\documentclass[12pt, draftclsnofoot, onecolumn]{IEEEtran}
\ifCLASSINFOpdf

\else

\fi
\usepackage{mathrsfs}

\usepackage{amsmath}
\usepackage{ntheorem}

\usepackage{makeidx}  
\usepackage{algorithm}
\usepackage{algorithmic}
\usepackage{graphicx}
\usepackage{subfigure}
\usepackage{epstopdf}
\usepackage{bm}
\usepackage{cite}
\usepackage{stfloats}

\usepackage{amssymb}
\usepackage{graphicx}

\usepackage{url}

\usepackage{tabularx,booktabs}
\usepackage{caption}

\newcolumntype{C}{>{\centering\arraybackslash}X} 
\usepackage{lipsum}

\usepackage{makecell} 

\usepackage{graphicx}
\usepackage{color}
\newtheorem{thm}{Theorem}
\newtheorem{lem}{Lemma}
\newtheorem{remark}{Remark}
\newtheorem{corollary}{Corollary}

\usepackage{diagbox}
\usepackage{multirow}

\definecolor{blue}{RGB}{0,0,0}

\begin{document}

\title {Two-Timescale Design for Reconfigurable Intelligent Surface-Aided Massive MIMO Systems with Imperfect CSI}

%
%
\author{Kangda Zhi, Cunhua Pan, Hong Ren, Kezhi Wang, Maged Elkashlan,  Marco Di Renzo, \IEEEmembership{Fellow, IEEE},  Robert Schober, \IEEEmembership{Fellow, IEEE}, H. Vincent Poor, \IEEEmembership{Fellow, IEEE}, Jiangzhou Wang, \IEEEmembership{Fellow, IEEE} and Lajos Hanzo, \IEEEmembership{Life Fellow, IEEE} 
	\thanks{(Corresponding author: Cunhua Pan). Part of this work has been presented in the IEEE SPAWC, 2021 \cite{zhi2021reconfigurable}. 
		 		
%
%
%
%
%
%

		K. Zhi, M Elkashlan are with the School of Electronic Engineering and Computer Science at Queen Mary University of London, London E1 4NS, U.K. (e-mail: k.zhi, maged.elkashlan@qmul.ac.uk).
C. Pan, H. Ren are with the National Mobile Communications Research Laboratory, Southeast University, Nanjing 210096, China. (e-mail: c.pan, hren@seu.edu.cn).
K. Wang is with Department of Computer and Information Sciences, Northumbria University, UK. (e-mail: kezhi.wang@northumbria.ac.uk).
M. Di Renzo is with Universit\'{e} Paris-Saclay, CNRS and CentraleSup\'{e}lec, Laboratoire des Signaux et Syst\`{e}mes, Gif-sur-Yvette, France. (e-mail: marco.direnzo@centralesupelec.fr).
R. Schober is with the Institute for Digital Communications, Friedrich-Alexander-University Erlangen-N\"{u}rnberg (FAU), Germany (e-mail: robert.schober@fau.de).	
H. V. Poor is with the Department of Electrical and Computer Engineering, Princeton University, Princeton, NJ 08544 USA (e-mail:
poor@princeton.edu). 
Jiangzhou Wang is with the School of Engineering and Digital Arts, University of Kent, UK. (e-mail: J.Z.Wang@kent.ac.uk).
Lajos Hanzo is with the School of Electronics and Computer Science, University of Southampton, Southampton, SO17 1BJ, U.K. (e-mail: lh@ecs.soton.ac.uk.).


} }

\maketitle
\begin{abstract}

		This paper investigates the two-timescale transmission scheme for reconfigurable intelligent surface (RIS)-aided massive multiple-input multiple-output (MIMO) systems, where the beamforming at the base station (BS) is adapted to the rapidly-changing instantaneous channel state information (CSI), while the nearly-passive beamforming at the RIS is adapted to the slowly-changing statistical CSI. Specifically, we first consider a system model with spatially-independent Rician fading channels, which leads to tractable expressions and offers analytical insights on the power scaling laws and on the impact of various system parameters. Then, we analyze a more general system model with spatially-correlated Rician fading channels and consider the impact of electromagnetic interference (EMI) caused by other devices present in the considered environment. For both case studies, we apply the linear minimum mean square error (LMMSE) estimator to estimate the aggregated channel from the users to the BS, utilize the low-complexity maximal ratio combining (MRC) detector, and derive a closed-form expression for a  lower bound of the achievable rate. Besides, {\color{blue}an accelerated} gradient ascent-based algorithm is proposed for solving the minimum user rate maximization problem.  Numerical results show that, {\color{blue}in the considered setup}, the spatially-independent model without EMI is sufficiently accurate when the inter-distance of the RIS elements is sufficiently large and the EMI is mild. In the presence of spatial correlation, we show that an RIS can better tailor the wireless environment. Furthermore, it is shown that deploying an RIS in a massive MIMO network brings significant gains when the RIS is deployed close to the cell-edge users. On the other hand, the gains obtained by the users distributed over a large area are shown to be modest.

\end{abstract}

\begin{IEEEkeywords}
Reconfigurable intelligent surface (RIS), massive MIMO, two-timescale transmission scheme, channel estimation, spatial correlation, electromagnetic interference (EMI). 
\end{IEEEkeywords}

\IEEEpeerreviewmaketitle

\section{Introduction}
As an emerging candidate for next-generation communication systems, reconfigurable intelligent surfaces (RISs), also termed intelligent reflecting surfaces (IRSs), have attracted significant interest from  academia and industry\cite{di2020smart,pan2020reconfigurable}. An RIS is a reconfigurable engineered surface that does not require active radio frequency (RF) chains, power amplifiers, and digital signal processing units, and is usually made of a large number of low cost and passive scattering elements that are coupled with simple low power electronic circuits. By intelligently tuning the phase shifts of the impinging waves with the aid of a controller, an RIS can constructively strengthen the desired signal or can deconstructively weaken the interference signals, which results in an appealing nearly-passive beamforming gain.

	Compared with existing multi-antenna systems\cite{Kammoun2019RZF,Couillet2011DE,wu2016secure,Wagner2012limite,Caire2010feedback,Adhikary2013LAR}, it has been demonstrated that RIS-aided systems have the potential to achieve better performance in terms of cost and energy consumption\cite{wu2019intelligent,huang2019reconfigu,pan2020multicell,zhang2019capacity,mei2021multibeam,dai2020experimental,tao2020analysis,sirojuddin2022low}. Recently, RISs have been considered for being integrated into various communication scenarios, such as terahertz, sub-terahertz, and millimeter-wave systems\cite{ning2021terherz,wang2021mmWave}, simultaneous wireless information and power transfer (SWIPT)\cite{pan2020intelligent}, unmanned aerial vehicle (UAV) communications\cite{lu2021UAV}, cell-free systems\cite{9352948}, physical-layer security\cite{hongsheng2020noise,9146177,chu2020secrecy}, mobile edge computing (MEC)\cite{bai2020latency,chen2021irsaided,chu2021MEC}, device-to-device (D2D) communications\cite{jia2021energy,9301375}. Furthermore, the effectiveness of RIS-aided systems in the presence of practical imperfections has been demonstrated in \cite{zhou2020framework,yuxianghao2020robost,hong2021hardware,zhou2020hardware}. Specifically, relying on imperfect instantaneous channel state information (CSI), the robust transmission design of RISs was studied in \cite{zhou2020framework,yuxianghao2020robost}. The authors of \cite{hong2021hardware}  studied the RIS beamforming design by considering transceiver hardware impairments. With the consideration of RF impairments and phase noises, the authors of \cite{zhou2020hardware} conducted a theoretical study on the fundamental tradeoffs between the spectral and energy efficiency of an RIS communication network. In addition, a valuable experimental investigation of  RIS-assisted channels was carried out in \cite{tang2021experimental}.

	While several benefits of RISs have been demonstrated in the above-mentioned contributions, most of them considered the design of the nearly-passive beamforming at the RIS under the assumption that the instantaneous CSI is estimated in each channel coherence interval. In practice, however, instantaneous CSI-based schemes face two challenges. The first one is the overhead for the acquisition of the instantaneous CSI.  Due to the absence of power amplifiers, digital signal processing units, and radio frequency chains at the RISs, many authors proposed to estimate the cascaded user-RIS-BS channels instead of the separated user-RIS and RIS-BS channels\cite{9130088,9087848}. The pilot overhead of these channel estimation schemes is  proportional to the number of RIS elements. However, an RIS generally consists of a large number of reflecting elements to ensure the desired coverage enhancement\cite{bigison2020IRS}, which incurs in a prohibitively high pilot overhead.
	Secondly, in each channel coherence time interval, the BS needs to calculate the optimal beamforming coefficients for the RIS, and needs to send them back to the RIS controller via dedicated feedback links. For instantaneous CSI-based schemes, therefore, the beamforming calculation and information feedback need to be executed frequently in each channel coherence interval, which results in a high computational complexity, feedback overhead, and energy consumption.

	To address these two practical challenges, recently, Han {\emph {et al.}}\cite{han2019large}  proposed a novel two-timescale based RIS scheme, which facilitates the deployment and operation of RIS-aided systems. This promising two-timescale scheme was further analyzed in recent research works \cite{zhao2021twoTimeScale,zhao2020twoQoS,guo2020historical,abrardo2020intelligent,gao2021LMMSE,zhang2021vehicle,kammoun2020asymptotic,9352967,jia2020analysis,zhi2020power,zhi2020directLinks,van2021reconfigurable}. 
	In the two-timescale scheme, the BS beamforming is designed based on the instantaneous aggregated CSI, which includes the direct and RIS-reflected links. The dimension of this aggregated channel is the same as for conventional RIS-free systems, which is independent of  the number of RIS elements.
	Hence, in the two-timescale scheme, the number of pilot signals  needs to be only larger than the number of users, which significantly reduces the channel estimation overhead. More importantly, the two-timescale scheme aims to optimize the RISs only based on long-term statistical CSI, such as the locations and the angles of arrival and departure of the users with respect to the BS and the RIS, which vary much slower than the instantaneous CSI, for typical applications in the sub-6 GHz bands. The phase shifts of the RIS elements need to be updated only when the large-scale channel information changes. Compared with instantaneous CSI-based designs that need to update the phase shifts of the RIS elements in each channel coherence interval, therefore, RIS-aided designs based on statistical CSI can significantly reduce the computational complexity, feedback overhead and energy consumption.

In addition, massive MIMO technology has been identified as the cornerstone of the fifth generation (5G) and future communication systems\cite{bjornson2019massive,bjornson2017massive}. Massive MIMO exploits tens or hundreds of BS antennas to serve multiple users simultaneously. Due to the complexity of wireless propagation environments, e.g., the presence of large blocking objects, however, the signal power received at the end-users may be still too weak, and it may be insufficient to support emerging applications that entail high date rate requirements, such as virtual reality (VR) or augmented reality (AR). Inspired by the capability of RISs to customize the wireless propagation environment, a natural idea is to integrate them into massive MIMO systems. By constructing alternative transmission paths, it is envisioned that RIS-aided massive MIMO systems can achieve significant performance gains, especially when the direct links between the BS and the users are blocked by obstacles. In RIS-aided massive MIMO systems, the transmission scheme needs to be carefully designed, and the channel estimation overhead needs to be taken into account considering the large channel dimension. The application of instantaneous CSI-assisted schemes, in particular, may lead to a prohibitive complexity and overhead. Instead, due to the reduced channel estimation and feedback overhead, the two-timescale scheme is deemed more suitable for RIS-aided massive MIMO systems.

  Even though RIS-aided massive MIMO systems have been investigated in some recent works \cite{demir2021channel,bijoson2020nearField,zhi2020power,zhi2020directLinks},   three key  issues are still not well understood. Firstly, it is crucial to identify the ultimate performance limits of RIS-aided massive MIMO systems based on the two-timescale scheme under imperfect CSI. In the presence of channel estimation errors, the impact of key system parameters, the achievable rate scaling law, and the power scaling law are  unknown. To tackle these open problems, it is necessary to derive explicit information-theoretic analytical frameworks that provide guidelines for system design. Secondly, it is essential to adopt realistic channel models that account for line-of-sight (LoS) and non-LoS (NLoS) components, so that the impact of the LoS and the scattered power can be appropriately modeled and analyzed. This enables one to provide guidelines for the deployment of RISs. Thirdly,  some unique and realistic characteristics need to be considered when analyzing RIS-aided systems, including the spatial correlation among the RIS elements and the electromagnetic interference (EMI). To date, the impact of spatial correlation and EMI have not been examined in RIS-aided massive MIMO systems based on the two-timescale scheme and in the presence of imperfect CSI. To be specific, due to the planar structure of the RIS, the channel spatial correlation among the RIS elements cannot be ignored \cite{Emil2021Correlation}. To model the LoS and NLoS channel components and the spatial correlation among the RIS elements, the correlated Rician fading model is considered an appropriate choice. Also, due to the large aperture, an RIS may be subject to a large amount of EMI, which is generated by any uncontrollable external sources (e.g., the signals from adjacent cells and the natural background radiation) \cite{Emil2022EMI,torres2022intelligent}. Therefore, the EMI re-radiated by a large RIS towards the intended receiver might deteriorate the channel estimation quality and reduce the end-to-end SINR, especially when the RIS is large and the useful signal power is weak. These three open research problems motivate the present research work.

In this paper, we analyze the uplink (UL) two-timescale transmission of an RIS-aided massive MIMO system that is subject to imperfect aggregated CSI. The Rician channel model is adopted to evaluate the impact of the LoS and NLoS channel components. To gain some initial design insights, we first analyze a channel model with spatial-independent Rician fading, which admits tractable expressions of the achievable rate, and enables us to develop a comprehensive theoretical framework to evaluate the impact of critical system parameters and power scaling laws. Then, we generalize our analysis to a channel model with spatially correlated Rician fading and EMI. In this context, we focus our attention on the impact of spatial correlation and EMI on the achievable rate and the power scaling laws. Finally, we propose a gradient ascent method to solve the minimum user rate maximization problem based only on statistical CSI.
The specific contributions of this paper are summarized as follows.

 		\begin{itemize}
 			\item To begin with, we consider the spatial-independent Rician fading model. The aggregated channel is estimated  by relying on the linear minimum mean square error (LMMSE)  method and its performance in terms of  mean square error (MSE) and normalized MSE (NMSE) is analyzed. Under the assumption of MRC detectors, we derive closed-form expressions for the use-and-then-forget (UatF) bound of the achievable rate. The derived results hold for an arbitrary number of BS antennas and RIS elements. Then, we analyze the impact of important system parameters, the asymptotic behavior of the rate, and the power scaling laws. We specialize our findings to the single-user case in order to obtain further engineering insights.

 			\item Next, we consider a more general system model that includes spatial correlation at the RIS and the EMI captured by the RIS. Also in this case, we compute the LMMSE channel estimates and formulate the UatF bound of the achievable rate in a closed-form expression. Our analysis shows that the presence of spatial correlation provides the RIS with an enhanced capability of customizing the wireless environment. On the other hand, the presence of severe EMI may result in different power scaling laws.
 			
 		\item For both the spatially-independent and spatially-correlated channel models, we propose {\color{blue}an accelerated gradient ascent-based  algorithm} to solve the minimum user rate maximization problem. We first apply a log-sum-exp approximation to obtain a smooth objective function. Then, we compute the  gradient vectors with respect to  the angle vectors. The performance loss in the projection is avoided since the objective function is periodic with the angles and the unit modulus constraint holds for all the angles. Besides, closed-form solutions are obtained in the special case of a single user.
 			
 			\item Numerical results validate the accuracy of  analytical insights derived by neglecting the spatial correlation and EMI. In the presence of spatial correlation and EMI, {\color{blue}the obtained numerical results} show that similar trends hold when the spatial correlation and the EMI are moderate. Specifically, our numerical study reveals that (i) an RIS with a large number of elements may benefit from the presence of spatial correlation; (ii) in the presence of severe EMI, an RIS-aided system may not offer better performance than a conventional massive MIMO system; (iii) the integration of RISs in massive MIMO systems is especially beneficial when the RISs are deployed near the cell edge users.


 		\end{itemize}

The remainder of this paper is organized as follows. The performance analysis based on spatially-independent channels without EMI is carried out in Section \ref{section_2}, \ref{section_3}, and \ref{section_4}. Specifically, the system model is introduced in Section \ref{section_2},   the LMMSE channel  estimator is derived and analyzed in    Section \ref{section_3}, and a closed-form lower bound expression of the achievable rate is obtained in Section   \ref{section_4}. The extension to spatially-correlated channels in the presence of EMI is discussed in Section \ref{section_cor_emi}. In Section \ref{section_optimization},  a gradient ascent-based algorithm for solving the minimum user rate maximization problem is introduced. Extensive numerical results are illustrated in Section \ref{section_6} and the conclusions are drawn in Section\ref{section_7}.

	\begin{table}[t]
		\renewcommand\arraystretch{0.75}
		\centering 
		\captionsetup{font={small}}
		\caption{List of Main Symbols}	
		\vspace{-5pt}
		\begin{tabular}{|c|c|c|c|c|c|c}
			\hline
			Symbol& Definition &Symbol& Definition   \\
			\hline
			$M$/$N$/$K$& Number of BS antennas/RIS elements/users& $p$& Transmit power for each user \\
			\hline
			$\theta_n$& Phase shift of the $n$-th RIS element&$\boldsymbol{\theta}$& Phase shift vector equal to $\left[{\theta_{1}}, {\theta_{2}}, \ldots, { \theta_{N}}\right]^{T}$\\
			\hline
			$\boldsymbol{c}$& Vector equal to $e^{j \boldsymbol{\theta}}$&$\bf\Phi$& RIS phase shifts matrix, $\boldsymbol{\Phi}=\operatorname{diag}(\mathbf{c})$\\
			\hline
			$\sigma^2$/$\sigma_e^2$, $\rho$&Power of thermal noise/EMI, $\rho=\frac{\sigma_e^2}{\sigma^2}$& $\mathbf{x}$/$\mathbf{n}$/$\boldsymbol{v}$ & Signal/noise/EMI vector\\
			\hline
			$d_{ris}$/$d_{bs}$&Element spacing of RIS/BS& $\lambda$& Wavelength\\
			\hline
			$\tau$/$\tau_c$&Lengths of pilot signal/coherence interval&$\mathbf{s}_k$, $\mathbf{S}$&User $k$'s pilot sequence, $\mathbf{S}=\left[\mathbf{s}_{1}, \mathbf{s}_{2}, \ldots, \mathbf{s}_{K}\right]$\\
			\hline
			$\mathbf{N}$/$\mathbf{V}$&Noise/EMI vectors over $\tau$ time slots&$\gamma_{k}$& Pathloss of user $k$'s direct link \\
			\hline
			$\alpha_k$&Pathloss of user $k$-RIS link& $\beta$&Pathloss of RIS-BS link \\
			\hline
			$\delta$& Rician factor of RIS-BS link&$\varepsilon_{k}$& Rician factor of user $k$-RIS link \\
			\hline
			$\mathbf{d}_{k}$, $\tilde{\mathbf{d}}_{k}$	&User $k$-BS direct link, $\mathbf{d}_{k}=\sqrt{\gamma_{k}} \tilde{\mathbf{d}}_{k}$&$\mathbf{h}_{k}$,  $\overline{\mathbf{h}}_{k}$, $\tilde{\mathbf{h}}_{k}$ & User $k$-RIS link, comprised of $\overline{\mathbf{h}}_{k}$ and $\tilde{\mathbf{h}}_{k}$ \\
			\hline
			$\mathbf{H}_{2}$/$\mathbf{H}_{c,2}$	& RIS-BS link without/with correlation&$\tilde{\mathbf{H}}_{2}$/$\tilde{\mathbf{H}}_{c,2}$& NLoS part of $\mathbf{H}_{2}$/$\mathbf{H}_{c,2}$	 \\
			\hline
			$\mathbf{q}_{k}$/$\mathbf{q}_{c,k}$&  Aggregated link without/with correlation &$\mathbf{Q}$/$\mathbf{Q}_c$& Matrix with the $k$-th column of $\mathbf{q}_{k}$/$\mathbf{q}_{c,k}$\\
			\hline
			$\hat{\mathbf{q}}_{k}$/$\hat{\mathbf{q}}_{c,k}$	&Channel estimate of $\mathbf{q}_{k}$/$\mathbf{q}_{c,k}$&$\hat{\mathbf{Q}}$/$\hat{\mathbf{Q}}_c$&Matrix with the $k$-th column of $\hat{\mathbf{q}}_{k}$/$\hat{\mathbf{q}}_{c,k}$	\\
			\hline
			$\mathbf{q}_{k}^{1}$-$\mathbf{q}_{k}^{4}$, $\underline{\mathbf{q}}_{k}$&Notations defined in (\ref{exact_channel}) &$\hat{\mathbf{q}}_{k}^{1}$-$\hat{\mathbf{q}}_{k}^{4}$, $\underline{\hat{\mathbf{q}}}_{k}$&Notations defined in (\ref{estimated_channel_detail})\\
			\hline
			$\mathbf{y}$/$\mathbf{y}_c$	&BS received signal without/with correlation &$\mathbf{r}$/$\mathbf{r}_c$	& Decoded symbols from $\mathbf{y}$/$\mathbf{y}_c$	\\
			\hline
			$\mathbf{Y}_P$/$\mathbf{Y}_{c,P}$&Received pilot signals at the BS&$\mathbf{y}_{p}^{k}$/$\mathbf{y}_{c,p}^{k}$	&Observation vector without/with correlation \\
			\hline
			$\mathbf{a}_{M}\!(.)$/$\mathbf{a}_{N}\!(.)$& Array response vector for BS/RIS&$E_u$& A constant used in the power scaling laws\\
			\hline
			$c_k$, $\widehat{c}_k$	& $c_{k}= \frac{\beta \alpha_{k}}{(\delta+1)\left(\varepsilon_{k}+1\right)}$, $\widehat{c}_k= \frac{\beta \alpha_{k}}{\delta+1}$&$a_{k1}$ - $a_{k4}$& Notations defined in Lemma \ref{lemma1} and Theorem \ref{theorem1} \\
			\hline
			$e_{k1}$ - $e_{k3}$&Notations defined in Lemma \ref{lemma_e1e2e3}& $f_{k}(\boldsymbol{\Phi})$& Scalar equal to $ \mathbf{a}_{N}^{H} \boldsymbol{\Phi} \overline{\mathbf{h}}_{k}$\\
			\hline
			$\mathbf{A}_k$, $\mathbf{B}_k$&Matrices defined in Theorem \ref{theorem1}& $\boldsymbol{\Upsilon}_k$& Matrix defined in Theorem \ref{theorem_mmse}\\
			\hline
			$\underline{R}_{k}$/$\underline{R}_{c,k}$& Rate of use $k$ without/with correlation & $f(\boldsymbol{\theta})$/$f_c(\boldsymbol{\theta})$& Approximated minimum user rate\\
			\hline
			$\mathbf{R}_{ris}$, $\!\!\mathbf{R}_{emi}$&Spatial correlation matrices&$\boldsymbol{{f}}_d(.)$, $\mathbf{z}_k(.)$& Function defined in Lemma \ref{lemma_apbp}, \ref{lemma_T_Upsilon}\\
			\hline
			\multicolumn{2}{|l |}{  $f_{c, 1}(\mathbf{\Phi})$, $f_{c,k, 2}(\mathbf{\Phi})$ - $f_{c,k, 7}(\mathbf{\Phi})$, $f_{c,ki, 8}(\mathbf{\Phi})$ - $f_{c,ki, 9}(\mathbf{\Phi})$ } &  \multicolumn{2}{l |}{ Scalar functions defined in (\ref{Notations}) } \\
			\hline
			\multicolumn{2}{|l |}{  			$\boldsymbol{f}_{c, 1}^{\prime}(\boldsymbol{\theta})$, $\boldsymbol{f}_{c, k, 2}^{\prime}(\boldsymbol{\theta})$ - $\boldsymbol{f}_{c, k, 7}^{\prime}(\boldsymbol{\theta})$, $\boldsymbol{f}_{c, ki, 8}^{\prime}(\boldsymbol{\theta})$ - $\boldsymbol{f}_{c, ki, 9}^{\prime}(\boldsymbol{\theta})$} &  \multicolumn{2}{l |}{ Gradient vectors defined in Lemma \ref{lemma_gradient}} \\
			\hline
			\multicolumn{2}{|l |}{  		$ E_{k}^{\text {signal }} $, $ I_{k i}$, $ E_{k}^{\text {leak }}$, $ E_{k}^{\text {noise }}$   } &  \multicolumn{2}{l |}{    Signal, interference, leakage, and noise in Theorem \ref{theorem2}} \\
			\hline
			\multicolumn{2}{|l |}{  		$ E_{c, k}^{\text {signal }} $, $ I_{c, k i}$, $ E_{c, k}^{\text {leak }}$, $ E_{c, k}^{\text {emi }}$, $ E_{c, k}^{\text {noise }}$   } &  \multicolumn{2}{l |}{    Signal, interference, leakage, EMI and noise in Theorem \ref{theorem4}} \\
			\hline
		\end{tabular}\label{tab2}
	\end{table}

\emph{Notations}: Vectors and matrices are denoted by boldface lower case and upper case letters, respectively. The transpose, conjugate, conjugate transpose, and inverse of matrix $\bf X$ are denoted by ${\bf X}^T$, ${\bf X}^*$, ${\bf X}^H$ and ${\bf X}^{-1}$, respectively. $\left[{\bf X}\right]_{m,n}$ denotes the $(m,n)$th entry of matrix $\bf X$. The real, imaginary, trace, expectation, and covariance operators are denoted by ${\rm Re}\left\{\cdot\right\}$, ${\rm Im}\left\{\cdot\right\}$, ${\rm Tr}\left\{\cdot\right\}$, ${\mathbb E}\left\{\cdot\right\} $, and ${\rm Cov}\left\{\cdot\right\} $, respectively. The $l_2$ norm of a vector and the absolute value of a complex number are denoted by $\left\|\cdot\right\|$ and $\left|\cdot\right|$, respectively. $\mathbb{C}^{M\times N}$ denotes the space of $M\times N$ complex matrices. ${\bf I}_{M}$ and $\bf 0$ denote the $M\times M$ identity matrix and all-zero matrix with appropriate dimension, respectively. The operator $\bmod$ returns the remainder after division, and $\lfloor x \rfloor$ denotes the nearest integer smaller than $x$. ${\bf x} \sim \mathcal{CN}\left({\bar{\bf x}},{\bf C}\right)$ is a complex Gaussian distributed vector with mean $\bar{\bf x}$ and covariance matrix ${\bf C}$. $\mathcal{O}$ denotes the standard big-O notation. Besides, for ease of reference, the main symbols used in this work are listed in Table \ref{tab2}.

\section{System Model}\label{section_2}
\begin{figure}[t]
	\setlength{\abovecaptionskip}{0pt}
	\setlength{\belowcaptionskip}{-20pt}
	\centering
	\includegraphics[width= 0.5\textwidth]{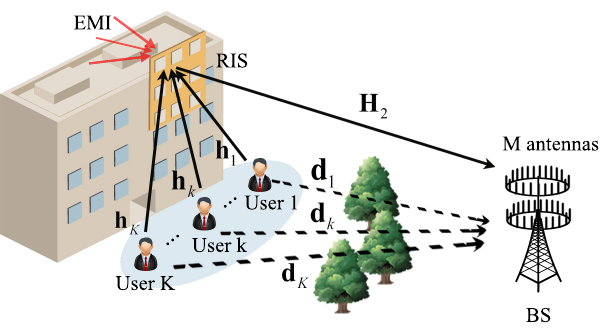}
	\DeclareGraphicsExtensions.
	\caption{An RIS-aided  massive MIMO system.}
	\label{figure1}
	\vspace{-10pt}
\end{figure}
To begin with, we consider an RIS-aided massive MIMO system under spatially-uncorrelated channels and in the absence of EMI. These two aspects will be analyzed in Section \ref{section_cor_emi}.
Specifically, as   illustrated in Fig. \ref{figure1}, we consider the UL transmission of an RIS-aided massive MIMO system, where an RIS is deployed in the
proximity of $K$  users to assist their UL transmissions to the BS. For convenience, we denote the set of users as $\mathcal{K}=\left\{1, 2,\ldots,K\right\}$. The BS is equipped with $M$ active antennas, the RIS comprises $N$ nearly-passive reflecting elements, and the $K$ users are equipped with a single transmit antenna. The channels from user $k$, $k \in \mathcal{K}$ to the BS, from user $k$, $k \in \mathcal{K}$ to the RIS, and from the RIS to the BS are denoted by $\mathbf {d}_k \in \mathbb{C}^{M\times 1}$,  $\mathbf {h}_k \in \mathbb{C}^{N\times 1}$, and $\mathbf {H}_2 \in \mathbb{C}^{M\times N}$, respectively. Additionally, we define $\mathbf{D}=\left[ \mathbf{d}_1, \mathbf{d}_2,\ldots,\mathbf{d}_K\right]$ and $\mathbf{H}_1=\left[ \mathbf{h}_1,\mathbf{h}_2,\ldots,\mathbf{h}_K\right]$. 

The RIS shapes the propagation environment by phase-shifting the impinging signals. Its phase shift matrix is denoted by $\mathbf{\Phi} = \mathrm{diag}\left\{e^{j\theta_1},e^{j\theta_2},\ldots,e^{j\theta_N}\right\}$, where $\theta_n\in [0,2\pi)$ represents the phase shift of the $n$th reflecting element. Based on these definitions, the cascaded user $k$-RIS-BS channel can be written as $\mathbf{g}_k = \mathbf{H}_2 {\bf \Phi} {\bf h}_k$, and the cascaded channels of the $K$ users are collected in the matrix $\mathbf{G} = [\mathbf{g}_1,\mathbf{g}_2,\ldots,\mathbf{g}_K]=\mathbf{H}_2 {\bf \Phi} {\bf H}_1   \in \mathbb{C}^{M\times K}$.

The $K$ users transmit their data in the same UL time-frequency resource. For ease of exposition, let $\mathbf{Q} = \mathbf{G}+\mathbf{D}=[{\bf q}_1,{\bf q}_2,\ldots,{\bf q}_K]  \in \mathbb{C}^{M\times K}$ denote the aggregated instantaneous channel matrix from the users to the BS. 
Thereby, the signal vector received at the BS is given by
\begin{align}\label{received_signal_model}
		\mathbf{y} = \sqrt{p} {\bf Q}\mathbf{x} +\mathbf{n} = \sqrt{p} \sum\nolimits_{k=1}^{K}{\bf q}_k x_k +\mathbf{n} ,
\end{align}
where $p$ is the average transmit power of each user, $\mathbf{x}=[x_1,x_2,\ldots,x_K]^T$ are the transmit symbols of the $K$ users, and $\mathbf{n}\sim \mathcal{CN}\left({\bf 0},\sigma^2\mathbf{I}_M\right)$ denotes the noise vector. 

The BS applies a low-complexity MRC receiver to detect the transmitted symbols. Before designing the MRC matrix, the channel $\bf Q$ has to be estimated at the BS. A standard LMMSE estimator is employed to obtain the estimated channel $\hat{\bf{Q}}$, as explained in the next section{\footnote{Given the LMMSE channel estimator, this work is focused on the  possible benefits of deploying RISs in massive MIMO systems. It is meaningful to investigate other channel estimators (such as the least-squares and element-wise MMSE\cite{bjornson2017massive,ozdogan2019massive,ozdogan2019performance}) and to evaluate the trade-off between estimation quality and implementation complexity. The comparison between different channel estimation strategies is postponed to a  future work.}.
Relying on the channel estimate, the BS performs MRC by multiplying the received signal $\bf y$ with $\hat{\bf Q}^H$, as follows
\begin{align}\label{MRC}
\mathbf{r} = \hat{\bf{Q}} ^H \mathbf{y} = \sqrt{p} \hat{\bf{Q}}  ^H \mathbf{Q} \mathbf{x} + \hat{\bf{Q}}   ^H \mathbf{n}.
\end{align}

Then, the $k$th element of the vector $\bf r$ can be expressed as
\begin{align}\label{rate_user_k}
r_{k}=\sqrt{p} \hat{\mathbf{q}}_{k}^{H} \mathbf{q}_{k} x_{k}+\sqrt{p} \sum\nolimits_{i=1, i \neq k}^{K} \hat{\mathbf{q}}_{k}^{H} \mathbf{q}_{i} x_{i}+\hat{\mathbf{q}}_{k}^{H} \mathbf{n}, \;\; k \in \mathcal{K},
\end{align}
where $\hat{\mathbf{q}}_{k}$ is the $k$th column of $\hat{\mathbf{Q}}$.

\subsection{Channel Model}
Since the users may be located far away from the BS and a large number of environmental blocking objects (i.e., blockages such as trees, vehicles, buildings) may exist in the area of interest, the LoS path between the users and the BS could be blocked. As in \cite{han2019large,9352967,jia2020analysis}, we adopt the Rayleigh fading model to describe the NLoS channel between the user $k$ and the BS, as follows
\begin{align}
\mathbf{d}_k = \sqrt{\gamma_k} \tilde{\mathbf{d}}_k, \;\; k \in \mathcal{K},
\end{align}
where $\gamma_k$ denotes the distance-dependent path-loss, and $\tilde{\mathbf{d}}_k$ denotes the fast fading NLoS channel.
The entries of $\tilde{\mathbf{d}}_k$ are independent and identically distributed (i.i.d.) complex Gaussian random variables, i.e., $ \tilde{\mathbf{d}}_k \sim \mathcal{CN}\left({\bf 0},\mathbf{I}_M\right)$.

Considering that the RIS is often installed on the facades of high-rise buildings and it could be placed near the users, the channels between the users and the RIS have a high LoS probability. In addition, the RIS and the BS are usually deployed at some heights above the ground, which implies that LoS paths are likely to exist between the RIS and the BS. Therefore, as in \cite{han2019large,9352967,jia2020analysis,zhi2020power,zhi2020directLinks}, we adopt the Rician fading model for the user-RIS and RIS-BS channels, as follows
\begin{align}
	&\mathbf{h}_{k}=\sqrt{\frac{\alpha_{k}}{\varepsilon_{k}+1}}\left(\sqrt{\varepsilon_{k}} \,\overline{\mathbf{h}}_{k}+\tilde{\mathbf{h}}_{k}\right), k \in \mathcal{K}, \\\label{H2}
	&\mathbf{H}_{2}=\sqrt{\frac{\beta}{\delta+1}}\left(\sqrt{\delta} \, \overline{\mathbf{H}}_{2}+\tilde{\mathbf{H}}_{2}\right),
\end{align}
where $\alpha_k$ and $\beta$ represent the path-loss coefficients, $\varepsilon_k$ and $\delta$ are the Rician factors that account for the ratio of the LoS power to the NLoS power of the corresponding propagation paths. Furthermore, $\overline{\mathbf{h}}_k $ and $\overline{\mathbf{H}}_2 $ denote the LoS components, whereas $\tilde{\mathbf{h}}_k $ and $\tilde{\mathbf{H}}_2 $ represent the NLoS components. For the NLoS paths, the components of $\tilde{\mathbf{h}}_k $ and $\tilde{\mathbf{H}}_2 $ are i.i.d. complex Gaussian random variables with zero mean and unit variance. For the LoS paths, the uniform linear array (ULA) and uniform squared planar array (USPA) models are adopted for the BS and the RIS, respectively. Hence, $\overline{\mathbf{h}}_k $ and $\overline{\mathbf{H}}_2 $ are, respectively, modelled as follows
\begin{align}\label{uspa_hk}
&{\overline {\bf h}_k} =  {\bf a}_N \left( {\varphi _{kr}^a,\varphi _{kr}^e} \right), k \in \mathcal{K},\\\label{uspa_H2}
&{\overline {\bf H}_2} = {\bf a}_M \left( {\phi _r^a,\phi _r^e} \right)    {\bf a}_N^H\left( {\varphi _t^a,\varphi _t^e} \right),
\end{align}
where $\varphi _{kr}^a$ ($\varphi _{kr}^e$) is the azimuth (elevation) angle of arrival (AoA) of the incident signal at the RIS from the user $k$, $\varphi _t^a$  ($\varphi _t^e$) is the azimuth (elevation) angle of departure (AoD) reflected by the RIS towards the BS, and $\phi _r^a$ ($\phi _r^e$) is the azimuth (elevation) AoA of the signal received at the BS from the RIS, respectively. Furthermore, $ {{\bf a}_X}\left( {\vartheta _{}^a,\vartheta _{}^e} \right) \in \mathbb{C}^{X\times 1}$ denotes the array response vector, whose $ x$-th entry is

	\begin{align}\label{uspa}
\begin{aligned}
&\left[ {{\bf a}_M}\left( {\vartheta _{}^a,\vartheta _{}^e} \right) \right]_{ x} = \exp\left\{j2\pi \frac{d_{bs}}{\lambda }
   (x-1)\sin \vartheta _{}^e\sin \vartheta _{}^a\right\},\\
&\left[ {{\bf a}_N}\left( {\vartheta _{}^a,\vartheta _{}^e} \right) \right]_{ x} = \exp\left\{j2\pi \frac{d_{ris}}{\lambda }
\left( {   \lfloor \left({ x} - 1 \right)/\sqrt{N}\rfloor \sin \vartheta _{}^e\sin \vartheta _{}^a
	+ \left(\left({x}-1\right)\bmod \sqrt{N}\right)  \cos \vartheta _{}^e} \right) \right\},
\end{aligned}
\end{align}
where $d_{bs}$, $d_{ris}$, and $\lambda$ denote the BS antenna spacing, the RIS element spacing, and the wavelength, respectively.

To simplify the notation, in the sequel, we denote ${\bf a}_M \left( {\phi _r^a,\phi _r^e} \right)$ and ${\bf a}_N\left( {\varphi _t^a,\varphi _t^e} \right)$ simply by ${\bf a}_M$ and ${\bf a}_N$, respectively. Then, the aggregated channel from the user $k$ to the BS can be expressed as
\begin{align}\label{exact_channel}
\mathbf{q}_{k}&=\mathbf{g}_{k}+\mathbf{d}_{k}=\mathbf{H}_2 {\bf \Phi} {\bf h}_k+\mathbf{d}_{k}  \nonumber\\
&={\underbrace{\sqrt{c_{k} \delta \varepsilon_{k}} \, \overline{\mathbf{H}}_{2} \mathbf{\Phi} \overline{\mathbf{h}}_{k}}_{\mathbf{q}_{k}^{1}}
	+\underbrace{\sqrt{c_{k} \delta} \, \overline{\mathbf{H}}_{2} \mathbf{\Phi} \tilde{\mathbf{h}}_{k}}_{\mathbf{q}_{k}^{2}}
	+\underbrace{\sqrt{c_{k} \varepsilon_{k}} \, \tilde{\mathbf{H}}_{2} \mathbf{\Phi} \overline{\mathbf{h}}_{k}}_{\mathbf{q}_{k}^{3}}
	+\underbrace{\sqrt{c_{k}} \, \tilde{\mathbf{H}}_{2} \mathbf{\Phi} \tilde{\mathbf{h}}_{k}}_{\mathbf{q}_{k}^{4}}} 
	+\sqrt{\gamma_{k}} \, \tilde{\mathbf{d}}_{k}\nonumber\\
	&\triangleq {\underline{\mathbf{q}}} _{k} + {\mathbf{d}}_{k},
\end{align}
where $c_{k}\triangleq \frac{\beta \alpha_{k}}{(\delta+1)\left(\varepsilon_{k}+1\right)}$, and ${\underline{\mathbf{q}}} _{k} =\sum_{\omega=1}^{4} {\bf q}^\omega_k$. Note that $\underline{\mathbf{q}} _{k} $ and ${\mathbf{d}}_{k}$ are mutually independent.


\section{Channel Estimation}\label{section_3}
	In this section, we use the LMMSE method to obtain the estimated aggregated instantaneous channel $\hat{\bf Q}$. Specifically, the BS estimates the aggregated channel matrix $\bf Q$ based on some predefined pilot signals.
	Let $\tau_c$ and $\tau$ denote the length of the channel coherence interval and the number of time slots used for channel estimation, respectively, where $\tau$ is no smaller than $K$, i.e., $\tau\geq K$. In each channel coherence interval, the $K$ users simultaneously transmit mutually orthogonal pilot sequences to the BS. The pilot sequence of  user $k$ is denoted by ${\bf s}_k \in \mathbb{C}^{\tau \times 1}$. By defining ${\bf S} = [{\bf s}_1,{\bf s}_2,\ldots,{\bf s}_K]$, we have ${\bf S}^H{\bf S} ={\bf I}_K$. Then, the $M\times \tau$ pilot signals received at the BS can be written as
\begin{align}\label{pilot_signal}
\mathbf{Y}_{p}=\sqrt{\tau p} \mathbf{Q }{\bf S}^H+\mathbf{N},
\end{align}
where $\tau p$ is the transmit pilot power, and $\bf N$ denotes the $M\times\tau$ noise matrix whose entries are i.i.d. complex Gaussian random variables with zero mean and variance $\sigma^2$. Multiplying (\ref{pilot_signal}) by $\frac{ {\bf s}_k} {\sqrt{\tau p}}$ and exploiting the orthogonality of the pilot signals, the BS obtains the following observation vector for user $k$
\begin{align}\label{y_p}
\mathbf{y}_{p}^{k}=\frac{1}{\sqrt{\tau p}} \mathbf{Y}_{p} \mathbf{s}_{k}=\mathbf{q}_{k}+\frac{1}{\sqrt{\tau p}} \mathbf{N s}_{k}.
\end{align}

The optimal estimate of the $k$-th user's channel based on the observation vector $\mathbf{y}_{p}^{k}$ can be determined based on the MMSE criterion, which has been widely utilized in conventional massive MIMO systems \cite{zhang2014power,ngo2013energy,Hassibi2003training}. 
In RIS-aided massive MIMO systems where Rician fading is considered for all RIS-aided channels, however, it is challenging to obtain the MMSE estimator. This is because the cascaded user-RIS-BS channel $\bf G$ in RIS-aided systems is not Gaussian distributed, but double Gaussian distributed \cite{DoubleGaussian2012}. To obtain closed-form channel estimates, as is needed to obtain useful design insights, we adopt the sub-optimal but tractable LMMSE estimator. This is because the LMMSE estimator only requires the knowledge of the first and second order statistics, and therefore it does not need to know the exact channel distributions. In the following lemma, we present the required statistics for the channel vector $\mathbf{q}_{k}$ and the observation vector $\mathbf{y}_{p}^{k}$.
\begin{lem}\label{lemma1}
	For $k \in \mathcal{K}$, the mean vectors and covariance matrices that are needed to compute the LMMSE estimator are given by
	\begin{align}
&\mathbb{E}\left\{\mathbf{q}_{k}\right\}=\mathbb{E}\left\{\mathbf{y}^k_{p}\right\}=\sqrt{c_{k} \delta \varepsilon_{k}} \,\overline{\mathbf{H}}_{2} \mathbf{\Phi} \overline{\mathbf{h}}_{k},\\
&\operatorname{Cov}\left\{\mathbf{q}_{k}, \mathbf{y}_{p}^{k}\right\}=\operatorname{Cov}\left\{\mathbf{y}_{p}^{k},\mathbf{q}_{k} \right\}=\operatorname{Cov}\left\{\mathbf{q}_{k}, \mathbf{q}_{k}\right\}=a_{k 1} \mathbf{a}_{M} \mathbf{a}_{M}^{H}+a_{k 2} \mathbf{I}_{M},\\
&\operatorname{Cov}\left\{\mathbf{y}_{p}^{k}, \mathbf{y}_{p}^{k}\right\}=\operatorname{Cov}\left\{\mathbf{q}_{k}, \mathbf{q}_{k}\right\}+\frac{\sigma^{2}}{\tau p} \mathbf{I}_{M}=a_{k 1} \mathbf{a}_{M} \mathbf{a}_{M}^{H}+\left(a_{k 2}+\frac{\sigma^{2}}{\tau p}\right) \mathbf{I}_{M},
	\end{align}
	where $a_{k1}=N c_{k} \delta$ and $a_{k2} = N c_{k}\left(\varepsilon_{k}+1\right)+\gamma_{k}$ are two auxiliary variables.
\end{lem}

\itshape {Proof:}  \upshape See Appendix \ref{appendix2}. \hfill $\blacksquare$

\begin{thm}\label{theorem1}
	Using the observation vector ${\mathbf{y}}^k_{p}$, the LMMSE estimate $\hat{\mathbf{q}}_{k}$ of the channel vector ${\mathbf{q}}_{k}$ is given by {\footnote{Note that $\hat{\mathbf{q}}_{k}^{1}$ and ${\mathbf{q}}_{k}^{1}$ are identical due to the unbiased estimation. However, we define two symbols in order to simplify the analytical formulation and make the derivations easier to understand (see (\ref{interference_expanded_sum}) and (\ref{leakage_firstTerm_expanded}) for example).}}
	\begin{align}\label{estimated_channel}
	 &\hat{\mathbf{q}}_{k}=\mathbf{A}_{k} \mathbf{y}_{p}^{k}+\mathbf{B}_{k}\\
	 &\quad=\underbrace{\underbrace{\sqrt{c_{k} \delta \varepsilon_{k}} \, \overline{\mathbf{H}}_{2} \mathbf{\Phi} \overline{\mathbf{h}}_{k}}_{\hat{\mathbf{q}}_{k}^{1}}
	 	+\underbrace{\left(M a_{k 3}+a_{k 4}\right) \sqrt{c_{k} \delta} \, \overline{\mathbf{H}}_{2} \mathbf{\Phi} \tilde{\mathbf{h}}_{k}}_{\hat{\mathbf{q}}_{k}^{2}}
	 	+\underbrace{\sqrt{c_{k}  \varepsilon_{k}} \mathbf{A}_{k}  \tilde{\mathbf{H}}_{2} \mathbf{\Phi} \overline{\mathbf{h}}_{k}}_{\hat{\mathbf{q}}_{k}^{3}}
	 	+\underbrace{\sqrt{c_{k}} \mathbf{A}_{k} \tilde{\mathbf{H}}_{2} \mathbf{\Phi} \tilde{\mathbf{h}}_{k}}_{\hat{\mathbf{q}}_{k}^{4}}}_{   \hat{\underline{\mathbf{q}}} _{k}  } \nonumber\\\label{estimated_channel_detail}
	 &\quad\quad+\sqrt{\gamma_{k}}\mathbf{A}_{k} \tilde{\mathbf{d}}_{k}+\frac{1}{\sqrt{\tau p}} \mathbf{A}_{k} \mathbf{N s}_{k},
	\end{align}
	where
	\begin{align}\label{Ak}
&\mathbf{A}_{k}=\mathbf{A}^H_{k}=a_{k 3} \mathbf{a}_{M} \mathbf{a}_{M}^{H}+a_{k 4} \mathbf{I}_{M},\\
&\mathbf{B}_{k}=\left(\mathbf{I}_{M}-\mathbf{A}_{k}\right) \sqrt{c_{k} \delta \varepsilon_{k}} \, \overline{\mathbf{H}}_{2} \mathbf{\Phi} \overline{\mathbf{h}}_{k},\\\label{ak3}
&a_{k 3}= \frac{a_{k 1} \frac{\sigma^{2}}{\tau p}}{\left(a_{k 2}+\frac{\sigma^{2}}{\tau p}\right)\left(a_{k 2}+\frac{\sigma^{2}}{\tau p}+M a_{k 1}\right)}, \\\label{ak4}
&a_{k 4}= \frac{a_{k 2}}{a_{k 2}+\frac{\sigma^{2}}{\tau p}},
	\end{align}
	and the NMSE of the estimate of $\mathbf{q} _{k}$ is   
	\begin{align}\label{NMSE}
\operatorname{NMSE}_{k}=\frac{    \operatorname{Tr}\left\{\operatorname{Cov}\left\{   {\bf q}_k - \hat{\bf q}_k, {\bf q}_k - \hat{\bf q}_k \right\}\right\}   }
{ \operatorname{Tr}\left\{\operatorname{Cov}\left\{\mathbf{q}_{k}, \mathbf{q}_{k}\right\}\right\}} =\frac{\frac{\sigma^{2}}{\tau p}\left(M a_{k 1} a_{k 2}+a_{k 2}^{2}+\left(a_{k 1}+a_{k 2}\right) \frac{\sigma^{2}}{\tau p}\right)}{\left(a_{k 2}+\frac{\sigma^{2}}{\tau p}\right)\left(a_{k 2}+\frac{\sigma^{2}}{\tau p}+M a_{k 1}\right)\left(a_{k 1}+a_{k 2}\right)}.
	\end{align}
	
\end{thm}

\itshape {Proof:}  \upshape See Appendix \ref{appendix3}. \hfill $\blacksquare$

As evident from Theorem \ref{theorem1}, we only estimate the aggregated channel matrix ${\bf Q} \in\mathbb{C}^{M\times K}$ including the reflected and direct channels, which has the same dimension as the user-BS channel matrix in conventional massive MIMO systems. Therefore, we only require that the length of the pilot sequences is no smaller than the number of users, i.e., $\tau\geq K$. Compared to  methods that estimate the $MN$ individual channels in RIS-aided communications\cite{9130088,9087848}, the proposed method has a lower overhead and computational complexity.


\begin{remark}\label{remark2}
	When $c_k=0,\forall k$, i.e., the RIS-assisted channels are absent, we have $a_{k1}=0$, $a_{k2} = \gamma_{k}$, $a_{k3}=0$, $a_{k4}=\frac{\gamma_{k}}{\gamma_{k}+\frac{\sigma^{2}}{\tau p}}$ and ${\bf B}_k={\bf 0}$. In this case, the estimate in (\ref{estimated_channel}) reduces to $\hat{\mathbf{q}}_{k}= \frac{\gamma_{k}}{\gamma_{k}+\frac{\sigma^{2}}{\tau p}} \mathbf{y}_{p}^{k}$ and the MSE matrix in (\ref{MSE_matrix}) reduces to $ {\rm \bf MSE}_k =\frac{  \gamma_{k} \frac{\sigma^{2}}{\tau p}}{\gamma_{k}+\frac{\sigma^{2}}{\tau p}} {\bf I}_M$, which, as expected, is the same as the MSE in conventional massive MIMO systems \cite{ngo2013energy}. If the RIS channels only have the LoS components, i.e., $\delta, \varepsilon_{k} \to\infty, \forall k$, we also obtain $a_{k1}\to 0$ and $a_{k2} \to \gamma_{k}$. In this case, the MSE matrix in (\ref{MSE_matrix}) is again the same as that in conventional massive MIMO systems. This is because the LoS channels are deterministic and known, and, thus, they do not introduce additional estimation errors.
\end{remark}

\begin{corollary}\label{corollary1}
	 In the low pilot power-to-noise ratio regime, high pilot power-to-noise ratio regime, and large $N$ regime, the asymptotic NMSE is, respectively, given by
	\begin{align}\label{NMSE_scale1}
	&\lim \nolimits_{\frac{\sigma^{2}}{\tau p} \rightarrow \infty} \mathrm{NMSE}_{k} \rightarrow 1, \\\label{NMSE_scale2}
	&\lim \nolimits_{  \frac{\sigma^{2}}{\tau p} \rightarrow 0} \mathrm{NMSE}_{k} \rightarrow 0, \\\label{NMSE_scale3}
	&\lim \nolimits_{N \rightarrow \infty} \mathrm{NMSE}_{k} \rightarrow 0.
	\end{align}
	Besides, assume that the power $p$ is scaled proportionally to $p=E_u/N$, where $E_u$ denotes a constant. As $N\to\infty$, we have
	\begin{align}\label{NMSE_scale4}
	\lim \nolimits_{p=\frac{E_u} {N}, N \rightarrow \infty} \operatorname{NMSE}_k <1.
	\end{align}	
\end{corollary}
\itshape {Proof:}  \upshape When $\frac{\sigma^{2}}{\tau p} \rightarrow \infty$ or $N \rightarrow \infty$, by selecting the dominant terms in (\ref{NMSE}), which scale with ${(\frac{\sigma^{2}}{\tau p})}^2$ or $N^3$, we arrive at (\ref{NMSE_scale1}) and (\ref{NMSE_scale3}), respectively. Substituting $\frac{\sigma^{2}}{\tau p} = 0$ into (\ref{NMSE}), its numerator reduces to zero, which leads to (\ref{NMSE_scale2}). Replacing the power $p$ in (\ref{NMSE}) with $p=E_u/N$, as $N \to \infty$, we can readily find that all the dominant terms in the numerator are present in the denominator as well, which results in (\ref{NMSE_scale4}).  We omit the specific limit of (\ref{NMSE_scale4}) since it is a complex expression but is simple to compute.
\hfill $\blacksquare$

It is worth noting that NMSE values between $ 0 $ (i.e., perfect estimation) and $ 1  $ (i.e., using the mean value of the variable as the estimate) quantify the relative estimation error \cite{bjornson2017massive}. In conventional massive MIMO systems, a common method for reducing the NMSE is to increase the length of the pilot sequence $\tau$. In RIS-aided massive MIMO systems, Corollary \ref{corollary1} indicates that increasing the number of RIS elements $N$ can play a similar role as increasing $\tau$. 
Therefore, increasing the number of RIS elements not only helps improve the system rate, but it also helps reduce the NMSE. Additionally, (\ref{NMSE_scale4}) reveals that an RIS equipped with a large number of reflecting elements $N$ can help the NMSE converge to a limit lower than one, even for low pilot powers.

To better understand the impact of increasing $N$ for channel estimation, we present the following asymptotic results.
\begin{corollary}\label{corollary2}
When $\tau\to\infty$, we have $\hat{\mathbf{q}}_{k} \to {\mathbf{q}}_{k}$, which implies ${\bf e}_k\to \bf 0$ and therefore the channel estimation is perfect. When $N\to\infty$, by contrast, we have 
\begin{align}\label{LS_estimator}
&\hat{\mathbf{q}}_{k}\to {\mathbf{q}}_{k} + \frac{1}{\sqrt{\tau p}}  \mathbf{N s}_{k},\\\label{error_LS}
&{\bf e}_k = {\bf q}_k - \hat{\bf q}_k \to \frac{-1}{\sqrt{\tau p}}  \mathbf{N s}_{k} ,\\\label{MSE_LS}
&{\rm \bf MSE}_k = \mathbb{E}\left\{  {\bf e}_k  {\bf e}_k^H    \right\} \to \frac{\sigma^2}{\tau p} {\bf I}_M.
\end{align}

\end{corollary}

\itshape {Proof:}  \upshape When $\tau\to\infty$ or $N\to\infty$, based on Theorem \ref{theorem1}, we have $a_{k3}\to 0$, $a_{k4}\to 1$, and $Ma_{k3}+a_{k4}\to 1$, which yields ${\bf A}_k \to {\bf I}_M$. If $\tau\to\infty$, we further get $\frac{1}{\sqrt{\tau p}}\to 0$, which completes the proof. \hfill $\blacksquare$

Although the NMSE converges to zero as $N\to\infty$ (see (\ref{NMSE_scale3})), Corollary \ref{corollary2} shows that, in contrast to increasing $\tau$, the MSE of the LMMSE estimator converges to a non-zero constant as $N\to\infty$.  
 If we estimate the channel $\mathbf{q}_{k}$ based on the least-squares (LS) estimator\cite[(3.35)]{bjornson2017massive}, it is interesting to note that we  obtain the same results as in (\ref{LS_estimator}) and (\ref{MSE_LS}).
  In general, the LS estimator, which does not exploit any prior channel statistics, has worse estimation performance (higher MSE) than the LMMSE estimator\cite{9087848,ozdogan2019massive,bjornson2017massive}. Therefore, Corollary \ref{corollary2} indicates that the MSE performance of the LMMSE estimation converges towards an upper bound, which is the MSE performance of the LS estimation, as $N\to\infty$. This result will be validated in Section \ref{section_6}.

\begin{corollary}\label{corollary3}
 
 When the RIS-BS channel reduces to the Rayleigh channel (i.e., $\delta=0$), the estimated channel vector, MSE, and NMSE, respectively, simplify to
 
\begin{align}
&{\hat{\bf q}}_k = a_{k4} \mathbf{I}_M \mathbf{y}^k_p=\frac{	N \beta \alpha_{k}+\gamma_{k}  }{N  \beta \alpha_{k}+\gamma_{k}  +  \frac{\sigma^{2}}{\tau p} } \mathbf{y}^k_{p},\\\label{MSE_rayleigh}
&{\rm \bf{MSE}}_{k}=\frac{    \left(  N \beta \alpha_{k}+\gamma_{k}    \right)    \frac{\sigma^{2}}{\tau p}    }{N \beta \alpha_{k}+\gamma_{k}+\frac{\sigma^{2}}{\tau p}} \mathbf{I}_{M},\\\label{NMSE_rayleigh}
&\operatorname{NMSE}_{k}=\frac{\frac{\sigma^{2}}{\tau p}}{N \beta \alpha_{k}+\gamma_{k}+\frac{\sigma^{2}}{\tau p}} .
\end{align}
\end{corollary}

\itshape {Proof:}  \upshape When $\delta=0$, we have $a_{k1}=0$, $a_{k2} = N  \beta\alpha_{k}+\gamma_{k}$, $a_{k3}=0$, $a_{k4}=\frac{N \beta \alpha_{k}+\gamma_{k}}{N \beta \alpha_{k}+\gamma_{k}+\frac{\sigma^{2}}{\tau p}}$ and ${\bf B}_k={\bf 0}$. The proof follows by inserting these results in Theorem \ref{theorem1} and (\ref{MSE_matrix}).
\hfill $\blacksquare$

	Corollary \ref{corollary3} corresponds to a scenario where a large number of scatterers exist nearby the RIS and the BS, and the LoS path between the RIS and the BS is negligible. Therefore, the RIS-BS channel is dominated by the NLoS paths. In this case, both the MSE and NMSE have simple analytical expressions, which help us better understand the conclusions drawn in Corollary \ref{corollary1} and Corollary \ref{corollary2}. It is apparent that the MSE (represented by the trace of ${\rm \bf MSE}_k$ in (\ref{MSE_rayleigh})) and the NMSE (represented by ${\rm NMSE}_k$ in (\ref{NMSE_rayleigh})) are decreasing functions of the pilot power $\tau p$. As a function of $N$, on the other hand, the MSE is an increasing function, while the NMSE is a decreasing function. When $N\to\infty$, we have ${\rm \bf MSE}_k \to \frac{\sigma^2}{\tau p} {\bf I}_M$ but ${\rm NMSE}_k \to { 0}$. 
Note that we can obtain the MSE and NMSE for conventional massive MIMO systems by setting $N=0$ in (\ref{MSE_rayleigh}) and (\ref{NMSE_rayleigh}). Therefore, the obtained result implies that the MSE of RIS-aided massive MIMO systems is worse than the MSE of massive MIMO systems without RISs, while the NMSE of RIS-aided massive MIMO systems is better than the NMSE of massive MIMO systems without RISs. 
	The reason is that an RIS introduces $N$ additional paths to the system, but the pilot length $\tau$ does not increase correspondingly, which increases the estimation error. However, the presence of an RIS results in better channel gains, which help decrease the normalized error.

 	Furthermore, if we reduce the power as $p=E_u/N$, as $N\to\infty$, the NMSE in (\ref{NMSE_rayleigh}) converges to a limit less than one, as follows
\begin{align}
\lim \nolimits_{\delta=0, \, p=\frac{E_u} {N}, \,N \rightarrow \infty      } \operatorname{NMSE}_k \to
\frac{{\sigma^{2}}}{ \tau E_u\beta \alpha_{k}+{\sigma^{2}}} <1.
\end{align}


\section{Analysis of the Achievable Rate}\label{section_4}
Based on the channel estimates provided in Theorem \ref{theorem1}, closed-form expressions for a lower bound of the achievable rate are derived and analyzed in this section{\footnote{To avoid verbose expressions, ``lower bound of the achievable rate'' is replaced with ``achievable rate" in the rest of this paper. It is, however, implied that we compute a lower bound.}}. In Section \ref{section_optimization}, the obtained analytical expressions are utilized for optimizing the phase shifts of the RIS based on statistical CSI.

\subsection{Derivation of the Rate}

	As in \cite{bjornson2014massiveHW,marzetta2016fundamentals,ozdogan2019massive,Nam2020correlate}, we utilize the so called UatF bound, which is a tractable lower bound, to characterize the ergodic rate of RIS-aided massive MIMO systems. First, we rewrite $r_k$ in (\ref{rate_user_k}) as
\begin{align}
r_{k}=\underbrace{\sqrt{p} \, \mathbb{E}\left\{\hat{\mathbf{q}}_{k}^{H} \mathbf{q}_{k}\right\} x_{k}}_{\text {Desired signal }}+\underbrace{\sqrt{p}\left(\hat{\mathbf{q}}_{k}^{H} \mathbf{q}_{k}-\mathbb{E}\left\{\hat{\mathbf{q}}_{k}^{H} \mathbf{q}_{k}\right\}\right) x_{k}}_{\text {Signal leakage }}+\underbrace{\sqrt{p} \sum\nolimits_{i=1, i \neq k}^{K} \hat{\mathbf{q}}_{k}^{H} \mathbf{q}_{i} x_{i}}_{\text {Multi-user interference }}+\underbrace{\hat{\mathbf{q}}_{k}^{H} \mathbf{n}}_{\text {Noise }}.
\end{align}

Then, we formulate the lower bound of the $k$-th user's ergodic rate as 
$ \underline{R}_{k}=\frac{\tau_c-\tau}{\tau_c}\log _{2}\left(1+\mathrm{SINR}_k\right) $,
where the pre-log factor $\frac{\tau_c-\tau}{\tau_c}$ represents the rate loss that originates from the pilot overhead, and 
the SINR is expressed as
\begin{align}\label{defination_rate}
\mathrm{SINR}_k
=\frac{  p   \left|\mathbb{E}\left\{\hat{\mathbf{q}}_{k}^{H} \mathbf{q}_{k}\right\}\right|^{2} }{p  \left(    \mathbb{E}\left\{\left|\hat{\mathbf{q}}_{k}^{H} \mathbf{q}_{k}\right|^{2}\right\}-\left|\mathbb{E}\left\{\hat{\mathbf{q}}_{k}^{H} \mathbf{q}_{k}\right\}\right|^{2}     \right)   +p \sum\limits_{i=1, i \neq k}^{K}    \mathbb{E}\left\{\left|\hat{\mathbf{q}}_{k}^{H} \mathbf{q}_{i}\right|^{2}\right\}   
	+\sigma^{2}   \mathbb{E}\left\{\left\|\hat{\mathbf{q}}_{k}\right\|^{2}\right\}  }.
\end{align}

To simplify the expression of $\underline{R}_k$, we define three auxiliary variables $e_{k1}$, $e_{k2}$, and $e_{k3}$. These variables capture the performance degradation due to the imperfect knowledge of the CSI.  
\begin{lem}\label{lemma_e1e2e3}
 For $k\in\mathcal{K}$, we have ${\rm Tr}\left\{{\bf A}_k\right\} = Me_{k1}$, 
  ${\bf A}_k \overline{\bf H}_2= e_{k2} \overline{\bf H}_2$ and
   ${\rm Tr}\left\{  {\bf A}_k{\bf A}_k\right\} = Me_{k3}$, where
   \begin{align}
   &e_{k 1} \triangleq a_{k 3}+a_{k 4} , \\
   &e_{k 2} \triangleq M a_{k3}+a_{k4}  , \\
   &e_{k 3} \triangleq M a_{k 3}^{2}+2 a_{k 3} a_{k 4}+a_{k 4}^{2} .
   \end{align}
   
   Furthermore, $ e_{k1}, e_{k2}$ and $ e_{k3}$ are bounded in $\left[0,1\right]$. When $\tau p\to\infty$ or $N\to\infty$, we have $e_{k1}, e_{k2}, e_{k3}\to1$. When $\tau p\to0$, by contrast, we have $e_{k1}, e_{k2}, e_{k3}\to 0$. 
\end{lem}

\itshape {Proof:}  \upshape See Appendix \ref{appendix_e1e2e3}. \hfill $\blacksquare$

In the following theorem, we derive a closed-form expression for the achievable rate.

\begin{thm}\label{theorem2}
A lower bound for the ergodic rate of the $k$-th user is given by{\footnote{The phase shift matrix $\bf\Phi$ is assumed to be fixed when deriving the achievable rate. After obtaining the achievable rate, we will design $\bf\Phi$ so that the derived rate is optimized.}}
\begin{align}\label{rate}
	\begin{aligned}
&	\underline{R}_{k}=\tau^{o}  \log _{2}\left(1+ \mathrm{S I N R}_k\right),\\
&	\mathrm{SINR}_k = \frac{  p E_{k}^{\rm s i g n a l}  \left(\mathbf{\Phi}\right)  }{p E_{k}^{\rm l e a k } \left(\mathbf{\Phi}\right)   +p \sum\limits_{i=1, i \neq k}^{K} I_{k i} \left(\mathbf{\Phi}\right) +\sigma^{2} E_{k}^{\rm {noise}}  \left(\mathbf{\Phi}\right)   },
	\end{aligned}
\end{align}
where $\tau^{o} = \frac{\tau_{c}-\tau}{\tau_{c}}$, $ E_{k}^{\rm signal} \left(\mathbf{\Phi}\right) =  \left\{E_{k}^{\rm {noise}} \left(\mathbf{\Phi}\right) \right\}^{2}$,
\begin{align}\label{signal_LMMSE}
E_{k}^{\rm {noise}} \left(\mathbf{\Phi}\right) =M\left\{    \left|f_{k}(\mathbf{\Phi})\right|^{2} c_{k} \delta \varepsilon_{k}+N c_{k} \delta e_{k 2}+\left(N c_{k}\left(\varepsilon_{k}+1\right)+\gamma_{k}\right) e_{k 1}  \right\},\qquad\quad\quad\quad\quad
\end{align}
\begin{align}
\begin{array}{l}
E_{k}^{\rm leak} \left(\mathbf{\Phi}\right) =M\left|f_{k}(\mathbf{\Phi})\right|^{2} c_{k}^{2} \delta \varepsilon_{k}\left\{N\left(M \delta+\varepsilon_{k}+1\right)\left(e_{k 2}^{2}+1\right)+2\left(M e_{k 1}+e_{k 2}\right)\left(e_{k 2}+1\right)\right\} \\
\qquad\qquad\;\;\quad+M\left|f_{k}(\mathbf{\Phi})\right|^{2} c_{k} \delta \varepsilon_{k}\left\{  \gamma_{k}+\left(\gamma_{k}+\frac{\sigma^{2}}{\tau p}\right) e_{k 2}^{2} \right\} \\
\qquad\qquad\;\;\quad+M^{2} N^{2} c_{k}^{2} \delta^{2} e_{k 2}^{2} +M N^{2} c_{k}^{2}\left\{2 \delta\left(\varepsilon_{k}+1\right) e_{k 2}^{2}+\left(\varepsilon_{k}+1\right)^{2} e_{k 3}\right\} \\
\qquad\qquad\;\;\quad+M^{2} N c_{k}^{2}\left\{\left(2 \varepsilon_{k}+1\right) e_{k 1}^{2}+2 \delta e_{k 1} e_{k 2}\right\} \\
\qquad\qquad\;\;\quad+M N c_{k}\left\{c_{k}\left(2 \delta e_{k 2}^{2}+\left(2 \varepsilon_{k}+1\right) e_{k 3}\right)+\left(2 \gamma_{k}+\frac{\sigma^{2}}{\tau p}\right)\left(\delta e_{k 2}^{2}+\left(\varepsilon_{k}+1\right) e_{k 3}\right)\right\} \\
\qquad\qquad\;\;\quad+M \gamma_{k}\left(\gamma_{k}+\frac{\sigma^{2}}{\tau p}\right) e_{k 3},
\end{array}
\end{align}
and
\begin{align}
\begin{array}{l}
I_{k i}\left(\mathbf{\Phi}\right) =M^{2}\left|f_{k}(\mathbf{\Phi})\right|^{2}\left|f_{i}(\mathbf{\Phi})\right|^{2} c_{k} c_{i} \delta^{2} \varepsilon_{k} \varepsilon_{i} \\
\qquad\quad+M\left|f_{k}(\mathbf{\Phi})\right|^{2} c_{k} \delta \varepsilon_{k}\left\{c_{i}\left(M N \delta+N \varepsilon_{i}+N+2 M e_{k 1}\right)+\gamma_{i}\right\} \\
\qquad\quad+M\left|f_{i}(\mathbf{\Phi})\right|^{2} c_{i} \delta \varepsilon_{i}\left\{c_{k} e_{k 2}\left(M N \delta e_{k 2}+N \varepsilon_{k} e_{k 2}+N e_{k 2}+2 M e_{k 1}\right)+\left(\gamma_{k}+\frac{\sigma^{2}}{\tau p}\right) e_{k 2}^{2}\right\} \\
\qquad\quad+M^{2} N^{2} c_{k} c_{i} \delta^{2} e_{k 2}^{2} \\
\qquad\quad+M N^{2} c_{k} c_{i}\left\{\delta\left(\varepsilon_{k}+\varepsilon_{i}+2\right) e_{k 2}^{2}+\left(\varepsilon_{k}+1\right)\left(\varepsilon_{i}+1\right) e_{k 3}\right\} \\
\qquad\quad+M^{2} N c_{k} c_{i} e_{k 1}\left\{\left(\varepsilon_{k}+\varepsilon_{i}+1\right) e_{k 1}+2 \delta e_{k 2}\right\} \\
\qquad\quad+M^{2} c_{k} c_{i} \varepsilon_{k} \varepsilon_{i} e_{k 1}\left(\left|\overline{\mathbf{h}}_{k}^{H} \overline{\mathbf{h}}_{i}\right|^{2} e_{k 1}+2 \delta \operatorname{Re}\left\{f_{k}^{H}(\mathbf{\Phi}) f_{i}(\mathbf{\Phi}) \overline{\mathbf{h}}_{i}^{H} \overline{\mathbf{h}}_{k}\right\}\right) \\
\qquad\quad+M N\left\{\left(\gamma_{k}+\frac{\sigma^{2}}{\tau p}\right) c_{i}\left(\delta e_{k 2}^{2}+\left(\varepsilon_{i}+1\right) e_{k 3}\right)+\gamma_{i} c_{k}\left(\delta e_{k 2}^{2}+\left(\varepsilon_{k}+1\right) e_{k 3}\right)\right\} \\
\qquad\quad+M \gamma_{i}\left(\gamma_{k}+\frac{\sigma^{2}}{\tau p}\right) e_{k 3},
\end{array}
\end{align}

with
\begin{align}\label{f_k_Phi_definition}
f_{k}({\bf\Phi}) \triangleq &\mathbf{a}_{N}^{H} {\bf\Phi} \overline{\mathbf{h}}_{k}=\sum\nolimits_{n=1}^{N} e^{j\left(\zeta_{n}^{k}+\theta_{n}\right)},\\
\zeta_{n}^{k}=&2 \pi \frac{d}{\lambda}  \left(   \lfloor(n-1) / \sqrt{N}\rfloor  \left(\sin \varphi_{kr}^{e} \sin \varphi_{kr}^{a}-\sin \varphi_{t}^{e} \sin \varphi_{t}^{a}\right)\right.\nonumber\\
&\left.+((n-1) \bmod \sqrt{N})\left( \cos \varphi_{kr}^{e}- \cos \varphi_{t}^{e}\right)\right).
\end{align}

\end{thm}

\itshape {Proof:}  \upshape See Appendix \ref{appendix4}. \hfill $\blacksquare$

	The closed-form expression in Theorem \ref{theorem2} does not involve the calculation of inverse matrices and the numerical computation of integrals. In contrast to time-consuming Monte Carlo simulations, the evaluation of the rate based on Theorem \ref{theorem2} has a low computational complexity even if $M$ and $N$ are large numbers, as usually is in RIS-aided massive MIMO systems. 
Besides, Theorem \ref{theorem2} only relies on statistical CSI. Therefore, by using the analytical expression of the rate in (\ref{rate}) as an objective function for system design, we are able to optimize the phase shifts of the RIS only based on long-term statistical CSI. {\color{blue}For clarity and analytical tractability, the statistical CSI is assumed to be perfectly known\cite{bjornson2017massive,demir2021channel,zhao2021twoTimeScale}. In practice, due to the user mobility, there may exist location and angular estimation errors based on, e.g., GPS (Global Positioning System) information, which could result  in some performance loss for the design of  receiver at the BS and passive beamforming at the RIS. The impact of imperfect statistical CSI can be analyzed by averaging the angular estimation error in  the expression of the achievable rate similar to  \cite{hu2020location}. This analysis is interesting and is left to a future research work.}


By comparing the formulation in Theorem \ref{theorem2} with that given in \cite[Theorem 1]{zhi2020power}, it can be seen that the impact of imperfect CSI is completely characterized by the parameters $e_{k1}, e_{k2}, e_{k3}$ and $\frac{\sigma^{2}}{\tau p}$. In the perfect CSI scenario, we have $\tau\to\infty$, which leads to $e_{k1}=e_{k2}=e_{k3}=1$ and $\frac{\sigma^{2}}{\tau p}=0$. 
 Based on Theorem \ref{theorem2}, we can analyze the performance of RIS-aided massive MIMO systems for arbitrary system parameters. 
Even though the obtained analytical expressions may look cumbersome at the first sight, they provide clear insights in terms of the key system parameters $M$, $N$, and $f_{k}(\mathbf{\Phi})$, $\forall k$. For example, since the interference term $I_{ki}$ scales as $\mathcal{O}(M^2)$, we infer that  RIS-aided massive MIMO systems suffer from stronger multi-user interference than conventional massive MIMO systems. In the following, we provide a comprehensive analysis of RIS-aided massive MIMO systems, including the asymptotic behavior of the rate for large values of $M$ and $N$, the power scaling laws, and the impact of the Rician factors.
To this end, we begin with a useful lemma.
\begin{lem}\label{f_i_Phi_is_bounded}
		\begin{itemize}
			
		\item If $N=1$, for arbitrary $\bf\Phi$, we have $\left|f_{k}(\mathbf{\Phi})\right| = 1$ in (\ref{f_k_Phi_definition}). 
		
		\item If $N>1$, by optimizing $\bf\Phi$, the range of values $ 0 \leq\left|f_{k}(\mathbf{\Phi})\right| \leq N$ is achievable in (\ref{f_k_Phi_definition}). 
		
		\item If we configure the phase shifts of the RIS to achieve $\left|f_{k}(\mathbf{\Phi})\right| = N$, unless the user $i,i\neq k$, has the same azimuth and elevation AoA as the user $k$, the function $\left|f_{i}(\mathbf{\Phi})\right| $ in (\ref{f_k_Phi_definition}) is bounded when $N\to\infty$.
		
		\item Unless the user $i,i\neq k$, has the same azimuth and elevation AoA as the user $k$, the term $\left|\overline{\mathbf{h}}_{k}^{H} \overline{\mathbf{h}}_{i}\right|^{2}$ is bounded when $N\to\infty$.
	\end{itemize}

\end{lem}

\itshape {Proof:}  \upshape See Appendix \ref{appendix5}. \hfill $\blacksquare$


\subsection{Multi-user Case}
In this section, we consider the general multi-user scenario, i.e., $K>1$. Since any two users are unlikely to be in the same location, we assume that the azimuth and elevation AoA of any two users are different, i.e., $\left( {\varphi _{kr}^a,\varphi _{kr}^e} \right) \neq \left( {\varphi _{ir}^a,\varphi _{ir}^e} \right)$. To begin with, we investigate the asymptotic behavior of the rate in (\ref{rate}) for large values of $M$ and $N$. 
\begin{remark}\label{remark_asymptotic_MN}
From Theorem \ref{theorem2}, we observe that, as a function of $M$, $ E_{k}^{\rm signal} \left(\mathbf{\Phi}\right)$, $ E_{k}^{\rm leak} \left(\mathbf{\Phi}\right)$ and $ I_{ki} \left(\mathbf{\Phi}\right)$ behave asymptotically as $\mathcal{O}\left(M^2\right)$. Therefore, the rate $\underline{R}_k$ converges to a finite limit when $M\to\infty$. If, on the other hand, we align the phase shifts of the RIS for maximizing the intended signal for the user $k$, i.e., we set $\left|f_{k}(\mathbf{\Phi})\right|=N$, then we have $\underline{R}_k\to\infty$ for user $k$, and $\underline{R}_i\to0$ for the other users $i \neq k$ as $N\to\infty$, based on Lemma \ref{f_i_Phi_is_bounded}. In a multi-user scenario, this implies that it is necessary to enforce some fairness requirements among the users when designing the phase shifts of the RIS.
\end{remark}

	Next, we study the power scaling laws of RIS-aided massive MIMO systems with different Rician factors. Specifically, the Rician factor characterizes the fading severity of the environment and the richness of scatterers in the environment.
	 The smaller the Rician factor, the larger the number of scatterers in the environment. 
	 If the Rician factor is zero, we retrieve the Rayleigh fading channel as a special case in which only the NLoS components exist.
	 If the Rician factor tends to infinity, the channel is deterministic and  is characterized only by the LoS component.  It is worth mentioning that, under the assumption of imperfect CSI, decreasing the transmit power $p$ results in a reduction of the power used for both the data and pilot signals.

	 We  analyze several scenarios for the RIS-BS and user-RIS channels. For ease of exposition, we summarize the obtained power scaling laws as a function of $M$ and $N$ in Table \ref{tab_M_N}.
	 \begin{table}[t]
	 		\renewcommand\arraystretch{0.8}
	 	\centering
	 	\captionsetup{font={small}}
	 	\caption{Power scaling laws in the multi-user case.}
	 	\vspace{-10pt}
	 	\begin{tabular}{c|c|c|c|c|c}
	 		\hline
	 		\multicolumn{2}{c| }{}&\multicolumn{4}{c }{$\left(\text{RIS-BS channel, user-RIS channels}\right)$}    \\
	 		\cline{3-6}
	 		\multicolumn{2}{c| }{}& $\left(\text{Rician, Rician}\right)$& $\left(\text{Rician, Rayleigh}\right)$& $\left(\text{Rayleigh, Rician}\right)$& $\left(\text{Rayleigh, Rayleigh}\right)$ \\
	 		\hline
	 		\multirow{2}*{Imperfect CSI}&  $ M $& ${1}/{M}$ & ${1}/{M}$ &   ${1}/{\sqrt{M}}$  &  ${1}/{\sqrt{M}}$ \\
	 		\cline{2-6}
	 		& $ N $& $\diagdown$ &  \multicolumn{3}{c }{$1/N$} \\
	 		\hline
	 		\multirow{2}*{Perfect CSI}& $ M $& \multicolumn{4}{c }{$1/M$} \\
	 		\cline{2-6}
	 		& $ N $& $\diagdown$ &  \multicolumn{3}{c }{$1/N$} \\
	 		\hline		
	 	\end{tabular}\label{tab_M_N}
	 	\label{tab_scale}
	 \end{table}
 Specifically, the following notations are used. ``Imperfect CSI'' and ``Perfect CSI'' are referred to the power scaling laws obtained for imperfect and perfect CSI, respectively. By setting $e_{k1}=e_{k2}=e_{k3}=1$ and $\frac{\sigma^2}{\tau p}=0$, which are obtained when $\tau\to\infty$, the imperfect CSI setup reduces to the perfect CSI setup. The notation ``$\left(\text{Rician, Rician}\right)$'' means that the RIS-BS channel and all the user-RIS channels are Rician distributed, i.e., $\delta>0$ and $\varepsilon_{k}>0,\forall k$. Similarly, the notation ``$\left(\text{Rician, Rayleigh}\right)$'' means that the RIS-BS channel is Rician distributed and all the user-RIS channels are Rayleigh distributed, i.e., $\delta>0$ and $\varepsilon_{k}=0,\forall k$. The notations ``$1/M$'', ``$1/\sqrt{M}$'' and ``$1/N$''  imply that the rate tends to a non-zero value if the transmit power scales  proportionally to $1/M$, $1/\sqrt{M}$ and $1/N$, respectively. We mention, for completeness, that the readers interested in the power scaling laws as a function of $M$ in conventional massive MIMO systems without RISs may refer to \cite{zhang2014power} and \cite{ngo2013energy}. Besides, we note that the rate does not depend on the RIS phase shift matrix $\bf\Phi$ if $\delta=0$ or $\varepsilon_{k}=0, \forall k$, which will be proved in Corollary \ref{corollary5_NLoS_expression}.
  In the following, we mainly consider the proof for the imperfect CSI case, since the perfect CSI setup can be obtained in a similar manner, by setting $e_{k1}=e_{k2}=e_{k3}=1$ and $\frac{\sigma^2}{\tau p}=0$. 
	 
\begin{corollary}\label{corollary4} (``$1/M$'' for ``(Rician, Rician)'' and ``(Rician, Rayleigh)'')
	Assume that the transmit power $p$ is scaled as $p={E_u}/{M}$. For $M\to\infty$, the rate of user $k$, $k\in\mathcal{K}$, is lower bounded by
	\begin{align}\label{rate_power_scaling_law_multipleUser}
\underline{R}_{k} \rightarrow \tau^{o}  \log _{2}\left(1+\frac{E_{u} c_{k}^{2} \delta^{2}\left(\left|f_{k}(\mathbf{\Phi})\right|^{2} \varepsilon_{k}+N e_{k 2}\right)^{2}}{E_{u} E_{k}^{\rm l e a k }\left(\mathbf{\Phi}\right)+E_{u} \sum\limits_{i=1, i \neq k}^{K} I_{k i} \left(\mathbf{\Phi}\right) +\sigma^{2} c_{k} \delta\left(\left|f_{k}(\mathbf{\Phi})\right|^{2} \varepsilon_{k}+N e_{k 2}\right)}\right),
	\end{align}
	where
	\begin{align}
E_{k}^{\rm {leak}}\left(\mathbf{\Phi}\right)=&N\left|f_{k}(\mathbf{\Phi})\right|^{2} c_{k}^{2} \delta^{2} \varepsilon_{k}\left(e_{k 2}^{2}+1\right)+\frac{\sigma^{2}}{\tau E_{u}}\left|f_{k}(\mathbf{\Phi})\right|^{2} c_{k} \delta \varepsilon_{k} e_{k 2}^{2} \nonumber\\
&+N^{2} c_{k}^{2} \delta^{2} e_{k 2}^{2}+\frac{\sigma^{2}}{\tau E_{u}} N c_{k} \delta e_{k 2}^{2} , \\
I_{k i}\left(\mathbf{\Phi}\right)=&\left|f_{k}(\mathbf{\Phi})\right|^{2}\left|f_{i}(\mathbf{\Phi})\right|^{2} c_{k} c_{i} \delta^{2} \varepsilon_{k} \varepsilon_{i}+N\left|f_{k}(\mathbf{\Phi})\right|^{2} c_{k} c_{i} \delta^{2} \varepsilon_{k} \nonumber\\
&+\left|f_{i}(\mathbf{\Phi})\right|^{2} c_{i} \delta \varepsilon_{i} e_{k 2}^{2}\left(N c_{k} \delta+\frac{\sigma^{2}}{\tau E_{u}}\right)+N^{2} c_{k} c_{i} \delta^{2} e_{k 2}^{2}+N \frac{\sigma^{2}}{\tau E_{u}} c_{i} \delta e_{k 2}^{2}, \\\label{e2_limit_in_powerScalingLaw}
e_{k 2}=&\frac{N c_{k} \delta}{\frac{\sigma^{2}}{\tau E_{u}}+N c_{k} \delta}.
	\end{align}
\end{corollary}

\itshape {Proof:}  \upshape  If $p=E_u/M$ and $M\to \infty$, we have $e_{k1}\to 0$, $e_{k3}\to 0$, and $e_{k2}$ tends to (\ref{e2_limit_in_powerScalingLaw}). 
The proof is completed by substituting $p=E_u/M$ into Theorem 2 and retaining the non-zero terms whose asymptotic behavior is $\mathcal{O}\left(M\right)$. \hfill $\blacksquare$


For a massive number of antennas, Corollary \ref{corollary4} shows that the rate of all the users tends to a non-zero value when the transmit power scales as $p=E_u/M$. 
From (\ref{rate_power_scaling_law_multipleUser}), we evince that the rate $\underline{R}_k$ is still non-zero if $\varepsilon_{k}=0,\forall k$, i.e., all the user-RIS channels are Rayleigh distributed. This proves the power scaling law ``$1/M$'' for the ``$\left(\text{Rician, Rayleigh}\right)$'' setup in Table \ref{tab_M_N}. 
However, the rate $\underline{R}_k$ in (\ref{rate_power_scaling_law_multipleUser}) reduces to zero if $c_k=0$ or $\delta=0$, i.e., the RIS-aided channels are absent or the RIS-BS channel is Rayleigh distributed. This indicates that the power scaling law ``$1/M$'' does not hold for these two case studies. Specifically, the considered system degenerates to an RIS-free massive MIMO system with Rayleigh fading if $c_k=0, \forall k$. In this case, it has been proven that the rate can maintain a non-zero value when the power scales as $p=E_u/\sqrt{M}$ \cite[(37)]{ngo2013energy}. As for the power scaling law for $\delta=0$, we first provide an analytical expression of the rate when $\delta=0$.



\begin{corollary}\label{corollary5_NLoS_expression}
If the RIS-BS channel is Rayleigh distributed ($\delta =0$), the rate of user $k$, $k\in\mathcal{K}$, is lower bounded by
\begin{align}\label{rate_NLoS}
\underline{R}_{k}^{(\rm{NL_1})}= \tau^{o}   \log _{2}\left(1+
\frac{p E_{k}^{\mathrm{signal}}}{p E_{k}^{\mathrm{leak}}+p \sum_{i=1, i \neq k}^{K} I_{k i}+\sigma^{2} E_{k}^{\mathrm{noise}}}\right),
\end{align}
where
\begin{align}\label{desired_signal_NLoS_in_corollary5}
E_{k}^{\rm {signal}}=&M\left(N c_{k}\left(\varepsilon_{k}+1\right)+\gamma_{k}\right)^{2} e_{k 1}, \\
E_{k}^{\rm {noise}}=&N c_{k}\left(\varepsilon_{k}+1\right)+\gamma_{k}, \\
E_{k}^{\rm {leak}}=&N^{2} c_{k}^{2}\left(\varepsilon_{k}+1\right)^{2} e_{k 1}+M N c_{k}^{2}\left(2 \varepsilon_{k}+1\right) e_{k 1} \nonumber\\
&+N c_{k}\left\{c_{k}\left(2 \varepsilon_{k}+1\right)+\left(2 \gamma_{k}+\frac{\sigma^{2}}{\tau p}\right)\left(\varepsilon_{k}+1\right)\right\} e_{k 1}+\gamma_{k}\left(\gamma_{k}+\frac{\sigma^{2}}{\tau p}\right) e_{k 1}, \\
I_{k i}=&N^{2} c_{k} c_{i}\left(\varepsilon_{k}+1\right)\left(\varepsilon_{i}+1\right) e_{k 1}+M N c_{k} c_{i}\left(\varepsilon_{k}+\varepsilon_{i}+1\right) e_{k 1}+M c_{k} c_{i} \varepsilon_{k} \varepsilon_{i}\left|\overline{\mathbf{h}}_{k}^{H} \overline{\mathbf{h}}_{i}\right|^{2} e_{k 1} \nonumber\\
&+N\left\{\left(\gamma_{k}+\frac{\sigma^{2}}{\tau p}\right) c_{i}\left(\varepsilon_{i}+1\right)+\gamma_{i} c_{k}\left(\varepsilon_{k}+1\right)\right\} e_{k 1}+\gamma_{i}\left(\gamma_{k}+\frac{\sigma^{2}}{\tau p}\right) e_{k 1},
\end{align}
and
\begin{align}\label{e_k1_NLoS}
e_{k1}=\frac{N \beta \alpha_{k}+\gamma_{k}}{N \beta \alpha_{k}+\gamma_{k}+\frac{\sigma^{2}}{\tau p}}.
\end{align}
\end{corollary}

\itshape {Proof:}  \upshape  When $\delta=0$, we have $a_{k1}=0$, $a_{k2}=N\beta\alpha_{k}+\gamma_{k}$, and $a_{k3}=0$. Thus, we obtain $e_{k3}=e_{k1}^2$, where $e_{k1}=a_{k4}$ is given in (\ref{e_k1_NLoS}). Substituting $\delta=0$ into Theorem \ref{theorem2} and using $e_{k3}=e_{k1}^2$, the proof follows with the aid of some algebraic simplifications. \hfill $\blacksquare$

It is observed that the rate  in Corollary \ref{corollary5_NLoS_expression} does not depend on $\bf\Phi$. Therefore, in a fully NLoS RIS-BS channel, any RIS phase shift matrix  results in the same ergodic rate. This is because the RIS phase shift matrix $\bf \Phi$ is a unitary matrix and the entries of the NLoS channel $\tilde{\bf H}_2$ are Gaussian distributed. Therefore, $\tilde{\bf H}_2{\bf\Phi}$ has the same statistical properties as $\tilde{\bf H}_2$. Likewise, there is no need to design the RIS phase shifts if all the user-RIS links are fully NLoS. This conclusion is apparent from (\ref{rate}) by setting $\varepsilon_{k}=0,\forall k$.

By analyzing the dominant terms of (\ref{rate_NLoS}) when $M,N\to\infty$, we evince that the rate increases without bound for all the users. This implies that fairness requirements among the users are implicitly guaranteed in this special case. As $N\to\infty$, specifically, the dominant terms in (\ref{rate_NLoS}) scale asymptotically as $\mathcal{O}\left(N^2\right)$, and the rate converges to
\begin{align}\label{limit_NLoS_delta}
\underline{R}_{k}^{(\rm{NL_1})}&\to  \tau^{o}   \log _{2}\left(1+\frac{ 
	M \alpha_k
}{ \sum_{i=1}^{K} 
	\alpha_i 
}\right), \text{ as } N\to\infty,\\
&= \tau^{o}   \log _{2}\left(1+{ 
	M 
}/{ K
}\right), \text{ if } \alpha_1=\ldots=\alpha_K.
\end{align}
From (\ref{limit_NLoS_delta}), we evince that the SINR, $  \frac{M \alpha_k}{ \sum_{i=1}^{K} \alpha_i } $, does not depend on the pilot power $\tau p$ and it increases linearly with $M$. Therefore, good performance can be obtained if $\delta=0$ and $N\to\infty$.




With the aid of Corollary 5, we investigate, in the following corollaries, the power scaling laws as a function of $M$ and $N$ when $\delta=0$.

\begin{corollary}\label{corollary6_NLoS_sqrtM} (``$1/\sqrt{M}$'' for ``(Rayleigh, Rician)'' and ``(Rayleigh, Rayleigh)'')
	If the RIS-BS channel is Rayleigh distributed ($\delta=0$), and the power is scaled as $p=E_u/\sqrt{M}$ with $M\to\infty$, the rate of user $k$, $k\in\mathcal{K}$ tends to $ \underline{R}_{k}^{(\mathrm{NL_1})} \rightarrow  \tau^{o}   \log _{2}\left(1+\mathrm{S I N R}_{k}\right) $, where the effective SINR is given by
	\begin{align}\label{SINR_NLOS_scaling_sqrtM}
\operatorname{SINR}_{k}=\frac{\tau E_{u}^{2}\left(N c_{k}\left(\varepsilon_{k}+1\right)+\gamma_{k}\right)^{2}}{\tau E_{u}^{2} N c_{k}^{2}\left(2 \varepsilon_{k}+1\right)+\sum_{i=1, i \neq k}^{K} \tau E_{u}^{2} c_{k} c_{i}\left\{N\left(\varepsilon_{k}+\varepsilon_{i}+1\right)+\varepsilon_{k} \varepsilon_{i}\left|\overline{\mathbf{h}}_{k}^{H} \overline{\mathbf{h}}_{i}\right|^{2}\right\}+\sigma^{4}}.
	\end{align}
\end{corollary}

\itshape {Proof:}  \upshape   First, we substitute $p=E_u/\sqrt{M}$ into Corollary \ref{corollary5_NLoS_expression} and ignore the terms that tend to zero as $M\to\infty$. Then, we divide the numerator and denominator of the SINR by $\frac{N\beta\alpha_{k}+\gamma_{k}}{\sigma^2}$. This yields (\ref{SINR_NLOS_scaling_sqrtM}) and the proof is completed.
 \hfill $\blacksquare$

From (\ref{SINR_NLOS_scaling_sqrtM}), we evince that the numerator of the SINR scales with $\mathcal{O}\left(N^2\right)$, but the denominator of the SINR only scales with $\mathcal{O}\left(N\right)$. Therefore, Corollary \ref{corollary6_NLoS_sqrtM} indicates that the rate scales logarithmically with $N$ if $p=E_u/\sqrt{M}$ and $M\to\infty$, which is a promising result for RIS-aided massive MIMO systems.
Besides, it is worth noting that (\ref{SINR_NLOS_scaling_sqrtM})  reduces to the same expression as in \cite[Eq. (37)]{ngo2013energy} when $c_k=0,\forall k$.

\begin{corollary}\label{corollary7_rate_NLoS_scaling_N} (``$1/N$'' for ``(Rayleigh, Rician)'' and ``(Rayleigh, Rayleigh)'')
	If the RIS-BS channel is Rayleigh distributed ($\delta=0$) and the power is scaled as $p=E_u/{N}$ with $N\to\infty$, the rate of user $k$, $k\in\mathcal{K}$, is lower bounded by
	\begin{align}\label{SINR_NLOS_scaling_N}
\underline{R}_{k}^{(\mathrm{NL_1})} \rightarrow \tau^{o}  \log _{2}\left(1+\frac{E_{u} M \beta \alpha_{k}}{\sum_{i=1}^{K}\left(E_{u} \beta \alpha_{i}+\frac{\alpha_{i}}{\alpha_{k}} \frac{\sigma^{2}}{\tau}\right)+\sigma^{2}\left(1+\frac{\sigma^{2}}{\tau E_{u} \beta \alpha_{k}}\right)}\right).
	\end{align}
\end{corollary}

\itshape {Proof:}  \upshape   First, we substitute $p=E_u/N$ into Corollary \ref{corollary5_NLoS_expression}. When $N\to\infty$, we have $e_{k 1} \rightarrow \frac{\beta \alpha_{k}}{\beta \alpha_{k}+\frac{\sigma^{2}}{\tau E_{u}}}  $. Then, we remove the non-dominant terms that do not scale as $\mathcal{O}\left({N}\right)$. By noting that $ c_{k}\left(\varepsilon_{k}+1\right)=\beta \alpha_{k},\forall k $, and dividing the numerator and denominator of the SINR by $ \beta\alpha_{k}$, we obtain (\ref{SINR_NLOS_scaling_N}). This completes the proof.
  \hfill $\blacksquare$

Corollary \ref{corollary7_rate_NLoS_scaling_N} sheds some interesting insights. Firstly, we note that Corollary \ref{corollary6_NLoS_sqrtM} has unveiled that the transmit power $p$ can only be reduced proportionally to $1/\sqrt{M}$, while maintaining a non-zero rate, when $\delta=0$. Corollary \ref{corollary7_rate_NLoS_scaling_N}, on the other hand, proves that the transmit power can be reduced proportionally to $1/N$, while maintaining a non-zero rate, when $\delta=0$. This reveals the positive role of deploying RISs in massive MIMO systems. Secondly, the obtained power scaling law does not depend on the Rician factors of the user-RIS links, i.e., $\varepsilon_{k},\forall k$. 
This implies that the rate in (\ref{SINR_NLOS_scaling_N}) is the same for LoS-only and NLoS-only user-RIS channels.
Thirdly, in (\ref{SINR_NLOS_scaling_N}), the desired signal term in (\ref{SINR_NLOS_scaling_N}) scales as $\mathcal{O}(M)$ and the interference term scales as $\mathcal{O}(1)$. 
As a result, the rate scales logarithmically with the number of BS antennas. 
When the number of antennas is large, the power of the interference is relatively small compared with the power of the desired signal, and then a good rate can be guaranteed with the setup stated in Corollary \ref{corollary7_rate_NLoS_scaling_N}.
Therefore, a rich-scattering environment between the RIS and the BS ($\delta=0$) is beneficial in RIS-aided massive MIMO systems, since it can provide sufficient spatial multiplexing gains and help mitigate the multi-user interference.
 Finally,  (\ref{SINR_NLOS_scaling_N}) unveils that, if the users are all located at the same distance from the RIS, i.e., $\alpha_{1}=\ldots=\alpha_{K}$, they all achieve the same rate. Therefore, fairness requirements can be guaranteed in this special case.

Corollary \ref{corollary7_rate_NLoS_scaling_N} sheds light on the achievable rate when the RIS-BS channel is Rayleigh distributed ($\delta=0$). In the next corollary, we analyze the opposite scenario in which the user-RIS channels are Rayleigh distributed ($\varepsilon_{k}=0,\forall k$).

\begin{corollary}\label{corollary8_rate_NLoS_varepusilon_scaling_N} (``$1/N$'' for ``(Rician, Rayleigh)'')
	Assume $\delta>0$. If the user-RIS channels are Rayleigh distributed ($\varepsilon_{k}=0,\forall k$) and the power is scaled as $p=E_u/{N}$ with $N\to\infty$, the rate of user $k$, $k\in\mathcal{K}$, is lower bounded by
	\begin{align}\label{SINR_NLOS_varepusilon_scaling_N}
\underline{R}_{k}^{({\rm NL_2})} \rightarrow \tau^{o}   \log _{2}\left(1+\frac{E_{u} M c_{k}^{2}\left(\delta e_{k 2}+e_{k 1}\right)^{2}}{E_{u}\left(E_{k}^{\mathrm {leak}}+\sum_{i=1, i \neq k}^{K} I_{k i}\right)+\sigma^{2} c_{k}\left(\delta e_{k 2}+e_{k 1}\right)}\right),
	\end{align}
	with
	\begin{align}
&E_{k}^{\mathrm {leak}}+\sum_{i=1, i \neq k}^{K} I_{k i}=\sum_{i=1}^{K} c_{i}\left\{M c_{k} \delta^{2} e_{k 2}^{2}+c_{k}\left(2 \delta e_{k 2}^{2}+e_{k 3}\right)+\frac{\sigma^{2}}{\tau E_{u}}\left(\delta e_{k 2}^{2}+e_{k 3}\right)\right\}, \\
&a_{k 3}=\frac{c_{k} \delta \frac{\sigma^{2}}{\tau E_{u}}}{\left(c_{k}+\frac{\sigma^{2}}{\tau E_{u}}\right)\left(c_{k}+\frac{\sigma^{2}}{\tau E_{u}}+M c_{k} \delta\right)}, \\
&a_{k 4}=\frac{c_{k}}{c_{k}+\frac{\sigma^{2}}{\tau E_{u}}}.
	\end{align}
\end{corollary}

\itshape {Proof:}  \upshape It follows from Theorem \ref{theorem2} by setting $\varepsilon_{k}=0,\forall k$ and  $p=E_u/N$, and by keeping only the dominant terms for $N\to\infty$. \hfill $\blacksquare$

Corollary \ref{corollary8_rate_NLoS_varepusilon_scaling_N} characterizes the achievable rate when the user-RIS channels are characterized by rich scattering. The obtained performance trends are different from those unveiled in Corollary \ref{corollary7_rate_NLoS_scaling_N} (i.e., the RIS-BS channel characterized by rich scattering).
In contrast to Corollary \ref{corollary7_rate_NLoS_scaling_N}, in particular, both the desired signal and the interference in (\ref{SINR_NLOS_varepusilon_scaling_N}) scale as $\mathcal{O}\left(M\right)$. As a result,
if the user-RIS channels are Rayleigh distributed, the rate in (\ref{SINR_NLOS_varepusilon_scaling_N}) is still bounded from above even if the number of BS antennas is very large. Besides, it is not hard to prove that the rate in (\ref{SINR_NLOS_varepusilon_scaling_N})  reduces to the same expression as (\ref{SINR_NLOS_scaling_N}) if we  set $\delta=0$. This result confirms the conclusion in Corollary \ref{corollary7_rate_NLoS_scaling_N} that the scaling law unrelated to the Rician factor $\varepsilon_{k}$ if $\delta=0$.

	From Corollary \ref{corollary7_rate_NLoS_scaling_N}  and Corollary \ref{corollary8_rate_NLoS_varepusilon_scaling_N}, we conclude that a small value of $\delta$ is beneficial in terms of power scaling laws. This is because a small $\delta$ corresponds to a high-rank RIS-BS channel, which provides sufficient spatial diversity for  multi-user communications. It is known that, due to the product pathloss law that characterizes RIS-aided links in the far-field region, it is better to deploy an RIS either close to the BS or close to the users\cite{wu2020survey,you2020deploy}. Our analysis  reveals that the best deployment for an RIS depends on the spatial diversity provided by the RIS-BS channel. When the RIS is deployed close to the users, $\delta$ could be small since the Rician factor commonly decreases with the communication distance\cite{3GPP}. Therefore, placing the RIS close to the users is still a good choice since this results in a high rank RIS-BS channel. If the RIS is deployed near the BS, $\delta$ could be large and the RIS-BS channel could become rank-deficient. In this context,  other methods are needed to improve the rank of the channel such as introducing some artificial scatterers between the BS and the RIS or placing the RIS very close to the BS\cite{kammoun2020asymptotic}.

\subsection{Single-user Case}\label{single_user_analysis}

In this subsection, we analyze the power scaling laws in the special case with only one user, i.e., $K=1$. Without loss of generality, the user is referred to as user $k$. Since no other user exists, the rate can be obtained from Theorem \ref{theorem2} by ignoring the multi-user interference term, i.e., by setting $I_{ki}\left(\mathbf{\Phi}\right)=0$. 
For analytical tractability, we further assume that the number of RIS elements is large. 
In this scenario (single-user and large $N$), it can be proved that the optimal phase shift matrix that maximizes the rate corresponds to the condition $\left|f_{k}(\mathbf{\Phi})\right|=N$. This statement is formally proved in the next section (Theorem \ref{theorem3}).

Therefore, by setting $I_{ki}=0$ and $\left|f_{k}(\mathbf{\Phi})\right|=N$ in Theorem \ref{theorem2}, we obtain that the power of the desired signal scales as $\mathcal{O}\left(M^2N^4\right)$, the power of the signal leakage scales as $\mathcal{O}\left(M^2N^3\right)$, and the power of the noise term scales as $\mathcal{O}\left(MN^2\right)$. Therefore, the rate is bounded for $M\to\infty$, but it can grow without bound for $N\to\infty$. 
For ease of exposition, similar to the multi-user case, we summarize the obtained power scaling laws in Table \ref{tab2_single_user_scaling}. 
In the following, we report the proofs only for some (those that lead to insightful design guidelines) system setups that are summarized in Table \ref{tab2_single_user_scaling}. The proof of each case study can, in fact, be obtained by using analytical steps similar to the multi-user case. Finally, we mention that the power scaling laws in the single-user case with perfect CSI can be derived readily based on \cite[Eq. (17)]{han2019large}.

	 \begin{table}[t]
	 		\renewcommand\arraystretch{0.8}
	\centering
	\captionsetup{font={small}}
	\caption{Power scaling laws in the single-user case.}
	\vspace{-10pt}
	\begin{tabular}{c|c|c|c|c|c}
		\hline
		\multicolumn{2}{c| }{}&\multicolumn{4}{c }{$\left(\text{RIS-BS channel, user-RIS channel}\right)$}    \\
		\cline{3-6}
		\multicolumn{2}{c| }{}& $\left(\text{Rician, Rician}\right)$& $\left(\text{Rician, Rayleigh}\right)$& $\left(\text{Rayleigh, Rician}\right)$& $\left(\text{Rayleigh, Rayleigh}\right)$ \\
		\hline
		\multirow{2}*{Imperfect CSI}& $ M $& ${1}/{M}$ & ${1}/{M}$ &   ${1}/{\sqrt{M}}$  &  ${1}/{\sqrt{M}}$ \\
		\cline{2-6}
		& $ N $& $1/N^2$ &  \multicolumn{3}{c }{$1/N$} \\
		\hline
		\multirow{2}*{Perfect CSI}& $ M $& \multicolumn{4}{c }{$1/M$} \\
		\cline{2-6}
		& $ N $& $1/N^2$ &  \multicolumn{3}{c }{$1/N$} \\
		\hline		
	\end{tabular}\label{tab2_single_user_scaling}
\end{table}

\begin{corollary}\label{corollary_single_user_MNN}
	Consider a single-user system with $\left|f_{k}(\mathbf{\Phi})\right|=N$. If the transmit power is scaled as $p={E_u}/{\left(MN^2\right)}$ with $M,N\to\infty$, the rate is lower bounded by
	\begin{align}\label{scaling_MNN}
\underline{R}_{k}\to \tau^{o}  \log _{2}\left(1+\frac{E_{u}}{\sigma^{2}} \frac{\beta \alpha_{k} \delta \varepsilon_{k}}{(\delta+1)\left(\varepsilon_{k}+1\right)}\right).
\end{align}

If the transmit power is scaled as $p={E_u}/{N^2}$ with $N\to\infty$, the rate is lower bounded by
\begin{align}\label{rate_singleUser_scale_NN}
\underline{R}_{k}^{} \rightarrow  \tau^{o}   \log _{2}\left(1+\frac{E_{u}}{\sigma^{2}} M c_{k} \delta \varepsilon_{k}\right).
\end{align}
\end{corollary}

\itshape {Proof:}  \upshape  Let us set $p={E_u}/{\left(MN^2\right)}$, $\left|f_{k}(\mathbf{\Phi})\right|=N$ and $I_{ki}=0$ in Theorem \ref{theorem2}. The rate in (\ref{scaling_MNN}) follows because $e_{k1},e_{k2},e_{k3}\to 0$ and by retaining the dominant terms that scale as $\mathcal{O}\left(MN^2\right)$ for $M,N\to\infty$. Similarly, let us set $p={E_u}/{N^2}$, $\left|f_{k}(\mathbf{\Phi})\right|=N$ and $I_{ki}=0$ in Theorem \ref{theorem2}. The rate in (\ref{rate_singleUser_scale_NN}) follows by retaining the dominant terms that scale as $\mathcal{O}\left(N^2\right)$ for $N\to\infty$.
\hfill $\blacksquare$

The SNRs in (\ref{scaling_MNN}) and (\ref{rate_singleUser_scale_NN}) do not depend on $\tau$, and except for a pre-log scaling factor, the same SNR as for perfect CSI-based systems can be obtained from \cite[Eq. (17)]{han2019large}. We evince, therefore, that $\tau=K=1$ is the optimal pilot length based on (\ref{scaling_MNN}) and (\ref{rate_singleUser_scale_NN}).
Therefore, the overhead for channel estimation is relatively low.
Furthermore, the rates in (\ref{scaling_MNN}) and (\ref{rate_singleUser_scale_NN}) are increasing functions with the Rician factors $\delta$ and $\varepsilon_{k}$, which unveils that LoS-dominated environments are favorable for RIS-aided single-user systems.
If both $\delta\to\infty$ and $\varepsilon_{k}\to\infty$, (\ref{scaling_MNN}) and (\ref{rate_singleUser_scale_NN}) are maximized. 
On the contrary, if $\delta=0$ or $\varepsilon_{k}=0$, we observe that (\ref{scaling_MNN}) and (\ref{rate_singleUser_scale_NN}) tend to zero. This implies that the power scaling law $1/N^2$ does not hold anymore.
In these two cases, the transmit power can be scaled only proportionally to $1/N$ to maintain a non-zero rate when $N\to\infty$. 
Mathematically, the corresponding power scaling laws can be proved from Corollary \ref{corollary7_rate_NLoS_scaling_N} and Corollary \ref{corollary8_rate_NLoS_varepusilon_scaling_N} by setting the multi-user interference to zero. As an example, the case study for $\delta=0$ is analyzed in the following corollary.
\begin{corollary}\label{corollary_single_user_rayleigh}
	Consider a single-user system with $\delta=0$. If the transmit power is scaled as $p={E_u}/{N}$ with $N\to\infty$, the rate is lower bounded by
	\begin{align}\label{scaling_N_rayleigh}
 \underline{R}_{k}^{(\mathrm{NL_1})} \rightarrow   \tau^{o}   \log _{2}\left(1+\frac{E_{u} M \beta \alpha_{k}}
 {  E_{u} \beta \alpha_{k}+ \frac{\sigma^{2}}{\tau}+\sigma^{2}\left(1+\frac{\sigma^{2}}{\tau E_{u} \beta \alpha_{k}}\right)}\right).
 \end{align}
\end{corollary}

As $\tau$ increases, the denominator of the SNR of (\ref{scaling_N_rayleigh}) decreases. Therefore, the SNR of (\ref{scaling_N_rayleigh}) is an increasing function of $\tau$. Therefore, $\tau=1$ is not guaranteed to be optimal in a rich-scattering environment ($\delta=0$), and a relatively large number of pilot signals may be needed. Thus, Corollary \ref{corollary_single_user_rayleigh} also unveils that LoS environments are favorable for RIS-aided single-user systems.

\section{Extension to Correlated Channels with EMI}\label{section_cor_emi}
In this section, we generalize the analysis in Section \ref{section_4} by considering the impact of spatial correlation at the RIS and the presence of EMI. 
We ignore the spatial correlation at the BS, since a ULA with half-wavelength antenna spacing is assumed at the BS. On the other hand, the RIS is usually modeled as a UPA and  the spatial correlation cannot be ignored in general \cite{Emil2021Correlation}.
Specifically, this section has two objectives: (1) to analyze the impact of spatial correlation and EMI in RIS-aided massive MIMO systems; and (2) to study to what extent the findings obtained in Section \ref{section_4}  hold in the presence of spatial correlation and EMI.

 \subsection{Channel Model with Spatial Correlation}
 The evaluation conducted in Section \ref{section_4} indicates that it is appropriate to place the RIS near the users. In this scenario, the LoS components  dominate the user-RIS channels, and therefore the Rician factor $\varepsilon_k$ is relatively large. For ease of analysis and brevity,  this section is focused on the scenario where the user-RIS channels are characterized only by the LoS component (i.e., $\varepsilon_{k}\to\infty$, $\forall k$).\footnote{ Many research works have revealed that the rate is marginally affected by the Rician factor when it is greater than 10\cite{han2019large,zhao2021twoTimeScale}. Thus, the considered scenario serves as a tractable approximation when $\varepsilon_k$ can be assumed to be relatively large. The analysis of arbitrary values for the Rician fading factors $\varepsilon_{k}, \forall k$, is postponed to a future research work.}  In the following, we  present the generalized system model in the presence of spatial correlation and EMI. For the avoidance of doubt, the subscript $c$ is utilized to indicate the existence of spatial correlation. 

In the presence of spatial correlation and EMI, the received signal at the BS is 
\begin{align}\label{model_EMI}
\mathbf{y}_{c}=\sqrt{p} \mathbf{Q}_{c} \mathbf{x}+\mathbf{H}_{c, 2} \mathbf{\Phi} \boldsymbol{v}+\mathbf{n},
\end{align}
where $\boldsymbol{v} \sim \mathcal{CN}\left(\mathbf{0}, \sigma_{e}^{2} \mathbf{R}_{e m i}\right)$ denotes the EMI received at the RIS whose spatial correlation matrix is $ \mathbf{R}_{e m i}$.  Specifically, the EMI  is  reflected by the RIS and reaches the BS through the RIS-BS channel $\mathbf{H}_{c, 2}$ resulting in  the term   $\mathbf{H}_{c, 2} \mathbf{\Phi} \boldsymbol{v}$ in (\ref{model_EMI}). 
The matrix $ \mathbf{Q}_{c}  =[\mathbf{q}_{c, 1},\mathbf{q}_{c, 2},\ldots,\mathbf{q}_{c, K} ] \in \mathbb{C}^{M \times K}$ denotes the spatially-correlated aggregated channel from the $K$ users to the BS, where $\mathbf{q}_{c, k}=\mathbf{H}_{c, 2} \boldsymbol{\Phi} \mathbf{h}_{k} + \mathbf{d}_{k}$ is the aggregated channel of user $k$. The user $k$-RIS channel $\mathbf{h}_{k}$ and the RIS-BS channel $\mathbf{H}_{c, 2}$ are, respectively, given by 
\begin{align}\label{hk_cor}
&\mathbf{h}_{k}=\sqrt{\alpha_{k}} \overline{\mathbf{h}}_{k}	,\\\label{H2_cor}
&\mathbf{H}_{c, 2}=\sqrt{\frac{\beta}{\delta+1}}\left(\sqrt{\delta} \overline{\mathbf{H}}_{2}+\tilde{\mathbf{H}}_{c, 2}\right),
\end{align}
where $\tilde{\mathbf{H}}_{c, 2}=\tilde{\mathbf{H}}_{2} \mathbf{R}_{r i s}^{1 / 2}$ and $\mathbf{R}_{ris}$ denotes the spatial correlation matrix of the NLoS channel components. Assuming an isotropic scattering environment for $\boldsymbol{v}$ and $\tilde{\mathbf{H}}_{c, 2}$, the spatial correlation matrices $\mathbf{R}_{e m i}$ and $ \mathbf{R}_{r i s}$ at the RIS can be formulated as $\mathbf{R}_{e m i}= \mathbf{R}_{r i s} = \mathbf{R}$ with\cite{Emil2021Correlation,Emil2022EMI}
\begin{align}
\left[\mathbf{R}\right]_{a, b}=\operatorname{sinc}\left(\frac{2\left\|\mathbf{u}_{a}-\mathbf{u}_{b}\right\|}{\lambda}\right), 1\leq a,b \leq N,
\end{align}
where $\left\|\mathbf{u}_{a}-\mathbf{u}_{b}\right\|$ denotes the distance between the $a$-th  and $b$-th elements of the RIS, which depends on the RIS element spacing $d_{ris}$. Since $\operatorname{sinc}(\cdot)$  is an even function, we have $\mathbf{R}=\mathbf{R}^H$. For ease of writing, we define $\widehat{c}_{k}=\frac{\alpha_{k} \beta}{\delta+1}$. Therefore,  based on (\ref{hk_cor}) and (\ref{H2_cor}), the spatially-correlated aggregated channel of user $k$ can be expressed as 
\begin{align}
\begin{aligned}
	&\mathbf{q}_{c, k}=\mathbf{H}_{c, 2} \boldsymbol{\Phi} \mathbf{h}_{k} + \mathbf{d}_{k}=\sqrt{\widehat{c}_{k} \delta} \overline{\mathbf{H}}_{2} \boldsymbol{\Phi} \overline{\mathbf{h}}_{k}+\sqrt{\widehat{c}_{k}} \tilde{\mathbf{H}}_{c, 2} \boldsymbol{\Phi} \overline{\mathbf{h}}_{k}+\sqrt{\gamma_{k}} \tilde{\mathbf{d}}_{k}.
\end{aligned}
\end{align}


\subsection{Channel Estimation}
In this section, we derive the LMMSE channel estimate $\hat{\mathbf{q}}_{c, k}$ for the aggregated channel of the $k$-th user.
During the channel estimation phase, the BS receives the $M \times \tau$ pilot signal as follows
\begin{align}
\mathbf{Y}_{c, P}=\sqrt{\tau p} \mathbf{Q}_{c} \mathbf{S}^H+\mathbf{H}_{c, 2} \mathbf{\Phi} \mathbf{V}+\mathbf{N},
\end{align} 
where $\mathbf{V}=\mathbf{R}_{e m i}^{1 / 2} \tilde{\mathbf{V}} \in \mathbb{C}^{N \times \tau}$ and each element of $\tilde{\mathbf{V}}\in \mathbb{C}^{N \times \tau}$ is independently distributed as $ \mathcal{CN}\left({0}, \sigma_{e}^{2} \right)$. After correlating $\mathbf{Y}_{c, P}$ with $\mathbf{s}_k$, the observation vector for the channel of the $k$-th user ${\mathbf{q}}_{c, k}$ is given by
\begin{align}
\begin{aligned}
	\mathbf{y}_{c, p}^{k}=\frac{1}{\sqrt{\tau p}} \mathbf{Y}_{c, P} \mathbf{s}_{k}=\mathbf{q}_{c, k}+\frac{\left(\mathbf{H}_{c, 2} \mathbf{\Phi} \mathbf{V}+\mathbf{N}\right) \mathbf{s}_{k}}{\sqrt{\tau p}}.
\end{aligned}
\end{align}

\begin{thm}\label{theorem_mmse}
Based on $\mathbf{y}_{c, p}^{k}$, the LMMSE channel estimate for $\mathbf{q}_{c, k}$ is given by
\begin{align}\label{channel_estimation_cor}
\hat{\mathbf{q}}_{c, k} = \sqrt{\widehat{c}_{k} \delta} \overline{\mathbf{H}}_{2} \mathbf{\Phi} \overline{\mathbf{h}}_{k}+\sqrt{\widehat{c}_{k}} \mathbf{\Upsilon}_{k} \tilde{\mathbf{H}}_{c, 2} \mathbf{\Phi} \overline{\mathbf{h}}_{k}+\sqrt{\gamma_{k}} \mathbf{\Upsilon}_{k} \tilde{\mathbf{d}}_{k}+\frac{\mathbf{\Upsilon}_{k} \mathbf{H}_{c, 2} \boldsymbol{\Phi} \mathbf{V} \mathbf{s}_{k}}{\sqrt{\tau p}}+\frac{\mathbf{\Upsilon}_{k} \mathbf{N } \mathbf{s} _ { k }}{\sqrt{\tau p}},
\end{align}
where
\begin{align}\label{Upsilon_k}
	 &\mathbf{\Upsilon}_{k}=\mathbf{\Upsilon}_{k}^H=\left(\widehat{c}_{k} \overline{\mathbf{h}}_{k}^{H} \boldsymbol{\Phi}^{H} \mathbf{R}_{ris} \mathbf{\Phi} \overline{\mathbf{h}}_{k}+\gamma_{k}\right)\times\nonumber\\
	 & \left\{\left(\widehat{c}_{k} \overline{\mathbf{h}}_{k}^{H} \boldsymbol{\Phi}^{H} \mathbf{R}_{ris} \boldsymbol{\Phi} \overline{\mathbf{h}}_{k}+\gamma_{k}+\frac{\sigma^{2}}{\tau p}+\frac{\sigma_{e}^{2} \beta \operatorname{Tr}\left\{\mathbf{R}_{e m i} \mathbf{\Phi}^{H} \mathbf{R}_{r i s} \boldsymbol{\Phi}\right\}}{\tau p(\delta+1)}\right) \mathbf{I}_{M}+\frac{\sigma_{e}^{2} \beta \delta \overline{\mathbf{H}}_{2} \boldsymbol{\Phi} \mathbf{R}_{e m i} \boldsymbol{\Phi}^{H} \overline{\mathbf{H}}_{2}^{H}}{\tau p(\delta+1)}\right\}^{-1}.
\end{align}
\end{thm}

\itshape {Proof:}  \upshape See Appendix \ref{appendix7}. \hfill $\blacksquare$

Besides, applying  \cite[Eq. (12.21)]{kay1993fundamentals}, the MSE matrix is given by
\begin{align}\label{MSE_cor}
\textbf{MSE}_{c,k} = \left(\widehat{c}_{k} \overline{\mathbf{h}}_{k}^{H} \boldsymbol{\Phi}^{H} \mathbf{R}_{ris} \mathbf{\Phi} \overline{\mathbf{h}}_{k}+\gamma_{k}\right)  \left( \mathbf{I}_M - \mathbf{\Upsilon}_{k}\right).
\end{align}

Equation (\ref{MSE_cor}) embodies the impact of spatial correlation and EMI on channel estimation. By the direct inspection of (\ref{MSE_cor}), we can make the following observations. On the one hand, the MSE may be degraded by the EMI power $\sigma_{e}^2$ through the term  $\mathbf{\Upsilon}_{k}$. On the other hand, the unitary matrices $\mathbf{\Phi}^H$ and $\mathbf{\Phi}$ do not cancel out in the presence of  spatial correlation, i.e., the matrices $\mathbf{R}_{ris}$ and $\mathbf{R}_{emi}$ are not identity matrices. This implies that an RIS can be utilized for improving the channel estimation accuracy for transmission over spatially-correlated channels. This is a benefit that spatial correlation brings in RIS-aided systems. 
If the spatial correlation is negligible, by contrast, we obtain $\mathbf{R}_{ris}=\mathbf{R}_{emi}=\mathbf{I}_N$ and the MSE matrix in (\ref{MSE_cor}) no longer depends on  $\bf\Phi$, and therefore we cannot optimize the phase shifts of the RIS to improve the quality of channel estimation.


\subsection{Achievable Rate}
Based on the estimated channel $ \hat{\mathbf{q}}_{c, k}$, the MRC detector can be obtained and the corresponding UatF bound of the achievable rate can be computed in the presence of spatial correlation and EMI as well. Specifically, by pre-multiplying the MRC decoding matrix $\hat{\mathbf{Q}}_{c}^H=\left[\hat{\mathbf{q}}_{c, 1}, \ldots, \hat{\mathbf{q}}_{c, K}\right]^H$ with the received signal $\mathbf{y}_c$ in (\ref{model_EMI}), the decoded symbols at the BS are given by
\begin{align}
\mathbf{r}_{c}=\hat{\mathbf{Q}}_{c}^{H} \mathbf{y}_{c}=\sqrt{p} \hat{\mathbf{Q}}_{c}^{H} \mathbf{Q}_{c} \mathbf{x}+\hat{\mathbf{Q}}_{c}^{H} \mathbf{H}_{c, 2} \mathbf{\Phi} \boldsymbol{v}+\hat{\mathbf{Q}}_{c}^{H} \mathbf{n}.
\end{align}

Then, the $k$-th entry of $\mathbf{r}_c$ can be expressed as follows
\begin{align}\label{r_ck}
	\begin{aligned}
r_{c,k}& = 
\sqrt{p} \mathbb{E}\left\{\hat{\mathbf{q}}_{c, k}^{H} \mathbf{q}_{c, k}\right\} x_{k}+\sqrt{p}\left(\hat{\mathbf{q}}_{c, k}^{H} \mathbf{q}_{c, k} -\mathbb{E}\left\{\hat{\mathbf{q}}_{c, k}^{H} \mathbf{q}_{c, k}\right\}  \right)x_{k}\\
&+\sqrt{p} \sum_{i=1, i \neq k}^{K} \hat{\mathbf{q}}_{c, k}^{H} \mathbf{q}_{c, i} x_{i}+\hat{\mathbf{q}}_{c, k}^{H} \mathbf{H}_{c, 2} \mathbf{\Phi} \boldsymbol{v}+\hat{\mathbf{q}}_{c, k}^{H} \mathbf{n}.
	\end{aligned}
\end{align}
Accordingly, the SINR of user $k$ can be written as
\begin{align}\label{SINR_cor_definition}
\operatorname{SINR}_{c, k}=\frac{p E_{c, k}^{\mathrm {signal }}}{p E_{c, k}^{\mathrm {leak }}+p \sum_{i=1, i \neq k}^{K} I_{c, k i}+\sigma_{e}^{2} E_{c, k}^{\mathrm{emi}}+\sigma^{2} E_{c, k}^{\mathrm{noise}}},
\end{align}
where the desired signal is $ E_{c, k}^{\mathrm {signal }}=\left|\mathbb{E}\left\{\hat{\mathbf{q}}_{c, k}^{H} \mathbf{q}_{c, k}\right\}\right|^{2} $, the signal leakage is $ E_{c, k}^{\mathrm {leak }}=\mathbb{E}\left\{\left|\hat{\mathbf{q}}_{c, k}^{H} \mathbf{q}_{c, k}\right|^{2}\right\}-\left|\mathbb{E}\left\{\hat{\mathbf{q}}_{c, k}^{H} \mathbf{q}_{c, k}\right\}\right|^{2} $,  the interference is $ I_{c, k i}=\mathbb{E}\left\{\left|\hat{\mathbf{q}}_{c, k}^{H} \mathbf{q}_{c, i}\right|^{2}\right\} $,  the EMI is $ E_{c, k}^{\mathrm{emi}}=\mathbb{E}\left\{\hat{\mathbf{q}}_{c, k}^{H} \mathbf{H}_{c, 2} \mathbf{\Phi} \mathbf{R}_{e m i} \mathbf{\Phi}^{H} \mathbf{H}_{c, 2}^{H} \hat{\mathbf{q}}_{c, k}\right\} $, and the noise is $ E_{c, k}^{\mathrm {noise }}=\mathbb{E}\left\{\left\|\hat{\mathbf{q}}_{c, k}\right\|^{2}\right\} $.

In order to obtain a compact expression for the UatF bound of the achievable rate, we introduce the following shorthand functions, for $1\leq k, i \leq K$
\begin{align}\label{Notations}
	\begin{aligned}
	&f_{c, 1}(\boldsymbol{\Phi})=\operatorname{Tr}\left\{\mathbf{R}_{r i s} \boldsymbol{\Phi} \mathbf{R}_{e m i} \boldsymbol{\Phi}^{H}\right\},\quad f_{c, k, 2}(\boldsymbol{\Phi})=\overline{\mathbf{h}}_{k}^{H} \boldsymbol{\Phi}^{H} \mathbf{R}_{r i s} \boldsymbol{\Phi}\overline{\mathbf{h}}_{k}, \\
	&f_{c, k, 3}(\boldsymbol{\Phi})=\operatorname{Tr}\left\{\mathbf{\Upsilon}_{k}^{2} \overline{\mathbf{H}}_{2} \boldsymbol{\Phi} \mathbf{R}_{e m i} \boldsymbol{\Phi}^{H} \overline{\mathbf{H}}_{2}^{H}\right\},\quad f_{c, k, 4}(\boldsymbol{\Phi})=\operatorname{Tr}\left\{   \boldsymbol{\Upsilon}_{k}^{2}\right\}, \\
	&f_{c, k, 5}(\boldsymbol{\Phi})=\left|\operatorname{Tr}\left\{  \boldsymbol{\Upsilon}_{k}   \right\}\right|^{2}, \quad f_{c, k, 6}(\boldsymbol{\Phi})=\overline{\mathbf{h}}_{k}^{H} \boldsymbol{\Phi}^{H} \mathbf{R}_{r i s} \boldsymbol{\Phi} \mathbf{R}_{e m i} \boldsymbol{\Phi}^{H} \mathbf{R}_{r i s} \boldsymbol{\Phi} \overline{\mathbf{h}}_{k},\\
	&f_{c, k, 7}(\boldsymbol{\Phi})=\left|f_{k}(\boldsymbol{\Phi})\right|^{2},\quad f_{c, ki, 8}(\boldsymbol{\Phi})=\overline{\mathbf{h}}_{i}^{H} \boldsymbol{\Phi}^{H} \overline{\mathbf{H}}_{2}^{H} \mathbf{\Upsilon}_{k}^{2} \overline{\mathbf{H}}_{2} \mathbf{\Phi} \overline{\mathbf{h}}_{i},\\
	&f_{c, ki, 9}(\boldsymbol{\Phi})=\overline{\mathbf{h}}_{i}^{H} \boldsymbol{\Phi}^{H} \overline{\mathbf{H}}_{2}^{H} \boldsymbol{\Upsilon}_{k} \overline{\mathbf{H}}_{2} \mathbf{\Phi} \mathbf{R}_{e m i} \boldsymbol{\Phi}^{H} \overline{\mathbf{H}}_{2}^{H} \boldsymbol{\Upsilon}_{k}^{H} \overline{\mathbf{H}}_{2} \boldsymbol{\Phi} \overline{\mathbf{h}}_{i}.
\end{aligned}
\end{align}

\begin{thm}\label{theorem4}
In the presence of spatial correlation and EMI, the UatF bound for the achievable rate of the $k$-th user is given by
\begin{align}\label{rate_cor}
	&\underline{R}_{c, k}=  \tau^{o}    \log _{2}\left(1+\operatorname{SINR}_{c, k}\right) ,\\\label{SINR_cor}
	&\operatorname{SINR}_{c, k}=\frac{p E_{c, k}^{\mathrm {signal }}}{p E_{c, k}^{\mathrm {leak }}+p \sum_{i=1, i \neq k}^{K} I_{c, k i}+\sigma_{e}^{2} E_{c, k}^{\mathrm{emi}}+\sigma^{2} E_{c, k}^{\mathrm {noise }}},
\end{align}
where the signal term is $E_{c, k}^{\mathrm {signal }} = \left( E_{c, k}^{\mathrm {noise }} \right)^2$ and the noise term is
\begin{align}\label{E_ck_noise}
	E_{c, k}^{\mathrm {noise }} = M \widehat{c}_{k} \delta\left|f_{k}(\boldsymbol{\Phi})\right|^{2}+\widehat{c}_{k} \operatorname{Tr}\left\{\mathbf{\Upsilon}_{k}\right\} \overline{\mathbf{h}}_{k}^{H} \boldsymbol{\Phi}^{H} \mathbf{R}_{r i s} \boldsymbol{\Phi} \overline{\mathbf{h}}_{k}+\gamma_{k} \operatorname{Tr}\left\{\mathbf{\Upsilon}_{k}\right\}.
\end{align}

The EMI term is given by $E_{c, k}^{\mathrm{e m i}}=\frac{\beta}{\delta+1} \sum_{\omega=1}^{8} E_{c, k}^{\omega, \mathrm{e m i}}$ where
\begin{align}\label{EMI_expec}
\begin{aligned}
	E_{c, k}^{1, \mathrm{e m i}} &=M^{2} \widehat{c}_{k} \delta^{2} f_{c, k, 7}(\boldsymbol{\Phi}) \mathbf{a}_{N}^{H} \boldsymbol{\Phi} \mathbf{R}_{e m i} \mathbf{\Phi}^{H} \mathbf{a}_{N} ,\\
	E_{c, k}^{2, \mathrm{e m i}} &=\left(\widehat{c}_{k} \delta f_{c, k, 2}(\boldsymbol{\Phi})+\frac{2 \beta \delta \sigma_{e}^{2}}{\tau p(\delta+1)} f_{c, 1}(\boldsymbol{\Phi})+\delta\left(\gamma_{k}+\frac{\sigma^{2}}{\tau p}\right)\right) f_{c, k, 3}(\boldsymbol{\Phi}) ,\\
	E_{c, k}^{3, \mathrm{e m i}} &=\left(M \widehat{c}_{k} \delta f_{c, k, 7}(\boldsymbol{\Phi})+\left(\frac{\sigma^{2}}{\tau p}+\gamma_{k}+\widehat{c}_{k} f_{c, k, 2}(\boldsymbol{\Phi})+\frac{\beta \sigma_{e}^{2}}{\tau p(\delta+1)} f_{c, 1}(\boldsymbol{\Phi})\right) f_{c, k, 4}(\boldsymbol{\Phi})\right) f_{c, 1}(\boldsymbol{\Phi}) ,\\
	E_{c, k}^{4, \mathrm{e m i}} &=\frac{\beta \delta^{2} \sigma_{e}^{2}}{\tau p(\delta+1)} \operatorname{Tr}\left\{\left(\mathbf{R}_{e m i} \boldsymbol{\Phi}^{H} \overline{\mathbf{H}}_{2}^{H}{ \bf\Upsilon}_{k} \overline{\mathbf{H}}_{2} \boldsymbol{\Phi}\right)^{2}\right\}, \\
	E_{c, k}^{5, \mathrm{e m i}} &=2 \widehat{c}_{k} \delta \operatorname{Tr}\left\{{\bf\Upsilon}_{k}\right\} \operatorname{Re}\left\{\overline{\mathbf{h}}_{k}^{H} \boldsymbol{\Phi}^{H} \overline{\mathbf{H}}_{2}^{H} \overline{\mathbf{H}}_{2} \boldsymbol{\Phi}_{e m i} \mathbf{\Phi}^{H} \mathbf{R}_{r i s} \boldsymbol{\Phi} \overline{\mathbf{h}}_{k}\right\}, \\
	E_{c, k}^{6, \mathrm{e m i}} &=\frac{2 \beta \delta \sigma_{e}^{2}}{\tau p(\delta+1)} \operatorname{Tr}\left\{{\bf\Upsilon}_{k}\right\} \operatorname{Tr}\left\{\mathbf{R}_{e m i} \mathbf{\Phi}^{H} \overline{\mathbf{H}}_{2}^{H} {\bf\Upsilon}_{k}^{H} \overline{\mathbf{H}}_{2} \mathbf{\Phi} \mathbf{R}_{e m i} \mathbf{\Phi}^{H} \mathbf{R}_{r i s} \boldsymbol{\Phi}\right\}, \\
	E_{c, k}^{7, \mathrm{e m i}} &=\widehat{c}_{k} f_{c, k, 5}(\boldsymbol{\Phi}) f_{c, k, 6}(\boldsymbol{\Phi}), \\
	E_{c, k}^{8, \mathrm{e m i}} &=\frac{\beta \sigma_{e}^{2}}{\tau p(\delta+1)} f_{c, k, 5}(\boldsymbol{\Phi}) \operatorname{Tr}\left\{\left(\mathbf{R}_{r i s} \boldsymbol{\Phi} \mathbf{R}_{e m i} \boldsymbol{\Phi}^{H}\right)^{2}\right\}.
\end{aligned}
\end{align}

The interference term is $I_{c, k i}=\sum_{\omega=1}^{8} I_{c, k i}^{\omega}$, where
\begin{align}
\begin{aligned}
	&I_{c, k i}^{1}=\gamma_{i}  E_{c, k}^{\mathrm {noise }} +M^{2} \widehat{c}_{k} \widehat{c}_{i} \delta^{2} f_{c, k, 7}(\mathbf{\Phi}) f_{c, i, 7}(\mathbf{\Phi}),\\
	&I_{c, k i}^{2}=\bigg\{
		M \widehat{c}_{k} \widehat{c}_{i} \delta f_{c, k, 7}(\boldsymbol{\Phi})+\left(\widehat{c}_{i}\left(\gamma_{k}+\frac{\sigma^{2}}{\tau p}\right)+\frac{\widehat{c}_{i} \beta \sigma_{e}^{2}}{\tau p(\delta+1)} f_{c, 1}(\boldsymbol{\Phi})\right) f_{c, k, 4}(\boldsymbol{\Phi}) \\
		&\qquad\quad+\frac{\widehat{c}_{i} \beta \delta \sigma_{e}^{2}}{\tau p(\delta+1)} f_{c, k, 3}(\boldsymbol{\Phi})\bigg\} f_{c, i, 2}(\boldsymbol{\Phi}),\\
	&I_{c, k i}^{3}=\left\{\widehat{c}_{k} \widehat{c}_{i} \delta f_{c, ki, 8}(\boldsymbol{\Phi})+\widehat{c}_{k} \widehat{c}_{i} f_{c, k, 4}(\boldsymbol{\Phi}) f_{c, i, 2}(\mathbf{\Phi})\right\} f_{c, k, 2}(\boldsymbol{\Phi}),\\
	&I_{c, k i}^{4}=\left\{\frac{\widehat{c}_{i} \beta \delta \sigma_{e}^{2}}{\tau p(\delta+1)} f_{c, 1}(\boldsymbol{\Phi})+\widehat{c}_{i} \delta\left(\gamma_{k}+\frac{\sigma^{2}}{\tau p}\right)\right\} f_{c, ki, 8}(\boldsymbol{\Phi}),\\
	&I_{c, k i}^{5}=\left\{\widehat{c}_{k} \widehat{c}_{i}\left|\overline{\mathbf{h}}_{k}^{H} \boldsymbol{\Phi}^{H} \mathbf{R}_{r i s} \boldsymbol{\Phi} \overline{\mathbf{h}}_{i}\right|^{2}+\frac{\widehat{c}_{i} \beta \sigma_{e}^{2}}{\tau p(\delta+1)} f_{c, i, 6}(\boldsymbol{\Phi})\right\} f_{c, k, 5}(\mathbf{\Phi}),\\
	&I_{c, k i}^{6}=2 \widehat{c}_{k} \widehat{c}_{i} \delta \operatorname{Tr}\left\{  {\bf\Upsilon}_{k}\right\} \operatorname{Re}\left\{\overline{\mathbf{h}}_{k}^{H} \boldsymbol{\Phi}^{H} \overline{\mathbf{H}}_{2}^{H} \overline{\mathbf{H}}_{2} \boldsymbol{\Phi} \overline{\mathbf{h}}_{i} \overline{\mathbf{h}}_{i}^{H} \boldsymbol{\Phi}^{H} \mathbf{R}_{r i s} \boldsymbol{\Phi} \overline{\mathbf{h}}_{k}\right\},\\
	&I_{c, k i}^{7}=\frac{\widehat{c}_{i} \beta \delta^{2} \sigma_{e}^{2}}{\tau p(\delta+1)} f_{c, ki, 9}(\boldsymbol{\Phi}),\\
	&I_{c, k i}^{8}=\frac{2 \widehat{c}_{i} \beta \delta \sigma_{e}^{2}}{\tau p(\delta+1)} \operatorname{Tr}\left\{{\bf\Upsilon}_{k}\right\} \operatorname{Re}\left\{\overline{\mathbf{h}}_{i}^{H} {\bf\Phi}^{H} \mathbf{R}_{r i s} \mathbf{\Phi} \mathbf{R}_{e m i} \mathbf{\Phi}^{H} \overline{\mathbf{H}}_{2}^{H} {\bf\Upsilon}_{k}^{H} \overline{\mathbf{H}}_{2} \boldsymbol{\Phi} \overline{\mathbf{h}}_{i}\right\}.
\end{aligned}
\end{align}

The signal leakage term is $E_{c, k}^{\mathrm {leak }}=\sum_{\omega=1}^{8} E_{c, k}^{\omega,\mathrm{ l e a k}}$, where
\begin{align}
\begin{aligned}
	E_{c, k}^{1, \mathrm { leak }} &=M \widehat{c}_{k} \delta \gamma_{k} f_{c, k, 7}(\boldsymbol{\Phi}), \\
	E_{c, k}^{2, \mathrm { leak }} &=\left\{M \widehat{c}_{k}^{2} \delta f_{c, k, 7}(\boldsymbol{\Phi})+\widehat{c}_{k}^{2} \delta f_{c, kk, 8}(\boldsymbol{\Phi})+\left(\widehat{c}_{k}^{2} f_{c, k, 2}(\boldsymbol{\Phi})+2 \widehat{c}_{k} \gamma_{k}+\frac{\widehat{c}_{k} \sigma^{2}}{\tau p}\right) f_{c, k, 4}(\boldsymbol{\Phi})\right\} f_{c, k, 2}(\boldsymbol{\Phi}), \\
	E_{c, k}^{3, \mathrm { leak }} &=\left\{\widehat{c}_{k} \delta \gamma_{k}+\frac{\widehat{c}_{k} \beta \delta \sigma_{e}^{2}}{\tau p(\delta+1)} f_{c, 1}(\boldsymbol{\Phi})+\frac{\widehat{c}_{k} \delta \sigma^{2}}{\tau p}\right\} f_{c, kk, 8}(\boldsymbol{\Phi}), \\
	E_{c, k}^{4, \mathrm { leak }} &=\left\{\gamma_{k}^{2}+\frac{\gamma_{k} \sigma^{2}}{\tau p}+\frac{\beta \sigma_{e}^{2}}{\tau p(\delta+1)}\left(\gamma_{k}+\widehat{c}_{k} f_{c, k, 2}(\boldsymbol{\Phi})\right) f_{c, 1}(\boldsymbol{\Phi})\right\} f_{c, k, 4}(\boldsymbol{\Phi}) ,\\
	E_{c, k}^{5, \mathrm { leak }} &=\frac{\widehat{c}_{k} \beta \delta^{2} \sigma_{e}^{2}}{\tau p(\delta+1)} f_{c, kk, 9}(\boldsymbol{\Phi}), \\
	E_{c, k}^{6, \mathrm { leak }} &=\frac{2 \widehat{c}_{k} \beta \delta \sigma_{e}^{2}}{\tau p(\delta+1)} \operatorname{Tr}\left\{\mathbf{\Upsilon}_{k}^{H}\right\} \operatorname{Re}\left\{\overline{\mathbf{h}}_{k}^{H} \boldsymbol{\Phi}^{H} \overline{\mathbf{H}}_{2}^{H} {\bf\Upsilon}_{k} \overline{\mathbf{H}}_{2} \mathbf{\Phi} \mathbf{R}_{e m i} \mathbf{\Phi}^{H} \mathbf{R}_{r i s} \mathbf{\Phi} \overline{\mathbf{h}}_{k}\right\} ,\\
	E_{c, k}^{7, \mathrm { leak }} &=\frac{\beta \delta \sigma_{e}^{2}}{\tau p(\delta+1)}\left\{\gamma_{k}+\widehat{c}_{k} f_{c, k, 2}(\boldsymbol{\Phi})\right\} f_{c, k, 3}(\boldsymbol{\Phi}), \\
	E_{c, k}^{8, \mathrm { leak }} &=\frac{\widehat{c}_{k} \beta \sigma_{e}^{2}}{\tau p(\delta+1)} f_{c, k, 5}(\boldsymbol{\Phi}) f_{c, k, 6}(\boldsymbol{\Phi}).
\end{aligned}
\end{align}
\end{thm}

\itshape {Proof:}  \upshape See Appendix \ref{appendix8}. \hfill $\blacksquare$

By comparing the rate $\underline{R}_{c,k}$ in Theorem \ref{theorem4} with the rate $\underline{R}_k$ in Theorem \ref{theorem2}, we can unveil the impact of spatial correlation and EMI. The impact of spatial correlation on the achievable rate is discussed in the following remark.

\begin{remark}\label{remark3}
As briefly mentioned for the MSE in (\ref{MSE_cor}), the presence of spatial correlation could enhance the capabilities of an RIS to tailor a wireless channel. This is apparent by the direct inspection of the rate in Theorem \ref{theorem4} as well.
To be specific, consider the term $\overline{\mathbf{h}}_{k}^{H} \boldsymbol{\Phi}^{H} \mathbf{R}_{r i s} \boldsymbol{\Phi} \overline{\mathbf{h}}_{k} $ as an example. If the spatial correlation is negligible, this term is fixed and equal to $N$ without any possibility to be adjusted by the RIS, since the matrix $ \boldsymbol{\Phi}$ is a unitary matrix and $ \boldsymbol{\Phi}^H  \boldsymbol{\Phi} = \mathbf{I}_N$. However, the same term can be shaped by an RIS in the presence of spatial correlation. For simplicity, let us assume the most severe setup in terms of spatial correlation, i.e., $\mathbf{R}_{ris}=\mathbf{1}_{N\times N}$ so that $\overline{\mathbf{h}}_{k}^{H} \boldsymbol{\Phi}^{H} \mathbf{R}_{r i s} \boldsymbol{\Phi} \overline{\mathbf{h}}_{k} = \left|      \overline{\mathbf{h}}_{k}^{H} \boldsymbol{\Phi}^{H} \mathbf{1}_{N\times 1}         \right|^2$. Based on the proof of Lemma \ref{f_i_Phi_is_bounded}, we have $0\leq  \left|      \overline{\mathbf{h}}_{k}^{H} \boldsymbol{\Phi}^{H} \mathbf{1}_{N\times 1}         \right|^2 \leq N^2$, which demonstrates the enhanced adjustment ability of an RIS to shape the channel in the presence of spatial correlation.

\end{remark}

Next, we discuss the impact of the  EMI on the power scaling laws.
Due to the complex expressions in (\ref{EMI_expec}) and the fact that the optimal design of the RIS phase shifts matrix $\bf\Phi$ cannot  be obtained in a closed-form expression, general conclusions cannot be drawn. However, some special cases are discussed in the following corollary based on the proof by contradiction method.
\begin{corollary}\label{corollary11}
 The power scaling laws summarized in Table \ref{tab_scale} are not guaranteed to hold in the presence of EMI.
\end{corollary}

\itshape {Proof:}  \upshape  
We first give a counterexample for the power scaling laws as a function of $M$. Specifically, we note that the desired signal $E_{c, k}^\mathrm{signal}$ and the EMI term $E_{c, k}^{1, \mathrm{e m i}}$ in (\ref{EMI_expec}) scale as $\mathcal{O}\left(M^2\right)$. If the power is scaled proportionally to $p=1/M$, therefore, the SINR in (\ref{SINR_cor}) tends to zero when $M\to\infty$.
Let us now give a    counterexample for the power scaling laws as a function of $N$. Consider the case study in which only the NLoS components of the channels are present, i.e., $\delta=0$, and no spatial correlation is present, i.e.,  $\mathbf{R}_{ris}=\mathbf{R}_{emi}=\mathbf{I}_N$. Accordingly, $\mathbf{\Upsilon}_k$ simplifies as follows
\begin{align} 
	&\operatorname{Tr}\{\mathbf{\Upsilon}_{k} \}= \frac{ M( N \widehat{c}_{k}  +\gamma_{k})     }{N \widehat{c}_{k}  + \gamma_{k}+\frac{\sigma^{2}}{\tau p}  +\frac{N \sigma_{e}^{2} \beta  }{\tau p}} .
\end{align}
Then, we have $E_{c, k}^{\mathrm {signal }} = \left( E_{c, k}^{\mathrm {noise }} \right)^2$ where
\begin{align}\label{E_ck_noise_special_case}
	E_{c, k}^{\mathrm {noise }} =\frac{ M( N \widehat{c}_{k}  +\gamma_{k})^2     }{N \widehat{c}_{k}  + \gamma_{k}+\frac{\sigma^{2}}{\tau p}  +\frac{N \sigma_{e}^{2} \beta  }{\tau p}} .
\end{align}
If the power is scaled proportionally to $p=1/N$ when $N\to\infty$, (\ref{E_ck_noise_special_case}) implies that $E_{c, k}^{\mathrm {signal }}\to (\frac{\tau M \widehat{c}^2_k}{\sigma_{e}^2 \beta})^2$, which implies $p E_{c, k}^{\mathrm {signal }}= E_{c, k}^{\mathrm {signal }}/N\to 0$. Therefore,  the SINR would tend to zero. This special case demonstrates that the power scaling laws with respect to $N$ are not guaranteed to hold in the presence of EMI.
 \hfill $\blacksquare$

A simple explanation for Corollary \ref{corollary11} is the following.  If the users' transmit power $p$ is scaled proportionally to $1/M$ or $1/N$, as $M$ or $N$ increases, the intended signal power received by the RIS becomes weaker and weaker while the power of the EMI received by the RIS is unaffected. Thus, the EMI becomes stronger and stronger as compared to the intended signal. In other words, as $M,N\to\infty$, the useful power becomes extremely weak and the EMI power dominates the received signal at the RIS.

Nevertheless, we note that the importance of the power scaling laws does not lie in the performance limits in the asymptotic regime for  $M,N\to\infty$. In practice, neither the number of BS antennas nor the number of RIS elements can be infinite. The analysis of the power scaling laws is insightful to understand whether the transmit power of the users can be reduced by increasing $M$ or $N$ while not significantly sacrificing the rate. Therefore, we are usually interested in the power scaling laws when $M$ or $N$ is large but finite. The considered channel model can, in addition, be applied in the far-field region of the BS and RIS, and hence it is not possible to consider an infinite number of BS antennas or RIS elements. Besides, the users share the same RIS-BS channel in RIS-aided systems, which results in strong multi-user interference when applying MRC, as noted in Remark \ref{remark_asymptotic_MN}. Even though the EMI re-radiated by an RIS may be stronger than the thermal noise, it may not necessarily be stronger than the multi-user interference when $M$ or $N$ is not very large. {\color{blue}Specifically, some numerical examples about the impact of the EMI on the achievable rate and power scaling laws are reported in Section VII.}


\section{Design of the RIS Phase Shifts}	\label{section_optimization}
In this section, we optimize the phase shifts of the RIS to maximize the achievable rate derived in Theorem \ref{theorem2} and Theorem \ref{theorem4}. Since the derived ergodic rate depends only on statistical CSI, we need to update the phase shifts of the RIS according to the time variations of the long-term CSI. This results in less frequent updates of the RIS phase shifts especially in the sub-6 GHz frequency range, which, in turn, reduces the channel acquisition overhead and the computational complexity.

\subsection{Single-user Case}
Before tackling the general optimization problem, we first justify the statement made in Section \ref{single_user_analysis} that the optimal phase shift matrix that maximizes the rate in the single-user case fulfills the condition $\left|f_{k}(\boldsymbol{\Phi})\right|=N$. To this end, this subsection aims to solve the phase shifts optimization problem in the  single-user case and in the absence of  spatial correlation and EMI.

	In the single-user case, only the user $k$ is present. We aim to find the phase shifts matrix $\bf \Phi$ that maximizes the lower bound of the ergodic rate $\underline{R}_{k}$ in Theorem \ref{theorem2} by setting $I_{ki}\left(\mathbf{\Phi}\right)=0$.  Only the scenarios with $N>1$, $\delta>0$ and $\varepsilon_k>0,\forall k$ are considered, since $\bf\Phi$ can be set arbitrarily otherwise. It can be observed that the phase shifts matrix $\bf\Phi$ appears only in the term $\left|f_k\left(\mathbf{\Phi}\right)\right|^2$. 
	{\color{blue}For clarity, we denote $x= \left|f_k\left(\mathbf{\Phi}\right)\right|^2$ as the optimization variable. Then, the rate $\underline{R}_{k}$ in Theorem \ref{theorem2} can be rewritten in form of (\ref{rate_single_user}) comprised of some constants $s_1$, $s_2$, $t_1$ and $t_2$ as follows} 
	\begin{align}\label{rate_single_user}
		\underline{R}_{k}&=\tau^{o} \log _{2}\left(1+\mathrm{SNR}_k\left(x\right)\right) \nonumber\\
		&=\tau^{o} \log _{2}\left(1+\frac{   E_{k}^{\rm s i g n a l}  \left(x\right)  }{ E_{k}^{\rm l e a k }\left(x\right) +\frac{\sigma^{2}}{p} E_{k}^{\rm {noise}}  \left(x\right)  }\right)  \nonumber\\
		&=  \tau^{o} \log _{2}\left(1+    \frac{\left(s_{1} x+s_{2}\right)^{2}}{t_{1} x+t_{2}}      \right).
	\end{align}
{\color{blue}The expressions of  $s_1$, $s_2$, $t_1$ and $t_2$ can be derived by direct inspection of Theorem \ref{theorem2} and therefore are omitted for brevity. Besides, it is readily to prove that  $s_1, s_2, t_1, t_2>0$.} From Lemma \ref{f_i_Phi_is_bounded}, we know that the domain of the variable $x$ is $0 \leq x \leq N^2$. Based on (\ref{rate_single_user}), therefore, the optimization problem can be formulated as follows
\begin{subequations}\label{Problem1}
	\begin{align}
		&\max _{x}\;   \;\mathrm{SNR}_{k}\left({x}\right) =\frac{\left(s_{1} x+s_{2}\right)^{2}}{t_{1} x+t_{2}}, \\\label{feasible_set_x}
		&\text { s.t. } \quad  0 \leq x \leq N^2.
	\end{align}
\end{subequations}
	

To solve the problem in (\ref{Problem1}), 
we compute the first-order derivative of $\mathrm{S N R}_k\left(x\right)$ with respect to $x$, as follows
\begin{align}\label{daoshu}
\frac{\partial \mathrm{SNR}_{k}\left(x\right)}{\partial x}=\frac{\left(s_{1} x+s_{2}\right) \left(s_{1} t_{1} x+2 s_{1} t_{2}-s_{2} t_{1}\right)}{\left(t_{1} x+t_{2}\right)^{2}}.
\end{align}

The first-order derivative of $\mathrm{S N R}_k\left(x\right)$ is positive or negative depending on the numerator in (\ref{daoshu}), which is a quadratic function of $x$, i.e., a parabola opening upward, with two roots. The two roots can be obtained by setting (\ref{daoshu}) equal to zero, which yields
\begin{align}\label{define_x0}
x_0^L = \frac{-s_{2}}{s_{1}},\quad x_0^R = \frac{s_{2} t_{1}-2 s_{1} t_{2}}{s_{1} t_{1}},
\end{align}
where $x_0^L<0$ while $ x_0^R $ can be positive.

We can design the optimal configuration of $\bf \Phi$ by analyzing the derivative $\frac{\partial \mathrm{SNR}_{k}\left(x\right)}{\partial x}$ in the domain of $x$, i.e., (\ref{feasible_set_x}), which depends on $ x_0^R $. For example, if $ x_0^R \leq 0$, for a parabola opening upward, we obtain $\frac{\partial \mathrm{SNR}_{k}\left(x\right)}{\partial x}\geq 0$ in the domain $0\leq x \leq N^2$. The complete optimal design criterion is summarized in the following theorem.


\begin{thm}\label{theorem3}
	For RIS-aided single-user systems subject to imperfect CSI, the optimal phase shift matrix $\bf\Phi$ obtained by maximizing the UatF bound of the achievable rate can be summarized as follows.
	\begin{itemize}
		{\color{blue}
			\item It is optimal to set $\left|f_k\left(\mathbf{\Phi}\right)\right|=N$ if (1) $x_0^R\leq 0$; or (2) $0<x_0^R< N^2$ and ${\rm SNR}_k\left(0\right)\leq{\rm SNR}_k\left(N^2\right)$; or (3) $N\to \infty$.
		
		\item It is optimal to set $\left|f_k\left(\mathbf{\Phi}\right)\right|=0$ if (4) $0<x_0^R< N^2$ and ${\rm SNR}_k\left(0\right)>{\rm SNR}_k\left(N^2\right)$; or (5) $x_0^R\ge N^2$.}
	\end{itemize}
\end{thm}

\itshape {Proof:}  \upshape  {\color{blue} It follows by direct inspection of $x_0^R$.  If $x_0^R\leq 0$,} we obtain $\frac{\partial \mathrm{SNR}_{k}\left(x\right)}{\partial x}\geq 0$ in the domain $0\leq x \leq N^2$. Thus, the SNR is an increasing function of $x$ in its domain, which implies that the maximum SNR is reached at the endpoint $x=N^2$. Therefore, it is optimal to set $\left|f_k\left(\mathbf{\Phi}\right)\right|=N$. {\color{blue}If $x_0^R\ge N^2$,} we obtain $\frac{\partial \mathrm{SNR}_{k}\left(x\right)}{\partial x}\leq 0$ in the domain of $x$. Thus, the SNR is a decreasing function of $x$, which implies that the maximum SNR is reached at the endpoint $x=0$. Therefore, it is optimal to set $\left|f_k\left(\mathbf{\Phi}\right)\right|=0$. {\color{blue}If $0<x_0^R< N^2$,} the SNR first decreases for $x<x_0^R$, and then increases for $x>x_0^R$. 
Therefore, the maximum SNR is obtained either at $x=0$ or at $x=N^2$. By comparing ${\rm SNR}_k\left(0\right)$ with ${\rm SNR}_k\left(N^2\right)$, we can identify the optimal design. {\color{blue}Finally, we focus on a special case of $N\to\infty$. In this context, we have ${\rm SNR}_k\left(0\right) < {\rm SNR}_k\left(N^2\right)$, since ${\rm SNR}_k\left(0\right)$ is bounded while ${\rm SNR}_k\left(N^2\right)\to\infty$. Therefore, it is optimal to set $\left|f_k\left(\mathbf{\Phi}\right)\right|=N$  if $N\to\infty$. }    \hfill $\blacksquare$


Finally, we note that the optimal design obtained in {\color{blue}the case of $N\to\infty$} substantiates the analysis reported in Section \ref{single_user_analysis} for large $N$.

\subsection{Multi-user Case}

In this subsection, we consider the design of the RIS phase shifts in the general multi-user scenario with $K>1$.
In the multi-user case, as mentioned in Remark \ref{remark_asymptotic_MN}, it is necessary to guarantee some fairness requirements among  the different users. To this end, we aim to maximize the minimum rate of the users. As a result, the optimization problem can be formulated as follows
\begin{subequations}\label{Problem2}
\begin{align}
&\max _{\bf\Phi}\; \min _{k\in\mathcal{K}} \;\;\underline{R}_{k}\left({\bf\Phi}\right)  \text{ or }  \underline{R}_{c,k}\left({\bf\Phi}\right)   , \\\label{sffggs}
&\text { s.t. } \quad  \left|\left[\mathbf{\Phi}\right]_{n,n}\right|=1, \forall n,
\end{align}
\end{subequations}
where $\underline{R}_k\left({\bf\Phi}\right)$ is given by (\ref{rate}) in Theorem \ref{theorem2} and $\underline{R}_{c,k}\left({\bf\Phi}\right)  $ is given by (\ref{rate_cor}) in Theorem \ref{theorem4}. Constraint (\ref{sffggs}) is the unit modulus constraint for the RIS phase shifts matrix.

For tractability, we introduce the vectors $\boldsymbol{\theta}=[\theta_{1},\theta_2,\ldots,\theta_N]^T$ and $\boldsymbol{c} = [e^{j \theta_{1}},e^{j \theta_{2}},\ldots,e^{j \theta_{N}}]^T$ so that  $\boldsymbol{c} = e^ {j \boldsymbol{\theta}}$ and $\mathbf{\Phi}=\mathrm{diag}  \left(  \boldsymbol{c}  \right)$. Then, the problem in (\ref{Problem2}) can be solved effectively based on the gradient ascent method with respect to the real variable $\boldsymbol{\theta}$.
It is worth noting that our proposed method is different from existing works which adopted the projected gradient ascent method with respect to complex variable $   \boldsymbol{c}$\cite{kammoun2020asymptotic}. To be specific, after updating $\boldsymbol{c}$, the projected gradient ascent method needs a projection operation to ensure that the updated variable $\boldsymbol{c}_{new}$ fulfills the unit modulus constraint $\left| \boldsymbol{c}_{new} \right| = \mathbf{1}$. By contrast, the proposed gradient ascent method avoids the suboptimality caused by the projection operation since the complex exponential functions  are periodic with $\boldsymbol{\theta}$ and the unit modulus constraint holds for every phase shifts vector $\boldsymbol \theta$. Besides, the performance of the gradient ascent method highly depends on the step size, and working with real variables makes the algorithm more robust to the choice of this tuning parameter\cite{qian2022joint}.

The gradient  with respect to $\boldsymbol{\theta}$ is given as follows. Since the objective function in (\ref{Problem2}) includes the min function, which is not differentiable, we first approximate the objective function in (\ref{Problem2})  as
\begin{align}
&\min _{k} \underline{R}_{ k} (\boldsymbol{\theta})\approx  -\frac{1}{\mu} \ln \left\{\sum_{k=1}^{K} \exp \left\{-\mu \underline{R}_{ k}(\mathbf{\boldsymbol{\theta}})\right\}\right\}  \triangleq   f(\mathbf{\boldsymbol{\theta}}),\\
&\min _{k} \underline{R}_{c, k} (\boldsymbol{\theta})   \approx  -\frac{1}{\mu} \ln \left\{\sum_{k=1}^{K} \exp \left\{-\mu \underline{R}_{c, k}(\mathbf{\boldsymbol{\theta}})\right\}\right\}  \triangleq   f_{c}(\mathbf{\boldsymbol{\theta}}),
\end{align}
where $\mu$ is a constant value for controlling the accuracy of the approximation. It can be proved that the approximation error is smaller than $\frac{\ln K}{\mu}$ based on the method in \cite{xingsi1992entropy}. Thus, the problem in (\ref{Problem2}) can be recast as
\begin{subequations}\label{Problem4}
	\begin{align}
		&\max _{   \boldsymbol{\theta}  }\;        f(\mathbf{\boldsymbol{\theta}})    \text{ or }     f_c(\mathbf{\boldsymbol{\theta}})   , \\\label{constraint_f}
		&\text { s.t. } \quad  0 \leq \theta_{n} < 2 \pi, \forall n.
	\end{align}
\end{subequations}

As mentioned, the constraint (\ref{constraint_f}) can be neglected thanks to the periodicity of the objective functions $ f(\mathbf{\boldsymbol{\theta}})$ and $ f_c(\mathbf{\boldsymbol{\theta}}) $ with respect to $\boldsymbol{\theta}$. Therefore, there is no need to perform any projection operation after updating variable $\boldsymbol{\theta}$.
Then, we need to calculate the gradient of $f(\mathbf{\boldsymbol{\theta}})  $ and $f_c(\mathbf{\boldsymbol{\theta}})$. Since these two gradients can be calculated in a similar way, we only provide the detailed process for $ \frac{\partial f_{c}(\boldsymbol{\theta})}{\partial \boldsymbol{\theta}} $.  Based on the chain rule, we have
\begin{align}\label{gradient_rate}
	&\frac{\partial f_{c}(\boldsymbol{\theta})}{\partial \boldsymbol{\theta}}=\frac{\tau^{o}  \sum_{k=1}^{K}\left\{\frac{\exp \left\{-\mu \underline{R}_{c, k}(\boldsymbol{\theta})\right\}}{1+\operatorname{SINR}_{c, k}(\boldsymbol{\theta})} \frac{\partial \operatorname{SINR}_{c, k}(\boldsymbol{\theta})}{\partial \boldsymbol{\theta}}\right\}}{(\ln 2)\left(\sum_{k=1}^{K} \exp \left\{-\mu \underline{R}_{c, k}(\boldsymbol{\theta})\right\}\right)},
\end{align}
and
\begin{align}\label{gradient_sinr}
\begin{aligned}
	&\frac{\partial \operatorname{SINR}_{c, k}( \boldsymbol{\theta} )}{\partial \boldsymbol{\theta}}=\frac{p \frac{\partial E_{c, k}^{\mathrm {signal }}}{\partial \boldsymbol{\theta}}}{p E_{c, k}^{\mathrm {leak }}+p \sum_{i=1, i \neq k}^{K} I_{c, k i}+\sigma_{e}^{2} E_{c, k}^{\mathrm {emi }}+\sigma^{2} E_{c, k}^{\mathrm {noise }}} \\
&-p E_{c, k}^{\mathrm {signal }} \frac{p \frac{\partial E_{c, k}^{\mathrm {leak }}}{\partial \boldsymbol{\theta}}+p \sum_{i=1, i \neq k}^{K} \frac{\partial I_{c, k i}}{\partial \boldsymbol{\theta}}+\sigma_{e}^{2} \frac{\partial E_{c, k}^{\mathrm {emi }}}{\partial \boldsymbol{\theta}}+\sigma^{2} \frac{\partial E_{c, k}^{\mathrm {noise }}}{\partial \boldsymbol{\theta}}}{\left(p E_{c, k}^{\mathrm {leak }}+p \sum_{i=1, i \neq k}^{K} I_{c, k i}+\sigma_{e}^{2} E_{c, k}^{\mathrm {emi }}+\sigma^{2} E_{c, k}^{\mathrm {noise }}\right)^{2}}.
\end{aligned}
\end{align}

Therefore, the gradient of $f_c(\boldsymbol{\theta})$ can be obtained after  calculating $\frac{\partial E_{c, k}^{\mathrm {signal }}}{\partial \boldsymbol{\theta}}$, $\frac{\partial E_{c, k}^{\mathrm {leak }}}{\partial \boldsymbol{\theta}}$, $\frac{\partial I_{c, k i}}{\partial \boldsymbol{\theta}}$, $\frac{\partial E_{c, k}^{\mathrm {emi }}}{\partial \boldsymbol{\theta}}$, and $\frac{\partial E_{c, k}^{\mathrm {noise }}}{\partial \boldsymbol{\theta}}$ in (\ref{gradient_sinr}). Based on Theorem \ref{theorem4}, we note that $ E_{c, k}^{\mathrm {signal }} $, $ I_{c, k i}$, $ E_{c, k}^{\mathrm {leak }}$, $ E_{c, k}^{\mathrm {emi }}$ and $ E_{c, k}^{\mathrm {noise }}$ can be computed from the functions in (\ref{Notations}). {\color{blue}For ease of following the key idea}, we first provide two useful lemmas and then use them to calculate the gradient of the terms in (\ref{Notations}).

\begin{lem}\label{lemma_apbp}
Given the deterministic matrices $\mathbf{A}$ and $\mathbf{B}$, the gradient of  $ \operatorname{Tr}\left\{\mathbf{A} \boldsymbol { \Phi } \mathbf{B} \boldsymbol{\Phi}^{H}\right\} $ with respect to $      \boldsymbol{\theta}      $ is given by
\begin{align}\label{apbp_vec}
\begin{aligned}
\frac{\partial \operatorname{Tr}\left\{\mathbf{A} \boldsymbol{\Phi  } \mathbf{B}  \boldsymbol{\Phi}^{H}\right\}}{\partial  \boldsymbol{\theta}}
&=j \boldsymbol{\Phi}^{T}\left(\mathbf{A}^{T} \odot \mathbf{B}\right) \boldsymbol{c}^{*}
-j \boldsymbol{\Phi}^{H}\left(\mathbf{A} \odot \mathbf{B}^{T}\right) \boldsymbol{c} \\
&\triangleq \boldsymbol{f}_{d}(\mathbf{A}, \mathbf{B}).
\end{aligned}
\end{align}
If $\mathbf{A}=\mathbf{A}^{H}, \mathbf{B}=\mathbf{B}^{H}$, we further have
\begin{align}\label{apbp_unitary}
\frac{\partial \operatorname{Tr}\left\{\mathbf{A} \boldsymbol{\Phi } \mathbf{B} \boldsymbol{\Phi}^{H}\right\}}{\partial  \boldsymbol{\theta}}=2 \operatorname{Im}\left\{\boldsymbol{\Phi}^{H}\left(\mathbf{A} \odot \mathbf{B}^{T}\right) \boldsymbol{c}\right\}.
\end{align}
\end{lem}

\itshape {Proof:}  \upshape See Appendix \ref{appendix9}. \hfill $\blacksquare$

\begin{lem}\label{lemma_T_Upsilon}
	Define $\psi_{k}^{1} = \widehat{c}_{k} \overline{\mathbf{h}}_{k}^{H} \boldsymbol{\Phi}^{H} \mathbf{R}_{ris} \mathbf{\Phi} \overline{\mathbf{h}}_{k}+\gamma_{k}$ and $\mathbf{\Upsilon}_{k}=\psi_{k}^{1} \mathbf{\Upsilon}_{k}^{1} $.
	Then, given the deterministic matrix $\bf T$, the gradient of  $ \operatorname{Tr}\left\{         \mathbf{T} \mathbf{\Upsilon}_{k}           \right\} $ with respect to $      \boldsymbol{\theta}      $ is given by
	\begin{align}\label{TUpsilon_vec}
		\begin{aligned}
			&\frac{\partial \operatorname{Tr}\left\{\mathbf{T} \mathbf{\Upsilon}_{k}\right\}}{\partial \boldsymbol{\theta}} \\
			&=2 \widehat{c}_{k}\left\{\operatorname{Tr}\left\{\mathbf{T }  \mathbf{\Upsilon}_{k}^{1}\right\}-\psi_{k}^{1} \operatorname{Tr}\left\{\mathbf{T}\left(\mathbf{\Upsilon}_{k}^{1}\right)^{2}\right\}\right\} \operatorname{Im}\left\{\boldsymbol{\Phi}^{H}\left(\mathbf{R}_{r i s} \odot\left(\overline{\mathbf{h}}_{k} \overline{\mathbf{h}}_{k}^{H}\right)^{T}\right) \boldsymbol{c}\right\} \\
			&- \frac{2\sigma_{e}^{2} \beta}{\tau p(\delta+1)} \psi_{k}^{1} \operatorname{Tr}\left\{\mathbf{T}\left(\mathbf{\Upsilon}_{k}^{1}\right)^{2}\right\} \operatorname{Im}\left\{\mathbf{\Phi}^{H}\left(\mathbf{R}_{r i s} \odot \mathbf{R}_{e m i}\right) \boldsymbol{c}\right\} \\
			&- \frac{\sigma_{e}^{2} \beta \delta}{\tau p(\delta+1)} \psi_{k}^{1}
			\boldsymbol{f}_d\left(        \overline{\mathbf{H}}_{2}^{H} \mathbf{\Upsilon}_{k}^{1} \mathbf{T} \mathbf{\Upsilon}_{k}^{1} \overline{\mathbf{H}}_{2},   \mathbf{R}_{e m i}   \right) \\
			&\triangleq \mathbf{z}_{k}(\mathbf{T})
		\end{aligned}
	\end{align}
\end{lem}

\itshape {Proof:} \upshape  The proof is similar to the proof of Lemma \ref{lemma_apbp}  after applying the chain rule to the inverse matrix  $\partial\left(\mathbf{X}^{-1}\right)=-\mathbf{X}^{-1}(\partial \mathbf{X}) \mathbf{X}^{-1}$. \hfill $\blacksquare$

With the aid of  Lemma \ref{lemma_apbp} and \ref{lemma_T_Upsilon}, we obtain the following lemma.
\begin{lem}\label{lemma_gradient}
The gradients of the functions defined in (\ref{Notations}) are given by
\begin{align}
	\begin{aligned}
\boldsymbol{f}_{c, 1}^{\prime}(  \boldsymbol{\theta}   )&=\frac{\partial f_{c, 1}(\boldsymbol{\Phi})}{\partial \boldsymbol{\theta}}=2 \operatorname{Im}\left\{\boldsymbol{\Phi}^{H}\left(\mathbf{R}_{r i s} \odot \mathbf{R}_{e m i}\right) \boldsymbol{c}\right\},\\
\boldsymbol{f}_{c, k, 2}^{\prime}( \boldsymbol{\theta} )&=\frac{\partial f_{c, k, 2}(\boldsymbol{\Phi})}{\partial \boldsymbol{\theta}}=2 \operatorname{Im}\left\{\boldsymbol{\Phi}^{H}\left(\mathbf{R}_{r i s} \odot\left(\overline{\mathbf{h}}_{k} \overline{\mathbf{h}}_{k}^{H}\right)^{T}\right) \boldsymbol{c}\right\},\\
 \boldsymbol{f}_{c, k, 3}^{\prime}(   \boldsymbol{\theta}   )&=\frac{\partial f_{c, k, 3}(\boldsymbol{\Phi})}{\partial \boldsymbol{\theta}}\\
&=\mathbf{z}_{k}\left(\overline{\mathbf{H}}_{2} \boldsymbol{\Phi} \mathbf{R}_{e m i} \boldsymbol{\Phi}^{H} \overline{\mathbf{H}}_{2}^{H} \boldsymbol{\Upsilon}_{k}\right)+2 \operatorname{Im}\left\{\boldsymbol{\Phi}^{H}\left(\overline{\mathbf{H}}_{2}^{H} \boldsymbol{\Upsilon}_{k}^{2} \overline{\mathbf{H}}_{2} \odot \mathbf{R}_{e m i}\right) \boldsymbol{c}\right\}\\
&+\mathbf{z}_{k}\left(\boldsymbol{\Upsilon}_{k} \overline{\mathbf{H}}_{2} \boldsymbol{\Phi} \mathbf{R}_{e m i} \boldsymbol{\Phi}^{H} \overline{\mathbf{H}}_{2}^{H}\right),
	\end{aligned}
\end{align}
\begin{align}
	\begin{aligned}
	\boldsymbol{f}_{c, k, 4}^{\prime}(   \boldsymbol{\theta}    )&=\frac{\partial f_{c, k, 4}(\boldsymbol{\Phi})}{\partial \boldsymbol{\theta}}=2 \mathbf{z}_{k}\left(\boldsymbol{\Upsilon}_{k}\right),\\
		\boldsymbol{f}_{c, k, 5}^{\prime}(    \boldsymbol{\theta}    )&=\frac{\partial f_{c, k, 5}(\boldsymbol{\Phi})}{\partial \boldsymbol{\theta}}=2 \operatorname{Tr}\left\{\mathbf{\Upsilon}_{k}\right\} \mathbf{z}_{k}\left(\mathbf{I}_{M}\right),\\
		\boldsymbol{f}_{c, k, 6}^{\prime}(   \boldsymbol{\theta}   )&=\frac{\partial f_{c, k, 6}(\boldsymbol{\Phi})}{\partial \boldsymbol{\theta}}\\
		&=2 \operatorname{Im}\left\{\boldsymbol{\Phi}^{H}\left(\mathbf{R}_{r i s} \boldsymbol{\Phi} \overline{\mathbf{h}}_{k} \overline{\mathbf{h}}_{k}^{H} \boldsymbol{\Phi}^{H} \mathbf{R}_{r i s} \odot \mathbf{R}_{e m i}\right) \boldsymbol{c}\right\}\\
		&+2 \operatorname{Im}\left\{\boldsymbol{\Phi}^{H}\left(\mathbf{R}_{r i s} \mathbf{\Phi} \mathbf{R}_{e m i} \boldsymbol{\Phi}^{H} \mathbf{R}_{r i s} \odot\left(\overline{\mathbf{h}}_{k} \overline{\mathbf{h}}_{k}^{H}\right)^{T}\right) \boldsymbol{c}\right\} ,\quad\qquad\qquad
	\end{aligned}
\end{align}
\begin{align}
	\begin{aligned}
		\boldsymbol{f}_{c, k, 7}^{\prime}(   \boldsymbol{\theta}    )&=\frac{\partial f_{c, k, 7}(\boldsymbol{\Phi})}{\partial \boldsymbol{\theta}}=2 \operatorname{Im}\left\{\boldsymbol{\Phi}^{H}\left(\mathbf{a}_{N} \mathbf{a}_{N}^{H} \odot\left(\overline{\mathbf{h}}_{k} \overline{\mathbf{h}}_{k}^{H}\right)^{T}\right) \boldsymbol{c}\right\} ,\\
		\boldsymbol{f}_{c, ki, 8}^{\prime}(   \boldsymbol{\theta}   )&=\frac{\partial f_{c, ki, 8}(\boldsymbol{\Phi})}{\partial \boldsymbol{\theta}}\\
		&=\mathbf{z}_{k}\left(\boldsymbol{\Upsilon}_{k} \overline{\mathbf{H}}_{2}  \boldsymbol{\Phi} \overline{\mathbf{h}}_{i} \overline{\mathbf{h}}_{i}^{H} \boldsymbol{\Phi}^{H} \overline{\mathbf{H}}_{2}^{H}\right)+\mathbf{z}_{k}\left(\overline{\mathbf{H}}_{2} \boldsymbol{\Phi} \overline{\mathbf{h}}_{i} \overline{\mathbf{h}}_{i}^{H} \boldsymbol{\Phi}^{H} \overline{\mathbf{H}}_{2}^{H} \boldsymbol{\Upsilon}_{k}\right)\\
		&+2 \operatorname{Im}\left\{\boldsymbol{\Phi}^{H}\left(\overline{\mathbf{H}}_{2}^{H} \boldsymbol{\Upsilon}_{k}^2 \overline{\mathbf{H}}_{2} \odot\left(\overline{\mathbf{h}}_{i} \overline{\mathbf{h}}_{i}^{H}\right)^{T}\right) \boldsymbol{c}\right\},\quad\qquad\qquad\qquad\qquad
	\end{aligned}
\end{align}
\begin{align}
	\begin{aligned}
\boldsymbol{f}_{c, k i, 9}^{\prime}(   \boldsymbol{\theta}   )&=\frac{\partial f_{c, k i, 9}(\boldsymbol{\Phi})}{\partial \boldsymbol{\theta}} \\
		&=\mathbf{z}_{k}\left(\overline{\mathbf{H}}_{2} \boldsymbol{\Phi} \mathbf{R}_{e m i} \boldsymbol{\Phi}^{H} \overline{\mathbf{H}}_{2}^{H} \boldsymbol{\Upsilon}_{k}^{H} \overline{\mathbf{H}}_{2} \boldsymbol{\Phi} \overline{\mathbf{h}}_{i} \overline{\mathbf{h}}_{i}^{H} \boldsymbol{\Phi}^{H} \overline{\mathbf{H}}_{2}^{H}\right)\\
		&+2 \operatorname{Im}\left\{\boldsymbol{\Phi}^{H}\left(\overline{\mathbf{H}}_{2}^{H} \boldsymbol{\Upsilon}_{k}^{H} \overline{\mathbf{H}}_{2} \boldsymbol{\Phi} \overline{\mathbf{h}}_{i} \overline{\mathbf{h}}_{i}^{H} \boldsymbol{\Phi}^{H} \overline{\mathbf{H}}_{2}^{H} \boldsymbol{\Upsilon}_{k} \overline{\mathbf{H}}_{2} \odot \mathbf{R}_{e m i}\right) \boldsymbol{c}\right\} \\
		&+\mathbf{z}_{k}\left(\overline{\mathbf{H}}_{2} \boldsymbol{\Phi} \overline{\mathbf{h}}_{i} \overline{\mathbf{h}}_{i}^{H} \boldsymbol{\Phi}^{H} \overline{\mathbf{H}}_{2}^{H} \boldsymbol{\Upsilon}_{k} \overline{\mathbf{H}}_{2} \boldsymbol{\Phi} \mathbf{R}_{e m i} \boldsymbol{\Phi}^{H} \overline{\mathbf{H}}_{2}^{H}\right)\\
		&+2 \operatorname{Im}\left\{\boldsymbol{\Phi}^{H}\left(\overline{\mathbf{H}}_{2}^{H} \boldsymbol{\Upsilon}_{k} \overline{\mathbf{H}}_{2} \mathbf{\Phi} \mathbf{R}_{e m i} \boldsymbol{\Phi}^{H} \overline{\mathbf{H}}_{2}^{H} \boldsymbol{\Upsilon}_{k}^{H} \overline{\mathbf{H}}_{2} \odot\left(\overline{\mathbf{h}}_{i} \overline{\mathbf{h}}_{i}^{H}\right)^{T}\right) \boldsymbol{c}\right\}.
	\end{aligned}
\end{align}
\end{lem}

\itshape {Proof:} \upshape  It follows by applying the chain rule to compute the derivatives and using Lemma \ref{lemma_apbp} and \ref{lemma_T_Upsilon}. Consider $\boldsymbol{f}_{c, k i, 3}^{\prime}(   \boldsymbol{\theta}   )$ as an example. By applying the chain rule, we have
\begin{align}
\begin{aligned}
&\boldsymbol{f}_{c, k, 3}^{\prime}(\boldsymbol{\theta})=\frac{\partial \operatorname{Tr}\left\{\boldsymbol{\Upsilon}_{k} \boldsymbol{\Upsilon}_{k} \overline{\mathbf{H}}_{2} \boldsymbol{\Phi} \mathbf{R}_{e m i} \mathbf{\Phi}^{H} \overline{\mathbf{H}}_{2}^{H}\right\}}{\partial \boldsymbol{\theta}}=\left.\frac{\partial \operatorname{Tr}\left\{\mathbf{T} \boldsymbol{\Upsilon}_{k}\right\}}{\partial \boldsymbol{\theta}}\right|_{\mathbf{T}=\mathbf{\Upsilon}_{k} \overline{\mathbf{H}}_{2} \boldsymbol{\Phi} \mathbf{R}_{e m i} \boldsymbol{\Phi}^{H} \overline{\mathbf{H}}_{2}^{H}}\\
&+\left.\frac{\partial \operatorname{Tr}\left\{\mathbf{T} \boldsymbol{\Upsilon}_{k}\right\}}{\partial \boldsymbol{\theta}}\right|_{\mathbf{T}=\overline{\mathbf{H}}_{2} \boldsymbol{\Phi} \mathbf{R}_{e mi} \boldsymbol{\Phi}^{H} \overline{\mathbf{H}}_{2}^{H} \boldsymbol{\Upsilon}_{k}}+\left.\frac{\partial \operatorname{Tr}\left\{\mathbf{A} \boldsymbol{\Phi} \mathbf{B} \boldsymbol{\Phi}^{H}\right\}}{\partial \boldsymbol{\theta}}\right|_{\mathbf{A}=\overline{\mathbf{H}}_{2}^{H} \boldsymbol{\Upsilon}_{k}^{2} \overline{\mathbf{H}}_{2}, \mathbf{B}=\mathbf{R}_{e m i}}.
\end{aligned}
\end{align}
The proof follows by applying Lemma \ref{lemma_apbp} and \ref{lemma_T_Upsilon}. The other terms can be obtained similarly.
\hfill $\blacksquare$

Therefore, the gradient of $ \frac{\partial f_{c}(\boldsymbol{\theta})}{\partial \boldsymbol{\theta}} $ in (\ref{gradient_rate}) follows from (\ref{gradient_sinr}), Lemmas \ref{lemma_apbp}, \ref{lemma_T_Upsilon}, \ref{lemma_gradient} and by applying the chain rule. For example, we have
\begin{align}\label{gra_signal}
\frac{\partial E_{c, k}^{\mathrm {signal }}}{\partial \boldsymbol{\theta}}=\frac{\partial\left\{\left(E_{c, k}^{\mathrm {noise }}\right)^{2}\right\}}{\partial \boldsymbol{\theta}}=2 E_{c, k}^{\mathrm{noise}} \frac{\partial E_{c, k}^{\mathrm{noise}}}{\partial \boldsymbol{\theta}},
\end{align}
and
\begin{align}
\frac{\partial E_{c, k}^{\mathrm {noise }}}{\partial \boldsymbol{\theta}}=M \widehat{c}_{k} \delta \boldsymbol{f}_{c, k, 7}^{\prime}(\boldsymbol{\theta})+\left\{\widehat{c}_{k} f_{c, k, 2}(\boldsymbol{\Phi})+\gamma_{k}\right\} \mathbf{z}_{k}\left(\mathbf{I}_{M}\right)+\widehat{c}_{k} \operatorname{Tr}\left\{\boldsymbol{\Upsilon}_{k}\right\} \boldsymbol{f}_{c, k, 2}^{\prime}(\boldsymbol{\theta}).
\end{align}

  \begin{algorithm}[t]
	\caption{\color{blue}Accelerated Gradient Ascent Algorithm}
	\color{blue}
	\begin{algorithmic}[1]\label{algorithm1}
		\STATE Initialize $\boldsymbol{\theta}_0$ randomly, $i=0$, $a_0=1$, $\boldsymbol{x}_{-1} =   \boldsymbol{\theta}_0  $;
		\WHILE{1}
		\STATE Calculate the gradient vector $  \boldsymbol{f}_{c }^{\prime}(   \boldsymbol{\theta}_i   )= \left.    \frac{\partial f_{c}(\boldsymbol{\theta})}{\partial \boldsymbol{\theta}}     \right| _  {\boldsymbol{\theta} = \boldsymbol{\theta}_i}   $;
		\STATE Obtain the step size $\kappa_i$ based on the backtracking line search;
		\STATE $\boldsymbol{x}_{i} =  \boldsymbol{\theta}_i +\kappa_i  \boldsymbol{f}_{c }^{\prime}(   \boldsymbol{\theta}_i   )  $;
		\STATE $a_{i+1}=(1+\sqrt{4 a_i^2+1}) / 2$;
		\STATE $\boldsymbol{\theta}_{i+1} =\boldsymbol{x}_{i}+\left(a_i-1\right)\left(     \boldsymbol{x}_{i}  - \boldsymbol{x}_{i-1}       \right) / a_{i+1}$;
		\IF {$f_c( \boldsymbol{\theta}_{i+1}  ) - f_c( \boldsymbol{\theta}_{i}  ) < 10^{-4}$}
		\STATE $\boldsymbol{\theta}^*= \boldsymbol{\theta}_{i+1}$, break;
		\ENDIF
		\STATE $i=i+1$;
		\ENDWHILE
	\end{algorithmic}
\end{algorithm}


All the other terms in $\frac{\partial f(\boldsymbol{\theta})}{\partial \boldsymbol{\theta}}$ and $\frac{\partial f_c(\boldsymbol{\theta})}{\partial \boldsymbol{\theta}}$ can be obtained similarly to (\ref{gra_signal}). The final analytical expressions of $\frac{\partial f(\boldsymbol{\theta})}{\partial \boldsymbol{\theta}}$ and $\frac{\partial f_c(\boldsymbol{\theta})}{\partial \boldsymbol{\theta}}$ are given in  Appendix \ref{appendix_K}. 
{\color{blue}It is known that gradient-based methods may have a slow convergence rate. To tackle this issue, we apply  Nesterov's accelerated gradient method, which  effectively increases the convergence speed of the gradient method\cite{nesterov1983method}.
For completeness, the  algorithm for optimizing $f_c(\boldsymbol{\theta})$ is presented in Algorithm \ref{algorithm1} where steps 6-7 correspond to Nesterov's acceleration method. }

\section{Numerical Results}\label{section_6}
In this section, we evaluate the performance of RIS-aided massive MIMO systems and validate the impact of key system parameters unveiled in the previous sections. We first consider a typical RIS-aided scenario where an RIS is deployed in close proximity to some cell-edge users. 
In this case, the direct links are relatively weak, and therefore an RIS may improve the end-to-end system performance.
Accordingly, we assume that $K=8$ users are evenly distributed on a semicircle centered at the RIS and of  radius $d_{UI}=20$ m. The distance between the RIS and the BS is $d_{IB}=700$ m. 
The distance between the user $k$ and the BS is obtained from the network topology, i.e., $\left(d_{k}^{\mathrm{UB}}\right)^{2}=\left(d_{\mathrm{IB}}-d_{\mathrm{UI}} \cos \left(\frac{\pi}{9} k\right)\right)^{2}+\left(d_{\mathrm{UI}} \sin \left(\frac{\pi}{9} k\right)\right)^{2}$. 
The path-loss exponent of the direct links is larger than the path-loss exponent of the RIS-assisted links in order to characterize the more severe signal attenuation due to the presence of blocking objects on the ground. Specifically, we set the distance-dependent path-loss factors equal to $ \alpha_{k}=10^{-3} d_{\mathrm{UI}}^{-2} $, $ \beta=10^{-3} d_{\mathrm{IB}}^{-2.5}  $ and $\gamma_{k}=10^{-3}\left(d_{k}^{\mathrm{UB}}\right)^{-4}, \forall k$. The number of symbols in each channel coherence time interval is $\tau_{c}=196$\cite{zhang2014power,ngo2013energy}, and $\tau=K=8$ symbols are utilized for channel estimation. The noise power is $\sigma^2=-104$ dBm (corresponding to a noise spectral density equal to $ -174 $ dBm/Hz over a bandwidth of 10 MHz).
The other simulation parameters (unless stated otherwise) are listed in Table \ref{tab1}.

\begin{table}[t]
		\renewcommand\arraystretch{0.6}
	\centering 
	\captionsetup{font={small}}
	\caption{Simulation parameters.}	
	\vspace{-5pt}
	\begin{tabular}{|c|c|c|c|c|c|c}
		\hline
		$\left( {\varphi _t^a,\varphi _t^e} \right)$ &$ (4.17, 0.09) $  &$ \left( {\phi _r^a,\phi _r^e} \right)$ &$ ( 6.28,4.21 ) $  \\
		\hline
		$ \left( {\varphi _{1r}^a,\varphi _{1r}^e} \right)$ &$ ( 5.20,4.32 )  $ & $ \left( {\varphi _{2r}^a,\varphi _{2r}^e} \right)$ &$ ( 0.41,2.52 ) $  \\
		\hline
		$ \left( {\varphi _{3r}^a,\varphi _{3r}^e} \right)$ &$ (3.84,1.78 ) $  & $ \left( {\varphi _{4r}^a,\varphi _{4r}^e} \right)$ &$ ( 1.35,4.15 )   $\\
		\hline
		
		$ \left( {\varphi _{5r}^a,\varphi _{5r}^e} \right)$ &$ ( 5.08,5.76 )  $ & $ \left( {\varphi _{6r}^a,\varphi _{6r}^e} \right)$ &$ ( 4.75,1.56 ) $  \\
		\hline
		$ \left( {\varphi _{7r}^a,\varphi _{7r}^e} \right)$ &$ (4.74,5.36 ) $  & $ \left( {\varphi _{8r}^a,\varphi _{8r}^e} \right)$ &$ ( 0.09,1.40 )   $\\
		\hline
		BS antennas & $M=64$ & RIS elements &	$N=64$\\	
		\hline
		Transmit power & $p=30$ dBm  & Antenna spacing& $d_{bs}=\lambda/2$\\
		\hline
		Rician factors &   {$\delta=1$, $\varepsilon_{k}=10,\forall k$}&Approximation factor & {$\mu=100$}\\
\hline
	\end{tabular}\label{tab1}
\end{table}

	\subsection{Spatial-independent Channels in the Absence of  EMI}
	We first validate the obtained analytical results by assuming that the channels are spatially independent and   the EMI is not present. This help us obtain  initial but useful insights on the performance offered by RIS-aided systems thanks to the simpler analytical expressions of the rate and the explicit analytical insights obtained in Section \ref{section_4}. Specifically, the analytical results are obtained by using Theorem \ref{theorem2} and related corollaries. The Monte Carlo simulations, which are referred to as ``Simulation'' in the legends of the figures, are obtained from (\ref{defination_rate}) by averaging over $10^5$ random channel realizations. The phase shifts are obtained by solving Problem (\ref{Problem4}) with respect to $f(\boldsymbol{\theta})$.

\subsubsection{Quality of the LMMSE Channel Estimation}

\begin{figure}[t]
	\setlength{\abovecaptionskip}{0pt}
	\setlength{\belowcaptionskip}{-20pt}
	\centering
	\includegraphics[width= 0.5\textwidth]{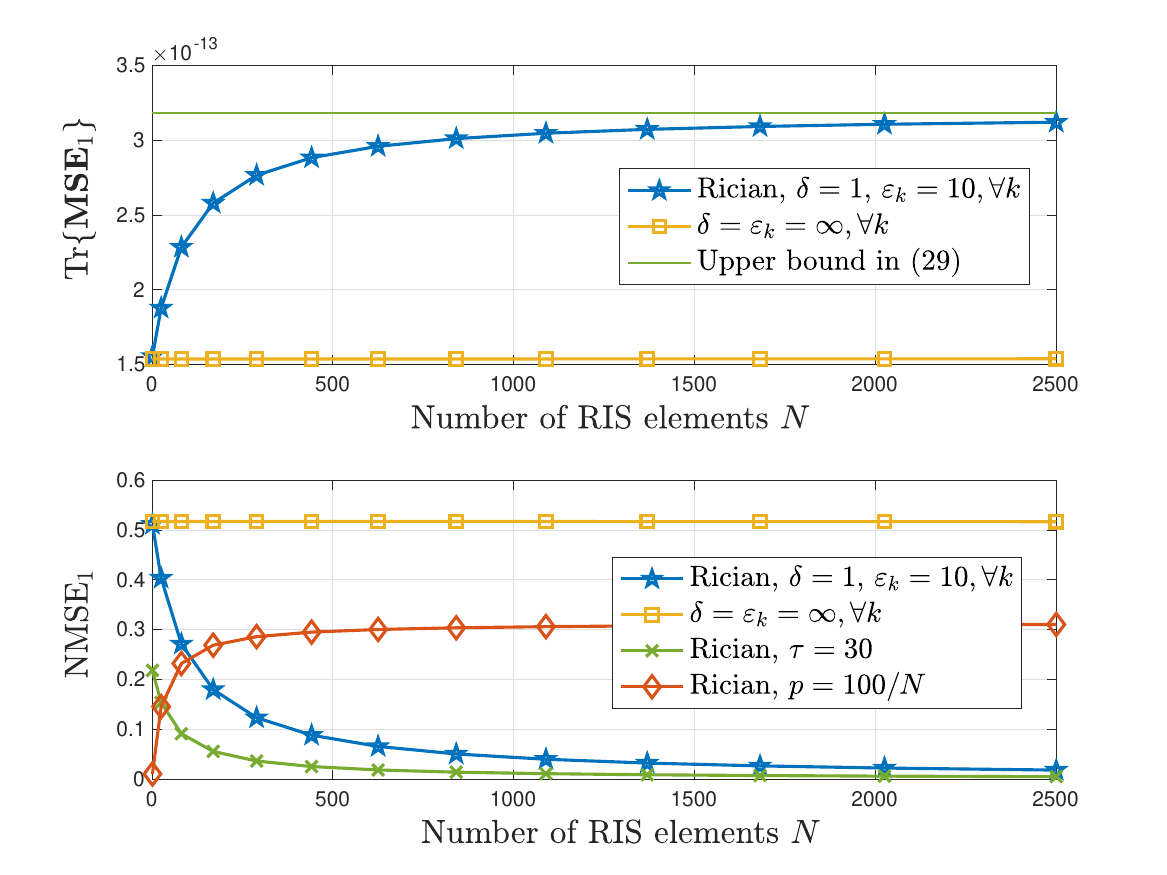}
	\DeclareGraphicsExtensions.
	\captionsetup{font={small}}
	\caption{MSE and NMSE of user $1$ versus the number of RIS elements.}
	\label{figure2}
\end{figure}

To begin with, we investigate the MSE and NMSE  of the proposed channel estimation scheme. The MSE and NMSE of the channel estimation algorithm of the $k$-th user are characterized through the functions $\mathrm{Tr}\left\{\mathbf{MSE}_k\right\}$ and $\mathrm{NMSE}_k$, respectively. 
Without loss of generality, Fig. \ref{figure2} illustrates the MSE and NMSE of user $1$ versus the number of RIS elements $N$.
In general Rician channels, we observe that the MSE is an increasing function of $N$ while the NMSE is a decreasing function of $N$, which is consistent with Corollaries \ref{corollary1}, \ref{corollary2} and \ref{corollary3}. This is because the number of communication paths increases with $N$, but the pilot length $\tau$ does not increase correspondingly, which increases the estimation error. However, the intensity of the channel gains increases with $N$, which, in turn, decreases the normalized errors. In purely LoS RIS-assisted channels ($\delta=\varepsilon_k\to\infty$), the MSE and NMSE are, on the other hand, independent of $N$. This is because  LoS channels are deterministic, and therefore do not introduce additional estimation errors. 
Also, we see that the MSE tends to an upper bound but the NMSE tends to zero when $N\to\infty$, which validates Corollary \ref{corollary1} and \ref{corollary2}. 
By increasing the length of the pilot signals from $8$ to $30$, we see that the NMSE decreases. However, the NMSE that is obtained for $\tau=30$ can also be obtained for $\tau=8$ but by using a larger value for $N$. This validates our remark that increasing the RIS elements can play a similar role as increasing $\tau$.
Finally, we see that the NMSE tends to a limit less than $1$ when the transmit power is scaled proportionally to $p=100/N$, as $N\to\infty$. This validates the correctness of (\ref{NMSE_scale4}).

\subsubsection{Single-user Case}
Next, we evaluate the ergodic achievable rate in the single-user scenario, where only user $1$ is present. 

\begin{figure*}
	\setlength{\abovecaptionskip}{-5pt}
	\setlength{\belowcaptionskip}{-15pt}
	\centering
	\begin{minipage}[t]{0.49\linewidth}
	\centering
	\includegraphics[width= 1\textwidth]{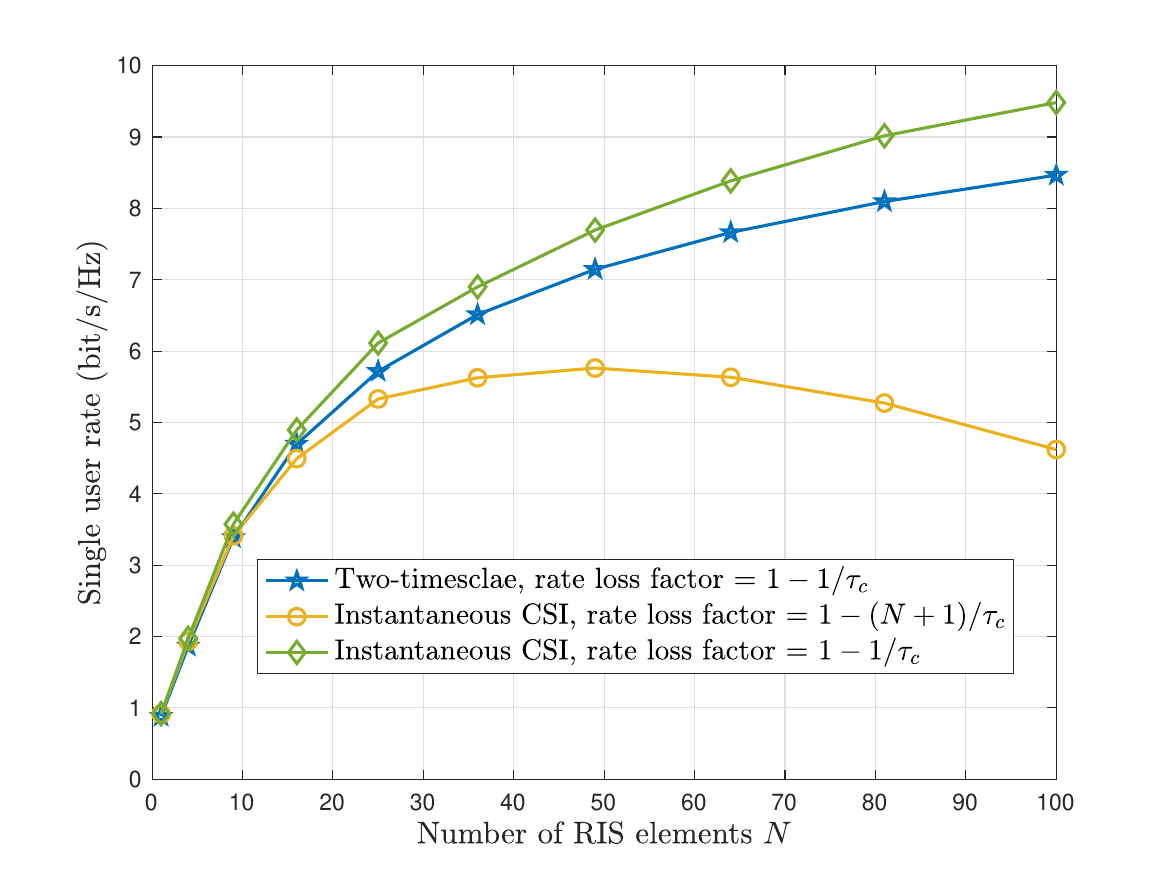}
	\DeclareGraphicsExtensions.
	\captionsetup{font={small}}
	\caption{Comparison of the two-timescale design\\ and instantaneous CSI-based design. }
	\label{figure2.5}
\end{minipage}
	\begin{minipage}[t]{0.49\linewidth}
		\centering
	\includegraphics[width= 1\textwidth]{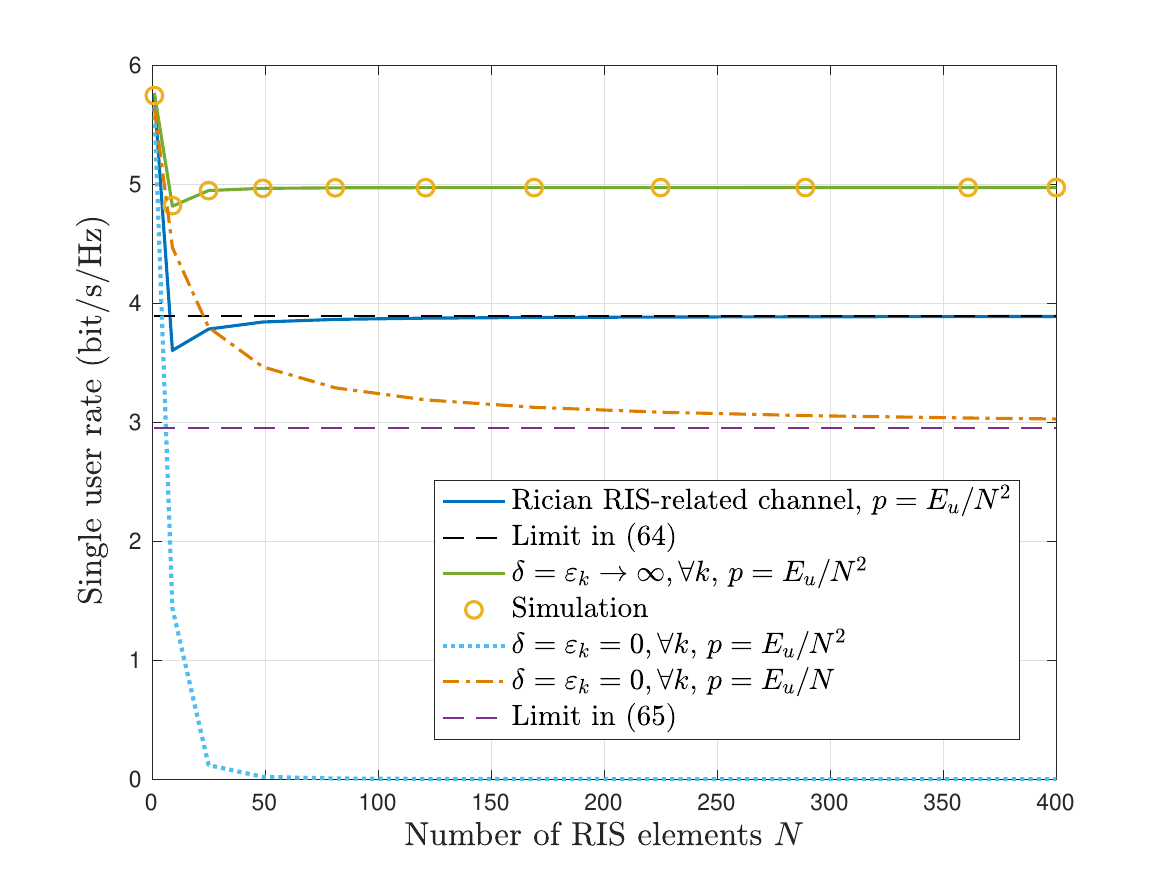}
	\DeclareGraphicsExtensions.
	\captionsetup{font={small}}
	\caption{ Rate versus $N$ in a single-user system. The transmit power is scaled as $p=E_u/N^2$ or $p=E_u/N$, where $E_u=20$ dB.}
	\label{figure3}
	\end{minipage}
	\vspace{-10pt}
\end{figure*}

In Fig. \ref{figure2.5}, we compare the proposed two-timescale scheme with the conventional instantaneous CSI-based scheme. The detailed implementation of the instantaneous CSI-based scheme is presented in Appendix \ref{appendix6}. By assuming the same rate loss factor (ideal but not achievable), it is seen that the instantaneous CSI-based scheme outperforms the proposed two-timescale scheme, especially when $N$ is large. This is because the LoS and NLoS channel components are both exploited in the instantaneous CSI-based RIS design. By contrast, the fast-fading NLoS channel information is averaged out in the proposed statistical CSI-based RIS design. When considering the actual channel estimation overhead, however, the proposed scheme outperforms the instantaneous CSI-based scheme. This is because the instantaneous CSI-based scheme requires a longer pilot length, which is proportional to $N$, even though it results in a higher SNR. When $N$ is large, the instantaneous CSI-based scheme needs a large number of time slots to transmit the pilot sequence, and then only a few symbols are left for data transmission. As a result of the high estimation overhead, the instantaneous CSI-based scheme incurs in a rate loss, which leads to a severe decrease of the rate in the large $N$ regime. Therefore, Fig. \ref{figure2.5} validates the effectiveness of the proposed two-timescale scheme.


In Fig. \ref{figure3}, we illustrate the power scaling law as a function of $N$ in a single-user scenario. In agreement with Corollary \ref{corollary_single_user_MNN}, the rate converges to a limit if we reduce the power proportionally to $1/N^2$ in Rician fading channels. Also, the limit is maximized in LoS-only RIS-assisted channels ($\delta=\varepsilon_k\to\infty$). In NLoS-only RIS-assisted channels ($\delta=\varepsilon_{k}=0$), scaling the power proportionally to $1/N^2$ reduces the rate to zero.
As proved in Corollary \ref{corollary_single_user_rayleigh}, in NLoS-only RIS-assisted channels, the power can only be scaled proportionally to $1/N$ for maintaining a non-zero rate.
These observations highlight that LoS environments are preferable for the deployment of RIS-aided single-user systems.

\subsubsection{Multi-user Case}
In Figs. \ref{figure8}-\ref{figure7}, we evaluate the performance of RIS-aided systems in the general multi-user scenario. 


\begin{figure*}
	\setlength{\abovecaptionskip}{-5pt}
	\setlength{\belowcaptionskip}{-15pt}
	\centering
	\begin{minipage}[t]{0.49\linewidth}
		\centering
	\includegraphics[width= 1\textwidth]{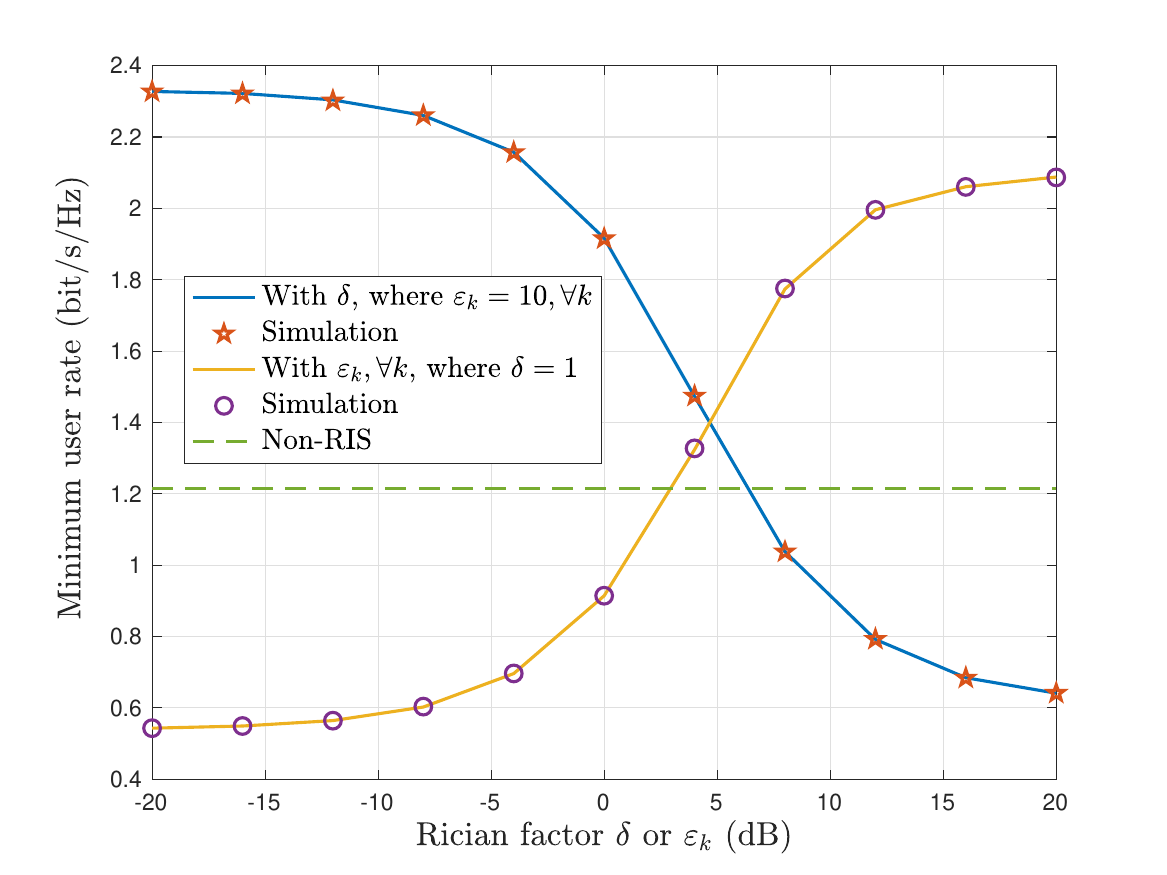}
\DeclareGraphicsExtensions.
\captionsetup{font={small}}
\caption{Minimum user rate versus the Rician factor $\delta$ or $\varepsilon_{k},\forall k$.}
\label{figure8}
	\end{minipage}%
	\begin{minipage}[t]{0.49\linewidth}
		\centering
	\includegraphics[width= 1\textwidth]{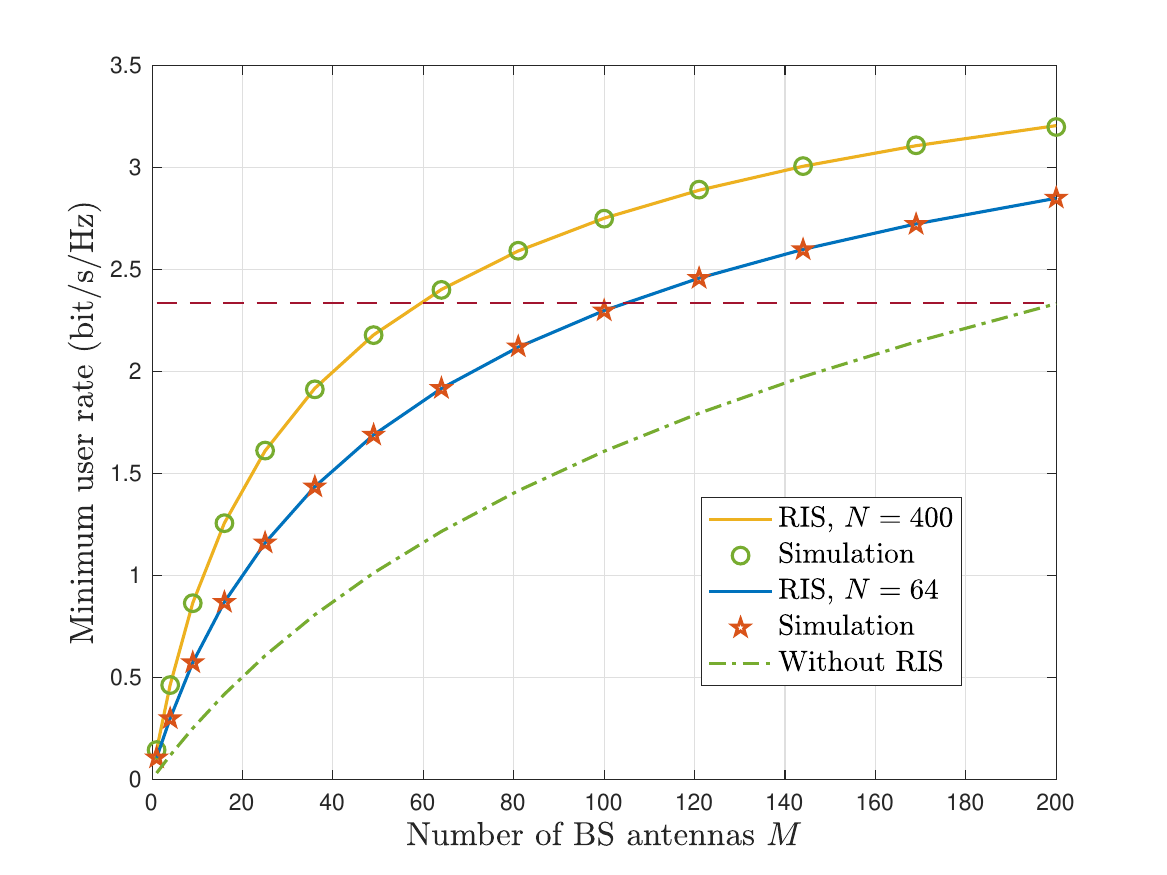}
\DeclareGraphicsExtensions.
\captionsetup{font={small}}
\caption{Minimum user rate versus $M$.}
\label{figure4_1}
	\end{minipage}
	\vspace{-10pt}
\end{figure*}

	Fig. \ref{figure8} shows the impact of the Rician factors. It can be observed that the achievable rate is a decreasing function of $\delta$ but an increasing function of $\varepsilon_{k},\forall k$. 
This is because the rank of the LoS component $\overline{\bf H}_2$ between the RIS and the BS is $ 1 $, while the rank of the LoS component $\overline{\mathbf{H}}_1$ between the users and the RIS is not. 
When $\delta\to\infty$, the rank of the RIS-BS channel tends to $1$, which leads to a rank-$1$ cascaded user-RIS-BS channel. As a result, the RIS-assisted channel becomes rank-deficient, which cannot effectively sustain the transmission of multiple users simultaneously. It is known that the RIS should be deployed either near the BS or near the users so that the product pathloss effect is mitigated\cite{wu2020survey}.  In addition, Fig. \ref{figure8} provides some suggestions with respect to the spatial diversity gain provided by the deployment of an RIS.  To increase $\varepsilon_{k}$, it is beneficial to install the RIS  at a certain height with respect to the ground, which results in increasing the strength of the LoS components of the RIS-user channels. Besides, it is necessary to guarantee a high-rank RIS-BS channel. This condition holds for small values of $\delta$ under the considered Rician fading model.  Since small values of $\delta$ are typically obtained when the RIS is deployed far away from the BS, it is still a good choice to place the RIS near the users after taking into consideration the impact of spatial diversity. On the contrary, if the RIS is deployed near the BS, $\delta$ could be large and the BS-RIS channel could be rank-deficiency under the considered Rician fading model. In this case, possible options for increasing the rank of the channel may be the deployment of artificial scatterers between the BS and the RIS or placing the RIS very close to the BS so that the spherical wave model is valid\cite{kammoun2020asymptotic}.


In Fig. \ref{figure4_1}, we evaluate the rate as a function of the number of BS antennas. The figure illustrates the impact of deploying an RIS in conventional massive MIMO systems. 
It is observed that the deployment of an RIS effectively improves the rate, and the improvement increases with the number of RIS elements. 
It is worth nothing that this performance gain is obtained by using a simple MRC receiver at the BS, and that the LMMSE channel estimator requires the same amount of overhead as conventional massive MIMO systems.
With the help of an RIS, we can achieve the same rate as conventional massive MIMO systems, but with a much smaller number of BS antennas. 
In particular, the rate obtained by a $ 200 $-antenna BS in conventional massive MIMO systems can be obtained by a $ 100 $-antenna BS in RIS-aided massive MIMO systems with $N=64$ RIS elements. 
The number of BS antennas can be further decreased to $M=64$ if the number of RIS elements is increased to $N=400$. Since the cost and energy consumption of one RIS element is much lower than that of one BS antenna, we conclude that the integration of RISs in conventional massive MIMO systems is a promising and cost-effective solution for future wireless communication systems.

\begin{figure*}
	\setlength{\abovecaptionskip}{-5pt}
	\setlength{\belowcaptionskip}{-15pt}
	\centering
	\begin{minipage}[t]{0.49\linewidth}
		\centering
	\includegraphics[width= 1\textwidth]{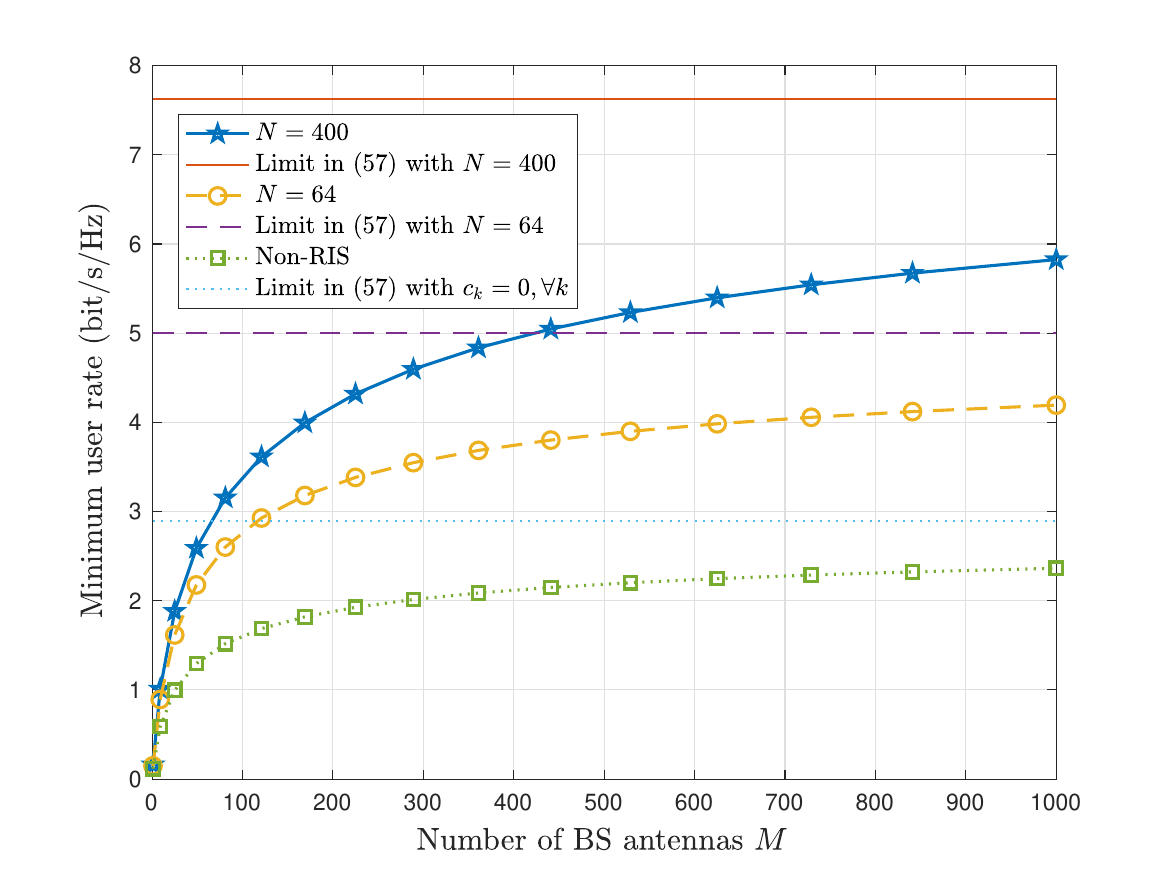}
\DeclareGraphicsExtensions.
\captionsetup{font={small}}
\caption{Minimum user rate versus $M$ when $\delta=0$. \\The transmit power is scaled as $p=E_u/\sqrt{M}$, \\where $E_u=10$ dB.}
\label{figure6}
	\end{minipage}%
	\begin{minipage}[t]{0.49\linewidth}
		\centering
	\includegraphics[width= 1\textwidth]{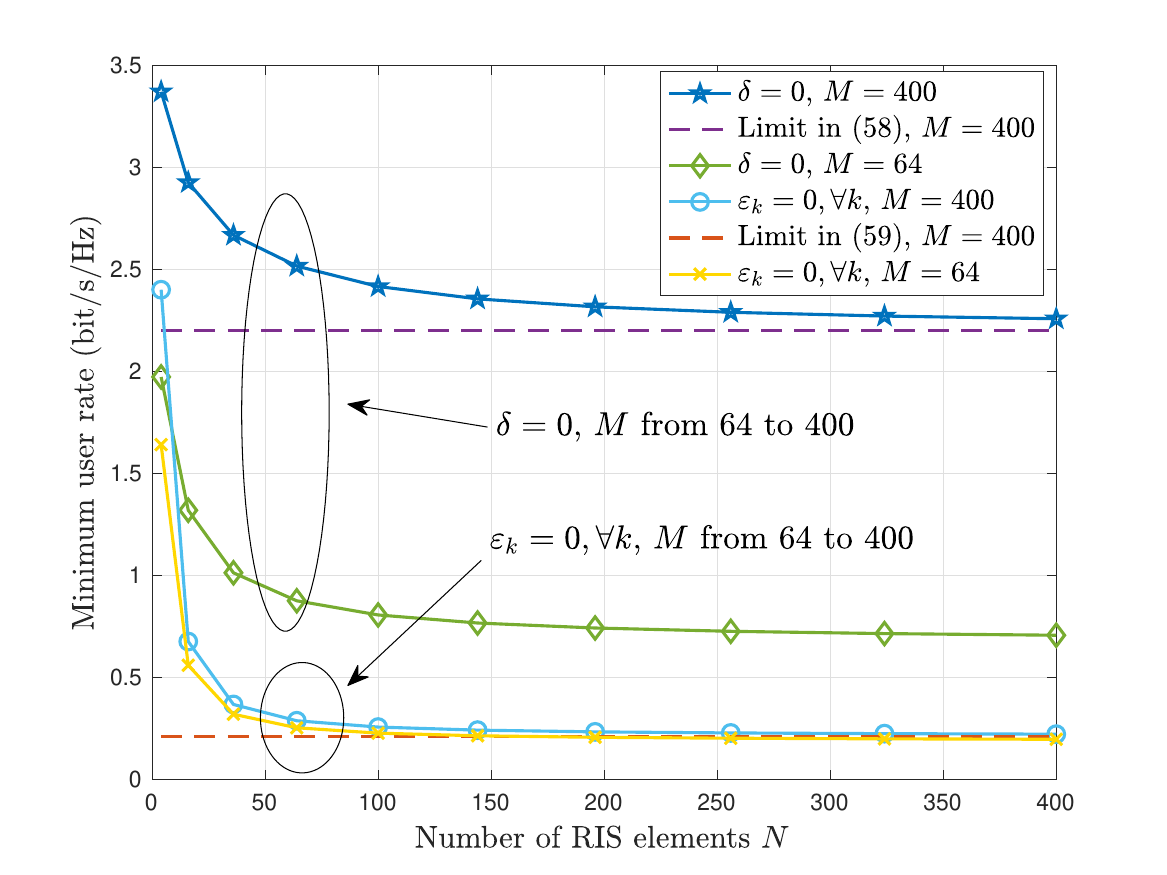}
\DeclareGraphicsExtensions.
\captionsetup{font={small}}
\caption{Minimum user rate versus $N$ when $\delta=0$ or $\varepsilon_{k}=0$. The transmit power is scaled as $p=E_u/N$, where $E_u=10$ dB.}
\label{figure7}
	\end{minipage}
	\vspace{-10pt}
\end{figure*}

In Fig. \ref{figure6} and Fig. \ref{figure7}, finally, we investigate the power scaling law over a purely NLoS RIS-BS channel ($\delta=0$) and a purely NLoS user-RIS channels ($\varepsilon_{k}=0, \forall k$). In Fig. \ref{figure6}, the transmit power is scaled proportionally to $1/\sqrt{M}$ for the NLoS RIS-BS channel ($\delta=0$). 
In agreement with Corollary \ref{corollary6_NLoS_sqrtM}, if $\delta=0$, the rate can be maintained to a non-zero value when the power is scaled proportionally to $1/\sqrt{M}$ as $M\to\infty$. Compared with conventional massive MIMO systems, the deployment of an RIS effectively improves the asymptotic limit when $M\to\infty$, and the rate gain could be further improved by increasing $N$.

In Fig. \ref{figure7}, the transmit power is scaled proportionally to $1/N$ over a purely NLoS RIS-BS channel ($\delta=0$) or purely NLoS user-RIS channels ($\varepsilon_{k}=0, \forall k$). For $N\to\infty$, the rate maintains a non-zero value, which is consistent with Corollaries \ref{corollary7_rate_NLoS_scaling_N} and \ref{corollary8_rate_NLoS_varepusilon_scaling_N}. Besides, in agreement with Corollary \ref{corollary7_rate_NLoS_scaling_N}, the asymptotic limit for $\delta=0$ when $N \to \infty$ can be significantly improved by increasing the number of BS antennas from $M=64$ to $M=400$. 
This is because the RIS-BS channel has a high rank if $\delta=0$, which decreases the spatial correlation among the users and mitigates the multi-user interference.
Furthermore, in agreement with Corollary \ref{corollary8_rate_NLoS_varepusilon_scaling_N}, the asymptotic limit for $\varepsilon_{k}=0, \forall k$ when $N\to\infty$ only marginally increases when increasing $M$ from $64$ to $400$. This observation confirms once again that guaranteeing the spatial diversity between the RIS and the BS could offer a good rate  in RIS-aided massive MIMO systems.

\subsection{Spatial-correlated Channels in the Presence of EMI}
The results illustrated in Figs. \ref{figure2}-\ref{figure7} have showcased the gain of RIS over spatially independent channels and in the absence of  EMI. 
In this section, {\color{blue}some numerical examples are presented to explore} the impact of spatial correlation and EMI and study under what conditions the spatial correlation and the EMI can be ignored as a function of the inter-distance between the RIS elements and the strength of the EMI. Specifically, the strength of EMI with respect to the thermal noise at the BS is characterized by the following ratio\cite{torres2022intelligent}
\begin{align}
\rho = \frac{\sigma_{e}^2}{\sigma^2}.
\end{align}

\begin{figure*}
	\setlength{\abovecaptionskip}{-5pt}
	\setlength{\belowcaptionskip}{-15pt}
	\centering
	\begin{minipage}[t]{0.49\linewidth}
	\centering
\includegraphics[width= 1\textwidth]{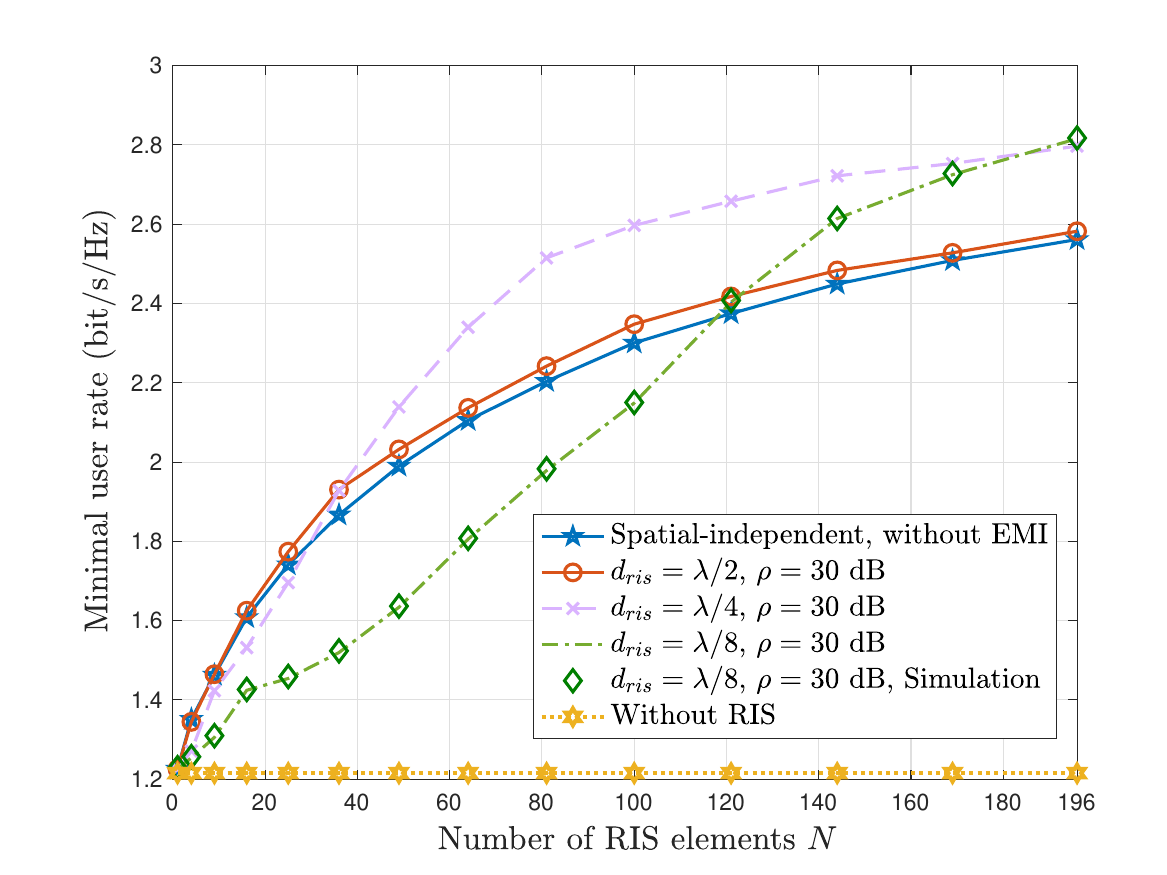}
\DeclareGraphicsExtensions.
\captionsetup{font={small}}
\caption{ Achievable rate versus $N$ for different values of the RIS element spacing $d_{ris}$.}
\label{figure11}
	\end{minipage}%
	\begin{minipage}[t]{0.49\linewidth}
	\centering
\includegraphics[width= 1\textwidth]{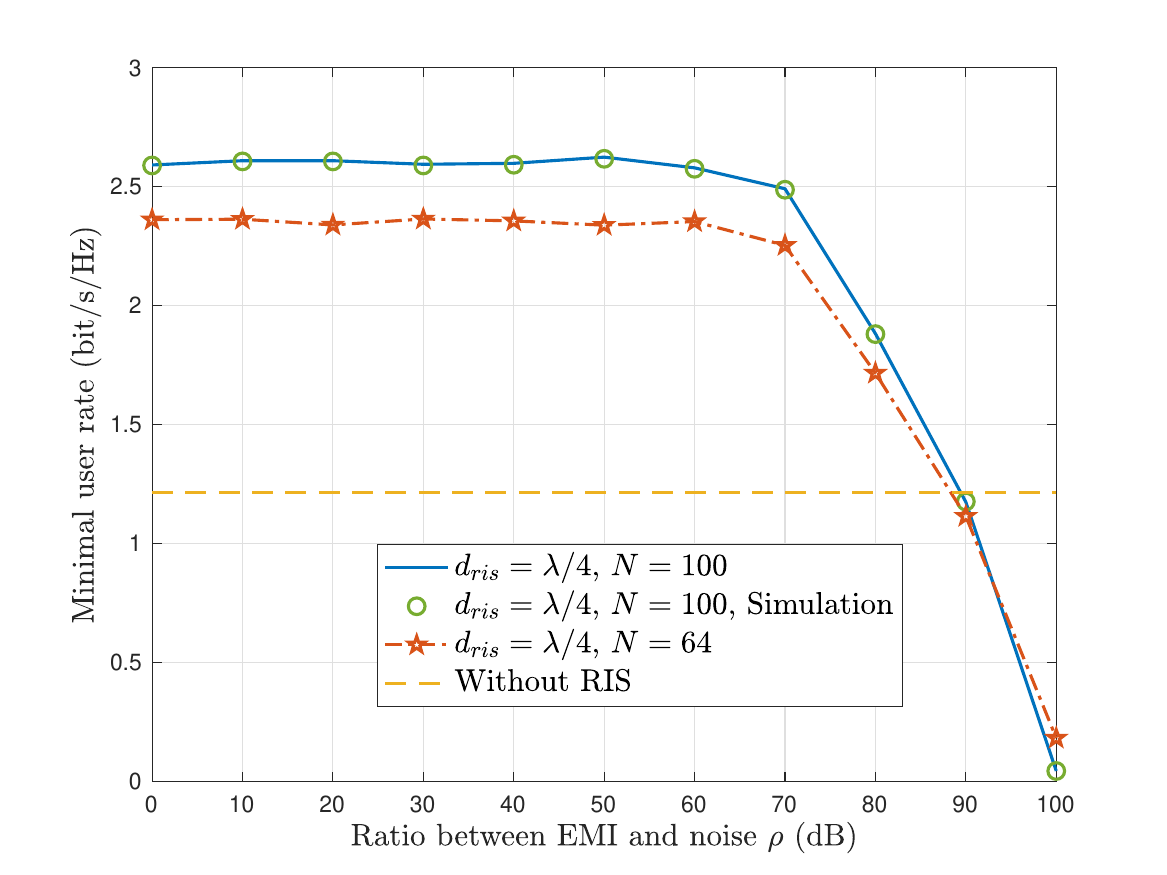}
\DeclareGraphicsExtensions.
\captionsetup{font={small}}
\caption{ Impact of the EMI.}
\label{figure12}
	\end{minipage}
	\vspace{-10pt}
\end{figure*}

Fig. \ref{figure11} illustrates the impact of channel spatial correlation, which is due to the sub-wavelength spacing between the RIS elements, on the achievable rate. In this context, the objective function of the optimization problem (\ref{Problem2}) is $\underline{R}_{c,k}\left({\bf\Phi}\right) $ where the  impact of spatial correlation is taken into account in the design of the RIS phase shifts.
First, as expected, we see that the impact of spatial correlation can be safely ignored when the inter-distance between the RIS elements is half of the wavelength ($d_{ris}=\lambda/2$) and the EMI is light ($\rho=30$ dB). This confirms that the analytical insights drawn in Section \ref{section_4} over spatially independent channels and in the absence of EMI are meaningful to understand the fundamental performance limits of RIS-aided systems in practically relevant scenarios. As the spacing between the RIS elements decreases ($d_{ris}=\lambda/4$, $\lambda/8$), however, the spatial correlation cannot be ignored and it has a non-negligible impact on the rate.  Specifically, we identify two operating regions: (i) small values of  RIS elements $N$ and (ii) large values of RIS elements $N$. For small values of $N$, the rate decreases as the inter-distance decreases. This is attributed to the decrease of the channel rank. For large values of $N$, the channel rank still decreases but we can leverage the large number of RIS elements and the greater ability of an RIS to customize the wireless channels in the presence of channel correlation, as discussed in Theorem \ref{theorem_mmse} and Remark \ref{remark3}. For large values of $N$, the beamforming gains provided by optimizing RIS outweigh the negative impact of spatial correlation, which in turn results in a better achievable rate. 


The impact of EMI is studied in Fig. \ref{figure12}. When the power of the EMI is sufficiently small with respect to the noise ($\rho<60$ dB), the impact of EMI on the achievable rate is negligible. This is attributed to the strong multi-user interference when using MRC. As a result, when the EMI is mild, its impact is negligible as compared with the multi-user interference. As $\rho$ increases, the EMI becomes more severe, and it eventually becomes the dominant contribution. For large values of the EMI,  RIS-aided systems may even perform worse than conventional massive MIMO systems.

\begin{figure*}
	\setlength{\abovecaptionskip}{-5pt}
	\setlength{\belowcaptionskip}{-15pt}
	\centering
	\begin{minipage}[t]{0.49\linewidth}
		\centering
		\includegraphics[width= 1\textwidth]{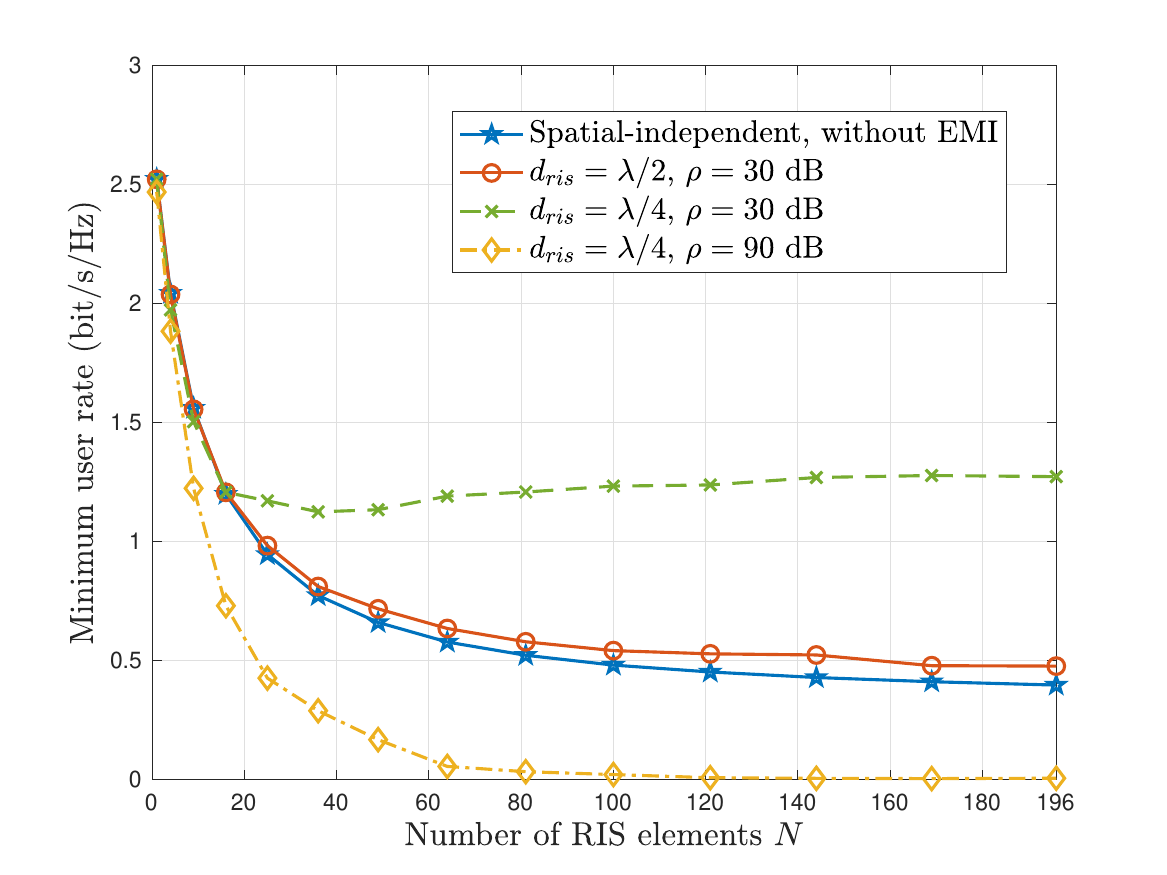}
		\DeclareGraphicsExtensions.
		\captionsetup{font={small}}
			\caption{  Achievable rate when the power is \\
				scaled  proportionally to $p=10/N$.}
		\label{figure13}
	\end{minipage}%
	\begin{minipage}[t]{0.49\linewidth}
	\centering
	\includegraphics[width= 1\textwidth]{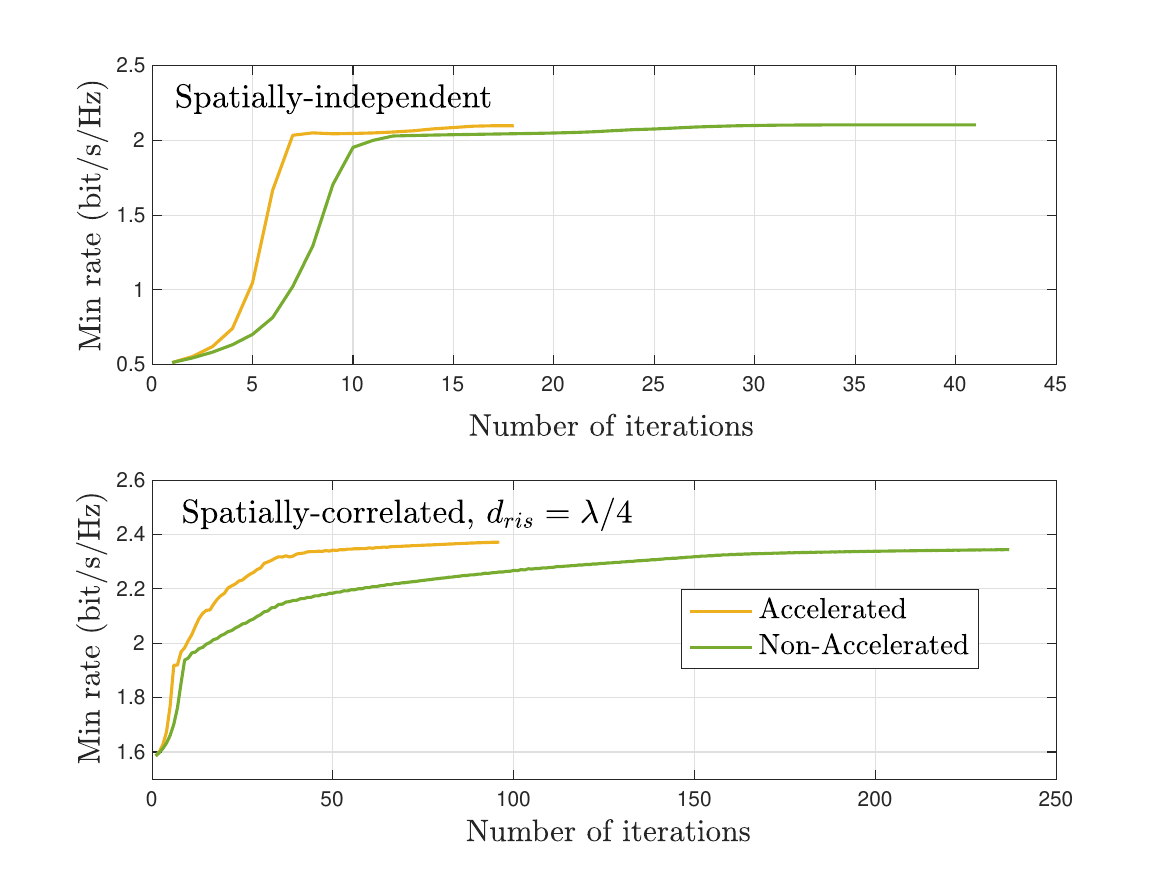}
	\DeclareGraphicsExtensions.
	\captionsetup{font={small}}
	\caption{\color{blue}Convergence behavior for spatially-independent and spatially-correlated cases, $M=N=64$.}
	\label{figure15}
\end{minipage}%
	\vspace{-10pt}
\end{figure*}
Fig. \ref{figure13} illustrates the power scaling laws as a function of the channel spatial correlation and EMI. Specifically, Fig.  \ref{figure13} shows the achievable rate when the power is scaled as $p = 10/N$. The figures validate Corollary \ref{corollary11}: if the EMI is mild, the power scaling law as a function of the transmit power is confirmed. On the other hand, it does not hold anymore in the presence of strong EMI. As a function of the inter-distance $d_{ris}$, Fig. \ref{corollary11} is in agreement with Fig. \ref{figure11}.

{\color{blue}In Fig. \ref{figure15}, we study the convergence behavior of the proposed accelerated gradient method compared with its non-accelerated counterpart. By applying the proposed acceleration method, it can be observed that the speed of convergence is effectively improved. In spatially-independent cases, the algorithm converges very quickly due to the  simple expression of the achievable rate. By contrast, when considering spatial correlation of $d_{ris}=\lambda/4$, the expression becomes more complex and the optimization variable $\bf \Phi$ appears more frequently, as discussed in Remark \ref{remark3}. As a result, the number of iterations needed for convergence increases. Nevertheless, it can  be observed that the accelerated gradient algorithm converges within $100$ iterations even though the number of optimization variables is $64$.}
 
\begin{figure*}
	\setlength{\abovecaptionskip}{-5pt}
	\setlength{\belowcaptionskip}{-15pt}
	\centering
	\begin{minipage}[t]{0.49\linewidth}
		\centering
		\includegraphics[width= 1\textwidth]{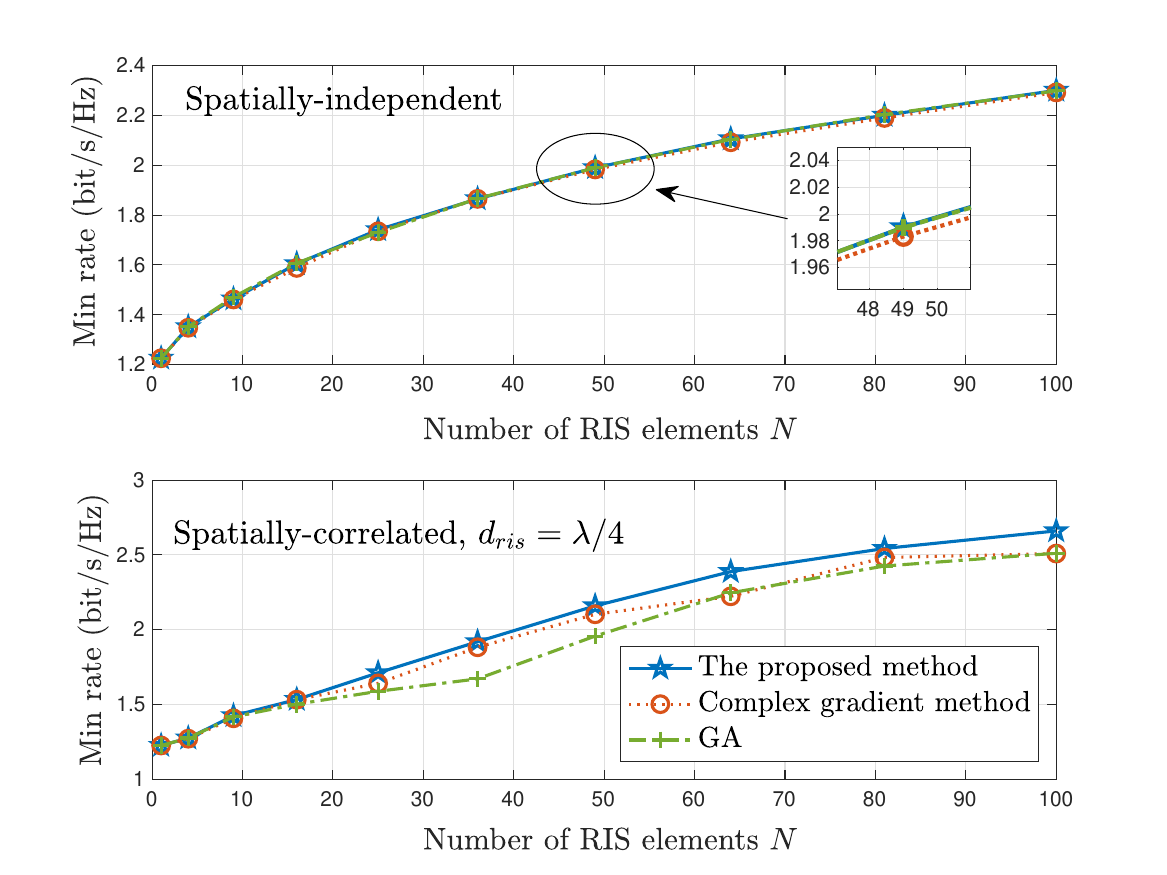}
		\DeclareGraphicsExtensions.
		\captionsetup{font={small}}
		\caption{\color{blue}Performance comparison between \\different optimization algorithms.}
		\label{figure16}
	\end{minipage}
	\begin{minipage}[t]{0.49\linewidth}
	\centering
	\includegraphics[width= 1\textwidth]{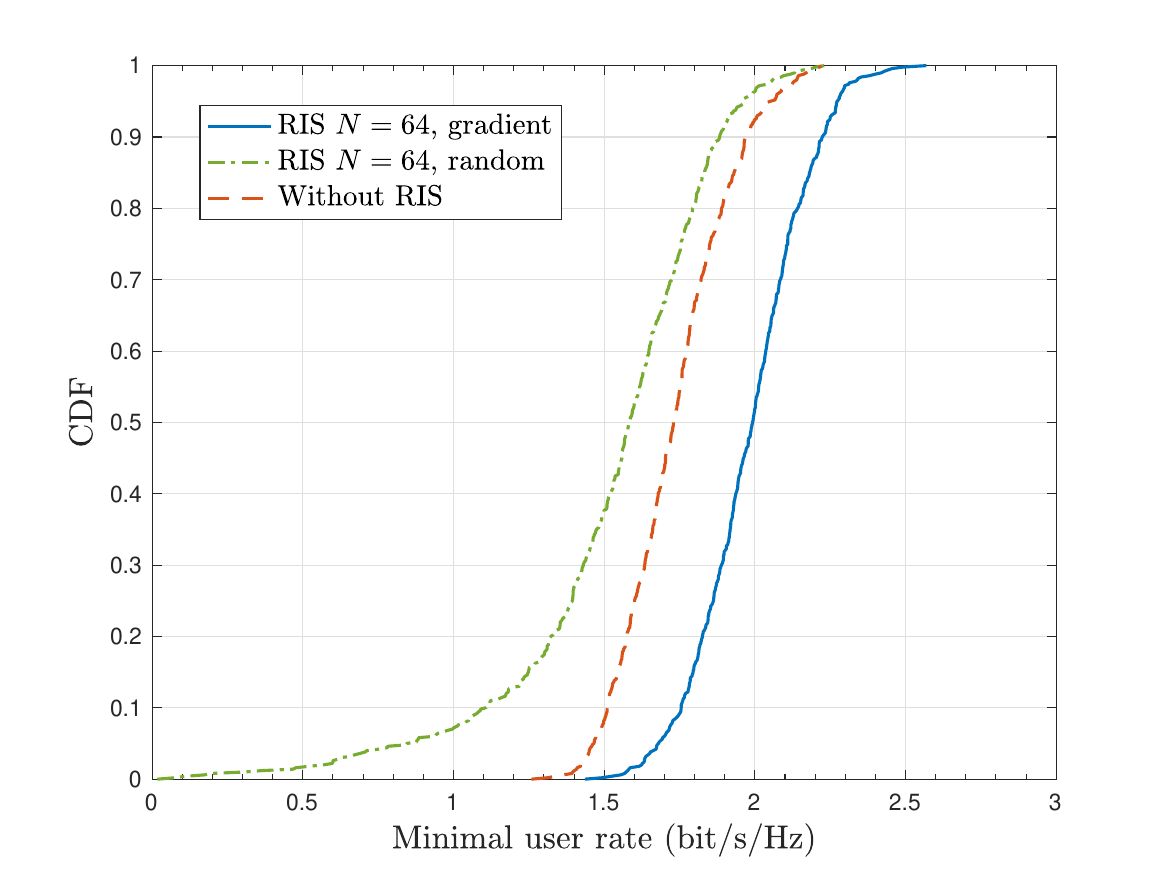}
	\DeclareGraphicsExtensions.
	\captionsetup{font={small}}
	\caption{CDF of the minimal user rate for RIS-aided and RIS-free systems, $d_{ris}=\lambda/4$, $\rho=30$ dB, $p=20$ dBm.}
	\label{figure14}
\end{minipage}
	\vspace{-10pt}
\end{figure*}

{\color{blue} Fig. \ref{figure16} compares the performance of the proposed method with two benchmark algorithms, i.e., the genetic algorithm (GA) \cite{zhi2020power}  and the gradient ascent method formulated in terms of the complex variables $ \boldsymbol{c} = e^ {j \boldsymbol{\theta}} $ \cite{kammoun2020asymptotic}. In spatially-independent cases, it can be observed that the three   algorithms provide almost  the same performance. This is because the objective function possesses a simple and tractable form.  Nevertheless, the proposed algorithm performs  slightly better than the gradient ascent method applied to complex-valued variables. This is because the proposed method treats the angles as optimization variables and therefore avoids the performance loss due to the projection operation. In the presence of spatial correlation, the objective function of the optimization problem becomes more complex. In this case, it can be seen that the proposed method outperforms the other two methods especially when $N$ is large. 
}

%

%

\subsection{Randomly Distributed Users in a Large Area}

The numerical results illustrated in the previous figures are obtained by assuming that the RIS is deployed near the cell-edge users. In this subsection we examine the case study in which the users are distributed over a large area and the transmit power may not be very high due to the deployment of many BS antennas. We set the transmit power equal to $p=20$ dBm and assume that the users are randomly distributed in a $100$ m $\times$ $100$ m area identified by the coordinates $ (200 \text{ m},0)  $ to $ (300\text{ m},100) $ \cite{demir2021channel}. The BS and the RIS are deployed in $ (0,0) $ and $ (200\text{ m},0) $, respectively. Also, we assume $d_{ris}=\lambda/4$ and $\rho=30$ dB for the spacing between the RIS elements and the EMI, respectively.


In Fig. \ref{figure14}, we illustrate the average rate of RIS-aided systems as a function of $500$ random locations of the user and compare it against the rate provided by conventional massive MIMO systems. We observe that the deployment of an RIS still provides some performance gains, but these are reduced as compared to the optimized deployment of the RIS near the cell-edge users. As expected, in addition, the achievable rate is low if the RIS phase shifts matrix is not optimized with the gradient ascent algorithm, but the phase shifts are randomly set.


\section{Conclusion}\label{section_7}
This paper investigated the two-timescale design for RIS-aided massive MIMO systems by taking into account the impact of channel estimation errors. 
We first considered a spatially-independent channel model in the absence of EMI, and we then extended the study to a spatially-correlated channel model in the presence of EMI.
In both cases, we obtained the LMMSE channel estimator for the user-BS aggregated channels, employed the MRC detector, derived the UatF bound of the achievable rate, and optimized the phase shifts of the RIS based on a gradient ascent method. To gain a better understanding of the performance offered by RIS-aided systems, we unveiled fundamental scaling laws over spatially-independent channel models. We demonstrated that the transmit power can be reduced proportionally to $1/M$, while maintaining a non-zero rate, as $M\to\infty$, over RIS-BS Rician channels. If the RIS-BS channel is Rayleigh distributed, on the other hand, a non-zero rate can be maintained when the power is reduced proportionally to $1/\sqrt{M}$ as $M\to\infty$ or proportionally to $1/N$ as $N\to\infty$. Over spatially-correlated channels and in the presence of EMI, we demonstrated that the presence of spatial correlation is beneficial in terms of shaping the wireless channels. We also found that it is beneficial to place the RIS close to the cell-edge users to compensate for the product path-loss behavior in the far-field region. Finally, we  proved that the scaling laws in the absence of EMI may not be preserved in the presence of EMI, especially if the EMI is strong enough.{\footnote{For brevity, the appendices provide a sketch of the proofs of the main results of the present paper. The interested readers may find the detailed proof in the companion extended version of the present paper \cite{zhi2021two}.}}


\begin{appendices}
	
	\section{Some Useful Results}\label{appendix1}
	\begin{lem}\label{lemma_HWH}
		Consider a matrix ${\bf X}  \in \mathbb{C}^{m\times n}  $, $m,n \ge 1$, whose entries are i.i.d. random variables with zero mean and ${ v}_{ x}$ variance. Consider a deterministic matrix ${\bf W} \in \mathbb{C}^{n\times n}$. Then, we have
		\begin{align}
			\mathbb{E}\left\{\mathbf{X} \mathbf{W} \mathbf{X}^{H}\right\}={v}_{{x}} \operatorname{Tr}\{\mathbf{W}\} \mathbf{I}_{m}.
		\end{align}
	\end{lem}
	
	\itshape {Proof:}  \upshape  Consider the matrix $\mathbf{X} \mathbf{W} \mathbf{X}^{H}$. The expectation of its $(i,j)$-th entry, where $i \neq j$, is given by
	\begin{align}
		\left[\mathbb{E}\left\{\mathbf{X} \mathbf{W} \mathbf{X}^{H}\right\}\right]_{i j}=\mathbb{E}\left\{\sum_{l=1}^{n} \sum_{k=1}^{n} \mathbf{X}_{i k} \mathbf{W}_{k l} [\mathbf{X}^{H}]_{l j}\right\}=\sum_{l=1}^{n} \sum_{k=1}^{n} \mathbb{E}\left\{\mathbf{X}_{i k} \mathbf{X}_{j l}^{*}\right\} \mathbf{W}_{k l}=0.
	\end{align}
	
	Similarly, the expectation of its $(i,i)$-th entry is 
	\begin{align}
		\left[\mathbb{E}\left\{\mathbf{X} \mathbf{W} \mathbf{X}^{H}\right\}\right]_{i i}=\sum_{l=1}^{n} \sum_{k=1}^{n} \mathbb{E}\left\{\mathbf{X}_{i k} \mathbf{X}_{i l}^{*}\right\} \mathbf{W}_{k l}=\sum_{k=1}^{n} \mathbb{E}\left\{\left|\mathbf{X}_{i k}\right|^{2}\right\} \mathbf{W}_{k k}={v}_{{x}} \operatorname{Tr}\{\mathbf{W}\}.
	\end{align}
	
	Therefore, the expectation of $\mathbf{X} \mathbf{W} \mathbf{X}^{H}$ is a diagonal matrix and its diagonal entries are all equal to ${v}_{{x}} \operatorname{Tr}\{\mathbf{W}\}$. This completes the proof. \hfill $\blacksquare$
	
	By letting $m=1$ or $n=1$, corresponding results for random vectors can be obtained.

	\begin{lem}\label{lemma4_HH_equal_0}
		Consider the deterministic matrices ${\bf W} \in \mathbb{C}^{N\times N}$ and vectors ${\bf w}_1,{\bf w}_2 \in \mathbb{C}^{N\times 1}$, and ${\bf w}_3,{\bf w}_4 \in \mathbb{C}^{M\times 1}$. Then, we have
		\begin{align}\label{HWH}
			&\mathbb{E}\left\{\tilde{\mathbf{H}}_{2} \mathbf{W} \tilde{\mathbf{H}}_{2}\right\}=\mathbb{E}\left\{\operatorname{Re}\left\{\tilde{\mathbf{H}}_{2} \mathbf{W} \tilde{\mathbf{H}}_{2}\right\}\right\}=\mathbf{0},\\\label{hW1hW2}
			&\mathbb{E}\left\{   \tilde{\mathbf{h}}_{k}^H \mathbf{w}_1 \tilde{\mathbf{h}}_{k}^H \mathbf{w}_2\right\}=
			\mathbb{E}\left\{\operatorname{Re}\left\{     \tilde{\mathbf{h}}_{k}^H \mathbf{w}_1 \tilde{\mathbf{h}}_{k}^H \mathbf{w}_2       \right\}\right\}=0,\\\label{wdwd}
			&\mathbb{E}\left\{  \mathbf{w}_3^H \tilde{\mathbf{d}}_{k}  \mathbf{w}_4^H \tilde{\mathbf{d}}_{k} \right\}=
			\mathbb{E}\left\{  {\rm Re}\left\{\mathbf{w}_3^H \tilde{\mathbf{d}}_{k}  \mathbf{w}_4^H \tilde{\mathbf{d}}_{k}\right\} \right\}=0.
		\end{align}
	\end{lem}
	
	\itshape {Proof:}  \upshape Let us consider a complex random variable $v=v_r +j v_i $ with $v_r, v_i\sim \mathcal{N}\left(0,1/2\right)$. Noting that for complex random variables, different from the result that $\mathbb{E}\left\{\left|v\right|^{2}\right\}=\mathbb{E}\left\{v_{r}^{2}\right\}+\mathbb{E}\left\{v_{i}^{2}\right\}=1$, we have
	\begin{align}\label{v_square}
		&\mathbb{E}\left\{v^{2}\right\}=\mathbb{E}\left\{v_{r}^{2}-v_{i}^{2}+2 j v_{r} v_{i}\right\}=\mathbb{E}\left\{v_{r}^{2}\right\}-\mathbb{E}\left\{v_{i}^{2}\right\}+2j \mathbb{E}\left\{v_r\right\}\mathbb{E}\left\{v_i\right\}=0,\\
		&\mathbb{E}\left\{\operatorname{Re}\left\{v^{2}\right\}\right\}=\mathbb{E}\left\{v_{r}^{2}\right\}-\mathbb{E}\left\{v_{i}^{2}\right\}=0.
	\end{align}
	
	The entries of $\tilde{\mathbf{H}}_2$ are i.i.d., each having the same distribution as $v$. Then, we have
	\begin{align}\label{HWH_expand}
		\mathbb{E}\left\{\left[\tilde{\mathbf{H}}_{2} \mathbf{W} \tilde{\mathbf{H}}_{2}\right]_{n 1, n 2}\right\}=\mathbb{E}\left\{\sum_{i=1}^{N} \sum_{m=1}^{N}\left[\tilde{\mathbf{H}}_{2}\right]_{n 1, m} \mathbf{W}_{m, i}\left[\tilde{\mathbf{H}}_{2}\right]_{i, n 2}\right\}.
	\end{align}
	
	For $(n1,m)\neq (i,n2)$  in (\ref{HWH_expand}), the expectation is zero, since the entries are independent and zero-mean. For $(n1,m)= (i,n2)$  in (\ref{HWH_expand}), the expectation is zero by using (\ref{v_square}). Therefore, (\ref{HWH}) is proved. Equations (\ref{hW1hW2}) and (\ref{wdwd}) can be proved \textit{mutatis mutandis}.
	\hfill $\blacksquare$

	{\color{blue}
		
		\begin{lem}\label{lemma_HCHWHCH}
			For deterministic matrices $ \mathbf{C}  \in \mathbb{C}^{M\times M}$ and $ \mathbf{W} \in \mathbb{C}^{N\times N}$, if $\mathbf{C} = \mathbf{C}^H$, there is
			\begin{align}\label{HCHWHCH}
				&\mathbb{E}\left\{\tilde{\mathbf{H}}_{2}^{H} \mathbf{C} \tilde{\mathbf{H}}_{2} \mathbf{W} \tilde{\mathbf{H}}_{2}^{H} \mathbf{C} \tilde{\mathbf{H}}_{2}\right\}=\operatorname{Tr}\{\mathbf{W}\} \operatorname{Tr}\left\{\mathbf{C}^{2}\right\} \mathbf{I}_{N}+|\operatorname{Tr}\{\mathbf{C}\}|^{2} \mathbf{W},\\\label{Hc2CHc2WHc2CHc2}
				&\mathbb{E}\left\{\tilde{\mathbf{H}}_{c, 2}^{H} \mathbf{C} \tilde{\mathbf{H}}_{c, 2} \mathbf{W} \tilde{\mathbf{H}}_{c, 2}^{H} \mathbf{C} \tilde{\mathbf{H}}_{c, 2}\right\} = \operatorname{Tr}\left\{\mathbf{R}_{r i s} \mathbf{W}\right\} \operatorname{Tr}\left\{\mathbf{C}^{2}\right\} \mathbf{R}_{r i s}+|\operatorname{Tr}\{\mathbf{C}\}|^{2} \mathbf{R}_{r i s} \mathbf{W R}_{r i s}.
			\end{align}
		\end{lem}
		
		\itshape {Proof:}  \upshape
		Define $\tilde{\mathbf{H}}_{2}=\left[{\bf J}_{1}, \ldots, {\bf J}_{N}\right]$, where ${\bf J}_{n} \in \mathbb{C}^{M \times 1}$, $1\leq n \leq N$, are independent of each other, and ${\bf J}_n \sim \mathcal{CN}\left({\bf 0}, {\bf I}_M\right)$. Denoting $[\mathbf{W}]_{m, n}={w}_{m n}$, then we have
		\begin{align}
			\begin{aligned}
				\left[\tilde{\mathbf{H}}_{2}^{H} \mathbf{C} \tilde{\mathbf{H}}_{2} \mathbf{W} \tilde{\mathbf{H}}_{2}^{H} \mathbf{C} \tilde{\mathbf{H}}_{2}\right]_{i, j} =\sum_{h=1}^{N} \sum_{m=1}^{N} \mathbf{J}_{i}^{H} \mathbf{C} \mathbf{J}_{m} w_{m h} \mathbf{J}_{h}^{H} \mathbf{C} \mathbf{J}_{j}.
			\end{aligned}
		\end{align}
		
		Note that $\mathbb{E}\left\{\left|\mathbf{J}_{i}^{H} \mathbf{C} \mathbf{J}_{i}\right|^{2}\right\}=|\operatorname{Tr}(\mathbf{C})|^{2}+\operatorname{Tr}\left(\mathbf{C}^{2}\right)$ \cite[(35)]{Emil2015circuit}. The expectation of the $i$-th diagonal term can be calculated as
		\begin{align}\label{diag_HCHWHCW}
			\begin{aligned}
				&\mathbb{E}\left\{\left[  \tilde{\mathbf{H}}_{2}^{H} \mathbf{C} \tilde{\mathbf{H}}_{2} \mathbf{W} \tilde{\mathbf{H}}_{2}^{H} \mathbf{C} \tilde{\mathbf{H}}_{2}  \right]_{i, i}\right\} \\
				&=\mathbb{E}\left\{\mathbf{J}_{i}^{H} \mathbf{C} \mathbf{J}_{i} w_{i i} \mathbf{J}_{i}^{H} \mathbf{C} \mathbf{J}_{i}\right\} +\mathbb{E}\left\{\sum_{m=1, m \neq i}^{N} \mathbf{J}_{i}^{H} \mathbf{C}\mathbf{J}_{m} w_{m m} \mathbf{J}_{m}^{H} \mathbf{C} \mathbf{J}_{i}\right\} \\
				&=w_{i i} \mathbb{E}\left\{\left|\mathbf{J}_{i}^{H} \mathbf{C} \mathbf{J}_{i}\right|^{2}\right\}+\mathbb{E}\left\{\sum_{m=1, m \neq i}^{N} w_{m m} \mathbf{J}_{i}^{H} \mathbf{C } \mathbb{E}\left\{\mathbf{J}_{m} \mathbf{J}_{m}^{H}\right\} \mathbf{C} \mathbf{J}_{i}\right\} \\
				&=w_{i i}|\operatorname{Tr}(\mathbf{C})|^{2}+\operatorname{Tr}\{\mathbf{W}\} \operatorname{Tr}\left\{\mathbf{C}^{2}\right\}.
			\end{aligned}
		\end{align}
		
		The expectation of the $(i,j)$-th  non-diagonal term is given by
		\begin{align}\label{non_diag_HCHWHCW}
			\begin{aligned}
				&\mathbb{E}\left\{\left[\tilde{\mathbf{H}}_{2}^{H} \mathbf{C} \tilde{\mathbf{H}}_{2} \mathbf{W} \tilde{\mathbf{H}}_{2}^{H} \mathbf{C}  \tilde{\mathbf{H}}_2\right]_{i, j}\right\}=\mathbb{E}\left\{\mathbf{J}_{i}^{H} \mathbf{C} \mathbf{J}_{i} w_{i j} \mathbf{J}_{j}^{H} \mathbf{C} \mathbf{J}_{j}\right\} \\
				&=w_{i j} \mathbb{E}\left\{\mathbf{J}_{i}^{H} \mathbf{C J}_{i}\right\} \mathbb{E}\left\{\mathbf{J}_{j}^{H} \mathbf{C} \mathbf{J}_{j}\right\} =w_{i j}|\operatorname{Tr}\{\mathbf{C}\}|^{2}.
			\end{aligned}
		\end{align}
		
		Combining (\ref{diag_HCHWHCW}) and (\ref{non_diag_HCHWHCW}) completes the proof of (\ref{HCHWHCH}). Then, we can prove (\ref{Hc2CHc2WHc2CHc2}) by using $\tilde{\mathbf{H}}_{c, 2}=\tilde{\mathbf{H}}_{2} \mathbf{R}_{r i s}^{1 / 2}$.
		\hfill $\blacksquare$

		Applying  Lemma \ref{lemma_HCHWHCH}, we can obtain some useful results as summarized in the following Lemma.
		\begin{lem}\label{lemma_HAHWHAH}
			For deterministic matrix $ \mathbf{W} \in \mathbb{C}^{N\times N}$, we have
			\begin{align}
				&\mathbb{E}\left\{\tilde{\mathbf{H}}_{2}^{H} \mathbf{A}_{k} \tilde{\mathbf{H}}_{2} \mathbf{W} \tilde{\mathbf{H}}_{2}^{H} \mathbf{A}_{k} \tilde{\mathbf{H}}_{2}\right\}=e_{k 1}^{2} M^{2} \mathbf{W}+e_{k 3} M \operatorname{Tr}\{\mathbf{W}\} \mathbf{I}_{N},\\\label{Wishart_H2}
				&\mathbb{E}\left\{  \tilde{\mathbf{H}}_{2}^{H} \tilde{\mathbf{H}}_{2} \mathbf{W} \tilde{\mathbf{H}}_{2}^{H} \tilde{\mathbf{H}}_{2}\right\}=M^{2} \mathbf{W}+M \operatorname{Tr}\{\mathbf{W}\} \mathbf{I}_{N},\\\label{Wishart_HaaH}
				&\mathbb{E}\left\{\tilde{\mathbf{H}}_{2}^{H} \mathbf{a}_{M} \mathbf{a}_{M}^{H} \tilde{\mathbf{H}}_{2} \mathbf{W} \tilde{\mathbf{H}}_{2}^{H} \mathbf{a}_{M} \mathbf{a}_{M}^{H} \tilde{\mathbf{H}}_{2}\right\}=M^{2} \mathbf{W}+M^{2} \operatorname{Tr}\{\mathbf{W}\} \mathbf{I}_{N}.
			\end{align}
			where $\mathbf{A}_k$ is defined in (\ref{Ak}), and $e_{k1}$ and $e_{k3}$ are defined in Lemma \ref{lemma_e1e2e3}. For deterministic matrix $ \mathbf{C}  \in \mathbb{C}^{M\times M}$ and random vector ${\bf u} \sim \mathcal{CN}\left({\bf 0}, {\bf I}_M\right)$, there are
			\begin{align}\label{Wishart_Jn}
				&\mathbb{E}\left\{\mathbf{u} \mathbf{u}^{H} \mathbf{C} \mathbf{u} \mathbf{u}^{H}\right\}=\mathbf{C}+\operatorname{Tr}\{\mathbf{C}\} \mathbf{I}_{M}, \\\label{Wishart_h4}
				&\mathbb{E}\left\{\left\|\mathbf{u} \right\|^{4}\right\}=\operatorname{Tr}\left\{\mathbb{E}\left\{\mathbf{u} \mathbf{u}^{H} \mathbf{u} \mathbf{u}^{H}\right\}\right\}=M^{2}+M.
			\end{align}
		\end{lem}
		
	}

	\section{}\label{appendix2}
	Recalling the definition of ${\bf q}_k$ in (\ref{exact_channel}), where $\tilde{\mathbf{H}}_{2}$, $\tilde{\mathbf{h}}_{k}$, $\tilde{\mathbf{d}}_{k}$, and $\bf N$ are independent of each other and composed of zero-mean entries, we have
	\begin{align}
		&\mathbb{E}\left\{\mathbf{y}_{p}^{k}\right\}=\mathbb{E}\left\{\mathbf{q}_{k}\right\}+\frac{1}{\sqrt{\tau p}} \mathbb{E}\left\{\mathbf{N }\right\} {\bf s}_{k}=\mathbb{E}\left\{\mathbf{q}_{k}\right\}=\sqrt{c_{k} \delta \varepsilon_{k}} \,\overline{\mathbf{H}}_{2} {\bf \Phi} \overline{\mathbf{h}}_{k}.
	\end{align}
	
	The covariance matrix between the unknown channel $\mathbf{q}_{k}$ and the observation vector $\mathbf{y}_{p}^{k}$ can be written as
	\begin{align}
		\begin{array}{l}
			\operatorname{Cov}\left\{\mathbf{q}_{k}, \mathbf{y}_{p}^{k}\right\}=\mathbb{E}\left\{\left(\mathbf{q}_{k}-\mathbb{E}\left\{\mathbf{q}_{k}\right\}\right)\left(\mathbf{y}_{p}^{k}-\mathbb{E}\left\{\mathbf{y}_{p}^{k}\right\}\right)^{H}\right\} \\
			=\mathbb{E}\left\{\left(\mathbf{q}_{k}-\mathbb{E}\left\{\mathbf{q}_{k}\right\}\right)\left(\mathbf{q}_{k}+\frac{1}{\sqrt{\tau p}} \mathbf{N s}_{k}-\mathbb{E}\left\{\mathbf{q}_{k}\right\}\right)^{H}\right\} \\
			=\mathbb{E}\left\{\left(\mathbf{q}_{k}-\mathbb{E}\left\{\mathbf{q}_{k}\right\}\right)\left(\mathbf{q}_{k}-\mathbb{E}\left\{\mathbf{q}_{k}\right\}\right)^{H}\right\} \\
			=\operatorname{Cov}\left\{\mathbf{q}_{k}, \mathbf{q}_{k}\right\},
		\end{array}
	\end{align}
	and
	\begin{align}
		\operatorname{Cov}\left\{\mathbf{y}_{p}^{k}, \mathbf{q}_{k}\right\}=\left(\operatorname{Cov}\left\{\mathbf{q}_{k}, \mathbf{y}_{p}^{k}\right\}\right)^{H}=\left(\operatorname{Cov}\left\{\mathbf{q}_{k}, \mathbf{q}_{k}\right\}\right)^{H}=\operatorname{Cov}\left\{\mathbf{q}_{k}, \mathbf{q}_{k}\right\}.
	\end{align}
	
	Invoking the definition of $\mathbf{q}_{k}$, we obtain
	\begin{align}
		\begin{array}{l}
			\operatorname{Cov}\left\{\mathbf{q}_{k}, \mathbf{q}_{k}\right\} =\mathbb{E}\left\{\left(\mathbf{q}_{k}-\mathbb{E}\left\{\mathbf{q}_{k}\right\}\right)\left(\mathbf{q}_{k}-\mathbb{E}\left\{\mathbf{q}_{k}\right\}\right)^{H}\right\} \\
			=\mathbb{E}\left\{\begin{array}{l}
				\left(\sqrt{c_{k} \delta} \,\overline{\mathbf{H}}_{2} {\bf \Phi} \tilde{\mathbf{h}}_{k}
				+\sqrt{c_{k} \varepsilon_{k}} \,\tilde{\mathbf{H}}_{2} {\bf \Phi} \overline{\mathbf{h}}_{k}
				+\sqrt{c_{k}} \,\tilde{\mathbf{H}}_{2} {\bf \Phi} \tilde{\mathbf{h}}_{k}+\sqrt{\gamma_{k}} \tilde{\mathbf{d}}_{k}\right) \\
				\times\left(\sqrt{c_{k} \delta} \, \tilde{\mathbf{h}}_{k}^{H} \mathbf{\Phi}^{H} \overline{\mathbf{H}}_{2}^{H}
				+\sqrt{c_{k} \varepsilon_{k}} \, \overline{\mathbf{h}}_{k}^{H} \mathbf{\Phi}^{H} \tilde{\mathbf{H}}_{2}^{H}
				+\sqrt{c_{k}} \,\tilde{\mathbf{h}}_{k}^{H} \mathbf{\Phi}^{H} \tilde{\mathbf{H}}_{2}^{H}+\sqrt{\gamma_{k}} \tilde{\mathbf{d}}_{k}^{H}\right)
			\end{array}\right\} \\
			=\mathbb{E}\left\{c_{k} \delta \overline{\mathbf{H}}_{2} {\bf \Phi} \tilde{\mathbf{h}}_{k} \tilde{\mathbf{h}}_{k}^{H} {\bf \Phi}^{H} \overline{\mathbf{H}}_{2}^{H}+c_{k} \varepsilon_{k} \tilde{\mathbf{H}}_{2} {\bf \Phi} \overline{\mathbf{h}}_{k} \overline{\mathbf{h}}_{k}^{H} {\bf \Phi}^{H} \tilde{\mathbf{H}}_{2}^{H}+c_{k} \tilde{\mathbf{H}}_{2} {\bf \Phi} \tilde{\mathbf{h}}_{k} \tilde{\mathbf{h}}_{k}^{H} {\bf \Phi}^{H} \tilde{\mathbf{H}}_{2}^{H}+\gamma_{k} \tilde{\mathbf{d}}_{k} \tilde{\mathbf{d}}_{k}^{H}\right\} \\
			{{\mathop  = \limits^{\left( b \right)} }}   N c_{k} \delta \mathbf{a}_{M} \mathbf{a}_{M}^{H}+\left(N c_{k}\left(\varepsilon_{k}+1\right)+\gamma_{k}\right) \mathbf{I}_{M},
		\end{array}
	\end{align}
	where ($ b $) exploits Lemma \ref{lemma_HWH} and the mutual independence of $\tilde{\mathbf{H}}_{2}$ and $\tilde{\mathbf{h}}_{k}$. 
	
	Similarly, we have
	\begin{align}
		\begin{array}{l}
			\operatorname{Cov}\left\{\mathbf{y}_{p}^{k}, \mathbf{y}_{p}^{k}\right\}=\mathbb{E}\left\{\left(\mathbf{y}_{p}^{k}-\mathbb{E}\left\{\mathbf{y}_{p}^{k}\right\}\right)\left(\mathbf{y}_{p}^{k}-\mathbb{E}\left\{\mathbf{y}_{p}^{k}\right\}\right)^{H}\right\} \\
			=\mathbb{E}\left\{\left(\mathbf{q}_{k}-\mathbb{E}\left\{\mathbf{q}_{k}\right\}+\frac{1}{\sqrt{\tau p}} \mathbf{N s}_{k}\right)\left(\mathbf{q}_{k}-\mathbb{E}\left\{\mathbf{q}_{k}\right\}+\frac{1}{\sqrt{\tau p}} \mathbf{N s}_{k}\right)^{H}\right\} \\
			=\mathbb{E}\left\{\left(\mathbf{q}_{k}-\mathbb{E}\left\{\mathbf{q}_{k}\right\}\right)\left(\mathbf{q}_{k}-\mathbb{E}\left\{\mathbf{q}_{k}\right\}\right)^{H}\right\}+\frac{1}{\tau p} \mathbb{E}\left\{\mathbf{N} \mathbf{s}_{k}\mathbf{s}_{k}^{H}  \mathbf{N}^{H}\right\} \\
			=\operatorname{Cov}\left\{\mathbf{q}_{k}, \mathbf{q}_{k}\right\}+\frac{\sigma^{2}}{\tau p} \mathbf{I}_{M}.
		\end{array} 
	\end{align}

	Finally, by introducing the auxiliary variables of $a_{k1}=N c_{k} \delta$ and $a_{k2} = N c_{k}\left(\varepsilon_{k}+1\right)+\gamma_{k}$, the proof is completed.

	\section{}\label{appendix3}
	The LMMSE estimate of the channel ${\bf q}_k$ based on the observation vector ${\bf y}^k_p$ can be written as\cite[Chapter 12.5]{ kay1993fundamentals}
	\begin{align}\label{estimate_define}
		\hat{\mathbf{q}}_{k}=\mathbb{E}\left\{\mathbf{q}_{k}\right\}+\operatorname{Cov}\left\{\mathbf{q}_{k}, \mathbf{y}_{p}^{k}\right\} \operatorname{Cov}^{-1}\left\{\mathbf{y}_{p}^{k}, \mathbf{y}_{p}^{k}\right\}\left(\mathbf{y}_{p}^{k}-\mathbb{E}\left\{\mathbf{y}_{p}^{k}\right\}\right),
	\end{align}
	where the mean and covariance matrices have been obtained in Lemma \ref{lemma1}.
	
	Let us compute $\operatorname{Cov}^{-1}\left\{\mathbf{y}_{p}^{k}, \mathbf{y}_{p}^{k}\right\}$. Using the Woodbury matrix identity\cite[Page 571]{ kay1993fundamentals}, we have
	\begin{align}
		\operatorname{Cov}^{-1}\left\{\mathbf{y}_{p}^{k}, \mathbf{y}_{p}^{k}\right\}&=\left(a_{k 1} \mathbf{a}_{M} \mathbf{a}_{M}^{H}+\left(a_{k 2}+\frac{\sigma^{2}}{\tau p}\right) \mathbf{I}_{M} \right)^{-1}\nonumber\\
		&=\left(a_{k 2}+\frac{\sigma^{2}}{\tau p}\right)^{-1} \mathbf{I}_{M}-\frac{a_{k 1}\left(a_{k 2}+\frac{\sigma^{2}}{\tau p}\right)^{-2}}{1+M a_{k 1}\left(a_{k 2}+\frac{\sigma^{2}}{\tau p}\right)^{-1}} \mathbf{a}_{M} \mathbf{a}_{M}^{H}.
	\end{align}
	
	As a result, we have
	\begin{align}\label{CC_inv}
		\begin{aligned}
			& \operatorname{Cov}\left\{\mathbf{q}_{k}, \mathbf{y}_{p}^{k}\right\} \operatorname{Cov}^{-1}\left\{\mathbf{y}_{p}^{k}, \mathbf{y}_{p}^{k}\right\} \\
			&=\left(a_{k 1} \mathbf{a}_{M} \mathbf{a}_{M}^{H}+a_{k 2} \mathbf{I}_{M}\right)\left\{\left(a_{k 2}+\frac{\sigma^{2}}{\tau p}\right)^{-1} \mathbf{I}_{M}-\frac{a_{k 1}\left(a_{k 2}+\frac{\sigma^{2}}{\tau p}\right)^{-2}}{1+M a_{k 1}\left(a_{k 2}+\frac{\sigma^{2}}{\tau p}\right)^{-1}} \mathbf{a}_{M} \mathbf{a}_{M}^{H}\right\} \\
			&= \frac{a_{k 1} \frac{\sigma^{2}}{\tau p}}{\left(a_{k 2}+\frac{\sigma^{2}}{\tau p}\right)\left\{\left(a_{k 2}+\frac{\sigma^{2}}{\tau p}\right)+M a_{k 1}\right\}} \mathbf{a}_{M} \mathbf{a}_{M}^{H}+\frac{a_{k 2}}{a_{k 2}+\frac{\sigma^{2}}{\tau p}} \mathbf{I}_{M} \\
			&\triangleq  a_{k 3} \mathbf{a}_{M} \mathbf{a}_{M}^{H}+a_{k 4} \mathbf{I}_{M} \triangleq \mathbf{A}_{k}=\mathbf{A}^H_{k}.
		\end{aligned}
	\end{align}
	
	Since we have $\mathbb{E}\left\{\mathbf{q}_{k}\right\}=\mathbb{E}\left\{\mathbf{y}^k_{p}\right\}=\sqrt{c_{k} \delta \varepsilon_{k}}\, \overline{\mathbf{H}}_{2} \mathbf{\Phi} \overline{\mathbf{h}}_{k}$, the LMMSE channel estimate in (\ref{estimate_define}) is calculated as
	\begin{align}
		\begin{array}{l}
			\hat{\mathbf{q}}_{k}=\sqrt{c_{k} \delta \varepsilon_{k}} \,\overline{\mathbf{H}}_{2} {\bf \Phi} \overline{\mathbf{h}}_{k}+\mathbf{A}_{k}\left(\mathbf{y}_{p}^{k}-\sqrt{c_{k} \delta \varepsilon_{k}} \,\overline{\mathbf{H}}_{2} {\bf \Phi} \overline{\mathbf{h}}_{k}\right) \\
			\,\quad=\mathbf{A}_{k} \mathbf{y}_{p}^{k}+\left(\mathbf{I}_{M}-\mathbf{A}_{k}\right) \sqrt{c_{k} \delta \varepsilon_{k}} \, \overline{\mathbf{H}}_{2} {\bf \Phi} \overline{\mathbf{h}}_{k} \\
			\,\quad\triangleq \mathbf{A}_{k} \mathbf{y}_{p}^{k}+\mathbf{B}_{k}.
		\end{array}
	\end{align}
	
	Additionally, we can expand the above linear expression and rewrite it as 
	\begin{align}
		\begin{aligned}
			\hat{\mathbf{q}}_{k}&=\mathbf{A}_{k}\left(\mathbf{q}_{k}+\frac{1}{\sqrt{\tau p}} \mathbf{N s}_{k}\right)+\mathbf{B}_{k} \\
			&=\mathbf{A}_{k}\left(\sum\nolimits_{\omega=1}^{4} \mathbf{q}_{k}^{\omega}+\sqrt{\gamma_{k}} \tilde{\mathbf{d}}_{k}+\frac{1}{\sqrt{\tau p}} \mathbf{N s}_{k}\right)+\left(\mathbf{I}_{M}-\mathbf{A}_{k}\right) \mathbf{q}_{k}^{1} \\
			&=\mathbf{q}_{k}^{1}+\sum\nolimits_{\omega=2}^{4} \mathbf{A}_{k} \mathbf{q}_{k}^{\omega}+\sqrt{\gamma_{k}} \mathbf{A}_{k} \tilde{\mathbf{d}}_{k}+\frac{1}{\sqrt{\tau p}} \mathbf{A}_{k} \mathbf{N s}_{k}.
		\end{aligned}
	\end{align}
	
	Then, by exploiting the property $\mathbf{A}_{k} \overline{\mathbf{H}}_{2}=\left(a_{k 3} \mathbf{a}_{M} \mathbf{a}_{M}^{H}+a_{k 4} \mathbf{I}_{M}\right) \mathbf{a}_{M} \mathbf{a}_{N}^{H}=\left(M a_{k 3}+a_{k 4}\right) \overline{\mathbf{H}}_{2}$, we arrive at (\ref{estimated_channel_detail}).
	
	Based on the estimate $\hat{\mathbf{q}}_{k}$, we can obtain the estimation error $\mathbf{e}_{k}=\mathbf{q}_{k}-\hat{\mathbf{q}}_{k}$. By direct inspection, the mean of $\mathbf{e}_{k}$ is zero. Exploiting \cite[Eq. (12.21)]{kay1993fundamentals}, Lemma \ref{lemma1} and (\ref{CC_inv}), the MSE matrix of the estimation error can be calculated as
	\begin{align}\label{MSE_matrix}
		\begin{array}{l}
			{\rm \bf{MSE}}_{k}=\mathbb{E}\left\{\mathbf{e}_{k} \mathbf{e}_{k}^{H}\right\} \\
			\qquad\,\quad=\operatorname{Cov}\left\{\mathbf{q}_{k}, \mathbf{q}_{k}\right\}-\operatorname{Cov}\left\{\mathbf{q}_{k}, \mathbf{y}_{p}^{k}\right\} \operatorname{Cov}^{-1}\left\{\mathbf{y}_{p}^{k}, \mathbf{y}_{p}^{k}\right\} \operatorname{Cov}\left\{\mathbf{y}_{p}^{k}, \mathbf{q}_{k}\right\} \\
			\qquad\,\quad= \operatorname{Cov}\left\{\mathbf{q}_{k}, \mathbf{q}_{k}\right\} - \mathbf{A}_k \operatorname{Cov}\left\{\mathbf{q}_{k}, \mathbf{q}_{k}\right\}\\
			\qquad\,\quad=\left(\mathbf{I}_{M}-\mathbf{A}_{k}\right) \operatorname{Cov}\left\{\mathbf{q}_{k}, \mathbf{q}_{k}\right\} \\
			\qquad\,\quad=\left(\left(1-a_{k 4}\right) \mathbf{I}_{M}-a_{k 3} \mathbf{a}_{M} \mathbf{a}_{M}^{H}\right)\left(a_{k 1} \mathbf{a}_{M} \mathbf{a}_{M}^{H}+a_{k 2} \mathbf{I}_{M}\right) \\
			\qquad\,\quad=\left(a_{k 1}\left(1-a_{k 4}\right)-M a_{k 1} a_{k 3}-a_{k 2} a_{k 3}\right) \mathbf{a}_{M} \mathbf{a}_{M}^{H}+a_{k 2}\left(1-a_{k 4}\right) \mathbf{I}_{M} \\
			\qquad\,\quad\triangleq a_{k5} \mathbf{a}_{M} \mathbf{a}_{M}^{H}+a_{k 6} \mathbf{I}_{M},
		\end{array}
	\end{align} 
	where
	\begin{align}
		\begin{aligned}
			a_{k 5}&=a_{k 1}\left(1-a_{k 4}\right)-\left(M a_{k 1}+a_{k 2}\right) a_{k 3} \\
			&=a_{k 1}\left(1-\frac{a_{k 2}}{a_{k 2}+\frac{\sigma^{2}}{\tau p}}\right)-\left(M a_{k 1}+a_{k 2}\right) \frac{a_{k 1} \frac{\sigma^{2}}{\tau p}}{\left(a_{k 2}+\frac{\sigma^{2}}{\tau p}\right)\left\{\left(a_{k 2}+\frac{\sigma^{2}}{\tau p}\right)+M a_{k 1}\right\}} \\
			&=\frac{a_{k 1}\left(\frac{\sigma^{2}}{\tau p}\right)^{2}}{\left(a_{k 2}+\frac{\sigma^{2}}{\tau p}\right)\left(a_{k 2}+\frac{\sigma^{2}}{\tau p}+M a_{k 1}\right)},
		\end{aligned}
	\end{align}
	and
	\begin{align}
		a_{k 6}=a_{k 2}\left(1-a_{k 4}\right) =a_{k 2}\left(1-\frac{a_{k 2}}{a_{k 2}+\frac{\sigma^{2}}{\tau p}}\right) =\frac{a_{k 2} \frac{\sigma^{2}}{\tau p}}{a_{k 2}+\frac{\sigma^{2}}{\tau p}}.
	\end{align}
	
	Based on the MSE matrix, the NMSE of the estimation error can be expressed as\cite[Eq. (3.20)]{bjornson2017massive}
	\begin{align}
		\operatorname{NMSE}_{k}&=\frac{\operatorname{Tr}\left\{{\rm\bf{MSE}}_{k}\right\}}{\operatorname{Tr}\left\{\operatorname{Cov}\left\{\mathbf{q}_{k}, \mathbf{q}_{k}\right\}\right\}} =\frac{M\left(a_{k 5}+a_{k 6}\right)}{M\left(a_{k 1}+a_{k 2}\right)} =\frac{a_{k 5}+a_{k 6}}{a_{k 1}+a_{k 2}}\nonumber\\
		&=\frac{\frac{\sigma^{2}}{\tau p}\left(M a_{k 1} a_{k 2}+a_{k 2}^{2}+\left(a_{k 1}+a_{k 2}\right) \frac{\sigma^{2}}{\tau p}\right)}{\left(a_{k 2}+\frac{\sigma^{2}}{\tau p}\right)\left(a_{k 2}+\frac{\sigma^{2}}{\tau p}+M a_{k 1}\right)\left(a_{k 1}+a_{k 2}\right)}.
	\end{align}
	
	Hence, the proof is completed.

	\section{}\label{appendix_e1e2e3}
	Recall that $\mathbf{A}_{k}=a_{k 3} \mathbf{a}_{M} \mathbf{a}_{M}^{H}+a_{k 4} \mathbf{I}_{M}$, and $ \overline{\mathbf{H}}_{2}=\mathbf{a}_{M} \mathbf{a}_{N}^{H} $. We can readily obtain
	\begin{align}
		\begin{array}{l}
			\operatorname{Tr}\left\{\mathbf{A}_{k}\right\}=M\left(a_{k 3}+a_{k 4}\right) \triangleq M e_{k 1}, \\
			\mathbf{A}_{k} \overline{\mathbf{H}}_{2}=a_{k 3} \mathbf{a}_{M} \mathbf{a}_{M}^{H} \mathbf{a}_{M} \mathbf{a}_{N}^{H}+a_{k 4} \mathbf{I}_{M} \mathbf{a}_{M} \mathbf{a}_{N}^{H}=\left(M a_{k 3}+a_{k 4}\right) \overline{\mathbf{H}}_{2} \triangleq e_{k 2} \overline{\mathbf{H}}_{2}, \\
			\mathbf{A}_{k} \mathbf{A}_{k}=a_{k 3} \mathbf{a}_{M} \mathbf{a}_{M}^{H}\left(a_{k 3} \mathbf{a}_{M} \mathbf{a}_{M}^{H}+a_{k 4} \mathbf{I}_{M}\right)+a_{k 4} \mathbf{I}_{M}\left(a_{k 3} \mathbf{a}_{M} \mathbf{a}_{M}^{H}+a_{k 4} \mathbf{I}_{M}\right) \\
			\qquad\quad=M a_{k 3}^{2} \mathbf{a}_{M} \mathbf{a}_{M}^{H}+2 a_{k 3} a_{k 4} \mathbf{a}_{M} \mathbf{a}_{M}^{H}+a_{k 4}^{2} \mathbf{I}_{M}, \\
			\operatorname{Tr}\left\{\mathbf{A}_{k} \mathbf{A}_{k}\right\}=M\left(M a_{k 3}^{2}+2 a_{k 3} a_{k 4}+a_{k 4}^{2}\right) \triangleq M e_{k 3}.
		\end{array}
	\end{align}
	
	By direct inspection of $e_{k1},e_{k2},e_{k3}$, we evince that they are composed of non-negative terms. Therefore, we have $e_{k1},e_{k2},e_{k3}\ge 0$. Then, we aim to prove $e_{k1},e_{k2},e_{k3}\leq1$. We first focus on the parameter $e_{k2}$. Using the expressions of  $a_{k3}$ and $a_{k4}$ in (\ref{ak3}) and (\ref{ak4}), we can expand $e_{k2}$ as
	\begin{align}\label{prove_ek2_small_1}
		\begin{aligned}
			e_{k 2}&=Ma_{k 3}+a_{k 4}=\frac{Ma_{k 1} \frac{\sigma^{2}}{\tau p}}{\left(a_{k 2}+\frac{\sigma^{2}}{\tau p}\right)\left(a_{k 2}+\frac{\sigma^{2}}{\tau p}+M a_{k 1}\right)}+\frac{a_{k 2}}{a_{k 2}+\frac{\sigma^{2}}{\tau p}}\\
			&=\frac{a_{k 2}\left(a_{k 2}+\frac{\sigma^{2}}{\tau p}+M a_{k 1}\right)+M a_{k 1}\frac{\sigma^{2}}{\tau p} }{a_{k 2}\left(a_{k 2}+\frac{\sigma^{2}}{\tau p}+M a_{k 1}\right)+M  a_{k 1} \frac{\sigma^{2}}{\tau p}+\frac{\sigma^{2}}{\tau p}\left(a_{k 2}+\frac{\sigma^{2}}{\tau p}\right)}.
		\end{aligned}
	\end{align}
	It is clear that the numerator in (\ref{prove_ek2_small_1}) is smaller than the denominator. Therefore, we proved that $e_{k2}\leq1$. Then, we can directly obtain
	\begin{align}
		&e_{k1}\leq e_{k2}\leq 1,\\
		&e_{k3}\leq e_{k2}^2\leq e_{k2} \leq 1.
	\end{align}
	
	Finally, when $\tau p\to\infty$ or $N\to\infty$, we have $a_{k3}\to 0$ and $a_{k4}\to 1$, which implies that $e_{k1}=e_{k2}=e_{k3}\to 1$. When $\tau p\to 0$, we have $a_{k3},a_{k4}\to 0$, which gives $e_{k1}=e_{k2}=e_{k3}\to 0$. This completes the proof.

	\section{}\label{appendix4}
	\subsection{Signal Term and Noise Term}
	According to the orthogonality property of the LMMSE estimator, we have $\mathbb{E}\left\{ \mathbf{e}_k \left(\mathbf{y}^k_{p}\right)^{H}\right\}={\bf 0}$. Besides, since ${\bf e}_k$ has zero mean, we obtain $\mathbb{E}\left\{\hat{\mathbf{q}}_{k}^{H} \mathbf{e}_k\right\}=\mathbb{E}\left\{\left(\mathbf{A}_{k} \mathbf{y}_{p}^{k}+\mathbf{B}_{k}\right)^H \mathbf{e}_k\right\}=0$.
	Therefore, we have
	\begin{align}\label{single=noise^2}
		\mathbb{E}\left\{\hat{\mathbf{q}}_{k}^{H} \mathbf{q}_{k}\right\}
		=\mathbb{E}\left\{\hat{\mathbf{q}}_{k}^{H} \hat{\mathbf{q}}_{k}\right\}+\mathbb{E}\left\{\hat{\mathbf{q}}_{k}^{H} \mathbf{e}_k\right\}=\mathbb{E}\left\{\left\|\hat{\mathbf{q}}_{k}\right\|^{2}\right\}.
	\end{align}
	
	Denote the signal term of (\ref{defination_rate}) as $   \left| \mathbb{E}\left\{\hat{\mathbf{q}}_{k}^{H} \mathbf{q}_{k}\right\} \right|^2 \triangleq E_{k}^{\rm s i g n a l} \left(\mathbf{\Phi}\right)$, and denote the noise term of (\ref{defination_rate})  as $  \mathbb{E}\left\{\left\|\hat{\mathbf{q}}_{k}\right\|^{2}\right\} \triangleq E_{k}^{\rm {noise}} \left(\mathbf{\Phi}\right)$. Clearly, $\mathbb{E}\left\{\left\|\hat{\mathbf{q}}_{k}\right\|^{2}\right\} $ is a real variable. Then, from (\ref{single=noise^2}), we obtain 
	\begin{align}\label{Eksignal}
		E_{k}^{\rm s i g n a l}  \left(\mathbf{\Phi}\right) =  \left| \mathbb{E}\left\{\hat{\mathbf{q}}_{k}^{H} \mathbf{q}_{k}\right\} \right|^2 = \left(  \mathbb{E}\left\{\left\|\hat{\mathbf{q}}_{k}\right\|^{2}\right\} \right)^2 =\left(E_{k}^{\rm {noise}} \left(\mathbf{\Phi}\right) \right)^2.
	\end{align}
	
	Let us now derive $E_{k}^{\rm {noise}} \left(\mathbf{\Phi}\right) $.
	Recall the expressions in (\ref{exact_channel}) and (\ref{estimated_channel_detail}). Since $\tilde{\bf H}_2$, $\tilde{\bf h}_k$, $\tilde{\bf d}_k$ and ${\bf N}$ are independent of each other and they all have a zero mean, we can derive the term $\mathbb{E}\left\{\hat{\mathbf{q}}_{k}^{H} \mathbf{q}_{k}\right\}$ by selecting the non-zero terms in the expansion as
	\begin{align}\label{derive_noise_and_signal}
		\begin{array}{l}
			E_{k}^{\rm {noise}} \left(\mathbf{\Phi}\right) = \mathbb{E}\left\{\left\|\hat{\mathbf{q}}_{k}\right\|^{2}\right\} =    \mathbb{E}\left\{\hat{\mathbf{q}}_{k}^{H} \mathbf{q}_{k}\right\}\\
			=\mathbb{E}\left\{\left(\sum_{\omega=1}^{4} \hat{\mathbf{q}}_{k}^{\omega}+\sqrt{\gamma_{k}} \mathbf{A}_{k} \tilde{\mathbf{d}}_{k}+\frac{1}{\sqrt{\tau p}} \mathbf{A}_{k} \mathbf{N s}_k\right)^{H}\left(\sum_{\psi=1}^{4} \mathbf{q}_{k}^{\psi}+\sqrt{\gamma_{k}} \tilde{\mathbf{d}}_{k}\right)\right\} \\
			=\sum_{\omega=1}^{4} \mathbb{E}\left\{\left(\hat{\mathbf{q}}_{k}^{\omega}\right)^{H} \mathbf{q}_{k}^{\omega}\right\}+\gamma_{k} \mathbb{E}\left\{\tilde{\mathbf{d}}_{k}^{H} \mathbf{A}_{k}^{H} \tilde{\mathbf{d}}_{k}\right\}\\
			=c_{k} \delta \varepsilon_{k} \overline{\mathbf{h}}_{k}^{H} \mathbf{\Phi}^{H} \overline{\mathbf{H}}_{2}^{H} \overline{\mathbf{H}}_{2} \mathbf{\Phi} \overline{\mathbf{h}}_{k}+e_{k2} c_{k} \delta \mathbb{E}\left\{\tilde{\mathbf{h}}_{k}^{H} \mathbf{\Phi}^{H} \overline{\mathbf{H}}_{2}^{H} \overline{\mathbf{H}}_{2} \mathbf{\Phi} \tilde{\mathbf{h}}_{k}\right\} \\
			\quad+c_{k} \varepsilon_{k} \overline{\mathbf{h}}_{k}^{H} \mathbf{\Phi}^{H} \mathbb{E}\left\{\tilde{\mathbf{H}}_{2}^{H} \mathbf{A}_{k}^{H} \tilde{\mathbf{H}}_{2}\right\} \mathbf{\Phi} \overline{\mathbf{h}}_{k}+c_{k} \mathbb{E}\left\{\tilde{\mathbf{h}}_{k}^{H} \mathbf{\Phi}^{H} \mathbb{E}\left\{\tilde{\mathbf{H}}_{2}^{H} \mathbf{A}_{k}^{H} \tilde{\mathbf{H}}_{2}\right\} \mathbf{\Phi} \tilde{\mathbf{h}}_{k}\right\}+\gamma_{k} \mathrm{Tr}\left\{ \mathbf{A}_{k}^{H} \right\} \\
			{{\mathop  = \limits^{\left( c \right)} }}c_{k} \delta \varepsilon_{k} M\left|f_{k}(\mathbf{\Phi})\right|^{2}+c_{k} \delta M N e_{k 2}+c_{k} \varepsilon_{k} M N e_{k 1}+c_{k} M N e_{k 1}+\gamma_{k} M e_{k 1} \\
			=M\left\{\left|f_{k}(\mathbf{\Phi})\right|^{2} c_{k} \delta \varepsilon_{k}+N c_{k} \delta e_{k 2}+\left(N c_{k}\left(\varepsilon_{k}+1\right)+\gamma_{k}\right) e_{k 1}\right\},
		\end{array}
	\end{align}
	where $(c)$ applies Lemma \ref{lemma_HWH} and exploits the identities ${\rm Tr}\{{\bf A}_k\}=Me_{k1}$, ${\bf \Phi}^H{\bf \Phi}={\bf I}_N$, and ${\rm Tr}\left\{\overline{\mathbf{H}}_{2}^{H} \overline{\mathbf{H}}_{2}\right\}=MN$. 
	Substituting (\ref{derive_noise_and_signal}) into (\ref{Eksignal}), we complete the calculation of $ E_{k}^{\rm s i g n a l}  \left(\mathbf{\Phi}\right) $ and $E_{k}^{\rm {noise}}\left(\mathbf{\Phi}\right)$.
	
	We conclude this subsection by deriving some useful results that are obtained by using a procedure similar to that used for obtaining (\ref{derive_noise_and_signal}). To be specific, we aim to derive $\mathbb{E}\left\{{\mathbf{q}}_{k}^{H} \mathbf{q}_{k}\right\}$, $\mathbb{E}\left\{  \hat{\underline{\mathbf{q}}}_{k}^{H} \hat{\underline{\mathbf{q}}}_{k}\right\}$, $\mathbb{E}\left\{{\underline{\mathbf{q}}}_{k}^{H} {\bf A}_k {\bf A}^H_k {\underline{\mathbf{q}}}_{k}\right\}$, $\mathbb{E}\left\{{\underline{\mathbf{q}}}_{i}^{H} {\bf A}_k {\bf A}^H_k {\underline{\mathbf{q}}}_{i}\right\}$, and $\mathbb{E}\left\{\hat{\underline{\mathbf{q}}}_{k}^{H} {\underline{\mathbf{q}}}_{k}\right\}$.
	
	Firstly, when ${\bf A}_k={\bf I}_M$ and $\tau\to\infty$, the imperfect estimate $\hat{\mathbf{q}}_{k}$ becomes the perfect estimate ${\mathbf{q}}_{k}$. Therefore, substituting ${\bf A}_k={\bf I}_M$ and $\tau\to\infty$ into (\ref{derive_noise_and_signal}), we have
	\begin{align}\label{perferct_gkgk}
		\mathbb{E}\left\{{\mathbf{q}}_{k}^{H} \mathbf{q}_{k}\right\}
		=M\left\{\left|f_{k}(\mathbf{\Phi})\right|^{2} c_{k} \delta \varepsilon_{k}+ N c_{k}\left(\delta +\varepsilon_{k}+1\right)+\gamma_{k} \right\}.
	\end{align}

	Secondly, by using the expression of $\hat{\underline{\mathbf{q}}}_{k}$ in (\ref{estimated_channel_detail}), we have
	\begin{align}\label{hat_un_hat_un}
		\begin{array}{l}
			\mathbb{E}\left\{  \hat{\underline{\mathbf{q}}}_{k}^{H} \hat{\underline{\mathbf{q}}}_{k}    \right\} =\mathbb{E}\left\{\sum_{\omega=1}^{4} \sum_{\psi=1}^{4}\left(\hat{\underline{\mathbf{q}}}_{k}^{\omega}\right)^{H} \hat{\underline{\mathbf{q}}}_{k}^{\psi}\right\}=\sum_{\omega=1}^{4} \mathbb{E}\left\{\left\|   \hat{\underline{\mathbf{q}}}_{k}^{\omega}\right\|^{2}\right\} \\
			{{\mathop  = \limits^{\left( d \right)} }}M\left\{\left|f_{k}(\mathbf{\Phi})\right|^{2} c_{k} \delta \varepsilon_{k}+N c_{k} \delta e_{k 2}^{2}+N c_{k}\left(\varepsilon_{k}+1\right) e_{k 3}\right\},
		\end{array}
	\end{align}
	where $(d)$ follows by applying the identity $ \operatorname{Tr}\left\{\mathbf{A}_{k}^{H} \mathbf{A}_{k}\right\}= M e_{k 3} $.
	
	Thirdly, using $\mathbf{A}_{k}^{H}=\mathbf{A}_{k}$ and $\mathbf{A}_{k}\overline{\mathbf{H}}_{2}=e_{k2}\overline{\mathbf{H}}_{2}$, we have
	\begin{align}\label{un_k_A_A_un_k}
		\begin{array}{l}
			\mathbb{E}\left\{\underline{\mathbf{q}}_{k}^{H}  \mathbf{A}_{k} \mathbf{A}_{k}^{H}\underline{\mathbf{q}}_{k}\right\}
			=\mathbb{E}\left\{\sum_{\omega=1}^{4} \sum_{\psi=1}^{4}\left(\mathbf{A}_{k} \mathbf{q}_{k}^{\omega}\right)^{H}\left(\mathbf{A}_{k} \mathbf{q}_{k}^{\psi}\right)\right\} \\
			=\left\|\sqrt{c_{k} \delta \varepsilon_{k}} \mathbf{A}_{k} \overline{\mathbf{H}}_{2} \mathbf{\Phi} \overline{\mathbf{h}}_{k}\right\|^{2}+\sum_{\omega=2}^{4} \mathbb{E}\left\{\left\|\hat{\mathbf{q}}_{k}^{\omega}\right\|^{2}\right\} \\
			=M\left\{\left|f_{k}(\mathbf{\Phi})\right|^{2} c_{k} \delta \varepsilon_{k} e_{k 2}^{2}+N c_{k} \delta e_{k 2}^{2}+N c_{k}\left(\varepsilon_{k}+1\right) e_{k 3}\right\}.
		\end{array}
	\end{align}
	Also, for $i\neq k$, we have
	\begin{align}\label{un_i_A_A_un_i}
		\begin{array}{l}
			\mathbb{E}\left\{\underline{\mathbf{q}}_{i}^{H}  \mathbf{A}_{k} \mathbf{A}_{k}^{H}\underline{\mathbf{q}}_{i}\right\}
			=M\left\{\left|f_{i}(\mathbf{\Phi})\right|^{2} c_{i} \delta \varepsilon_{i} e_{k 2}^{2}+N c_{i} \delta e_{k 2}^{2}+N c_{i}\left(\varepsilon_{i}+1\right) e_{k 3}\right\}.
		\end{array}
	\end{align}

	Finally, by substituting $\gamma_{k}=0$ into (\ref{derive_noise_and_signal}), we arrive at
	\begin{align}\label{hat_un_un}
		\mathbb{E}\left\{  \hat{\underline{\mathbf{q}}}_{k}^{H} {\underline{\mathbf{q}}}_{k}    \right\} =M\left\{\left|f_{k}(\mathbf{\Phi})\right|^{2} c_{k} \delta \varepsilon_{k}+N c_{k} \delta e_{k 2}+N c_{k}\left(\varepsilon_{k}+1\right) e_{k 1}\right\}.
	\end{align}
	
	\subsection{Interference Term}
	In this subsection, we derive the interference term of (\ref{defination_rate}). The interference term is denoted by $\mathbb{E}\left\{\left|\hat{\mathbf{q}}_{k}^{H} \mathbf{q}_{i}\right|^{2}\right\} \triangleq I_{ki} \left(\mathbf{\Phi}\right) $. 
	First, it is worth noting that the derivation of the interference term in the presence of imperfect CSI and double-Rician channels in RIS-aided massive MIMO systems has two main differences compared to conventional massive MIMO systems. 
	Firstly,
	the channel ${\bf q}_k$ and ${\bf q}_i$ are not independent, since different users experience the same RIS-BS channel. This can be readily validated by examining that $\mathbb{E}\left\{\mathbf{q}_{k}^{H} \mathbf{q}_{i}\right\} \neq \mathbb{E}\left\{\mathbf{q}_{k}^{H}\right\} \mathbb{E}\left\{\mathbf{q}_{i}\right\}$. 
	Secondly,
	the LMMSE error ${\bf e}_k$ is uncorrelated with but dependent on the estimate $\hat{\bf q}_k$, since the cascaded channel is not Gaussian distributed. 
	To tackle these two challenges, we derive the interference term by decomposing it as
	\begin{align}\label{expanded_interference_term}
		I_{ki} \left(\mathbf{\Phi}\right)=& \mathbb{E}\left\{\left|\hat{\mathbf{q}}_{k}^{H} \mathbf{q}_{i}\right|^{2}\right\}
		=\mathbb{E}\left\{\left|\left(  \underline{ \hat{\mathbf{q}} }_k  + \mathbf{A}_{k} {\mathbf{d}}_{k}+\frac{1}{\sqrt{\tau p}} \mathbf{A}_{k} \mathbf{N s}_{k}\right)^{H}\left(  \underline{\mathbf{q}}_{i}  +{\mathbf{d}}_{i}\right)\right|^{2}\right\} \nonumber\\
		=&\mathbb{E}\left\{\left| \underline{\hat{\mathbf{q}}}_{k}^{H} \underline{\mathbf{q}}_{i}+ \underline{\hat{\mathbf{q}}}_{k}^{H} {\mathbf{d}}_{i}+ {\mathbf{d}}_{k}^{H} \mathbf{A}_{k}^{H} \underline{\mathbf{q}}_{i}+ {\mathbf{d}}_{k}^{H} \mathbf{A}_{k}^{H} {\mathbf{d}}_{i}+\frac{1}{\sqrt{\tau p}} \mathbf{s}_{k}^H \mathbf{N}^{H} \mathbf{A}_{k}^{H} \underline{\mathbf{q}}_{i}+\frac{1}{\sqrt{\tau p}} \mathbf{s}_{k}^H \mathbf{N}^{H} \mathbf{A}_{k}^{H} {\mathbf{d}}_{i}\right|^{2}\right\} \nonumber\\
		=&\mathbb{E}\left\{\left| \underline{\hat{\mathbf{q}}}_{k}^{H} \underline{\mathbf{q}}_{i}\right|^{2}\right\}+ \mathbb{E}\left\{\left| \underline{\hat{\mathbf{q}}}_{k}^{H} {\mathbf{d}}_{i}\right|^{2}\right\}+ \mathbb{E}\left\{\left|{\mathbf{d}}_{k}^{H} \mathbf{A}_{k}^{H} \underline{\mathbf{q}}_{i}\right|^{2}\right\}+\mathbb{E}\left\{\left|{\mathbf{d}}_{k}^{H} \mathbf{A}_{k}^{H} {\mathbf{d}}_{i}\right|^{2}\right\} \nonumber\\
		&+\frac{1}{\tau p} \mathbb{E}\left\{\left|\mathbf{s}_{k}^H \mathbf{N}^{H} \mathbf{A}_{k}^{H} \underline{\mathbf{q}}_{i}\right|^{2}\right\}+\frac{1}{\tau p} \mathbb{E}\left\{\left|\mathbf{s}_{k}^H \mathbf{N}^{H} \mathbf{A}_{k}^{H} {\mathbf{d}}_{i}\right|^{2}\right\}.
	\end{align}
	
	We aim to derive the six expectations in (\ref{expanded_interference_term}) one by one, but the first one $\mathbb{E}\left\{\left| \underline{\hat{\mathbf{q}}}_{k}^{H} \underline{\mathbf{q}}_{i}\right|^{2}\right\}$ will be derived last.  The second term in (\ref{expanded_interference_term}) is
	\begin{align}\label{interference_5terms_1}
		\mathbb{E}\left\{\left|\hat{\underline{\mathbf{q}}}_{k}^{H} \mathbf{d}_{i}\right|^{2}\right\}=\mathbb{E}\left\{  \hat{\underline{\mathbf{q}}}_{k}^{H} \mathbf{d}_{i} \mathbf{d}_{i}^{H} \hat{\underline{\mathbf{q}}}_{k}\right\}=\mathbb{E}\left\{  \hat{\underline{\mathbf{q}}}_{k}^{H} \mathbb{E}\left\{\mathbf{d}_{i} \mathbf{d}_{i}^{H}\right\} \hat{\underline{\mathbf{q}}}_{k}\right\}
		=\gamma_{i} \mathbb{E}\left\{\hat{\underline{\mathbf{q}}}_{k}^{H} \hat{\underline{\mathbf{q}}}_{k}\right\},
	\end{align}
	where $  \mathbb{E}\left\{\hat{\underline{\mathbf{q}}}_{k}^{H} \hat{\underline{\mathbf{q}}}_{k}\right\}$ is given in (\ref{hat_un_hat_un}).
	
	The third term in (\ref{expanded_interference_term}) is
	\begin{align}\label{interference_5terms_2}
		\mathbb{E}\left\{\left|\mathbf{d}_{k}^{H} \mathbf{A}_{k}^{H} \underline{\mathbf{q}}_{i}\right|^{2}\right\}=\mathbb{E}\left\{\underline{\mathbf{q}}_{i}^{H} \mathbf{A}_{k} \mathbb{E}\left\{\mathbf{d}_{k} \mathbf{d}_{k}^{H}\right\} \mathbf{A}_{k}^{H} \underline{\mathbf{q}}_{i}\right\}
		=\gamma_{k} \mathbb{E}\left\{\underline{\mathbf{q}}_{i}^{H} \mathbf{A}_{k} \mathbf{A}_{k}^{H} \underline{\mathbf{q}}_{i}\right\},
	\end{align}
	where $\mathbb{E}\left\{\underline{\mathbf{q}}_{i}^{H} \mathbf{A}_{k} \mathbf{A}_{k}^{H} \underline{\mathbf{q}}_{i}\right\}$ is given in (\ref{un_i_A_A_un_i}).
	
	The fourth term in (\ref{expanded_interference_term}) is
	\begin{align}\label{interference_5terms_3}
		\begin{array}{l}
			\mathbb{E}\left\{\left|\mathbf{d}_{k}^{H} \mathbf{A}_{k}^{H} \mathbf{d}_{i}\right|^{2}\right\}=\mathbb{E}\left\{\mathbf{d}_{k}^{H} \mathbf{A}_{k}^{H} \mathbb{E}\left\{\mathbf{d}_{i} \mathbf{d}_{i}^{H}\right\} \mathbf{A}_{k} \mathbf{d}_{k}\right\} \\
			=\gamma_{i} \mathbb{E}\left\{\mathbf{d}_{k}^{H} \mathbf{A}_{k}^{H} \mathbf{A}_{k} \mathbf{d}_{k}\right\}=\gamma_{k} \gamma_{i} \operatorname{Tr}\left\{\mathbf{A}_{k}^{H} \mathbf{A}_{k}\right\} \\
			=\gamma_{k} \gamma_{i} M e_{k 3}.
		\end{array}
	\end{align}
	
	The fifth term in (\ref{expanded_interference_term}) is
	\begin{align}\label{interference_5terms_4}
		\frac{1}{{\tau p}}\mathbb{E}\left\{\left| \mathbf{s}_{k}^H \mathbf{N}^{H} \mathbf{A}_{k}^{H} \underline{\mathbf{q}}_{i}\right|^{2}\right\}=\frac{1}{\tau p} \mathbb{E}\left\{\underline{\mathbf{q}}_{i}^{H} \mathbf{A}_{k} \mathbb{E}\left\{\mathbf{N} \mathbf{s}_{k}\mathbf{s}_{k}^H \mathbf{N}^{H}\right\} \mathbf{A}_{k}^{H} \underline{\mathbf{q}}_{i}\right\}=\frac{\sigma^{2}}{\tau p} \mathbb{E}\left\{\underline{\mathbf{q}}_{i}^{H} \mathbf{A}_{k} \mathbf{A}_{k}^{H} \underline{\mathbf{q}}_{i}\right\},
	\end{align}
	where $\mathbb{E}\left\{\underline{\mathbf{q}}_{i}^{H} \mathbf{A}_{k} \mathbf{A}_{k}^{H} \underline{\mathbf{q}}_{i}\right\}$ is given in (\ref{un_i_A_A_un_i}).
	
	The sixth term in (\ref{expanded_interference_term}) is
	\begin{align}\label{interference_5terms_5}
		\frac{1}{{\tau p}}\mathbb{E}\left\{\left| \mathbf{s}_{k}^H \mathbf{N}^{H} \mathbf{A}_{k}^{H} \mathbf{d}_{i}\right|^{2}\right\}=\frac{\sigma^{2}}{\tau p} \gamma_{i} \operatorname{Tr}\left\{ \mathbf{A}_{k} \mathbf{A}_{k}^{H} \right\}=\frac{\sigma^{2}}{\tau p} \gamma_{i} M e_{k 3}.
	\end{align}
	
	Finally, we derive the first term $\mathbb{E}\left\{\left| \underline{\hat{\mathbf{q}}}_{k}^{H} \underline{\mathbf{q}}_{i}\right|^{2}\right\}$, which can be expanded as
	\begin{align}\label{interference_expanded_sum}
		\begin{array}{l}
			\mathbb{E}\left\{\left| \underline{\hat{\mathbf{q}}}_{k}^{H} \underline{\mathbf{q}}_{i}\right|^{2}\right\}
			=\mathbb{E}\left\{\left|\sum_{\omega=1}^{4} \sum_{\psi=1}^{4}\left(\hat{\mathbf{q}}_{k}^{\omega}\right)^{H} \mathbf{q}_{i}^{\psi}\right|^{2}\right\} \\
			=\sum_{\omega=1}^{4} \sum_{\psi=1}^{4} \mathbb{E}\left\{\left|\left(\hat{\mathbf{q}}_{k}^{\omega}\right)^{H} \mathbf{q}_{i}^{\psi}\right|^{2}\right\}+\sum_{\omega 1, \psi 1, \omega 2, \psi 2, \atop(\omega 1, \psi 1) \neq(\omega 2, \psi 2)}^{4} \mathbb{E}\left\{\left(\left(\hat{\mathbf{q}}_{k}^{\omega 1}\right)^{H} \mathbf{q}_{i}^{\psi 1}\right)\left(\left(\hat{\mathbf{q}}_{k}^{\omega 2}\right)^{H} \mathbf{q}_{i}^{\psi 2}\right)^{H}\right\},
		\end{array}
	\end{align}
	where ${\bf q}^1_k$ - ${\bf q}^4_k$ are defined in (\ref{exact_channel}), and $\hat{\bf q}^1_k$ - $\hat{\bf q}^4_k$ are defined in (\ref{estimated_channel_detail}).
	
	Equation (\ref{interference_expanded_sum}) can be derived by calculating the expectations of the $16$ modulus-square terms and the expectations of the other cross-terms. We first calculate the former $16$ modulus-square terms in (\ref{interference_expanded_sum}) one by one. The derivation utilizes Lemma \ref{lemma_HWH}, Lemma \ref{lemma_HAHWHAH}, and the independence between $\tilde{\bf H}_2$, $\tilde{\bf h}_k$, and $\tilde{\bf h}_i$.
	
	Firstly, we consider the terms with $\omega=1$. When $\psi=1$, we have
	\begin{align}\label{interference_16terms_1}
		\begin{array}{l}
			\mathbb{E}\left\{\left|\sqrt{c_{k} \delta \varepsilon_{k}} \sqrt{c_{i} \delta \varepsilon_{i}} \,\overline{\mathbf{h}}_{k}^{H} \mathbf{\Phi}^{H} \overline{\mathbf{H}}_{2}^{H} \overline{\mathbf{H}}_{2} \mathbf{\Phi} \overline{\mathbf{h}}_{i}\right|^{2}\right\} \\
			=c_{k} c_{i} \delta^{2} \varepsilon_{k} \varepsilon_{i} \overline{\mathbf{h}}_{k}^{H} \mathbf{\Phi}^{H} \mathbf{a}_{N} \mathbf{a}_{M}^{H} \mathbf{a}_{M} \mathbf{a}_{N}^{H} \mathbf{\Phi} \overline{\mathbf{h}}_{i} \overline{\mathbf{h}}_{i}^{H} \mathbf{\Phi}^{H} \mathbf{a}_{N} \mathbf{a}_{M}^{H} \mathbf{a}_{M} \mathbf{a}_{N}^{H} \mathbf{\Phi} \overline{\mathbf{h}}_{k} \\
			=c_{k} c_{i} \delta^{2} \varepsilon_{k} \varepsilon_{i} M^{2}\left|f_{k}(\mathbf{\Phi})\right|^{2}\left|f_{i}(\mathbf{\Phi})\right|^{2}.
		\end{array}
	\end{align}
	
	When $\psi=2$, we have 
	\begin{align}\label{interference_16terms_2}
		\begin{array}{l}
			\mathbb{E}\left\{\left|\sqrt{c_{k} \delta \varepsilon_{k}} \sqrt{c_{i} \delta} \,\overline{\mathbf{h}}_{k}^{H} \mathbf{\Phi}^{H} \overline{\mathbf{H}}_{2}^{H} \overline{\mathbf{H}}_{2} \mathbf{\Phi} \tilde{\mathbf{h}}_{i}\right|^{2}\right\} \\
			=c_{k} c_{i} \delta^{2} \varepsilon_{k} \overline{\mathbf{h}}_{k}^{H} \mathbf{\Phi}^{H} \mathbf{a}_{N} \mathbf{a}_{M}^{H} \mathbf{a}_{M} \mathbf{a}_{N}^{H} \mathbf{\Phi} \mathbb{E}\left\{\tilde{\mathbf{h}}_{i} \tilde{\mathbf{h}}_{i}^{H}\right\} \mathbf{\Phi}^{H} \mathbf{a}_{N} \mathbf{a}_{M}^{H} \mathbf{a}_{M} \mathbf{a}_{N}^{H} \mathbf{\Phi} \overline{\mathbf{h}}_{k} \\
			=c_{k} c_{i} \delta^{2} \varepsilon_{k} M^{2} N\left|f_{k}(\mathbf{\Phi})\right|^{2}.
		\end{array}
	\end{align}
	
	When $\psi=3$, using Lemma \ref{lemma_HWH}, we arrive at 
	\begin{align}\label{interference_16terms_3}
		\begin{array}{l}
			\mathbb{E}\left\{\left|\sqrt{c_{k} \delta \varepsilon_{k}} \sqrt{c_{i} \varepsilon_{i}} \, \overline{\mathbf{h}}_{k}^{H} \mathbf{\Phi}^{H} \overline{\mathbf{H}}_{2}^{H} \tilde{\mathbf{H}}_{2} \mathbf{\Phi} \overline{\mathbf{h}}_{i}\right|^{2}\right\} \\
			=c_{k} c_{i} \delta \varepsilon_{k} \varepsilon_{i} \overline{\mathbf{h}}_{k}^{H} \mathbf{\Phi}^{H} \overline{\mathbf{H}}_{2}^{H} \mathbb{E}\left\{\tilde{\mathbf{H}}_{2} \mathbf{\Phi} \overline{\mathbf{h}}_{i} \overline{\mathbf{h}}_{i}^{H} \mathbf{\Phi}^{H} \tilde{\mathbf{H}}_{2}^{H}\right\} \overline{\mathbf{H}}_{2} \mathbf{\Phi} \overline{\mathbf{h}}_{k} \\
			=c_{k} c_{i} \delta \varepsilon_{k} \varepsilon_{i} N \overline{\mathbf{h}}_{k}^{H} \mathbf{\Phi}^{H} \overline{\mathbf{H}}_{2}^{H} \overline{\mathbf{H}}_{2} \mathbf{\Phi} \overline{\mathbf{h}}_{k} \\
			=c_{k} c_{i} \delta \varepsilon_{k} \varepsilon_{i} M N\left|f_{k}(\mathbf{\Phi})\right|^{2}.
		\end{array}
	\end{align}
	
	When $\psi=4$, we have 
	\begin{align}\label{interference_16terms_4}
		\begin{array}{l}
			\mathbb{E}\left\{\left|\sqrt{c_{k} \delta \varepsilon_{k}} \sqrt{c_{i}} \,\overline{\mathbf{h}}_{k}^{H} \mathbf{\Phi}^{H} \overline{\mathbf{H}}_{2}^{H} \tilde{\mathbf{H}}_{2} \mathbf{\Phi} \tilde{\mathbf{h}}_{i}\right|^{2}\right\} \\
			=c_{k} c_{i} \delta \varepsilon_{k} \overline{\mathbf{h}}_{k}^{H} \mathbf{\Phi}^{H} \overline{\mathbf{H}}_{2}^{H} \mathbb{E}\left\{\tilde{\mathbf{H}}_{2} \mathbf{\Phi} \mathbb{E}\left\{\tilde{\mathbf{h}}_{i} \tilde{\mathbf{h}}_{i}^{H}\right\} \mathbf{\Phi}^{H} \tilde{\mathbf{H}}_{2}^{H}\right\} \overline{\mathbf{H}}_{2} \mathbf{\Phi} \overline{\mathbf{h}}_{k} \\
			=c_{k} c_{i} \delta \varepsilon_{k} \overline{\mathbf{h}}_{k}^{H} \mathbf{\Phi}^{H} \overline{\mathbf{H}}_{2}^{H} \mathbb{E}\left\{\tilde{\mathbf{H}}_{2} \tilde{\mathbf{H}}_{2}^{H}\right\} \overline{\mathbf{H}}_{2} \mathbf{\Phi} \overline{\mathbf{h}}_{k} \\
			=c_{k} c_{i} \delta \varepsilon_{k} M N\left|f_{k}(\mathbf{\Phi})\right|^{2}.
		\end{array}
	\end{align}
	
	Secondly, we consider the terms with $\omega=2$. When $\psi=1$, we have 
	\begin{align}\label{interference_16terms_5}
		\begin{array}{l}
			\mathbb{E}\left\{\left|e_{k 2} \sqrt{c_{k} \delta} \sqrt{c_{i} \delta \varepsilon_{i}} \,\tilde{\mathbf{h}}_{k}^{H} \mathbf{\Phi}^{H} \overline{\mathbf{H}}_{2}^{H} \overline{\mathbf{H}}_{2} \mathbf{\Phi} \overline{\mathbf{h}}_{i}\right|^{2}\right\} \\
			=e_{k 2}^{2} c_{k} c_{i} \delta^{2} \varepsilon_{i} \mathbb{E}\left\{\tilde{\mathbf{h}}_{k}^{H} \mathbf{\Phi}^{H} \overline{\mathbf{H}}_{2}^{H} \overline{\mathbf{H}}_{2} \mathbf{\Phi} \overline{\mathbf{h}}_{i} \overline{\mathbf{h}}_{i}^{H} \mathbf{\Phi}^{H} \overline{\mathbf{H}}_{2}^{H} \overline{\mathbf{H}}_{2} \mathbf{\Phi} \tilde{\mathbf{h}}_{k}\right\} \\
			=e_{k 2}^{2} c_{k} c_{i} \delta^{2} \varepsilon_{i} \operatorname{Tr}\left\{\overline{\mathbf{H}}_{2}^{H} \overline{\mathbf{H}}_{2} \mathbf{\Phi} \overline{\mathbf{h}}_{i} \overline{\mathbf{h}}_{i}^{H} \mathbf{\Phi}^{H} \overline{\mathbf{H}}_{2}^{H} \overline{\mathbf{H}}_{2}\right\} \\
			=e_{k 2}^{2} c_{k} c_{i} \delta^{2} \varepsilon_{i} \operatorname{Tr}\left\{\mathbf{a}_{N}^{H} \mathbf{\Phi} \overline{\mathbf{h}}_{i} \overline{\mathbf{h}}_{i}^{H} \mathbf{\Phi}^{H} \mathbf{a}_{N} \mathbf{a}_{M}^{H} \mathbf{a}_{M} \mathbf{a}_{N}^{H} \mathbf{a}_{N} \mathbf{a}_{M}^{H} \mathbf{a}_{M}\right\} \\
			=e_{k 2}^{2} c_{k} c_{i} \delta^{2} \varepsilon_{i} M^{2} N\left|f_{i}(\mathbf{\Phi})\right|^{2}.
		\end{array}
	\end{align}
	
	When $\psi=2$, we arrive at 
	\begin{align}\label{interference_16terms_6}
		\begin{array}{l}
			\mathbb{E}\left\{\left|e_{k 2} \sqrt{c_{k} \delta} \sqrt{c_{i} \delta} \,\tilde{\mathbf{h}}_{k}^{H} \mathbf{\Phi}^{H} \overline{\mathbf{H}}_{2}^{H} \overline{\mathbf{H}}_{2} \boldsymbol{\Phi} \tilde{\mathbf{h}}_{i}\right|^{2}\right\} \\
			=e_{k 2}^{2} c_{k} c_{i} \delta^{2} \mathbb{E}\left\{\tilde{\mathbf{h}}_{k}^{H} \mathbf{\Phi}^{H} \overline{\mathbf{H}}_{2}^{H} \overline{\mathbf{H}}_{2} \mathbf{\Phi} \mathbb{E}\left\{\tilde{\mathbf{h}}_{i} \tilde{\mathbf{h}}_{i}^{H}\right\} \mathbf{\Phi}^{H} \overline{\mathbf{H}}_{2}^{H} \overline{\mathbf{H}}_{2} \mathbf{\Phi} \tilde{\mathbf{h}}_{k}\right\} \\
			=e_{k 2}^{2} c_{k} c_{i} \delta^{2} \operatorname{Tr}\left\{\overline{\mathbf{H}}_{2}^{H} \overline{\mathbf{H}}_{2} \overline{\mathbf{H}}_{2}^{H} \overline{\mathbf{H}}_{2}\right\} \\
			=e_{k 2}^{2} c_{k} c_{i} \delta^{2} M^{2} N^{2}.
		\end{array}
	\end{align}
	
	When $\psi=3$, we get 
	\begin{align}\label{interference_16terms_7}
		\begin{array}{l}
			\mathbb{E}\left\{\left|e_{k 2} \sqrt{c_{k} \delta} \sqrt{c_{i} \varepsilon_{i}} \,\tilde{\mathbf{h}}_{k}^{H} \mathbf{\Phi}^{H} \overline{\mathbf{H}}_{2}^{H} \tilde{\mathbf{H}}_{2} \mathbf{\Phi} \overline{\mathbf{h}}_{i}\right|^{2}\right\} \\
			=e_{k 2}^{2} c_{k} c_{i} \delta \varepsilon_{i} \mathbb{E}\left\{\tilde{\mathbf{h}}_{k}^{H} \mathbf{\Phi}^{H} \overline{\mathbf{H}}_{2}^{H} \mathbb{E}\left\{\tilde{\mathbf{H}}_{2} \mathbf{\Phi} \overline{\mathbf{h}}_{i} \overline{\mathbf{h}}_{i}^{H} \mathbf{\Phi}^{H} \tilde{\mathbf{H}}_{2}^{H}\right\} \overline{\mathbf{H}}_{2} \mathbf{\Phi} \tilde{\mathbf{h}}_{k}\right\} \\
			=e_{k 2}^{2} c_{k} c_{i} \delta \varepsilon_{i} N \mathbb{E}\left\{\tilde{\mathbf{h}}_{k}^{H} \mathbf{\Phi}^{H} \overline{\mathbf{H}}_{2}^{H} \overline{\mathbf{H}}_{2} \boldsymbol{\Phi} \tilde{\mathbf{h}}_{k}\right\} \\
			=e_{k 2}^{2} c_{k} c_{i} \delta \varepsilon_{i} N \operatorname{Tr}\left\{\overline{\mathbf{H}}_{2}^{H} \overline{\mathbf{H}}_{2}\right\} \\
			=e_{k 2}^{2} c_{k} c_{i} \delta \varepsilon_{i} M N^{2}.
		\end{array}
	\end{align}
	
	When $\psi=4$, we have 
	\begin{align}\label{interference_16terms_8}
		\begin{array}{l}
			\mathbb{E}\left\{\left|e_{k 2} \sqrt{c_{k} \delta} \sqrt{c_{i}} \,\tilde{\mathbf{h}}_{k}^{H} \mathbf{\Phi}^{H} \overline{\mathbf{H}}_{2}^{H} \tilde{\mathbf{H}}_{2} \mathbf{\Phi} \tilde{\mathbf{h}}_{i}\right|^{2}\right\} \\
			=e_{k 2}^{2} c_{k} c_{i} \delta \mathbb{E}\left\{\tilde{\mathbf{h}}_{k}^{H} \mathbf{\Phi}^{H} \overline{\mathbf{H}}_{2}^{H} \mathbb{E}\left\{\tilde{\mathbf{H}}_{2} \boldsymbol{\Phi} \mathbb{E}\left\{\tilde{\mathbf{h}}_{i} \tilde{\mathbf{h}}_{i}^{H}\right\} \mathbf{\Phi}^{H} \tilde{\mathbf{H}}_{2}^{H}\right\} \overline{\mathbf{H}}_{2} \mathbf{\Phi} \tilde{\mathbf{h}}_{k}\right\} \\
			=e_{k 2}^{2} c_{k} c_{i} \delta \mathbb{E}\left\{\tilde{\mathbf{h}}_{k}^{H} \mathbf{\Phi}^{H} \overline{\mathbf{H}}_{2}^{H} \mathbb{E}\left\{\tilde{\mathbf{H}}_{2} \tilde{\mathbf{H}}_{2}^{H}\right\} \overline{\mathbf{H}}_{2} \mathbf{\Phi} \tilde{\mathbf{h}}_{k}\right\} \\
			=e_{k 2}^{2} c_{k} c_{i} \delta N \operatorname{Tr}\left\{\overline{\mathbf{H}}_{2}^{H} \overline{\mathbf{H}}_{2}\right\} \\
			=e_{k 2}^{2} c_{k} c_{i} \delta M N^{2}.
		\end{array}
	\end{align}

	Thirdly, we consider the terms with $\omega=3$. When $\psi=1$, using $\mathbf{A}_{k} \overline{\mathbf{H}}_{2}=e_{k2} \overline{\mathbf{H}}_{2} $, we have 
	\begin{align}\label{interference_16terms_9}
		\begin{array}{l}
			\mathbb{E}\left\{\left|\sqrt{c_{k} \varepsilon_{k}} \sqrt{c_{i} \delta \varepsilon_{i}} \, \overline{\mathbf{h}}_{k}^{H} \mathbf{\Phi}^{H} \tilde{\mathbf{H}}_{2}^{H} \mathbf{A}_{k}^{H} \overline{\mathbf{H}}_{2} \mathbf{\Phi} \overline{\mathbf{h}}_{i}\right|^{2}\right\} \\
			=c_{k} c_{i} \delta \varepsilon_{k} \varepsilon_{i} \overline{\mathbf{h}}_{k}^{H} \mathbf{\Phi}^{H} \mathbb{E}\left\{\tilde{\mathbf{H}}_{2}^{H} \mathbf{A}_{k}^{H} \overline{\mathbf{H}}_{2} \mathbf{\Phi} \overline{\mathbf{h}}_{i} \overline{\mathbf{h}}_{i}^{H} \mathbf{\Phi}^{H} \overline{\mathbf{H}}_{2}^{H} \mathbf{A}_{k} \tilde{\mathbf{H}}_{2}\right\} \mathbf{\Phi} \overline{\mathbf{h}}_{k} \\
			=c_{k} c_{i} \delta \varepsilon_{k} \varepsilon_{i} \overline{\mathbf{h}}_{k}^{H} \mathbf{\Phi}^{H} \operatorname{Tr}\left\{e_{k 2} \overline{\mathbf{H}}_{2} \mathbf{\Phi} \overline{\mathbf{h}}_{i} \overline{\mathbf{h}}_{i}^{H} \mathbf{\Phi}^{H} \overline{\mathbf{H}}_{2}^{H} e_{k 2}\right\} \mathbf{\Phi} \overline{\mathbf{h}}_{k} \\
			=e_{k 2}^{2} c_{k} c_{i} \delta \varepsilon_{k} \varepsilon_{i} M\left|f_{i}(\mathbf{\Phi})\right|^{2} \overline{\mathbf{h}}_{k}^{H} \mathbf{\Phi}^{H} \mathbf{\Phi} \overline{\mathbf{h}}_{k} \\
			=e_{k 2}^{2} c_{k} c_{i} \delta \varepsilon_{k} \varepsilon_{i} M N\left|f_{i}(\mathbf{\Phi})\right|^{2}.
		\end{array}
	\end{align}
	
	When $\psi=2$, we arrive at 
	\begin{align}\label{interference_16terms_10}
		\begin{array}{l}
			\mathbb{E}\left\{\left|\sqrt{c_{k} \varepsilon_{k}} \sqrt{c_{i} \delta} \,\overline{\mathbf{h}}_{k}^{H} \mathbf{\Phi}^{H} \tilde{\mathbf{H}}_{2}^{H} \mathbf{A}_{k}^{H} \overline{\mathbf{H}}_{2} \mathbf{\Phi} \tilde{\mathbf{h}}_{i}\right|^{2}\right\} \\
			=c_{k} c_{i} \delta \varepsilon_{k} \overline{\mathbf{h}}_{k}^{H} \mathbf{\Phi}^{H} \mathbb{E}\left\{\tilde{\mathbf{H}}_{2}^{H} \mathbf{A}_{k}^{H} \overline{\mathbf{H}}_{2} \mathbf{\Phi} \mathbb{E}\left\{\tilde{\mathbf{h}}_{i} \tilde{\mathbf{h}}_{i}^{H}\right\} \mathbf{\Phi}^{H} \overline{\mathbf{H}}_{2}^{H} \mathbf{A}_{k} \tilde{\mathbf{H}}_{2}\right\} \mathbf{\Phi} \overline{\mathbf{h}}_{k} \\
			=e_{k 2}^{2} c_{k} c_{i} \delta \varepsilon_{k} \overline{\mathbf{h}}_{k}^{H} \mathbf{\Phi}^{H} \operatorname{Tr}\left\{\overline{\mathbf{H}}_{2} \overline{\mathbf{H}}_{2}^{H}\right\} \mathbf{\Phi} \overline{\mathbf{h}}_{k} \\
			=e_{k 2}^{2} c_{k} c_{i} \delta \varepsilon_{k} M N^{2}.
		\end{array}
	\end{align}
	
	When $\psi=3$, using Lemma \ref{lemma_HAHWHAH}, we have 
	\begin{align}\label{interference_16terms_11}
		\begin{array}{l}
			\mathbb{E}\left\{\left|\sqrt{c_{k} \varepsilon_{k}} \sqrt{c_{i} \varepsilon_{i}} \,\overline{\mathbf{h}}_{k}^{H} \mathbf{\Phi}^{H} \tilde{\mathbf{H}}_{2}^{H} \mathbf{A}_{k}^{H} \tilde{\mathbf{H}}_{2} \mathbf{\Phi} \overline{\mathbf{h}}_{i}\right|^{2}\right\} \\
			=c_{k} c_{i} \varepsilon_{k} \varepsilon_{i} \overline{\mathbf{h}}_{k}^{H} \mathbf{\Phi}^{H} \mathbb{E}\left\{\tilde{\mathbf{H}}_{2}^{H} \mathbf{A}_{k}^{H} \tilde{\mathbf{H}}_{2} \mathbf{\Phi} \overline{\mathbf{h}}_{i} \overline{\mathbf{h}}_{i}^{H} \mathbf{\Phi}^{H} \tilde{\mathbf{H}}_{2}^{H} \mathbf{A}_{k} \tilde{\mathbf{H}}_{2}\right\} \mathbf{\Phi} \overline{\mathbf{h}}_{k} \\
			=c_{k} c_{i} \varepsilon_{k} \varepsilon_{i} \overline{\mathbf{h}}_{k}^{H} \mathbf{\Phi}^{H}\left(e_{k 1}^{2} M^{2} \mathbf{\Phi} \overline{\mathbf{h}}_{i} \overline{\mathbf{h}}_{i}^{H} \mathbf{\Phi}^{H}+e_{k 3} M\operatorname{Tr}\left\{\mathbf{\Phi} \overline{\mathbf{h}}_{i} \overline{\mathbf{h}}_{i}^{H} \mathbf{\Phi}^{H}\right\} \mathbf{I}_{N}\right) \mathbf{\Phi} \overline{\mathbf{h}}_{k} \\
			=c_{k} c_{i} \varepsilon_{k} \varepsilon_{i}\left(e_{k 1}^{2} M^{2} \overline{\mathbf{h}}_{k}^{H} \overline{\mathbf{h}}_{i} \overline{\mathbf{h}}_{i}^{H} \overline{\mathbf{h}}_{k}+e_{k 3} M\operatorname{Tr}\left\{\overline{\mathbf{h}}_{i} \overline{\mathbf{h}}_{i}^{H}\right\} \overline{\mathbf{h}}_{k}^{H} \overline{\mathbf{h}}_{k}\right) \\
			=c_{k} c_{i} \varepsilon_{k} \varepsilon_{i}\left(e_{k 1}^{2} M^{2}\left|\overline{\mathbf{h}}_{k}^{H} \overline{\mathbf{h}}_{i}\right|^{2}+e_{k 3} M N^{2}\right).
		\end{array}
	\end{align}
	
	When $\psi=4$, using Lemma \ref{lemma_HAHWHAH} with ${\bf W}={\bf I}_N$, we get 
	\begin{align}\label{interference_16terms_12}
		\begin{array}{l}
			\mathbb{E}\left\{\left|\sqrt{c_{k} \varepsilon_{k}} \sqrt{c_{i}} \,\overline{\mathbf{h}}_{k}^{H} \mathbf{\Phi}^{H} \tilde{\mathbf{H}}_{2}^{H} \mathbf{A}_{k}^{H} \tilde{\mathbf{H}}_{2} \mathbf{\Phi} \tilde{\mathbf{h}}_{i}\right|^{2}\right\} \\
			=c_{k} c_{i} \varepsilon_{k} \overline{\mathbf{h}}_{k}^{H} \mathbf{\Phi}^{H} \mathbb{E}\left\{\tilde{\mathbf{H}}_{2}^{H} \mathbf{A}_{k}^{H} \tilde{\mathbf{H}}_{2} \mathbf{\Phi} \mathbb{E}\left\{\tilde{\mathbf{h}}_{i} \tilde{\mathbf{h}}_{i}^{H}\right\} \mathbf{\Phi}^{H} \tilde{\mathbf{H}}_{2}^{H} \mathbf{A}_{k} \tilde{\mathbf{H}}_{2}\right\} \mathbf{\Phi} \overline{\mathbf{h}}_{k} \\
			=c_{k} c_{i} \varepsilon_{k} \overline{\mathbf{h}}_{k}^{H} \mathbf{\Phi}^{H} \mathbb{E}\left\{\tilde{\mathbf{H}}_{2}^{H} \mathbf{A}_{k}^{H} \tilde{\mathbf{H}}_{2} \tilde{\mathbf{H}}_{2}^{H} \mathbf{A}_{k} \tilde{\mathbf{H}}_{2}\right\} \mathbf{\Phi} \overline{\mathbf{h}}_{k} \\
			=c_{k} c_{i} \varepsilon_{k}\left(e_{k 1}^{2} M^{2}+e_{k 3} M N\right) \overline{\mathbf{h}}_{k}^{H} \mathbf{\Phi}^{H} \mathbf{I}_{N} \mathbf{\Phi} \overline{\mathbf{h}}_{k} \\
			=c_{k} c_{i} \varepsilon_{k}\left(e_{k 1}^{2} M^{2}+e_{k 3} M N\right) N.
		\end{array}
	\end{align}
	
	Fourthly, we consider the terms with $\omega=4$. When $\psi=1$, we have 
	\begin{align}\label{interference_16terms_13}
		\begin{array}{l}
			\mathbb{E}\left\{\left|\sqrt{c_{k}} \sqrt{c_{i} \delta \varepsilon_{i}} \,\tilde{\mathbf{h}}_{k}^{H} \mathbf{\Phi}^{H} \tilde{\mathbf{H}}_{2}^{H} \mathbf{A}_{k}^{H} \overline{\mathbf{H}}_{2} \mathbf{\Phi} \overline{\mathbf{h}}_{i}\right|^{2}\right\} \\
			=c_{k} c_{i} \delta \varepsilon_{i} \mathbb{E}\left\{\tilde{\mathbf{h}}_{k}^{H} \mathbf{\Phi}^{H} \mathbb{E}\left\{\tilde{\mathbf{H}}_{2}^{H} \mathbf{A}_{k}^{H} \overline{\mathbf{H}}_{2} \mathbf{\Phi} \overline{\mathbf{h}}_{i} \overline{\mathbf{h}}_{i}^{H} \mathbf{\Phi}^{H} \overline{\mathbf{H}}_{2}^{H} \mathbf{A}_{k} \tilde{\mathbf{H}}_{2}\right\} \mathbf{\Phi} \tilde{\mathbf{h}}_{k}\right\} \\
			=e_{k 2}^{2} c_{k} c_{i} \delta \varepsilon_{i} \mathbb{E}\left\{\tilde{\mathbf{h}}_{k}^{H} \mathbf{\Phi}^{H} \operatorname{Tr}\left\{\overline{\mathbf{H}}_{2} \mathbf{\Phi} \overline{\mathbf{h}}_{i} \overline{\mathbf{h}}_{i}^{H} \mathbf{\Phi}^{H} \overline{\mathbf{H}}_{2}^{H}\right\} \mathbf{\Phi} \tilde{\mathbf{h}}_{k}\right\} \\
			=e_{k 2}^{2} c_{k} c_{i} \delta \varepsilon_{i} M\left|f_{i}(\mathbf{\Phi})\right|^{2} \mathbb{E}\left\{\tilde{\mathbf{h}}_{k}^{H} \tilde{\mathbf{h}}_{k}\right\} \\
			=e_{k 2}^{2} c_{k} c_{i} \delta \varepsilon_{i} M N\left|f_{i}(\mathbf{\Phi})\right|^{2}.
		\end{array}
	\end{align}
	
	When $\psi=2$, we have 
	\begin{align}\label{interference_16terms_14}
		\begin{array}{l}
			\mathbb{E}\left\{\left|\sqrt{c_{k}} \sqrt{c_{i} \delta} \,\tilde{\mathbf{h}}_{k}^{H} \mathbf{\Phi}^{H} \tilde{\mathbf{H}}_{2}^{H} \mathbf{A}_{k}^{H} \overline{\mathbf{H}}_{2} \mathbf{\Phi} \tilde{\mathbf{h}}_{i}\right|^{2}\right\} \\
			=c_{k} c_{i} \delta \mathbb{E}\left\{\tilde{\mathbf{h}}_{k}^{H} \mathbf{\Phi}^{H} \mathbb{E}\left\{\tilde{\mathbf{H}}_{2}^{H} \mathbf{A}_{k}^{H} \overline{\mathbf{H}}_{2} \mathbf{\Phi} \mathbb{E}\left\{\tilde{\mathbf{h}}_{i} \tilde{\mathbf{h}}_{i}^{H}\right\} \mathbf{\Phi}^{H} \overline{\mathbf{H}}_{2}^{H} \mathbf{A}_{k} \tilde{\mathbf{H}}_{2}\right\} \mathbf{\Phi} \tilde{\mathbf{h}}_{k}\right\} \\
			=e_{k 2}^{2} c_{k} c_{i} \delta \mathbb{E}\left\{\tilde{\mathbf{h}}_{k}^{H} \mathbf{\Phi}^{H} \mathbb{E}\left\{\tilde{\mathbf{H}}_{2}^{H} \overline{\mathbf{H}}_{2} \overline{\mathbf{H}}_{2}^{H} \tilde{\mathbf{H}}_{2}\right\} \mathbf{\Phi} \tilde{\mathbf{h}}_{k}\right\} \\
			=e_{k 2}^{2} c_{k} c_{i} \delta M N \mathbb{E}\left\{\tilde{\mathbf{h}}_{k}^{H} \tilde{\mathbf{h}}_{k}\right\} \\
			=e_{k 2}^{2} c_{k} c_{i} \delta M N^{2}.
		\end{array}
	\end{align}
	
	When $\psi=3$, we have 
	\begin{align}\label{interference_16terms_15}
		\begin{array}{l}
			\mathbb{E}\left\{\left|\sqrt{c_{k}} \sqrt{c_{i} \varepsilon_{i}} \,\tilde{\mathbf{h}}_{k}^{H} \mathbf{\Phi}^{H} \tilde{\mathbf{H}}_{2}^{H} \mathbf{A}_{k}^{H} \tilde{\mathbf{H}}_{2} \mathbf{\Phi} \overline{\mathbf{h}}_{i}\right|^{2}\right\} \\
			=c_{k} c_{i} \varepsilon_{i} \mathbb{E}\left\{\tilde{\mathbf{h}}_{k}^{H} \mathbf{\Phi}^{H} \mathbb{E}\left\{\tilde{\mathbf{H}}_{2}^{H} \mathbf{A}_{k}^{H} \tilde{\mathbf{H}}_{2} \mathbf{\Phi} \overline{\mathbf{h}}_{i} \overline{\mathbf{h}}_{i}^{H} \mathbf{\Phi}^{H} \tilde{\mathbf{H}}_{2}^{H} \mathbf{A}_{k} \tilde{\mathbf{H}}_{2}\right\} \mathbf{\Phi} \tilde{\mathbf{h}}_{k}\right\} \\
			=c_{k} c_{i} \varepsilon_{i} \mathbb{E}\left\{\tilde{\mathbf{h}}_{k}^{H} \mathbf{\Phi}^{H}\left(e_{k 1}^{2} M^{2} \mathbf{\Phi} \overline{\mathbf{h}}_{i} \overline{\mathbf{h}}_{i}^{H} \mathbf{\Phi}^{H}+e_{k 3} M\operatorname{Tr}\left\{\mathbf{\Phi} \overline{\mathbf{h}}_{i} \overline{\mathbf{h}}_{i}^{H} \mathbf{\Phi}^{H}\right\} \mathbf{I}_{N}\right) \mathbf{\Phi} \tilde{\mathbf{h}}_{k}\right\} \\
			=c_{k} c_{i} \varepsilon_{i} \mathbb{E}\left\{       e_{k 1}^{2} M^{2} \tilde{\mathbf{h}}_{k}^{H} \overline{\mathbf{h}}_{i} \overline{\mathbf{h}}_{i}^{H} \tilde{\mathbf{h}}_{k}+e_{k 3} M N \tilde{\mathbf{h}}_{k}^{H} \tilde{\mathbf{h}}_{k}      \right\} \\
			=c_{k} c_{i} \varepsilon_{i}\left(e_{k 1}^{2} M^{2} N+e_{k 3} M N^{2}\right).
		\end{array}
	\end{align}
	
	When $\psi=4$, we get 
	\begin{align}\label{interference_16terms_16}
		\begin{array}{l}
			\mathbb{E}\left\{\left|\sqrt{c_{k}} \sqrt{c_{i}} \tilde{\mathbf{h}}_{k}^{H} \mathbf{\Phi}^{H} \tilde{\mathbf{H}}_{2}^{H} \mathbf{A}_{k}^{H} \tilde{\mathbf{H}}_{2} \mathbf{\Phi} \tilde{\mathbf{h}}_{i}\right|^{2}\right\} \\
			=c_{k} c_{i} \mathbb{E}\left\{\tilde{\mathbf{h}}_{k}^{H} \mathbf{\Phi}^{H} \mathbb{E}\left\{\tilde{\mathbf{H}}_{2}^{H} \mathbf{A}_{k}^{H} \tilde{\mathbf{H}}_{2} \mathbf{\Phi} \mathbb{E}\left\{\tilde{\mathbf{h}}_{i} \tilde{\mathbf{h}}_{i}^{H}\right\} \mathbf{\Phi}^{H} \tilde{\mathbf{H}}_{2}^{H} \mathbf{A}_{k} \tilde{\mathbf{H}}_{2}\right\} \mathbf{\Phi} \tilde{\mathbf{h}}_{k}\right\} \\
			=c_{k} c_{i} \mathbb{E}\left\{\tilde{\mathbf{h}}_{k}^{H} \mathbf{\Phi}^{H} \mathbb{E}\left\{\tilde{\mathbf{H}}_{2}^{H} \mathbf{A}_{k}^{H} \tilde{\mathbf{H}}_{2} \tilde{\mathbf{H}}_{2}^{H} \mathbf{A}_{k} \tilde{\mathbf{H}}_{2}\right\} \mathbf{\Phi} \tilde{\mathbf{h}}_{k}\right\} \\
			=c_{k} c_{i} \mathbb{E}\left\{\left(e_{k 1}^{2} M^{2}+e_{k 3} M N\right) \tilde{\mathbf{h}}_{k}^{H} \tilde{\mathbf{h}}_{k}\right\} \\
			=c_{k} c_{i}\left(e_{k 1}^{2} M^{2} N+e_{k 3} M N^{2}\right).
		\end{array}
	\end{align}
	
	The calculation of the expectations of the $16$ modulus-square terms in (\ref{interference_expanded_sum}) is completed. Then, we focus on the remaining cross-terms in (\ref{interference_expanded_sum}). Even though the total number of cross-terms is $16\times 15$, only a few terms are non-zero. To help identify the non-zero cross-terms, we expand $ \underline{\hat{\mathbf{q}}}_{k}^{H} \underline{\mathbf{q}}_{i} $ as
	\begin{align}\label{Expansion_ki}
		\begin{array}{l}
			\underline{\hat{\mathbf{q}}}_{k}^{H} \underline{\mathbf{q}}_{i} 
			=\sum_{\omega=1}^{4} \sum_{\psi=1}^{4}\left(\hat{\mathbf{q}}_{k}^{\omega}\right)^{H} \mathbf{q}_{i}^{\psi} \\
			=\eta_{11} \overline{\mathbf{h}}_{k}^{H} \mathbf{\Phi}^{H} \overline{\mathbf{H}}_{2}^{H} \overline{\mathbf{H}}_{2} \mathbf{\Phi} \overline{\mathbf{h}}_{i}+\eta_{12} \overline{\mathbf{h}}_{k}^{H} \mathbf{\Phi}^{H} \overline{\mathbf{H}}_{2}^{H} \overline{\mathbf{H}}_{2} \mathbf{\Phi} \tilde{\mathbf{h}}_{i}+\eta_{13} \overline{\mathbf{h}}_{k}^{H} \mathbf{\Phi}^{H} \overline{\mathbf{H}}_{2}^{H} \tilde{\mathbf{H}}_{2} \mathbf{\Phi} \overline{\mathbf{h}}_{i}+\eta_{14} \overline{\mathbf{h}}_{k}^{H} \mathbf{\Phi}^{H} \overline{\mathbf{H}}_{2}^{H} \tilde{\mathbf{H}}_{2} \mathbf{\Phi} \tilde{\mathbf{h}}_{i} \\
			+\eta_{21} \tilde{\mathbf{h}}_{k}^{H} \mathbf{\Phi}^{H} \overline{\mathbf{H}}_{2}^{H} \overline{\mathbf{H}}_{2} \mathbf{\Phi} \overline{\mathbf{h}}_{i}+\eta_{22} \tilde{\mathbf{h}}_{k}^{H} \mathbf{\Phi}^{H} \overline{\mathbf{H}}_{2}^{H} \overline{\mathbf{H}}_{2} \mathbf{\Phi} \tilde{\mathbf{h}}_{i}+\eta_{23} \tilde{\mathbf{h}}_{k}^{H} \mathbf{\Phi}^{H} \overline{\mathbf{H}}_{2}^{H} \tilde{\mathbf{H}}_{2} \mathbf{\Phi} \overline{\mathbf{h}}_{i}+\eta_{24} \tilde{\mathbf{h}}_{k}^{H} \mathbf{\Phi}^{H} \overline{\mathbf{H}}_{2}^{H} \tilde{\mathbf{H}}_{2} \mathbf{\Phi} \tilde{\mathbf{h}}_{i} \\
			+\eta_{31} \overline{\mathbf{h}}_{k}^{H} \mathbf{\Phi}^{H} \tilde{\mathbf{H}}_{2}^{H} \mathbf{A}_{k}^{H} \overline{\mathbf{H}}_{2} \mathbf{\Phi} \overline{\mathbf{h}}_{i}+\eta_{32} \overline{\mathbf{h}}_{k}^{H} \mathbf{\Phi}^{H} \tilde{\mathbf{H}}_{2}^{H} \mathbf{A}_{k}^{H} \overline{\mathbf{H}}_{2} \mathbf{\Phi} \tilde{\mathbf{h}}_{i} \\
			+\eta_{33} \overline{\mathbf{h}}_{k}^{H} \mathbf{\Phi}^{H} \tilde{\mathbf{H}}_{2}^{H} \mathbf{A}_{k}^{H} \tilde{\mathbf{H}}_{2} \mathbf{\Phi} \overline{\mathbf{h}}_{i}+\eta_{34} \overline{\mathbf{h}}_{k}^{H} \mathbf{\Phi}^{H} \tilde{\mathbf{H}}_{2}^{H} \mathbf{A}_{k}^{H} \tilde{\mathbf{H}}_{2} \mathbf{\Phi} \tilde{\mathbf{h}}_{i} \\
			+\eta_{41} \tilde{\mathbf{h}}_{k}^{H} \mathbf{\Phi}^{H} \tilde{\mathbf{H}}_{2}^{H} \mathbf{A}_{k}^{H} \overline{\mathbf{H}}_{2} \mathbf{\Phi} \overline{\mathbf{h}}_{i}+\eta_{42} \tilde{\mathbf{h}}_{k}^{H} \mathbf{\Phi}^{H} \tilde{\mathbf{H}}_{2}^{H} \mathbf{A}_{k}^{H} \overline{\mathbf{H}}_{2} \mathbf{\Phi} \tilde{\mathbf{h}}_{i} \\
			+\eta_{43} \tilde{\mathbf{h}}_{k}^{H} \mathbf{\Phi}^{H} \tilde{\mathbf{H}}_{2}^{H} \mathbf{A}_{k}^{H} \tilde{\mathbf{H}}_{2} \mathbf{\Phi} \overline{\mathbf{h}}_{i}+\eta_{44} \tilde{\mathbf{h}}_{k}^{H} \mathbf{\Phi}^{H} \tilde{\mathbf{H}}_{2}^{H} \mathbf{A}_{k}^{H} \tilde{\mathbf{H}}_{2} \mathbf{\Phi} \tilde{\mathbf{h}}_{i}.
		\end{array}
	\end{align}
	
	For brevity, we use the notation $\eta_{11}$ - $ \eta_{44}$ to identify the coefficients (path-loss factors and Rician factors) in front of the product of vectors and matrices, since these coefficients are deterministic and do not determine whether the expectations of the cross-terms are zero or not. Then, we can calculate the cross-terms in (\ref{interference_expanded_sum}) by calculating the expectation of the product of one term in (\ref{Expansion_ki}) with the conjugate transpose of another term in (\ref{Expansion_ki}). Therefore, by exploiting Lemma \ref{lemma4_HH_equal_0}, the independence and the zero-mean properties of $\tilde{\bf H}_2$, $\tilde{\bf h}_k$ and $\tilde{\bf h}_i$, we find that only $8$ cross-terms have a non-zero expectation. They can be combined as
	\begin{align}\label{interference_cross_sum}
		\begin{array}{l}
			\sum_{\omega 1, \psi 1, \omega 2, \psi 2, \atop(\omega 1, \psi 1) \neq(\omega 2, \psi 2)}^{4} \mathbb{E}\left\{\left(\left(\hat{\mathbf{q}}_{k}^{\omega 1}\right)^{H} \mathbf{q}_{i}^{\psi 1}\right)\left(\left(\hat{\mathbf{q}}_{k}^{\omega 2}\right)^{H} \mathbf{q}_{i}^{\psi 2}\right)^{H}\right\} \\
			=2 \operatorname{Re}\left\{\mathbb{E}\left\{\left(\left(\hat{\mathbf{q}}_{k}^{1}\right)^{H} \mathbf{q}_{i}^{1}\right)\left(\left(\hat{\mathbf{q}}_{k}^{3}\right)^{H} \mathbf{q}_{i}^{3}\right)^{H}\right\}\right\}
			+2 \operatorname{Re}\left\{\mathbb{E}\left\{\left(\left(\hat{\mathbf{q}}_{k}^{1}\right)^{H} \mathbf{q}_{i}^{2}\right)\left(\left(\hat{\mathbf{q}}_{k}^{3}\right)^{H} \mathbf{q}_{i}^{4}\right)^{H}\right\}\right\} \\
			+2 \operatorname{Re}\left\{\mathbb{E}\left\{\left(\left(\hat{\mathbf{q}}_{k}^{2}\right)^{H} \mathbf{q}_{i}^{1}\right)\left(\left(\hat{\mathbf{q}}_{k}^{4}\right)^{H} \mathbf{q}_{i}^{3}\right)^{H}\right\}\right\}
			+2 \operatorname{Re}\left\{\mathbb{E}\left\{\left(\left(\hat{\mathbf{q}}_{k}^{2}\right)^{H} \mathbf{q}_{i}^{2}\right)\left(\left(\hat{\mathbf{q}}_{k}^{4}\right)^{H} \mathbf{q}_{i}^{4}\right)^{H}\right\}\right\}.
		\end{array}
	\end{align}
	
	Then, we calculate these $4$ terms in (\ref{interference_cross_sum}) one by one. Using ${\rm Tr}\left\{{\bf A}_k\right\}=Me_{k1}$, the first cross-term is derived as
	\begin{align}\label{interference_4cross_1}
		\begin{array}{l}
			2 \operatorname{Re}\left\{\mathbb{E}\left\{\left(\left(\hat{\mathbf{q}}_{k}^{1}\right)^{H} \mathbf{q}_{i}^{1}\right)\left(\left(\hat{\mathbf{q}}_{k}^{3}\right)^{H} \mathbf{q}_{i}^{3}\right)^{H}\right\}\right\}\\
			=2 \operatorname{Re}\left\{\mathbb{E}\left\{\sqrt{c_{k} \delta \varepsilon_{k}} \sqrt{c_{i} \delta \varepsilon_{i}}\, \overline{\mathbf{h}}_{k}^{H} \mathbf{\Phi}^{H} \overline{\mathbf{H}}_{2}^{H} \overline{\mathbf{H}}_{2} \boldsymbol{\Phi} \overline{\mathbf{h}}_{i}\left(\sqrt{c_{k} \varepsilon_{k}} \sqrt{c_{i} \varepsilon_{i}} \, \overline{\mathbf{h}}_{k}^{H} \mathbf{\Phi}^{H} \tilde{\mathbf{H}}_{2}^{H} \mathbf{A}_{k}^{H} \tilde{\mathbf{H}}_{2} \mathbf{\Phi} \overline{\mathbf{h}}_{i}\right)^{H}\right\}\right\} \\
			=2 c_{k} c_{i} \delta \varepsilon_{k} \varepsilon_{i} \operatorname{Re}\left\{\overline{\mathbf{h}}_{k}^{H} \mathbf{\Phi}^{H} \overline{\mathbf{H}}_{2}^{H} \overline{\mathbf{H}}_{2} \mathbf{\Phi} \overline{\mathbf{h}}_{i} \overline{\mathbf{h}}_{i}^{H} \mathbf{\Phi}^{H} \mathbb{E}\left\{\tilde{\mathbf{H}}_{2}^{H} \mathbf{A}_{k} \tilde{\mathbf{H}}_{2}\right\} \mathbf{\Phi} \overline{\mathbf{h}}_{k}\right\} \\
			=2 c_{k} c_{i} \delta \varepsilon_{k} \varepsilon_{i} e_{k1} M \operatorname{Re}\left\{\overline{\mathbf{h}}_{k}^{H} \mathbf{\Phi}^{H} \overline{\mathbf{H}}_{2}^{H} \overline{\mathbf{H}}_{2} \mathbf{\Phi} \overline{\mathbf{h}}_{i} \overline{\mathbf{h}}_{i}^{H} \overline{\mathbf{h}}_{k}\right\} \\
			=2 c_{k} c_{i} \delta \varepsilon_{k} \varepsilon_{i} e_{k1} M \operatorname{Re}\left\{\overline{\mathbf{h}}_{k}^{H} \mathbf{\Phi}^{H} \mathbf{a}_{N} \mathbf{a}_{M}^{H} \mathbf{a}_{M} \mathbf{a}_{N}^{H} \mathbf{\Phi} \overline{\mathbf{h}}_{i} \overline{\mathbf{h}}_{i}^{H} \overline{\mathbf{h}}_{k}\right\} \\
			=2 c_{k} c_{i} \delta \varepsilon_{k} \varepsilon_{i} e_{k 1} M^{2} \operatorname{Re}\left\{f_{k}^{H}(\mathbf{\Phi}) f_{i}(\mathbf{\Phi}) \overline{\mathbf{h}}_{i}^{H} \overline{\mathbf{h}}_{k}\right\}.
		\end{array}
	\end{align}
	
	Note that the real operator ${\rm Re}\left\{\cdot\right\}$ will be omitted for the calculation of the 2nd, 3rd, and 4th cross-terms, since the results derived will show that they only have real parts. Then, the second cross-term is
	\begin{align}\label{interference_4cross_2}
		\begin{array}{l}
			2 \operatorname{Re}\left\{\mathbb{E}\left\{\left(\left(\hat{\mathbf{q}}_{k}^{1}\right)^{H} \mathbf{q}_{i}^{2}\right)\left(\left(\hat{\mathbf{q}}_{k}^{3}\right)^{H} \mathbf{q}_{i}^{4}\right)^{H}\right\}\right\}\\
			=2 \mathbb{E}\left\{\sqrt{c_{k} \delta \varepsilon_{k}} \sqrt{c_{i} \delta}\, \overline{\mathbf{h}}_{k}^{H} \mathbf{\Phi}^{H} \overline{\mathbf{H}}_{2}^{H} \overline{\mathbf{H}}_{2} \mathbf{\Phi} \tilde{\mathbf{h}}_{i}\left(\sqrt{c_{k} \varepsilon_{k}} \sqrt{c_{i}} \,\overline{\mathbf{h}}_{k}^{H} \mathbf{\Phi}^{H} \tilde{\mathbf{H}}_{2}^{H} \mathbf{A}_{k}^{H} \tilde{\mathbf{H}}_{2} \mathbf{\Phi} \tilde{\mathbf{h}}_{i}\right)^{H}\right\} \\
			=2 c_{k} c_{i} \delta \varepsilon_{k} \overline{\mathbf{h}}_{k}^{H} \mathbf{\Phi}^{H} \overline{\mathbf{H}}_{2}^{H} \overline{\mathbf{H}}_{2} \mathbf{\Phi} \mathbb{E}\left\{\tilde{\mathbf{h}}_{i} \tilde{\mathbf{h}}_{i}^{H}\right\} \mathbf{\Phi}^{H} \mathbb{E}\left\{\tilde{\mathbf{H}}_{2}^{H} \mathbf{A}_{k} \tilde{\mathbf{H}}_{2}\right\} \mathbf{\Phi} \overline{\mathbf{h}}_{k} \\
			=2 c_{k} c_{i} \delta \varepsilon_{k} e_{k1} M \overline{\mathbf{h}}_{k}^{H} \mathbf{\Phi}^{H} \overline{\mathbf{H}}_{2}^{H} \overline{\mathbf{H}}_{2} \mathbf{\Phi} \overline{\mathbf{h}}_{k} \\
			=2 c_{k} c_{i} \delta \varepsilon_{k} e_{k 1} M^{2}\left|f_{k}(\mathbf{\Phi})\right|^{2}.
		\end{array}
	\end{align}
	
	The third cross-term is
	\begin{align}\label{interference_4cross_3}
		\begin{array}{l}
			2 \operatorname{Re}\left\{\mathbb{E}\left\{\left(\left(\hat{\mathbf{q}}_{k}^{2}\right)^{H} \mathbf{q}_{i}^{1}\right)\left(\left(\hat{\mathbf{q}}_{k}^{4}\right)^{H} \mathbf{q}_{i}^{3}\right)^{H}\right\}\right\}\\
			=2 \mathbb{E}\left\{e_{k 2} \sqrt{c_{k} \delta} \sqrt{c_{i} \delta \varepsilon_{i}} \,\tilde{\mathbf{h}}_{k}^{H} \mathbf{\Phi}^{H} \overline{\mathbf{H}}_{2}^{H} \overline{\mathbf{H}}_{2} \mathbf{\Phi} \overline{\mathbf{h}}_{i}\left(\sqrt{c_{k}} \sqrt{c_{i} \varepsilon_{i}} \,\tilde{\mathbf{h}}_{k}^{H} \mathbf{\Phi}^{H} \tilde{\mathbf{H}}_{2}^{H} \mathbf{A}_{k}^{H} \tilde{\mathbf{H}}_{2} \mathbf{\Phi} \overline{\mathbf{h}}_{i}\right)^{H}\right\} \\
			=2 c_{k} c_{i} \delta \varepsilon_{i} e_{k 2} \mathbb{E}\left\{\tilde{\mathbf{h}}_{k}^{H} \mathbf{\Phi}^{H} \overline{\mathbf{H}}_{2}^{H} \overline{\mathbf{H}}_{2} \mathbf{\Phi} \overline{\mathbf{h}}_{i} \overline{\mathbf{h}}_{i}^{H} \mathbf{\Phi}^{H} \mathbb{E}\left\{\tilde{\mathbf{H}}_{2}^{H} \mathbf{A}_{k} \tilde{\mathbf{H}}_{2}\right\} \mathbf{\Phi} \tilde{\mathbf{h}}_{k}\right\} \\
			=2 c_{k} c_{i} \delta \varepsilon_{i} e_{k 1} e_{k 2} M \mathbb{E}\left\{\tilde{\mathbf{h}}_{k}^{H} \mathbf{\Phi}^{H} \overline{\mathbf{H}}_{2}^{H} \overline{\mathbf{H}}_{2} \mathbf{\Phi} \overline{\mathbf{h}}_{i} \overline{\mathbf{h}}_{i}^{H} \tilde{\mathbf{h}}_{k}\right\} \\
			=2 c_{k} c_{i} \delta \varepsilon_{i} e_{k 1} e_{k 2} M\operatorname{Tr}\left\{\overline{\mathbf{h}}_{i}^{H} \mathbf{\Phi}^{H} \overline{\mathbf{H}}_{2}^{H} \overline{\mathbf{H}}_{2} \mathbf{\Phi} \overline{\mathbf{h}}_{i}\right\} \\
			=2 c_{k} c_{i} \delta \varepsilon_{i} e_{k 1} e_{k 2} M^{2}\left|f_{i}(\mathbf{\Phi})\right|^{2}.
		\end{array}
	\end{align}
	
	The fourth cross-term is
	\begin{align}\label{interference_4cross_4}
		\begin{array}{l}
			2 \operatorname{Re}\left\{\mathbb{E}\left\{\left(\left(\hat{\mathbf{q}}_{k}^{2}\right)^{H} \mathbf{q}_{i}^{2}\right)\left(\left(\hat{\mathbf{q}}_{k}^{4}\right)^{H} \mathbf{q}_{i}^{4}\right)^{H}\right\}\right\}\\
			=2 \mathbb{E}\left\{e_{k 2} \sqrt{c_{k} \delta} \sqrt{c_{i} \delta} \,\tilde{\mathbf{h}}_{k}^{H} \mathbf{\Phi}^{H} \overline{\mathbf{H}}_{2}^{H} \overline{\mathbf{H}}_{2} \mathbf{\Phi} \tilde{\mathbf{h}}_{i}\left(\sqrt{c_{k}} \sqrt{c_{i}}\, \tilde{\mathbf{h}}_{k}^{H} \mathbf{\Phi}^{H} \tilde{\mathbf{H}}_{2}^{H} \mathbf{A}_{k}^{H} \tilde{\mathbf{H}}_{2} \mathbf{\Phi} \tilde{\mathbf{h}}_{i}\right)^{H}\right\} \\
			=2 c_{k} c_{i} \delta e_{k 2} \mathbb{E}\left\{\tilde{\mathbf{h}}_{k}^{H} \mathbf{\Phi}^{H} \overline{\mathbf{H}}_{2}^{H} \overline{\mathbf{H}}_{2} \mathbf{\Phi} \mathbb{E}\left\{\tilde{\mathbf{h}}_{i} \tilde{\mathbf{h}}_{i}^{H}\right\} \mathbf{\Phi}^{H} \mathbb{E}\left\{\tilde{\mathbf{H}}_{2}^{H} \mathbf{A}_{k} \tilde{\mathbf{H}}_{2}\right\} \mathbf{\Phi} \tilde{\mathbf{h}}_{k}\right\} \\
			=2 c_{k} c_{i} \delta e_{k 1} e_{k 2} M \mathbb{E}\left\{\tilde{\mathbf{h}}_{k}^{H} \mathbf{\Phi}^{H} \overline{\mathbf{H}}_{2}^{H} \overline{\mathbf{H}}_{2} \mathbf{\Phi} \tilde{\mathbf{h}}_{k}\right\} \\
			=2 c_{k} c_{i} \delta e_{k 1} e_{k 2} M^{2} N.
		\end{array}
	\end{align}
	
	We have completed the calculation of the expectation of $8$ cross-terms. 
	Finally, the interference term ${ I}_{ki} \left(\mathbf{\Phi}\right)$ is obtained by combining (\ref{interference_5terms_1}) - (\ref{interference_5terms_5}), (\ref{interference_16terms_1}) - (\ref{interference_16terms_16}) and (\ref{interference_4cross_1}) - (\ref{interference_4cross_4}) to (\ref{expanded_interference_term})  with some direct simplifications.

	\subsection{Signal Leakage}
	In this subsection, we derive the signal leakage term of (\ref{defination_rate}) as
	\begin{align}\label{leakage_make_difference}
		E_{k}^{(\text {leakage})} \left(\mathbf{\Phi}\right) =\mathbb{E}\left\{\left|\hat{\mathbf{q}}_{k}^{H} \mathbf{q}_{k}\right|^{2}\right\}-\left|\mathbb{E}\left\{\hat{\mathbf{q}}_{k}^{H} \mathbf{q}_{k}\right\}\right|^{2},
	\end{align}
	where $ \mathbb{E}\left\{\hat{\mathbf{q}}_{k}^{H} \mathbf{q}_{k}\right\} $ is given in (\ref{derive_noise_and_signal}). Therefore, we only need to derive the expectation $\mathbb{E}\left\{\left|\hat{\mathbf{q}}_{k}^{H} \mathbf{q}_{k}\right|^{2}\right\}$. By exploiting the zero-mean properties of  ${\bf d}_k$ and $\bf N$, and exploiting the independence between the cascaded channel, the direct channel, and the noise, we can expand this term  and remove the terms with zero expectation as
	\begin{align}\label{leakage_expanded-sum}
		\begin{array}{l}
			\mathbb{E}\left\{\left|\hat{\mathbf{q}}_{k}^{H} \mathbf{q}_{k}\right|^{2}\right\}
			=\mathbb{E}\left\{   \left|\left( \hat{\underline{\mathbf{q}}}_{k}^{H}+\mathbf{d}_{k}^{H} \mathbf{A}_{k}^{H}+\frac{1}{\sqrt{\tau p}} \mathbf{s}_{k}^H \mathbf{N}^{H} \mathbf{A}_{k}^{H}\right)\left(\underline{\mathbf{q}}_{k}+\mathbf{d}_{k}\right)\right|^{2}\right\} \\
			=\mathbb{E}\left\{\left|  \hat{\underline{\mathbf{q}}}_{k}   ^{H} \underline{\mathbf{q}}_{k}+\hat{\underline{\mathbf{q}}}_{k}^{H} \mathbf{d}_{k}+  \mathbf{d}_{k}^{H} \mathbf{A}_{k}^{H} \underline{\mathbf{q}}_{k}+\mathbf{d}_{k}^{H} \mathbf{A}_{k}^{H} \mathbf{d}_{k}+\frac{1}{\sqrt{\tau p}} \mathbf{s}_{k}^H \mathbf{N}^{H} \mathbf{A}_{k}^{H} \underline{\mathbf{q}}_{k}+\frac{1}{\sqrt{\tau p}} \mathbf{s}_{k}^H \mathbf{N}^{H} \mathbf{A}_{k}^{H} \mathbf{d}_{k}\right|^{2}\right\} \\
			{{\mathop  = \limits^{\left( d \right)} }}\mathbb{E}\left\{\left|   \hat{\underline{\mathbf{q}}}_{k}^{H} \underline{\mathbf{q}}_{k}\right|^{2}\right\}+\mathbb{E}\left\{\left|  \hat{\underline{\mathbf{q}}}_{k}^{H} \mathbf{d}_{k}\right|^{2}\right\}+\mathbb{E}\left\{\left|\mathbf{d}_{k}^{H} \mathbf{A}_{k}^{H} \underline{\mathbf{q}}_{k}\right|^{2}\right\}+\mathbb{E}\left\{\left|\mathbf{d}_{k}^{H} \mathbf{A}_{k}^{H} \mathbf{d}_{k}\right|^{2}\right\} \\
			+\mathbb{E}\left\{\left|\frac{1}{\sqrt{\tau p}} \mathbf{s}_{k}^H \mathbf{N}^{H} \mathbf{A}_{k}^{H} \underline{\mathbf{q}}_{k}\right|^{2}\right\}+\mathbb{E}\left\{\left|\frac{1}{\sqrt{\tau p}} \mathbf{s}_{k}^H \mathbf{N}^{H} \mathbf{A}_{k}^{H} \mathbf{d}_{k}\right|^{2}\right\} +2 \operatorname{Re}\left\{\mathbb{E}\left\{  \hat{\underline{\mathbf{q}}}_{k}^{H} \underline{\mathbf{q}}_{k}\left(\mathbf{d}_{k}^{H} \mathbf{A}_{k}^{H} \mathbf{d}_{k}\right)^{H}\right\}\right\},
		\end{array}
	\end{align}
	where in $ (d) $ the cross-term $\mathbb{E}\left\{\hat{\underline{\mathbf{q}}}_{k}^{H} \mathbf{d}_{k}\left(\mathbf{d}_{k}^{H} \mathbf{A}_{k}^{H} \underline{\mathbf{q}}_{k}\right)^{H}\right\}$ is zero due to Lemma \ref{lemma4_HH_equal_0}, and the cross-term $ \mathbb{E}\left\{\hat{\underline{\mathbf{q}}}_{k}^{H} \mathbf{d}_{k}\left(\mathbf{d}_{k}^{H} \mathbf{A}_{k}^{H} \mathbf{d}_{k}\right)^{H}\right\} $ is zero because the odd-order central moments of a zero-mean Gaussian variable are zero\cite[Eq. (12)]{winkelbauer2012moments}. 
	
	Next, we derive the $2$nd - $7$th terms in (\ref{leakage_expanded-sum}), but the first term in (\ref{leakage_expanded-sum}) is calculated at the end. The second term in (\ref{leakage_expanded-sum}) is
	\begin{align}\label{leakage_7terms_2}
		\mathbb{E}   \left\{\left|\hat{\underline{\mathbf{q}}}_{k}^{H} \mathbf{d}_{k}\right|^{2}\right\}
		=\mathbb{E} \left\{\hat{\underline{\mathbf{q}}}_{k}^{H} \mathbb{E}\left\{\mathbf{d}_{k} \mathbf{d}_{k}^{H}\right\} \hat{\underline{\mathbf{q}}}_{k}\right\}
		=\gamma_{k} \mathbb{E} \left\{\hat{\underline{\mathbf{q}}}_{k}^{H} \hat{\underline{\mathbf{q}}}_{k}\right\},
	\end{align}
	where $ \mathbb{E} \left\{\hat{\underline{\mathbf{q}}}_{k}^{H} \hat{\underline{\mathbf{q}}}_{k}\right\} $ is given in (\ref{hat_un_hat_un}).
	
	The third term in (\ref{leakage_expanded-sum}) is
	\begin{align}\label{leakage_7terms_3}
		\mathbb{E}\left\{\left|\mathbf{d}_{k}^{H} \mathbf{A}_{k}^{H} \underline{\mathbf{q}}_{k}\right|^{2}\right\}=\mathbb{E}\left\{\underline{\mathbf{q}}_{k}^{H} \mathbf{A}_{k} \mathbb{E}\left\{\mathbf{d}_{k} \mathbf{d}_{k}^{H}\right\} \mathbf{A}_{k}^{H} \underline{\mathbf{q}}_{k}\right\}
		=\gamma_{k} \mathbb{E}\left\{\underline{\mathbf{q}}_{k}^{H} \mathbf{A}_{k} \mathbf{A}_{k}^{H} \underline{\mathbf{q}}_{k}\right\},
	\end{align}
	where $ \mathbb{E}\left\{\underline{\mathbf{q}}_{k}^{H} \mathbf{A}_{k} \mathbf{A}_{k}^{H} \underline{\mathbf{q}}_{k}\right\} $ is given in (\ref{un_k_A_A_un_k}).
	
	Since $ \tilde{\mathbf{d}}_{k} \tilde{\mathbf{d}}_{k}^{H} \sim \mathcal{C} \mathcal{W}_{M}\left(\mathbf{I}_{M}, 1\right)$, using the property of the Wishart distribution (\ref{Wishart_Jn}), the fourth term  in (\ref{leakage_expanded-sum}) can be obtained as 
	\begin{align}\label{leakage_7terms_4}
		\begin{array}{l}
			\mathbb{E}\left\{\left|\mathbf{d}_{k}^{H} \mathbf{A}_{k}^{H} \mathbf{d}_{k}\right|^{2}\right\}=\mathbb{E}\left\{\mathbf{d}_{k}^{H} \mathbf{A}_{k}^{H} \mathbf{d}_{k} \mathbf{d}_{k}^{H} \mathbf{A}_{k} \mathbf{d}_{k}\right\} \\
			=\operatorname{Tr}\left\{\mathbb{E}\left\{\mathbf{A}_{k}^{H} \mathbf{d}_{k} \mathbf{d}_{k}^{H} \mathbf{A}_{k} \mathbf{d}_{k} \mathbf{d}_{k}^{H}\right\}\right\} \\
			=\gamma_{k}^{2} \operatorname{Tr}\left\{\mathbf{A}_{k}^{H}\left(\mathbf{A}_{k}+\operatorname{Tr}\left\{\mathbf{A}_{k}\right\} \mathbf{I}_{M}\right)\right\} \\
			=\gamma_{k}^{2} \operatorname{Tr}\left\{\mathbf{A}_{k}^{H} \mathbf{A}_{k}\right\}+\gamma_{k}^{2}\left(\operatorname{Tr}\left\{\mathbf{A}_{k}\right\}\right)^{2} \\
			=\gamma_{k}^{2} M e_{k 3}+\gamma_{k}^{2} M^{2} e_{k 1}^{2}.
		\end{array}
	\end{align}
	
	The fifth term in (\ref{leakage_expanded-sum}) is calculated as
	\begin{align}\label{leakage_7terms_5}
		\mathbb{E}\left\{\left|\frac{1}{\sqrt{\tau p}} \mathbf{s}_{k}^H \mathbf{N}^{H} \mathbf{A}_{k}^{H} \underline{\mathbf{q}}_{k}\right|^{2}\right\}
		=\frac{1}{\tau p} \mathbb{E}\left\{\underline{\mathbf{q}}_{k}^{H} \mathbf{A}_{k} \mathbb{E}\left\{\mathbf{N} \mathbf{s}_{k} \mathbf{s}_{k}^{H} \mathbf{N}^{H}\right\} \mathbf{A}_{k}^{H} \underline{\mathbf{q}}_{k}\right\}
		=\frac{\sigma^{2}}{\tau p} \mathbb{E}\left\{\underline{\mathbf{q}}_{k}^{H} \mathbf{A}_{k} \mathbf{A}_{k}^{H} \underline{\mathbf{q}}_{k}\right\},
	\end{align}
	where $ \mathbb{E}\left\{\underline{\mathbf{q}}_{k}^{H} \mathbf{A}_{k} \mathbf{A}_{k}^{H} \underline{\mathbf{q}}_{k}\right\} $ is given in (\ref{un_k_A_A_un_k}).
	
	The sixth term in (\ref{leakage_expanded-sum}) is
	\begin{align}\label{leakage_7terms_6}
		\begin{array}{l}
			\mathbb{E}\left\{\left|\frac{1}{\sqrt{\tau p}} \mathbf{s}_{k}^H \mathbf{N}^{H} \mathbf{A}_{k}^{H} \mathbf{d}_{k}\right|^{2}\right\}
			=\frac{1}{\tau p} \mathbb{E}\left\{\mathbf{d}_{k}^{H} \mathbf{A}_{k} \mathbb{E}\left\{\mathbf{N} \mathbf{s}_{k} \mathbf{s}_{k}^H \mathbf{N}^{H}\right\} \mathbf{A}_{k}^{H} \mathbf{d}_{k}\right\} \\
			=\frac{\sigma^{2}}{\tau p} \gamma_{k} \operatorname{Tr}\left\{\mathbf{A}_{k} \mathbf{A}_{k}^{H}\right\}=\frac{\sigma^{2}}{\tau p} \gamma_{k} M e_{k 3}.
		\end{array}
	\end{align}
	
	The seventh term in (\ref{leakage_expanded-sum}) is
	\begin{align}\label{leakage_7terms_7}
		\begin{array}{l}
			2 \operatorname{Re}\left\{\mathbb{E}\left\{  \underline{\hat{\mathbf{q}}}_{k}^{H} \underline{\mathbf{q}}_{k}\left(\mathbf{d}_{k}^{H} \mathbf{A}_{k}^{H} \mathbf{d}_{k}\right)^{H}\right\}\right\}
			=2 \operatorname{Re}\left\{\mathbb{E}\left\{   \underline{\hat{\mathbf{q}}}_{k}^{H} \underline{\mathbf{q}}_{k}\right\} \mathbb{E}\left\{\mathbf{d}_{k}^{H} \mathbf{A}_{k} \mathbf{d}_{k}\right\}\right\}\\
			=2 \gamma_{k} M e_{k 1} \operatorname{Re}\left\{ \mathbb{E}\left\{\underline{\hat{\mathbf{q}}}_{k}^{H} \underline{\mathbf{q}}_{k}\right\}\right\}
			=2 \gamma_{k} M e_{k 1} \mathbb{E}\left\{\underline{\hat{\mathbf{q}}}_{k}^{H} \underline{\mathbf{q}}_{k}\right\},
		\end{array}
	\end{align}
	where $\mathbb{E}\left\{\underline{\hat{\mathbf{q}}}_{k}^{H} \underline{\mathbf{q}}_{k}\right\}$ is given in (\ref{hat_un_un}).
	
	Finally, we derive the first term $ \mathbb{E}\left\{\left|   \hat{\underline{\mathbf{q}}}_{k}^{H} \underline{\mathbf{q}}_{k}\right|^{2}\right\} $ in (\ref{leakage_expanded-sum}), which can be expanded as
	\begin{align}\label{leakage_firstTerm_expanded}
		\begin{array}{l}
			\mathbb{E}\left\{\left|\underline{\hat{\mathbf{q}}}_{k}^{H} \underline{\mathbf{q}}_{k}\right|^{2}\right\}=\mathbb{E}\left\{\left|\sum_{\omega=1}^{4} \sum_{\psi=1}^{4}\left(\hat{\mathbf{q}}_{k}^{\omega}\right)^{H} \mathbf{q}_{k}^{\psi}\right|^{2}\right\} \\
			=\sum_{\omega=1}^{4} \sum_{\psi=1}^{4} \mathbb{E}_{\boldsymbol{}}\left\{\left|\left(\hat{\mathbf{q}}_{k}^{\omega}\right)^{H} \mathbf{q}_{k}^{\psi}\right|^{2}\right\}
			+ \sum_{\omega 1, \psi 1, \omega 2, \psi 2 \atop(\omega 1, \psi 1) \neq(\omega 2, \psi 2)}^{4} \mathbb{E}\left\{\left(\left(\hat{\mathbf{q}}_{k}^{\omega 1}\right)^{H} \mathbf{q}_{k}^{\psi 1}\right)\left(\left(\hat{\mathbf{q}}_{k}^{\omega 2}\right)^{H} \mathbf{q}_{k}^{\psi 2}\right)^{H}\right\}.
		\end{array}
	\end{align}
	
	In the following, we first calculate the $16$ modulus-square terms in (\ref{leakage_firstTerm_expanded}), and then calculate the remaining cross-terms.
	
	Firstly, we consider the terms with $\omega=1$. When $\psi=1$, we have
	\begin{align}\label{leakage_16terms_1}
		\begin{array}{l}
			\mathbb{E}\left\{\left|\sqrt{c_{k} \delta \varepsilon_{k}} \sqrt{c_{k} \delta \varepsilon_{k}} \,\overline{\mathbf{h}}_{k}^{H} \mathbf{\Phi}^{H} \overline{\mathbf{H}}_{2}^{H} \overline{\mathbf{H}}_{2} \mathbf{\Phi} \overline{\mathbf{h}}_{k}\right|^{2}\right\} \\
			=c_{k}^{2} \delta^{2} \varepsilon_{k}^{2}   \left|\left(\overline{\mathbf{h}}_{k}^{H} \mathbf{\Phi}^{H} \mathbf{a}_{N}\right) \mathbf{a}_{M}^{H} \mathbf{a}_{M}\left(\mathbf{a}_{N}^{H} \mathbf{\Phi} \overline{\mathbf{h}}_{k}\right)\right|^{2} \\
			=c_{k}^{2} \delta^{2} \varepsilon_{k}^{2} M^{2}\left|f_{k}(\mathbf{\Phi})\right|^{4}.
		\end{array} 
	\end{align}
	
	When $\psi=2$, we arrive at
	\begin{align}\label{leakage_16terms_2}
		\begin{array}{l}
			\mathbb{E}\left\{\left|\sqrt{c_{k} \delta \varepsilon_{k}} \sqrt{c_{k} \delta} \,\overline{\mathbf{h}}_{k}^{H} \mathbf{\Phi}^{H} \overline{\mathbf{H}}_{2}^{H} \overline{\mathbf{H}}_{2} \mathbf{\Phi} \tilde{\mathbf{h}}_{k}\right|^{2}\right\} \\
			=c_{k}^{2} \delta^{2} \varepsilon_{k} \overline{\mathbf{h}}_{k}^{H} \mathbf{\Phi}^{H} \overline{\mathbf{H}}_{2}^{H} \overline{\mathbf{H}}_{2} \mathbf{\Phi} \mathbb{E}\left\{\tilde{\mathbf{h}}_{k} \tilde{\mathbf{h}}_{k}^{H}\right\} \mathbf{\Phi}^{H} \overline{\mathbf{H}}_{2}^{H} \overline{\mathbf{H}}_{2} \mathbf{\Phi} \overline{\mathbf{h}}_{k} \\
			=c_{k}^{2} \delta^{2} \varepsilon_{k} \overline{\mathbf{h}}_{k}^{H} \mathbf{\Phi}^{H} \overline{\mathbf{H}}_{2}^{H} \overline{\mathbf{H}}_{2} \overline{\mathbf{H}}_{2}^{H} \overline{\mathbf{H}}_{2} \mathbf{\Phi} \overline{\mathbf{h}}_{k} \\
			=c_{k}^{2} \delta^{2} \varepsilon_{k} \overline{\mathbf{h}}_{k}^{H} \mathbf{\Phi}^{H} \mathbf{a}_{N} \mathbf{a}_{M}^{H} \mathbf{a}_{M} \mathbf{a}_{N}^{H} \mathbf{a}_{N} \mathbf{a}_{M}^{H} \mathbf{a}_{M} \mathbf{a}_{N}^{H} \mathbf{\Phi} \overline{\mathbf{h}}_{k} \\
			=c_{k}^{2} \delta^{2} \varepsilon_{k} M^{2} N \overline{\mathbf{h}}_{k}^{H} \mathbf{\Phi}^{H} \mathbf{a}_{N} \mathbf{a}_{N}^{H} \mathbf{\Phi} \overline{\mathbf{h}}_{k} \\
			=c_{k}^{2} \delta^{2} \varepsilon_{k} M^{2} N\left|f_{k}(\mathbf{\Phi})\right|^{2}.
		\end{array}
	\end{align}
	
	When $\psi=3$, we have
	\begin{align}\label{leakage_16terms_3}
		\begin{array}{l}
			\mathbb{E}\left\{\left| \sqrt{c_{k} \delta \varepsilon_{k}} \sqrt{c_{k} \varepsilon_{k}}  \, \overline{\mathbf{h}}_{k}^{H} \mathbf{\Phi}^{H} \overline{\mathbf{H}}_{2}^{H} \tilde{\mathbf{H}}_{2} \mathbf{\Phi} \overline{\mathbf{h}}_{k}\right|^{2}\right\} \\
			=c_{k}^{2} \delta \varepsilon_{k}^{2} \overline{\mathbf{h}}_{k}^{H} \mathbf{\Phi}^{H} \overline{\mathbf{H}}_{2}^{H} \mathbb{E}\left\{\tilde{\mathbf{H}}_{2} \mathbf{\Phi} \overline{\mathbf{h}}_{k} \overline{\mathbf{h}}_{k}^{H} \mathbf{\Phi}^{H} \tilde{\mathbf{H}}_{2}^{H}\right\} \overline{\mathbf{H}}_{2} \mathbf{\Phi} \overline{\mathbf{h}}_{k} \\
			=c_{k}^{2} \delta \varepsilon_{k}^{2} N \overline{\mathbf{h}}_{k}^{H} \mathbf{\Phi}^{H} \overline{\mathbf{H}}_{2}^{H} \overline{\mathbf{H}}_{2} \mathbf{\Phi} \overline{\mathbf{h}}_{k} \\
			=c_{k}^{2} \delta \varepsilon_{k}^{2} M N\left|f_{k}(\mathbf{\Phi})\right|^{2}.
		\end{array}
	\end{align}
	
	When $\psi=4$, we get
	\begin{align}\label{leakage_16terms_4}
		\begin{array}{l}
			\mathbb{E}\left\{\left|   \sqrt{c_{k} \delta \varepsilon_{k}}  \sqrt{c_{k}}  \, \overline{\mathbf{h}}_{k}^{H} \mathbf{\Phi}^{H} \overline{\mathbf{H}}_{2}^{H} \tilde{\mathbf{H}}_{2} \mathbf{\Phi} \tilde{\mathbf{h}}_{k}\right|^{2}\right\} \\
			=c_{k}^{2} \delta \varepsilon_{k} \overline{\mathbf{h}}_{k}^{H} \mathbf{\Phi}^{H} \overline{\mathbf{H}}_{2}^{H} \mathbb{E}\left\{\tilde{\mathbf{H}}_{2} \mathbf{\Phi} \mathbb{E}\left\{\tilde{\mathbf{h}}_{k} \tilde{\mathbf{h}}_{k}^{H}\right\} \mathbf{\Phi}^{H} \tilde{\mathbf{H}}_{2}^{H}\right\} \overline{\mathbf{H}}_{2} \mathbf{\Phi} \overline{\mathbf{h}}_{k} \\
			=c_{k}^{2} \delta \varepsilon_{k} M N\left|f_{k}(\mathbf{\Phi})\right|^{2}.
		\end{array}
	\end{align}

	Secondly, we consider the terms with $\omega=2$. When $\psi=1$, we have
	\begin{align}\label{leakage_16terms_5}
		\begin{array}{l}
			\mathbb{E}\left\{\left|e_{k 2} \sqrt{c_{k} \delta} \sqrt{c_{k} \delta \varepsilon_{k}} \,\tilde{\mathbf{h}}_{k}^{H} \mathbf{\Phi}^{H} \overline{\mathbf{H}}_{2}^{H} \overline{\mathbf{H}}_{2} \mathbf{\Phi} \overline{\mathbf{h}}_{k}\right|^{2}\right\} \\
			=c_{k}^{2} \delta^{2} \varepsilon_{k} e_{k 2}^{2} \mathbb{E}\left\{\tilde{\mathbf{h}}_{k}^{H} \mathbf{\Phi}^{H} \overline{\mathbf{H}}_{2}^{H} \overline{\mathbf{H}}_{2} \mathbf{\Phi} \overline{\mathbf{h}}_{k} \overline{\mathbf{h}}_{k}^{H} \mathbf{\Phi}^{H} \overline{\mathbf{H}}_{2}^{H} \overline{\mathbf{H}}_{2} \mathbf{\Phi} \tilde{\mathbf{h}}_{k}\right\} \\
			=c_{k}^{2} \delta^{2} \varepsilon_{k} e_{k 2}^{2} \operatorname{Tr}\left\{\overline{\mathbf{H}}_{2}^{H} \overline{\mathbf{H}}_{2} \mathbf{\Phi} \overline{\mathbf{h}}_{k} \overline{\mathbf{h}}_{k}^{H} \mathbf{\Phi}^{H} \overline{\mathbf{H}}_{2}^{H} \overline{\mathbf{H}}_{2}\right\} \\
			=c_{k}^{2} \delta^{2} \varepsilon_{k} e_{k 2}^{2} \operatorname{Tr}\left\{\mathbf{a}_{M}^{H} \mathbf{a}_{M}\left(\mathbf{a}_{N}^{H} \mathbf{\Phi} \overline{\mathbf{h}}_{k} \overline{\mathbf{h}}_{k}^{H} \mathbf{\Phi}^{H} \mathbf{a}_{N}\right) \mathbf{a}_{M}^{H} \mathbf{a}_{M} \mathbf{a}_{N}^{H} \mathbf{a}_{N}\right\} \\
			=c_{k}^{2} \delta^{2} \varepsilon_{k} e_{k 2}^{2} M^{2} N\left|f_{k}(\mathbf{\Phi})\right|^{2}.
		\end{array}
	\end{align}
	
	Since $\tilde{\mathbf{h}}_{k} \tilde{\mathbf{h}}_{k}^{H} \sim \mathcal{C} \mathcal{W}_{N}\left(\mathbf{I}_{N}, 1\right)$, using (\ref{Wishart_Jn}), when $\psi=2$, we arrive at
	\begin{align}\label{leakage_16terms_6}
		\begin{array}{l}
			\mathbb{E}\left\{\left|e_{k 2} \sqrt{c_{k} \delta} \sqrt{c_{k} \delta} \,\tilde{\mathbf{h}}_{k}^{H} \mathbf{\Phi}^{H} \overline{\mathbf{H}}_{2}^{H} \overline{\mathbf{H}}_{2} \mathbf{\Phi} \tilde{\mathbf{h}}_{k}\right|^{2}\right\} \\
			=c_{k}^{2} \delta^{2} e_{k 2}^{2} \mathbb{E}\left\{\tilde{\mathbf{h}}_{k}^{H} \mathbf{\Phi}^{H} \overline{\mathbf{H}}_{2}^{H} \overline{\mathbf{H}}_{2} \mathbf{\Phi} \tilde{\mathbf{h}}_{k} \tilde{\mathbf{h}}_{k}^{H} \mathbf{\Phi}^{H} \overline{\mathbf{H}}_{2}^{H} \overline{\mathbf{H}}_{2} \mathbf{\Phi} \tilde{\mathbf{h}}_{k}\right\} \\
			=c_{k}^{2} \delta^{2} e_{k 2}^{2} \operatorname{Tr}\left\{\mathbf{\Phi}^{H} \overline{\mathbf{H}}_{2}^{H} \overline{\mathbf{H}}_{2} \mathbf{\Phi} \mathbb{E}\left\{\tilde{\mathbf{h}}_{k} \tilde{\mathbf{h}}_{k}^{H} \mathbf{\Phi}^{H} \overline{\mathbf{H}}_{2}^{H} \overline{\mathbf{H}}_{2} \mathbf{\Phi} \tilde{\mathbf{h}}_{k} \tilde{\mathbf{h}}_{k}^{H}\right\}\right\} \\
			=c_{k}^{2} \delta^{2} e_{k 2}^{2} \operatorname{Tr}\left\{\mathbf{\Phi}^{H} \overline{\mathbf{H}}_{2}^{H} \overline{\mathbf{H}}_{2} \mathbf{\Phi}\left(\mathbf{\Phi}^{H} \overline{\mathbf{H}}_{2}^{H} \overline{\mathbf{H}}_{2} \mathbf{\Phi}+\operatorname{Tr}\left\{\mathbf{\Phi}^{H} \overline{\mathbf{H}}_{2}^{H} \overline{\mathbf{H}}_{2} \mathbf{\Phi}\right\} \mathbf{I}_{N}\right)\right\} \\
			=c_{k}^{2} \delta^{2} e_{k 2}^{2} \operatorname{Tr}\left\{\mathbf{\Phi}^{H} \overline{\mathbf{H}}_{2}^{H} \overline{\mathbf{H}}_{2} \mathbf{\Phi} \mathbf{\Phi}^{H} \overline{\mathbf{H}}_{2}^{H} \overline{\mathbf{H}}_{2} \mathbf{\Phi}+M N \mathbf{\Phi}^{H} \overline{\mathbf{H}}_{2}^{H} \overline{\mathbf{H}}_{2} \mathbf{\Phi}\right\} \\
			=c_{k}^{2} \delta^{2} e_{k 2}^{2}\left(\operatorname{Tr}\left\{\mathbf{a}_{N} \mathbf{a}_{M}^{H} \mathbf{a}_{M} \mathbf{a}_{N}^{H} \mathbf{a}_{N} \mathbf{a}_{M}^{H} \mathbf{a}_{M} \mathbf{a}_{N}^{H}\right\}+\operatorname{Tr}\left\{M N \mathbf{a}_{N} \mathbf{a}_{M}^{H} \mathbf{a}_{M} \mathbf{a}_{N}^{H}\right\}\right) \\
			=2 c_{k}^{2} \delta^{2} e_{k 2}^{2} M^{2} N^{2}.
		\end{array}
	\end{align}
	
	When $\psi=3$, we have
	\begin{align}\label{leakage_16terms_7}
		\begin{array}{l}
			\mathbb{E}\left\{\left|e_{k 2} \sqrt{c_{k} \delta} \sqrt{c_{k} \varepsilon_{k}} \,\tilde{\mathbf{h}}_{k}^{H} \mathbf{\Phi}^{H} \overline{\mathbf{H}}_{2}^{H} \tilde{\mathbf{H}}_{2} \mathbf{\Phi} \overline{\mathbf{h}}_{k}\right|^{2}\right\} \\
			=c_{k}^{2} \delta \varepsilon_{k} e_{k 2}^{2} \mathbb{E}\left\{\tilde{\mathbf{h}}_{k}^{H} \mathbf{\Phi}^{H} \overline{\mathbf{H}}_{2}^{H} \mathbb{E}\left\{\tilde{\mathbf{H}}_{2} \mathbf{\Phi} \overline{\mathbf{h}}_{k} \overline{\mathbf{h}}_{k}^{H} \mathbf{\Phi}^{H} \tilde{\mathbf{H}}_{2}^{H}\right\} \overline{\mathbf{H}}_{2} \mathbf{\Phi} \tilde{\mathbf{h}}_{k}\right\} \\
			=c_{k}^{2} \delta \varepsilon_{k} e_{k 2}^{2} N \mathbb{E}\left\{\tilde{\mathbf{h}}_{k}^{H} \mathbf{\Phi}^{H} \overline{\mathbf{H}}_{2}^{H} \overline{\mathbf{H}}_{2} \mathbf{\Phi} \tilde{\mathbf{h}}_{k}\right\} \\
			=c_{k}^{2} \delta \varepsilon_{k} e_{k 2}^{2} M N^{2}.
		\end{array}
	\end{align}
	
	When $\psi=4$, we arrive at
	\begin{align}\label{leakage_16terms_8}
		\begin{array}{l}
			\mathbb{E}\left\{\left|e_{k 2} \sqrt{c_{k} \delta} \sqrt{c_{k}} \,\tilde{\mathbf{h}}_{k}^{H} \mathbf{\Phi}^{H} \overline{\mathbf{H}}_{2}^{H} \tilde{\mathbf{H}}_{2} \mathbf{\Phi} \tilde{\mathbf{h}}_{k}\right|^{2}\right\} \\
			=c_{k}^{2} \delta e_{k 2}^{2} \mathbb{E}\left\{\tilde{\mathbf{h}}_{k}^{H} \mathbf{\Phi}^{H} \overline{\mathbf{H}}_{2}^{H} \tilde{\mathbf{H}}_{2} \mathbf{\Phi} \tilde{\mathbf{h}}_{k} \tilde{\mathbf{h}}_{k}^{H} \mathbf{\Phi}^{H} \tilde{\mathbf{H}}_{2}^{H} \overline{\mathbf{H}}_{2} \mathbf{\Phi} \tilde{\mathbf{h}}_{k}\right\} \\
			{{\mathop  = \limits^{\left( e \right)} }}c_{k}^{2} \delta e_{k 2}^{2} \mathbb{E}_{\tilde{\mathbf{h}}_{k}}\left\{\tilde{\mathbf{h}}_{k}^{H} \mathbf{\Phi}^{H} \overline{\mathbf{H}}_{2}^{H} \mathbb{E}_{\tilde{\mathbf{H}}_{2}}\left\{\tilde{\mathbf{H}}_{2} \mathbf{\Phi} \tilde{\mathbf{h}}_{k} \tilde{\mathbf{h}}_{k}^{H} \mathbf{\Phi}^{H} \tilde{\mathbf{H}}_{2}^{H}\right\} \overline{\mathbf{H}}_{2} \mathbf{\Phi} \tilde{\mathbf{h}}_{k} \mid \tilde{\mathbf{h}}_{k}\right\} \\
			=c_{k}^{2} \delta e_{k 2}^{2} \mathbb{E}_{\tilde{\mathbf{h}}_{k}}\left\{\tilde{\mathbf{h}}_{k}^{H} \mathbf{\Phi}^{H} \overline{\mathbf{H}}_{2}^{H} \operatorname{Tr}\left\{\tilde{\mathbf{h}}_{k} \tilde{\mathbf{h}}_{k}^{H}\right\} \overline{\mathbf{H}}_{2} \mathbf{\Phi} \tilde{\mathbf{h}}_{k} \mid \tilde{\mathbf{h}}_{k}\right\} \\
			{{\mathop  = \limits^{\left( f \right)} }}c_{k}^{2} \delta e_{k 2}^{2} \mathbb{E}\left\{\tilde{\mathbf{h}}_{k}^{H} \mathbf{\Phi}^{H} \overline{\mathbf{H}}_{2}^{H} \overline{\mathbf{H}}_{2} \mathbf{\Phi} \tilde{\mathbf{h}}_{k}\left(\tilde{\mathbf{h}}_{k}^{H} \tilde{\mathbf{h}}_{k}\right)\right\} \\
			=c_{k}^{2} \delta e_{k 2}^{2} \operatorname{Tr}\left\{\mathbf{\Phi}^{H} \overline{\mathbf{H}}_{2}^{H} \overline{\mathbf{H}}_{2} \mathbf{\Phi} \mathbb{E}\left\{\tilde{\mathbf{h}}_{k} \tilde{\mathbf{h}}_{k}^{H} \tilde{\mathbf{h}}_{k} \tilde{\mathbf{h}}_{k}^{H}\right\}\right\} \\
			{{\mathop  = \limits^{\left( g \right)} }}c_{k}^{2} \delta e_{k 2}^{2} \operatorname{Tr}\left\{\mathbf{\Phi}^{H} \overline{\mathbf{H}}_{2}^{H} \overline{\mathbf{H}}_{2} \mathbf{\Phi}(N+1) \mathbf{I}_{N}\right\} \\
			=c_{k}^{2} \delta e_{k 2}^{2} M N(N+1),
		\end{array}
	\end{align}
	where $(e)$ utilizes the law of total expectation, which calculates the conditional expectation of $\tilde{\mathbf{H}}_2$ given $\tilde{\mathbf{h}}_k$, and then calculates the expectation of $\tilde{\mathbf{h}}_k$. Since $\tilde{\mathbf{H}}_2$ is independent of $\tilde{\mathbf{h}}_k$, the conditional expectation of $\tilde{\mathbf{H}}_2$ given $\tilde{\mathbf{h}}_k$ is the same as its unconditional expectation; $(f)$ comes from $\operatorname{Tr}\left\{\tilde{\mathbf{h}}_{k} \tilde{\mathbf{h}}_{k}^{H}\right\}=\tilde{\mathbf{h}}_{k}^{H} \tilde{\mathbf{h}}_{k}$ which is a scalar number and its place can be arbitrarily moved; and $(g)$ applies a special case of (\ref{Wishart_Jn}).

	Thirdly, we consider the terms with $\omega=3$. When $\psi=1$, we have
	\begin{align}\label{leakage_16terms_9}
		\begin{array}{l}
			\mathbb{E}\left\{\left|\sqrt{c_{k} \varepsilon_{k}} \sqrt{c_{k} \delta \varepsilon_{k}} \,\overline{\mathbf{h}}_{k}^{H} \mathbf{\Phi}^{H} \tilde{\mathbf{H}}_{2}^{H} \mathbf{A}_{k}^{H} \overline{\mathbf{H}}_{2} \mathbf{\Phi} \overline{\mathbf{h}}_{k}\right|^{2}\right\} \\
			=c_{k}^{2} \delta \varepsilon_{k}^{2} \overline{\mathbf{h}}_{k}^{H} \mathbf{\Phi}^{H} \mathbb{E}\left\{\tilde{\mathbf{H}}_{2}^{H} \mathbf{A}_{k}^{H} \overline{\mathbf{H}}_{2} \mathbf{\Phi} \overline{\mathbf{h}}_{k} \overline{\mathbf{h}}_{k}^{H} \mathbf{\Phi}^{H} \overline{\mathbf{H}}_{2}^{H} \mathbf{A}_{k} \tilde{\mathbf{H}}_{2}\right\} \mathbf{\Phi} \overline{\mathbf{h}}_{k} \\
			=c_{k}^{2} \delta \varepsilon_{k}^{2} \overline{\mathbf{h}}_{k}^{H} \mathbf{\Phi}^{H} \operatorname{Tr}\left\{\mathbf{A}_{k}^{H} \overline{\mathbf{H}}_{2} \mathbf{\Phi} \overline{\mathbf{h}}_{k} \overline{\mathbf{h}}_{k}^{H} \mathbf{\Phi}^{H} \overline{\mathbf{H}}_{2}^{H} \mathbf{A}_{k}\right\} \mathbf{\Phi} \overline{\mathbf{h}}_{k} \\
			=c_{k}^{2} \delta \varepsilon_{k}^{2} e_{k 2}^{2} \overline{\mathbf{h}}_{k}^{H} \mathbf{\Phi}^{H} \operatorname{Tr}\left\{\mathbf{a}_{N}^{H} \mathbf{\Phi} \overline{\mathbf{h}}_{k} \overline{\mathbf{h}}_{k}^{H} \mathbf{\Phi}^{H} \mathbf{a}_{N} \mathbf{a}_{M}^{H} \mathbf{a}_{M}\right\} \mathbf{\Phi} \overline{\mathbf{h}}_{k} \\
			=c_{k}^{2} \delta \varepsilon_{k}^{2} e_{k 2}^{2} M N\left|f_{k}(\mathbf{\Phi})\right|^{2}.
		\end{array}
	\end{align}
	
	When $\psi=2$, we have
	\begin{align}\label{leakage_16terms_10}
		\begin{array}{l}
			\mathbb{E}\left\{\left|\sqrt{c_{k} \varepsilon_{k}} \sqrt{c_{k} \delta}\, \overline{\mathbf{h}}_{k}^{H} \mathbf{\Phi}^{H} \tilde{\mathbf{H}}_{2}^{H} \mathbf{A}_{k}^{H} \overline{\mathbf{H}}_{2} \mathbf{\Phi} \tilde{\mathbf{h}}_{k}\right|^{2}\right\} \\
			=c_{k}^{2} \delta \varepsilon_{k} \overline{\mathbf{h}}_{k}^{H} \mathbf{\Phi}^{H} \mathbb{E}\left\{\tilde{\mathbf{H}}_{2}^{H} \mathbf{A}_{k}^{H} \overline{\mathbf{H}}_{2} \mathbf{\Phi} \mathbb{E}\left\{\tilde{\mathbf{h}}_{k} \tilde{\mathbf{h}}_{k}^{H}\right\} \mathbf{\Phi}^{H} \overline{\mathbf{H}}_{2}^{H} \mathbf{A}_{k} \tilde{\mathbf{H}}_{2}\right\} \mathbf{\Phi} \overline{\mathbf{h}}_{k} \\
			=c_{k}^{2} \delta \varepsilon_{k} \overline{\mathbf{h}}_{k}^{H} \mathbf{\Phi}^{H} \operatorname{Tr}\left\{\mathbf{A}_{k}^{H} \overline{\mathbf{H}}_{2} \overline{\mathbf{H}}_{2}^{H} \mathbf{A}_{k}\right\} \mathbf{\Phi} \overline{\mathbf{h}}_{k} \\
			=c_{k}^{2} \delta \varepsilon_{k} e_{k 2}^{2} M N^{2}.
		\end{array}
	\end{align}
	
	When $\psi=3$, using Lemma \ref{lemma_HAHWHAH}, we have
	\begin{align}\label{leakage_16terms_11}
		\begin{array}{l}
			\mathbb{E}\left\{\left|\sqrt{c_{k} \varepsilon_{k}} \sqrt{c_{k} \varepsilon_{k}} \,\overline{\mathbf{h}}_{k}^{H} \mathbf{\Phi}^{H} \tilde{\mathbf{H}}_{2}^{H} \mathbf{A}_{k}^{H} \tilde{\mathbf{H}}_{2} \mathbf{\Phi} \overline{\mathbf{h}}_{k}\right|^{2}\right\} \\
			=c_{k}^{2} \varepsilon_{k}^{2} \overline{\mathbf{h}}_{k}^{H} \mathbf{\Phi}^{H} \mathbb{E}\left\{\tilde{\mathbf{H}}_{2}^{H} \mathbf{A}_{k}^{H} \tilde{\mathbf{H}}_{2} \mathbf{\Phi} \overline{\mathbf{h}}_{k} \overline{\mathbf{h}}_{k}^{H} \mathbf{\Phi}^{H} \tilde{\mathbf{H}}_{2}^{H} \mathbf{A}_{k} \tilde{\mathbf{H}}_{2}\right\} \mathbf{\Phi} \overline{\mathbf{h}}_{k} \\
			=c_{k}^{2} \varepsilon_{k}^{2} \overline{\mathbf{h}}_{k}^{H} \mathbf{\Phi}^{H}\left(e_{k 1}^{2} M^{2} \mathbf{\Phi} \overline{\mathbf{h}}_{k} \overline{\mathbf{h}}_{k}^{H} \mathbf{\Phi}^{H}+e_{k 3} M\operatorname{Tr}\left\{\mathbf{\Phi} \overline{\mathbf{h}}_{k} \overline{\mathbf{h}}_{k}^{H} \mathbf{\Phi}^{H}\right\} \mathbf{I}_{N}\right) \mathbf{\Phi} \overline{\mathbf{h}}_{k} \\
			=c_{k}^{2} \varepsilon_{k}^{2}\left(e_{k 1}^{2} M^{2} \overline{\mathbf{h}}_{k}^{H} \overline{\mathbf{h}}_{k} \overline{\mathbf{h}}_{k}^{H} \overline{\mathbf{h}}_{k}+e_{k 3} MN \overline{\mathbf{h}}_{k}^{H} \overline{\mathbf{h}}_{k}\right) \\
			=c_{k}^{2} \varepsilon_{k}^{2}\left(e_{k 1}^{2} M^{2} N^{2}+e_{k 3} M N^{2}\right).
		\end{array}
	\end{align}
	
	When $\psi=4$, using Lemma \ref{lemma_HAHWHAH} with ${\bf W}={\bf I}_N$, we have
	\begin{align}\label{leakage_16terms_12}
		\begin{array}{l}
			\mathbb{E}\left\{\left|\sqrt{c_{k} \varepsilon_{k}} \sqrt{c_{k}} \,\overline{\mathbf{h}}_{k}^{H} \mathbf{\Phi}^{H} \tilde{\mathbf{H}}_{2}^{H} \mathbf{A}_{k}^{H} \tilde{\mathbf{H}}_{2} \mathbf{\Phi} \tilde{\mathbf{h}}_{k}\right|^{2}\right\} \\
			=c_{k}^{2} \varepsilon_{k} \overline{\mathbf{h}}_{k}^{H} \mathbf{\Phi}^{H} \mathbb{E}\left\{\tilde{\mathbf{H}}_{2}^{H} \mathbf{A}_{k}^{H} \tilde{\mathbf{H}}_{2} \mathbf{\Phi} \mathbb{E}\left\{\tilde{\mathbf{h}}_{k} \tilde{\mathbf{h}}_{k}^{H}\right\} \mathbf{\Phi}^{H} \tilde{\mathbf{H}}_{2}^{H} \mathbf{A}_{k} \tilde{\mathbf{H}}_{2}\right\} \mathbf{\Phi} \overline{\mathbf{h}}_{k} \\
			=c_{k}^{2} \varepsilon_{k} \overline{\mathbf{h}}_{k}^{H} \mathbf{\Phi}^{H} \mathbb{E}\left\{\tilde{\mathbf{H}}_{2}^{H} \mathbf{A}_{k}^{H} \tilde{\mathbf{H}}_{2} \tilde{\mathbf{H}}_{2}^{H} \mathbf{A}_{k} \tilde{\mathbf{H}}_{2}\right\} \mathbf{\Phi} \overline{\mathbf{h}}_{k} \\
			=c_{k}^{2} \varepsilon_{k}\left(e_{k 1}^{2} M^{2}+e_{k 3} M N\right) \overline{\mathbf{h}}_{k}^{H} \mathbf{\Phi}^{H} \mathbf{I}_{N} \mathbf{\Phi} \overline{\mathbf{h}}_{k} \\
			=c_{k}^{2} \varepsilon_{k}\left(e_{k 1}^{2} M^{2} N+e_{k 3} M N^{2}\right).
		\end{array}
	\end{align}

	Fourthly, we consider the terms with $\omega=4$. When $\psi=1$, we have
	\begin{align}\label{leakage_16terms_13}
		\begin{array}{l}
			\mathbb{E}\left\{\left|\sqrt{c_{k}} \sqrt{c_{k} \delta \varepsilon_{k}} \,\tilde{\mathbf{h}}_{k}^{H} \mathbf{\Phi}^{H} \tilde{\mathbf{H}}_{2}^{H} \mathbf{A}_{k}^{H} \overline{\mathbf{H}}_{2} \mathbf{\Phi} \overline{\mathbf{h}}_{k}\right|^{2}\right\} \\
			=c_{k}^{2} \delta \varepsilon_{k} \mathbb{E}\left\{\tilde{\mathbf{h}}_{k}^{H} \mathbf{\Phi}^{H} \mathbb{E}\left\{\tilde{\mathbf{H}}_{2}^{H} \mathbf{A}_{k}^{H} \overline{\mathbf{H}}_{2} \mathbf{\Phi} \overline{\mathbf{h}}_{k} \overline{\mathbf{h}}_{k}^{H} \mathbf{\Phi}^{H} \overline{\mathbf{H}}_{2}^{H} \mathbf{A}_{k} \tilde{\mathbf{H}}_{2}\right\} \mathbf{\Phi} \tilde{\mathbf{h}}_{k}\right\} \\
			=c_{k}^{2} \delta \varepsilon_{k} e_{k 2}^{2} \mathbb{E}\left\{\tilde{\mathbf{h}}_{k}^{H} \mathbf{\Phi}^{H} \operatorname{Tr}\left\{\overline{\mathbf{H}}_{2} \mathbf{\Phi} \overline{\mathbf{h}}_{k} \overline{\mathbf{h}}_{k}^{H} \mathbf{\Phi}^{H} \overline{\mathbf{H}}_{2}^{H}\right\} \mathbf{\Phi} \tilde{\mathbf{h}}_{k}\right\} \\
			=c_{k}^{2} \delta \varepsilon_{k} e_{k 2}^{2} M\left|f_{k}(\mathbf{\Phi})\right|^{2} \mathbb{E}\left\{\tilde{\mathbf{h}}_{k}^{H} \tilde{\mathbf{h}}_{k}\right\} \\
			=c_{k}^{2} \delta \varepsilon_{k} e_{k 2}^{2} M N\left|f_{k}(\mathbf{\Phi})\right|^{2}.
		\end{array}
	\end{align}
	
	When $\psi=2$, we have
	\begin{align}\label{leakage_16terms_14}
		\begin{array}{l}
			\mathbb{E}\left\{\left|\sqrt{c_{k}} \sqrt{c_{k} \delta} \,\tilde{\mathbf{h}}_{k}^{H} \mathbf{\Phi}^{H} \tilde{\mathbf{H}}_{2}^{H} \mathbf{A}_{k}^{H} \overline{\mathbf{H}}_{2} \mathbf{\Phi} \tilde{\mathbf{h}}_{k}\right|^{2}\right\} \\
			=c_{k}^{2} \delta \mathbb{E}_{\tilde{\mathbf{h}}_{k}}\left\{\tilde{\mathbf{h}}_{k}^{H} \mathbf{\Phi}^{H} \mathbb{E}_{\tilde{\mathbf{H}}_{2}}\left\{\tilde{\mathbf{H}}_{2}^{H} \mathbf{A}_{k}^{H} \overline{\mathbf{H}}_{2} \mathbf{\Phi} \tilde{\mathbf{h}}_{k} \tilde{\mathbf{h}}_{k}^{H} \mathbf{\Phi}^{H} \overline{\mathbf{H}}_{2}^{H} \mathbf{A}_{k} \tilde{\mathbf{H}}_{2}\right\} \mathbf{\Phi} \tilde{\mathbf{h}}_{k} \mid \tilde{\mathbf{h}}_{k}\right\} \\
			=c_{k}^{2} \delta e_{k 2}^{2} \mathbb{E}\left\{\tilde{\mathbf{h}}_{k}^{H} \mathbf{\Phi}^{H} \operatorname{Tr}\left\{\overline{\mathbf{H}}_{2} \mathbf{\Phi} \tilde{\mathbf{h}}_{k} \tilde{\mathbf{h}}_{k}^{H} \mathbf{\Phi}^{H} \overline{\mathbf{H}}_{2}^{H}\right\} \mathbf{\Phi} \tilde{\mathbf{h}}_{k}\right\} \\
			{{\mathop  = \limits^{\left( h \right)} }}c_{k}^{2} \delta e_{k 2}^{2} \mathbb{E}\left\{\tilde{\mathbf{h}}_{k}^{H} \mathbf{\Phi}^{H} \mathbf{\Phi} \tilde{\mathbf{h}}_{k}\left(\tilde{\mathbf{h}}_{k}^{H} \mathbf{\Phi}^{H} \overline{\mathbf{H}}_{2}^{H} \overline{\mathbf{H}}_{2} \mathbf{\Phi} \tilde{\mathbf{h}}_{k}\right)\right\} \\
			=c_{k}^{2} \delta e_{k 2}^{2} \operatorname{Tr}\left\{\mathbb{E}\left\{\tilde{\mathbf{h}}_{k} \tilde{\mathbf{h}}_{k}^{H} \mathbf{\Phi}^{H} \overline{\mathbf{H}}_{2}^{H} \overline{\mathbf{H}}_{2} \mathbf{\Phi} \tilde{\mathbf{h}}_{k} \tilde{\mathbf{h}}_{k}^{H}\right\}\right\} \\
			=c_{k}^{2} \delta e_{k 2}^{2} \operatorname{Tr}\left\{\mathbf{\Phi}^{H} \overline{\mathbf{H}}_{2}^{H} \overline{\mathbf{H}}_{2} \mathbf{\Phi}+\operatorname{Tr}\left\{\mathbf{\Phi}^{H} \overline{\mathbf{H}}_{2}^{H} \overline{\mathbf{H}}_{2} \mathbf{\Phi}\right\} \mathbf{I}_{N}\right\} \\
			=c_{k}^{2} \delta e_{k 2}^{2} \operatorname{Tr}\left\{\overline{\mathbf{H}}_{2}^{H} \overline{\mathbf{H}}_{2}+MN\mathbf{I}_{N}\right\} \\
			=c_{k}^{2} \delta e_{k 2}^{2}\left(M N+M N^{2}\right),
		\end{array}
	\end{align}
	where $(h)$ comes from $  \operatorname{Tr}\left\{\overline{\mathbf{H}}_{2} \mathbf{\Phi} \tilde{\mathbf{h}}_{k} \tilde{\mathbf{h}}_{k}^{H} \mathbf{\Phi}^{H} \overline{\mathbf{H}}_{2}^{H}\right\}=\tilde{\mathbf{h}}_{k}^{H} \mathbf{\Phi}^{H} \overline{\mathbf{H}}_{2}^{H} \overline{\mathbf{H}}_{2} \mathbf{\Phi} \tilde{\mathbf{h}}_{k}$, which is a $1\times 1$ number and can be moved to the end of the equation.

	When $\psi=3$, using Lemma \ref{lemma_HWH} and Lemma \ref{lemma_HAHWHAH}, we have
	\begin{align}\label{leakage_16terms_15}
		\begin{array}{l}
			\mathbb{E}\left\{\left|\sqrt{c_{k}} \sqrt{c_{k} \varepsilon_{k}} \,\tilde{\mathbf{h}}_{k}^{H} \mathbf{\Phi}^{H} \tilde{\mathbf{H}}_{2}^{H} \mathbf{A}_{k}^{H} \tilde{\mathbf{H}}_{2} \mathbf{\Phi} \overline{\mathbf{h}}_{k}\right|^{2}\right\} \\
			=c_{k}^{2} \varepsilon_{k} \mathbb{E}\left\{\tilde{\mathbf{h}}_{k}^{H} \mathbf{\Phi}^{H} \mathbb{E}\left\{\tilde{\mathbf{H}}_{2}^{H} \mathbf{A}_{k}^{H} \tilde{\mathbf{H}}_{2} \mathbf{\Phi} \overline{\mathbf{h}}_{k} \overline{\mathbf{h}}_{k}^{H} \mathbf{\Phi}^{H} \tilde{\mathbf{H}}_{2}^{H} \mathbf{A}_{k} \tilde{\mathbf{H}}_{2}\right\} \mathbf{\Phi} \tilde{\mathbf{h}}_{k}\right\} \\
			=c_{k}^{2} \varepsilon_{k} \mathbb{E}\left\{\tilde{\mathbf{h}}_{k}^{H} \mathbf{\Phi}^{H}\left(e_{k 1}^{2} M^{2} \mathbf{\Phi} \overline{\mathbf{h}}_{k} \overline{\mathbf{h}}_{k}^{H} \mathbf{\Phi}^{H}+e_{k 3} M\operatorname{Tr}\left\{\mathbf{\Phi} \overline{\mathbf{h}}_{k} \overline{\mathbf{h}}_{k}^{H} \mathbf{\Phi}^{H}\right\} \mathbf{I}_{N}\right) \mathbf{\Phi} \tilde{\mathbf{h}}_{k}\right\} \\
			=c_{k}^{2} \varepsilon_{k} \mathbb{E}\left\{e_{k 1}^{2} M^{2} \tilde{\mathbf{h}}_{k}^{H} \mathbf{\Phi}^{H} \mathbf{\Phi} \overline{\mathbf{h}}_{k} \overline{\mathbf{h}}_{k}^{H} \mathbf{\Phi}^{H} \mathbf{\Phi} \tilde{\mathbf{h}}_{k}+e_{k 3} M N \tilde{\mathbf{h}}_{k}^{H} \mathbf{\Phi}^{H} \mathbf{\Phi} \tilde{\mathbf{h}}_{k}\right\} \\
			=c_{k}^{2} \varepsilon_{k} \mathbb{E}\left\{e_{k 1}^{2} M^{2} \tilde{\mathbf{h}}_{k}^{H} \overline{\mathbf{h}}_{k} \overline{\mathbf{h}}_{k}^{H} \tilde{\mathbf{h}}_{k}+e_{k 3} M N \tilde{\mathbf{h}}_{k}^{H} \tilde{\mathbf{h}}_{k}\right\} \\
			=c_{k}^{2} \varepsilon_{k}\left(e_{k 1}^{2} M^{2} N+e_{k 3} M N^{2}\right).
		\end{array}
	\end{align}
	
	When $\psi=4$, we get
	\begin{align}\label{leakage_16terms_16}
		\begin{array}{l}
			\mathbb{E}\left\{\left|\sqrt{c_{k}} \sqrt{c_{k}} \,\tilde{\mathbf{h}}_{k}^{H} \mathbf{\Phi}^{H} \tilde{\mathbf{H}}_{2}^{H} \mathbf{A}_{k}^{H} \tilde{\mathbf{H}}_{2} \mathbf{\Phi} \tilde{\mathbf{h}}_{k}\right|^{2}\right\} \\
			=c_{k}^{2} \mathbb{E}_{\tilde{\mathbf{h}}_{k}}\left\{\tilde{\mathbf{h}}_{k}^{H} \mathbf{\Phi}^{H} \mathbb{E}_{\tilde{\mathbf{H}}_{2}}\left\{\tilde{\mathbf{H}}_{2}^{H} \mathbf{A}_{k}^{H} \tilde{\mathbf{H}}_{2} \mathbf{\Phi} \tilde{\mathbf{h}}_{k} \tilde{\mathbf{h}}_{k}^{H} \mathbf{\Phi}^{H} \tilde{\mathbf{H}}_{2}^{H} \mathbf{A}_{k} \tilde{\mathbf{H}}_{2}\right\} \mathbf{\Phi} \tilde{\mathbf{h}}_{k} \mid \tilde{\mathbf{h}}_{k}\right\} \\
			=c_{k}^{2} \mathbb{E}_{\tilde{\mathbf{h}}_{k}}\left\{\tilde{\mathbf{h}}_{k}^{H} \mathbf{\Phi}^{H}\left(e_{k 1}^{2} M^{2} \mathbf{\Phi} \tilde{\mathbf{h}}_{k} \tilde{\mathbf{h}}_{k}^{H} \mathbf{\Phi}^{H}+e_{k 3}M \operatorname{Tr}\left\{\mathbf{\Phi} \tilde{\mathbf{h}}_{k} \tilde{\mathbf{h}}_{k}^{H} \mathbf{\Phi}^{H}\right\} \mathbf{I}_{N}\right) \mathbf{\Phi} \tilde{\mathbf{h}}_{k}\right\} \\
			=c_{k}^{2} \mathbb{E}\left\{e_{k 1}^{2} M^{2} \tilde{\mathbf{h}}_{k}^{H} \tilde{\mathbf{h}}_{k} \tilde{\mathbf{h}}_{k}^{H} \tilde{\mathbf{h}}_{k}+e_{k 3} M \tilde{\mathbf{h}}_{k}^{H} \tilde{\mathbf{h}}_{k} \tilde{\mathbf{h}}_{k}^{H} \tilde{\mathbf{h}}_{k}\right\} \\
			{{\mathop  = \limits^{\left( i \right)} }}c_{k}^{2}\left\{e_{k 1}^{2} M^{2} N(N+1)+e_{k 3} M N(N+1)\right\},
		\end{array}
	\end{align}
	where $(i)$ uses (\ref{Wishart_h4}).
	
	Herein, the calculation of the $16$ modulus-square terms are completed. Now, we focus on the expectation of the remaining cross-terms. To better understand the form of the cross-terms, we  give the expansion of $ \underline{\hat{\mathbf{q}}}_{k}^{H} \underline{\mathbf{q}}_{k} $ as
	\begin{align}\label{Expansion_kk}
		\begin{array}{l}
			\underline{\hat{\mathbf{q}}}_{k}^{H} \underline{\mathbf{q}}_{k}
			=\sum_{\omega=1}^{4} \sum_{\psi=1}^{4}\left(\hat{\mathbf{q}}_{k}^{\omega}\right)^{H} \mathbf{q}_{k}^{\psi} \\
			=\eta^{11} \overline{\mathbf{h}}_{k}^{H} \mathbf{\Phi}^{H} \overline{\mathbf{H}}_{2}^{H} \overline{\mathbf{H}}_{2} \mathbf{\Phi} \overline{\mathbf{h}}_{k}+\eta^{12} \overline{\mathbf{h}}_{k}^{H} \mathbf{\Phi}^{H} \overline{\mathbf{H}}_{2}^{H} \overline{\mathbf{H}}_{2} \mathbf{\Phi} \tilde{\mathbf{h}}_{k}+\eta^{13} \overline{\mathbf{h}}_{k}^{H} \mathbf{\Phi}^{H} \overline{\mathbf{H}}_{2}^{H} \tilde{\mathbf{H}}_{2} \mathbf{\Phi} \overline{\mathbf{h}}_{k}+\eta^{14} \overline{\mathbf{h}}_{k}^{H} \mathbf{\Phi}^{H} \overline{\mathbf{H}}_{2}^{H} \tilde{\mathbf{H}}_{2} \mathbf{\Phi} \tilde{\mathbf{h}}_{k} \\
			+\eta^{21} \tilde{\mathbf{h}}_{k}^{H} \mathbf{\Phi}^{H} \overline{\mathbf{H}}_{2}^{H} \overline{\mathbf{H}}_{2} \mathbf{\Phi} \overline{\mathbf{h}}_{k}+\eta^{22} \tilde{\mathbf{h}}_{k}^{H} \mathbf{\Phi}^{H} \overline{\mathbf{H}}_{2}^{H} \overline{\mathbf{H}}_{2} \mathbf{\Phi} \tilde{\mathbf{h}}_{k}+\eta^{23} \tilde{\mathbf{h}}_{k}^{H} \mathbf{\Phi}^{H} \overline{\mathbf{H}}_{2}^{H} \tilde{\mathbf{H}}_{2} \mathbf{\Phi} \overline{\mathbf{h}}_{k}+\eta^{24} \tilde{\mathbf{h}}_{k}^{H} \mathbf{\Phi}^{H} \overline{\mathbf{H}}_{2}^{H} \tilde{\mathbf{H}}_{2} \mathbf{\Phi} \tilde{\mathbf{h}}_{k} \\
			+\eta^{31} \overline{\mathbf{h}}_{k}^{H} \mathbf{\Phi}^{H} \tilde{\mathbf{H}}_{2}^{H} \mathbf{A}_{k}^{H} \overline{\mathbf{H}}_{2} \mathbf{\Phi} \overline{\mathbf{h}}_{k}+\eta^{32} \overline{\mathbf{h}}_{k}^{H} \mathbf{\Phi}^{H} \tilde{\mathbf{H}}_{2}^{H} \mathbf{A}_{k}^{H} \overline{\mathbf{H}}_{2} \mathbf{\Phi} \tilde{\mathbf{h}}_{k} \\
			+\eta^{33} \overline{\mathbf{h}}_{k}^{H} \mathbf{\Phi}^{H} \tilde{\mathbf{H}}_{2}^{H} \mathbf{A}_{k}^{H} \tilde{\mathbf{H}}_{2} \mathbf{\Phi} \overline{\mathbf{h}}_{k}+\eta^{34} \overline{\mathbf{h}}_{k}^{H} \mathbf{\Phi}^{H} \tilde{\mathbf{H}}_{2}^{H} \mathbf{A}_{k}^{H} \tilde{\mathbf{H}}_{2} \mathbf{\Phi} \tilde{\mathbf{h}}_{k} \\
			+\eta^{41} \tilde{\mathbf{h}}_{k}^{H} \mathbf{\Phi}^{H} \tilde{\mathbf{H}}_{2}^{H} \mathbf{A}_{k}^{H} \overline{\mathbf{H}}_{2} \mathbf{\Phi} \overline{\mathbf{h}}_{k}+\eta^{42} \tilde{\mathbf{h}}_{k}^{H} \mathbf{\Phi}^{H} \tilde{\mathbf{H}}_{2}^{H} \mathbf{A}_{k}^{H} \overline{\mathbf{H}}_{2} \mathbf{\Phi} \tilde{\mathbf{h}}_{k} \\
			+\eta^{43} \tilde{\mathbf{h}}_{k}^{H} \mathbf{\Phi}^{H} \tilde{\mathbf{H}}_{2}^{H} \mathbf{A}_{k}^{H} \tilde{\mathbf{H}}_{2} \mathbf{\Phi} \overline{\mathbf{h}}_{k}+\eta^{44} \tilde{\mathbf{h}}_{k}^{H} \mathbf{\Phi}^{H} \tilde{\mathbf{H}}_{2}^{H} \mathbf{A}_{k}^{H} \tilde{\mathbf{H}}_{2} \mathbf{\Phi} \tilde{\mathbf{h}}_{k}.
		\end{array}
	\end{align}
	
	We use the notation $\eta^{11} $ - $ \eta^{44}$ to identify the coefficients in front of the product of vectors and matrices. We can calculate the cross-terms in (\ref{leakage_firstTerm_expanded}) by calculating the expectation of the product of one term in (\ref{Expansion_kk}) with the conjugate transpose of another term in (\ref{Expansion_kk}). There exist $16\times 15$ cross-terms, but only $20$ of them are non-zero. Using Lemma \ref{lemma4_HH_equal_0}, the independence and zero-mean properties of $\tilde{\bf H}_2$ and $\tilde{\bf h}_k$, we can filter the $20$ non-zero cross-terms, and combine them into the following $10$ terms:
	\begin{align}\label{leakage_10Cross_terms_sum}
		\begin{aligned}
			&\sum\nolimits_{\omega 1, \psi 1, \omega 2, \psi 2, \atop(\omega 1, \psi 1) \neq(\omega 2, \psi 2)}^{4} \mathbb{E}\left\{\left(\left(\hat{\mathbf{q}}_{k}^{\omega 1}\right)^{H} \mathbf{q}_{k}^{\psi 1}\right)\left(\left(\hat{\mathbf{q}}_{k}^{\omega 2}\right)^{H} \mathbf{q}_{k}^{\psi 2}\right)^{H}\right\}\\
			&=2 \operatorname{Re}\left\{\mathbb{E}\left\{\left(\left(\hat{\mathbf{q}}_{k}^{1}\right)^{H} \mathbf{q}_{k}^{1}\right)\left(\left(\hat{\mathbf{q}}_{k}^{2}\right)^{H} \mathbf{q}_{k}^{2}\right)^{H}\right\}\right\}+2 \operatorname{Re}\left\{\mathbb{E}\left\{\left(\left(\hat{\mathbf{q}}_{k}^{1}\right)^{H} \mathbf{q}_{k}^{1}\right)\left(\left(\hat{\mathbf{q}}_{k}^{3}\right)^{H} \mathbf{q}_{k}^{3}\right)^{H}\right\}\right\}\\
			&+2 \operatorname{Re}\left\{\mathbb{E}\left\{\left(\left(\hat{\mathbf{q}}_{k}^{1}\right)^{H} \mathbf{q}_{k}^{1}\right)\left(\left(\hat{\mathbf{q}}_{k}^{4}\right)^{H} \mathbf{q}_{k}^{4}\right)^{H}\right\}\right\}+2 \operatorname{Re}\left\{\mathbb{E}\left\{\left(\left(\hat{\mathbf{q}}_{k}^{1}\right)^{H} \mathbf{q}_{k}^{2}\right)\left(\left(\hat{\mathbf{q}}_{k}^{3}\right)^{H} \mathbf{q}_{k}^{4}\right)^{H}\right\}\right\}\\
			&+2 \operatorname{Re}\left\{\mathbb{E}\left\{\left(\left(\hat{\mathbf{q}}_{k}^{1}\right)^{H} \mathbf{q}_{k}^{3}\right)\left(\left(\hat{\mathbf{q}}_{k}^{2}\right)^{H} \mathbf{q}_{k}^{4}\right)^{H}\right\}\right\}+2 \operatorname{Re}\left\{\mathbb{E}\left\{\left(\left(\hat{\mathbf{q}}_{k}^{2}\right)^{H} \mathbf{q}_{k}^{1}\right)\left(\left(\hat{\mathbf{q}}_{k}^{4}\right)^{H} \mathbf{q}_{k}^{3}\right)^{H}\right\}\right\}\\
			&+2 \operatorname{Re}\left\{\mathbb{E}\left\{\left(\left(\hat{\mathbf{q}}_{k}^{2}\right)^{H} \mathbf{q}_{k}^{2}\right)\left(\left(\hat{\mathbf{q}}_{k}^{3}\right)^{H} \mathbf{q}_{k}^{3}\right)^{H}\right\}\right\}+2 \operatorname{Re}\left\{\mathbb{E}\left\{\left(\left(\hat{\mathbf{q}}_{k}^{2}\right)^{H} \mathbf{q}_{k}^{2}\right)\left(\left(\hat{\mathbf{q}}_{k}^{4}\right)^{H} \mathbf{q}_{k}^{4}\right)^{H}\right\}\right\}\\
			&+2 \operatorname{Re}\left\{\mathbb{E}\left\{\left(\left(\hat{\mathbf{q}}_{k}^{3}\right)^{H} \mathbf{q}_{k}^{1}\right)\left(\left(\hat{\mathbf{q}}_{k}^{4}\right)^{H} \mathbf{q}_{k}^{2}\right)^{H}\right\}\right\}+2 \operatorname{Re}\left\{\mathbb{E}\left\{\left(\left(\hat{\mathbf{q}}_{k}^{3}\right)^{H} \mathbf{q}_{k}^{3}\right)\left(\left(\hat{\mathbf{q}}_{k}^{4}\right)^{H} \mathbf{q}_{k}^{4}\right)^{H}\right\}\right\}.
		\end{aligned}
	\end{align}
	
	Now, we derive these $10$ terms in sequence. Note that the real operator ${\rm Re}\left\{\cdot\right\}$ is omitted, since the results show that they only have real parts.
	
	Let us begin with the calculation of the first term as follows
	\begin{align}\label{leakage_10Cross_terms_1}
		\begin{array}{l}
			2 \mathbb{E}\left\{\sqrt{c_{k} \delta \varepsilon_{k}} \sqrt{c_{k} \delta \varepsilon_{k}}\, \overline{\mathbf{h}}_{k}^{H} \mathbf{\Phi}^{H} \overline{\mathbf{H}}_{2}^{H} \overline{\mathbf{H}}_{2} \mathbf{\Phi} \overline{\mathbf{h}}_{k}
			\left(e_{k 2} \sqrt{c_{k} \delta} \sqrt{c_{k} \delta} \,\tilde{\mathbf{h}}_{k}^{H} \mathbf{\Phi}^{H} \overline{\mathbf{H}}_{2}^{H} \overline{\mathbf{H}}_{2} \mathbf{\Phi} \tilde{\mathbf{h}}_{k}\right)^{H}\right\} \\
			=2 c_{k}^{2} \delta^{2} \varepsilon_{k} e_{k 2} \overline{\mathbf{h}}_{k}^{H} \mathbf{\Phi}^{H} \overline{\mathbf{H}}_{2}^{H} \overline{\mathbf{H}}_{2} \mathbf{\Phi} \overline{\mathbf{h}}_{k} \mathbb{E}\left\{\tilde{\mathbf{h}}_{k}^{H} \mathbf{\Phi}^{H} \overline{\mathbf{H}}_{2}^{H} \overline{\mathbf{H}}_{2} \mathbf{\Phi} \tilde{\mathbf{h}}_{k}\right\} \\
			=2 c_{k}^{2} \delta^{2} \varepsilon_{k} e_{k 2}\left(\overline{\mathbf{h}}_{k}^{H} \mathbf{\Phi}^{H} \mathbf{a}_{N}\right) \mathbf{a}_{M}^{H} \mathbf{a}_{M}\left(\mathbf{a}_{N}^{H} \mathbf{\Phi} \overline{\mathbf{h}}_{k}\right) \operatorname{Tr}\left\{\mathbf{\Phi}^{H} \mathbf{a}_{N} \mathbf{a}_{M}^{H} \mathbf{a}_{M} \mathbf{a}_{N}^{H} \mathbf{\Phi}\right\} \\
			=2 c_{k}^{2} \delta^{2} \varepsilon_{k} e_{k 2} M^{2} N\left|f_{k}(\mathbf{\Phi})\right|^{2}.
		\end{array}
	\end{align}
	
	The second term is
	\begin{align}\label{leakage_10Cross_terms_2}
		\begin{array}{l}
			2 \mathbb{E}\left\{\sqrt{c_{k} \delta \varepsilon_{k}} \sqrt{c_{k} \delta \varepsilon_{k}} \, \overline{\mathbf{h}}_{k}^{H} \mathbf{\Phi}^{H} \overline{\mathbf{H}}_{2}^{H} \overline{\mathbf{H}}_{2} \mathbf{\Phi} \overline{\mathbf{h}}_{k}
			\left(\sqrt{c_{k} \varepsilon_{k}} \sqrt{c_{k} \varepsilon_{k}} \,\overline{\mathbf{h}}_{k}^{H} \mathbf{\Phi}^{H} \tilde{\mathbf{H}}_{2}^{H} \mathbf{A}_{k}^{H} \tilde{\mathbf{H}}_{2} \mathbf{\Phi} \overline{\mathbf{h}}_{k}\right)^{H}\right\} \\
			=2 c_{k}^{2} \delta \varepsilon_{k}^{2} \overline{\mathbf{h}}_{k}^{H} \mathbf{\Phi}^{H} \overline{\mathbf{H}}_{2}^{H} \overline{\mathbf{H}}_{2} \mathbf{\Phi} \overline{\mathbf{h}}_{k} \overline{\mathbf{h}}_{k}^{H} \mathbf{\Phi}^{H} \mathbb{E}\left\{\tilde{\mathbf{H}}_{2}^{H} \mathbf{A}_{k} \tilde{\mathbf{H}}_{2}\right\} \mathbf{\Phi} \overline{\mathbf{h}}_{k} \\
			=2 c_{k}^{2} \delta \varepsilon_{k}^{2} e_{k 1} M\left(\overline{\mathbf{h}}_{k}^{H} \mathbf{\Phi}^{H} \mathbf{a}_{N}\right) \mathbf{a}_{M}^{H} \mathbf{a}_{M}\left(\mathbf{a}_{N}^{H} \mathbf{\Phi} \overline{\mathbf{h}}_{k}\right) \overline{\mathbf{h}}_{k}^{H} \overline{\mathbf{h}}_{k} \\
			=2 c_{k}^{2} \delta \varepsilon_{k}^{2} e_{k 1} M^{2} N\left|f_{k}(\mathbf{\Phi})\right|^{2}.
		\end{array}
	\end{align}
	
	The third term is
	\begin{align}\label{leakage_10Cross_terms_3}
		\begin{array}{l}
			2 \mathbb{E}\left\{\sqrt{c_{k} \delta \varepsilon_{k}} \sqrt{c_{k} \delta \varepsilon_{k}} \, \overline{\mathbf{h}}_{k}^{H} \mathbf{\Phi}^{H} \overline{\mathbf{H}}_{2}^{H} \overline{\mathbf{H}}_{2} \mathbf{\Phi} \overline{\mathbf{h}}_{k}
			\left(\sqrt{c_{k}} \sqrt{c_{k}} \, \tilde{\mathbf{h}}_{k}^{H} \mathbf{\Phi}^{H} \tilde{\mathbf{H}}_{2}^{H} \mathbf{A}_{k}^{H} \tilde{\mathbf{H}}_{2} \mathbf{\Phi} \tilde{\mathbf{h}}_{k}\right)^{H}\right\} \\
			=2 c_{k}^{2} \delta \varepsilon_{k} \overline{\mathbf{h}}_{k}^{H} \mathbf{\Phi}^{H} \overline{\mathbf{H}}_{2}^{H} \overline{\mathbf{H}}_{2} \mathbf{\Phi} \overline{\mathbf{h}}_{k} \mathbb{E}\left\{\tilde{\mathbf{h}}_{k}^{H} \mathbf{\Phi}^{H} \mathbb{E}\left\{\tilde{\mathbf{H}}_{2}^{H} \mathbf{A}_{k} \tilde{\mathbf{H}}_{2}\right\} \mathbf{\Phi} \tilde{\mathbf{h}}_{k}\right\} \\
			=2 c_{k}^{2} \delta \varepsilon_{k} e_{k 1} M\left(\overline{\mathbf{h}}_{k}^{H} \mathbf{\Phi}^{H} \mathbf{a}_{N}\right) \mathbf{a}_{M}^{H} \mathbf{a}_{M}\left(\mathbf{a}_{N}^{H} \mathbf{\Phi} \overline{\mathbf{h}}_{k}\right) \mathbb{E}\left\{\tilde{\mathbf{h}}_{k}^{H} \tilde{\mathbf{h}}_{k}\right\} \\
			=2 c_{k}^{2} \delta \varepsilon_{k} e_{k 1} M^{2} N\left|f_{k}(\mathbf{\Phi})\right|^{2}.
		\end{array}
	\end{align}
	
	The fourth term is
	\begin{align}\label{leakage_10Cross_terms_4}
		\begin{array}{l}
			2 \mathbb{E}\left\{\sqrt{c_{k} \delta \varepsilon_{k}} \sqrt{c_{k} \delta} \,\overline{\mathbf{h}}_{k}^{H} \mathbf{\Phi}^{H} \overline{\mathbf{H}}_{2}^{H} \overline{\mathbf{H}}_{2} \mathbf{\Phi} \tilde{\mathbf{h}}_{k}
			\left(\sqrt{c_{k} \varepsilon_{k}} \sqrt{c_{k}} \,\overline{\mathbf{h}}_{k}^{H} \mathbf{\Phi}^{H} \tilde{\mathbf{H}}_{2}^{H} \mathbf{A}_{k}^{H} \tilde{\mathbf{H}}_{2} \mathbf{\Phi} \tilde{\mathbf{h}}_{k}\right)^{H}\right\} \\
			=2 c_{k}^{2} \delta \varepsilon_{k} \overline{\mathbf{h}}_{k}^{H} \mathbf{\Phi}^{H} \overline{\mathbf{H}}_{2}^{H} \overline{\mathbf{H}}_{2} \mathbf{\Phi} \mathbb{E}\left\{\tilde{\mathbf{h}}_{k} \tilde{\mathbf{h}}_{k}^{H}\right\} \mathbf{\Phi}^{H} \mathbb{E}\left\{\tilde{\mathbf{H}}_{2}^{H} \mathbf{A}_{k} \tilde{\mathbf{H}}_{2}\right\} \mathbf{\Phi} \overline{\mathbf{h}}_{k} \\
			=2 c_{k}^{2} \delta \varepsilon_{k} e_{k 1} M \overline{\mathbf{h}}_{k}^{H} \mathbf{\Phi}^{H} \overline{\mathbf{H}}_{2}^{H} \overline{\mathbf{H}}_{2} \mathbf{\Phi} \overline{\mathbf{h}}_{k} \\
			=2 c_{k}^{2} \delta \varepsilon_{k} e_{k 1} M^{2}\left|f_{k}(\mathbf{\Phi})\right|^{2}.
		\end{array}
	\end{align}
	
	The fifth term is
	\begin{align}\label{leakage_10Cross_terms_5}
		\begin{array}{l}
			2 \mathbb{E}\left\{\sqrt{c_{k} \delta \varepsilon_{k}} \sqrt{c_{k} \varepsilon_{k}} \,\overline{\mathbf{h}}_{k}^{H} \mathbf{\Phi}^{H} \overline{\mathbf{H}}_{2}^{H} \tilde{\mathbf{H}}_{2} \mathbf{\Phi} \overline{\mathbf{h}}_{k}
			\left(e_{k 2} \sqrt{c_{k} \delta} \sqrt{c_{k}} \,\tilde{\mathbf{h}}_{k}^{H} \mathbf{\Phi}^{H} \overline{\mathbf{H}}_{2}^{H} \tilde{\mathbf{H}}_{2} \mathbf{\Phi} \tilde{\mathbf{h}}_{k}\right)^{H}\right\} \\
			=2 c_{k}^{2} \delta \varepsilon_{k} e_{k 2} \overline{\mathbf{h}}_{k}^{H} \mathbf{\Phi}^{H} \overline{\mathbf{H}}_{2}^{H} \mathbb{E}\left\{\tilde{\mathbf{H}}_{2} \mathbf{\Phi} \overline{\mathbf{h}}_{k} \tilde{\mathbf{h}}_{k}^{H} \mathbf{\Phi}^{H} \tilde{\mathbf{H}}_{2}^{H} \overline{\mathbf{H}}_{2} \mathbf{\Phi} \tilde{\mathbf{h}}_{k}\right\} \\
			=2 c_{k}^{2} \delta \varepsilon_{k} e_{k 2} \overline{\mathbf{h}}_{k}^{H} \mathbf{\Phi}^{H} \overline{\mathbf{H}}_{2}^{H} \mathbb{E}_{\tilde{\mathbf{h}}_{k}}\left\{\mathbb{E}_{\tilde{\mathbf{H}}_{2}}\left\{\tilde{\mathbf{H}}_{2} \mathbf{\Phi} \overline{\mathbf{h}}_{k} \tilde{\mathbf{h}}_{k}^{H} \mathbf{\Phi}^{H} \tilde{\mathbf{H}}_{2}^{H}\right\} \overline{\mathbf{H}}_{2} \mathbf{\Phi} \tilde{\mathbf{h}}_{k} \mid \tilde{\mathbf{h}}_{k}\right\} \\
			=2 c_{k}^{2} \delta \varepsilon_{k} e_{k 2} \overline{\mathbf{h}}_{k}^{H} \mathbf{\Phi}^{H} \overline{\mathbf{H}}_{2}^{H} \mathbb{E}\left\{\operatorname{Tr}\left\{\boldsymbol{\Phi} \overline{\mathbf{h}}_{k} \tilde{\mathbf{h}}_{k}^{H} \mathbf{\Phi}^{H}\right\} \overline{\mathbf{H}}_{2} \mathbf{\Phi} \tilde{\mathbf{h}}_{k}\right\} \\
			{{\mathop  = \limits^{\left( j \right)} }}2 c_{k}^{2} \delta \varepsilon_{k} e_{k 2} \overline{\mathbf{h}}_{k}^{H} \mathbf{\Phi}^{H} \overline{\mathbf{H}}_{2}^{H} \mathbb{E}\left\{\overline{\mathbf{H}}_{2} \boldsymbol{\Phi} \tilde{\mathbf{h}}_{k}\left(\tilde{\mathbf{h}}_{k}^{H} \overline{\mathbf{h}}_{k}\right)\right\} \\
			=2 c_{k}^{2} \delta \varepsilon_{k} e_{k 2} \overline{\mathbf{h}}_{k}^{H} \mathbf{\Phi}^{H} \overline{\mathbf{H}}_{2}^{H} \overline{\mathbf{H}}_{2} \boldsymbol{\Phi} \mathbb{E}\left\{\tilde{\mathbf{h}}_{k} \tilde{\mathbf{h}}_{k}^{H}\right\} \overline{\mathbf{h}}_{k} \\
			=2 c_{k}^{2} \delta \varepsilon_{k} e_{k 2} \overline{\mathbf{h}}_{k}^{H} \mathbf{\Phi}^{H} \overline{\mathbf{H}}_{2}^{H} \overline{\mathbf{H}}_{2} \mathbf{\Phi} \overline{\mathbf{h}}_{k} \\
			=2 c_{k}^{2} \delta \varepsilon_{k} e_{k 2} M\left|f_{k}(\mathbf{\Phi})\right|^{2},
		\end{array}
	\end{align}
	where $(j)$ uses $\operatorname{Tr}\left\{\mathbf{\Phi} \overline{\mathbf{h}}_{k} \tilde{\mathbf{h}}_{k}^{H} \mathbf{\Phi}^{H}\right\}=\tilde{\mathbf{h}}_{k}^{H} \overline{\mathbf{h}}_{k}$ and then places it at the end of the equation.
	
	The sixth term is
	\begin{align}\label{leakage_10Cross_terms_6}
		\begin{array}{l}
			2 \mathbb{E}\left\{e_{k 2} \sqrt{c_{k} \delta} \sqrt{c_{k} \delta \varepsilon_{k}}\, \tilde{\mathbf{h}}_{k}^{H} \mathbf{\Phi}^{H} \overline{\mathbf{H}}_{2}^{H} \overline{\mathbf{H}}_{2} \mathbf{\Phi} \overline{\mathbf{h}}_{k}
			\left(\sqrt{c_{k}} \sqrt{c_{k} \varepsilon_{k}} \, \tilde{\mathbf{h}}_{k}^{H} \mathbf{\Phi}^{H} \tilde{\mathbf{H}}_{2}^{H} \mathbf{A}_{k}^{H} \tilde{\mathbf{H}}_{2} \mathbf{\Phi} \overline{\mathbf{h}}_{k}\right)^{H}\right\} \\
			=2 c_{k}^{2} \delta \varepsilon_{k} e_{k 2} \mathbb{E}\left\{\tilde{\mathbf{h}}_{k}^{H} \mathbf{\Phi}^{H} \overline{\mathbf{H}}_{2}^{H} \overline{\mathbf{H}}_{2} \mathbf{\Phi} \overline{\mathbf{h}}_{k} \overline{\mathbf{h}}_{k}^{H} \mathbf{\Phi}^{H} \mathbb{E}\left\{\tilde{\mathbf{H}}_{2}^{H} \mathbf{A}_{k} \tilde{\mathbf{H}}_{2}\right\} \mathbf{\Phi} \tilde{\mathbf{h}}_{k}\right\} \\
			=2 c_{k}^{2} \delta \varepsilon_{k} e_{k 1} e_{k 2} M \mathbb{E}\left\{\tilde{\mathbf{h}}_{k}^{H} \mathbf{\Phi}^{H} \overline{\mathbf{H}}_{2}^{H} \overline{\mathbf{H}}_{2} \mathbf{\Phi} \overline{\mathbf{h}}_{k} \overline{\mathbf{h}}_{k}^{H} \tilde{\mathbf{h}}_{k}\right\} \\
			=2 c_{k}^{2} \delta \varepsilon_{k} e_{k 1} e_{k 2} M \operatorname{Tr}\left\{\overline{\mathbf{h}}_{k}^{H} \mathbf{\Phi}^{H} \mathbf{a}_{N} \mathbf{a}_{M}^{H} \mathbf{a}_{M} \mathbf{a}_{N}^{H} \mathbf{\Phi} \overline{\mathbf{h}}_{k}\right\} \\
			=2 c_{k}^{2} \delta \varepsilon_{k} e_{k 1} e_{k 2} M^{2}\left|f_{k}(\mathbf{\Phi})\right|^{2}.
		\end{array}
	\end{align}
	
	The seventh term is
	\begin{align}\label{leakage_10Cross_terms_7}
		\begin{array}{l}
			2 \mathbb{E}\left\{e_{k 2} \sqrt{c_{k} \delta} \sqrt{c_{k} \delta} \,\tilde{\mathbf{h}}_{k}^{H} \mathbf{\Phi}^{H} \overline{\mathbf{H}}_{2}^{H} \overline{\mathbf{H}}_{2} \mathbf{\Phi} \tilde{\mathbf{h}}_{k}
			\left(\sqrt{c_{k} \varepsilon_{k}} \sqrt{c_{k} \varepsilon_{k}} \,\overline{ \mathbf{h}}_{k}^{H} \mathbf{\Phi}^{H} \tilde{\mathbf{H}}_{2}^{H} \mathbf{A}_{k}^{H} \tilde{\mathbf{H}}_{2} \mathbf{\Phi} \overline{\mathbf{h}}_{k}\right)^{H}\right\} \\
			=2 c_{k}^{2} \delta\varepsilon_{k}  e_{k 2} \mathbb{E}\left\{\tilde{\mathbf{h}}_{k}^{H} \mathbf{\Phi}^{H} \overline{\mathbf{H}}_{2}^{H} \overline{\mathbf{H}}_{2} \mathbf{\Phi} \tilde{\mathbf{h}}_{k}\right\} \overline{\mathbf{h}}_{k}^{H} \mathbf{\Phi}^{H} \mathbb{E}\left\{\tilde{\mathbf{H}}_{2}^{H} \mathbf{A}_{k} \tilde{\mathbf{H}}_{2}\right\} \mathbf{\Phi} \overline{\mathbf{h}}_{k} \\
			=2 c_{k}^{2} \delta \varepsilon_{k} e_{k 1} e_{k 2} M \mathbb{E}\left\{\tilde{\mathbf{h}}_{k}^{H} \mathbf{\Phi}^{H} \overline{\mathbf{H}}_{2}^{H} \overline{\mathbf{H}}_{2} \mathbf{\Phi} \tilde{\mathbf{h}}_{k}\right\} \overline{\mathbf{h}}_{k}^{H} \overline{\mathbf{h}}_{k} \\
			=2 c_{k}^{2} \delta \varepsilon_{k} e_{k 1} e_{k 2}M \operatorname{Tr}\left\{\overline{\mathbf{H}}_{2}^{H} \overline{\mathbf{H}}_{2}\right\} \overline{\mathbf{h}}_{k}^{H} \overline{\mathbf{h}}_{k} \\
			=2 c_{k}^{2} \delta \varepsilon_{k} e_{k 1} e_{k 2} M^{2} N^{2}.
		\end{array}
	\end{align}
	
	The eighth term is
	\begin{align}\label{leakage_10Cross_terms_8}
		\begin{array}{l}
			2 \mathbb{E}\left\{e_{k 2} \sqrt{c_{k} \delta} \sqrt{c_{k} \delta} \,\tilde{\mathbf{h}}_{k}^{H} \mathbf{\Phi}^{H} \overline{\mathbf{H}}_{2}^{H} \overline{\mathbf{H}}_{2} \mathbf{\Phi} \tilde{\mathbf{h}}_{k}
			\left(\sqrt{c_{k}} \sqrt{c_{k}} \,\tilde{\mathbf{h}}_{k}^{H} \mathbf{\Phi}^{H} \tilde{\mathbf{H}}_{2}^{H} \mathbf{A}_{k}^{H} \tilde{\mathbf{H}}_{2} \mathbf{\Phi} \tilde{\mathbf{h}}_{k}\right)^{H}\right\} \\
			=2 c_{k}^{2} \delta e_{k 2} \mathbb{E}\left\{\tilde{\mathbf{h}}_{k}^{H} \mathbf{\Phi}^{H} \overline{\mathbf{H}}_{2}^{H} \overline{\mathbf{H}}_{2} \mathbf{\Phi} \tilde{\mathbf{h}}_{k} \tilde{\mathbf{h}}_{k}^{H} \mathbf{\Phi}^{H} \mathbb{E}\left\{\tilde{\mathbf{H}}_{2}^{H} \mathbf{A}_{k} \tilde{\mathbf{H}}_{2}\right\} \mathbf{\Phi} \tilde{\mathbf{h}}_{k}\right\} \\
			=2 c_{k}^{2} \delta e_{k 1} e_{k 2} M \mathbb{E}\left\{\tilde{\mathbf{h}}_{k}^{H} \mathbf{\Phi}^{H} \overline{\mathbf{H}}_{2}^{H} \overline{\mathbf{H}}_{2} \mathbf{\Phi} \tilde{\mathbf{h}}_{k} \tilde{\mathbf{h}}_{k}^{H} \tilde{\mathbf{h}}_{k}\right\} \\
			=2 c_{k}^{2} \delta e_{k 1} e_{k 2} M\operatorname{Tr}\left\{\mathbf{\Phi}^{H} \overline{\mathbf{H}}_{2}^{H} \overline{\mathbf{H}}_{2} \mathbf{\Phi} \mathbb{E}\left\{\tilde{\mathbf{h}}_{k} \tilde{\mathbf{h}}_{k}^{H} \tilde{\mathbf{h}}_{k} \tilde{\mathbf{h}}_{k}^{H}\right\}\right\} \\
			=2 c_{k}^{2} \delta e_{k 1} e_{k 2} M\operatorname{Tr}\left\{\mathbf{\Phi}^{H} \overline{\mathbf{H}}_{2}^{H} \overline{\mathbf{H}}_{2} \mathbf{\Phi}(N+1) \mathbf{I}_{N}\right\} \\
			=2 c_{k}^{2} \delta e_{k 1} e_{k 2} M^{2} N(N+1).
		\end{array}
	\end{align}
	
	The ninth term is
	\begin{align}\label{leakage_10Cross_terms_9}
		\begin{array}{l}
			2 \mathbb{E}\left\{\sqrt{c_{k} \varepsilon_{k}} \sqrt{c_{k} \delta \varepsilon_{k}} \, \overline{\mathbf{h}}_{k}^{H} \mathbf{\Phi}^{H} \tilde{\mathbf{H}}_{2}^{H} \mathbf{A}_{k}^{H} \overline{\mathbf{H}}_{2} \mathbf{\Phi} \overline{\mathbf{h}}_{k}
			\left(\sqrt{c_{k}} \sqrt{c_{k} \delta} \,\tilde{\mathbf{h}}_{k}^{H} \mathbf{\Phi}^{H} \tilde{\mathbf{H}}_{2}^{H} \mathbf{A}_{k}^{H} \overline{\mathbf{H}}_{2} \mathbf{\Phi} \tilde{\mathbf{h}}_{k}\right)^{H}\right\} \\
			=2 c_{k}^{2} \delta \varepsilon_{k} \overline{\mathbf{h}}_{k}^{H} \mathbf{\Phi}^{H} \mathbb{E}_{\tilde{\mathbf{h}}_{k}}\left\{\mathbb{E}_{\tilde{\mathbf{H}}_{2}}\left\{\tilde{\mathbf{H}}_{2}^{H} \mathbf{A}_{k}^{H} \overline{\mathbf{H}}_{2} \mathbf{\Phi} \overline{\mathbf{h}}_{k} \tilde{\mathbf{h}}_{k}^{H} \mathbf{\Phi}^{H} \overline{\mathbf{H}}_{2}^{H} \mathbf{A}_{k} \tilde{\mathbf{H}}_{2}\right\} \mathbf{\Phi} \tilde{\mathbf{h}}_{k} \mid \tilde{\mathbf{h}}_{k}\right\} \\
			=2 c_{k}^{2} \delta \varepsilon_{k} \overline{\mathbf{h}}_{k}^{H} \mathbf{\Phi}^{H} \mathbb{E}\left\{\operatorname{Tr}\left\{\mathbf{A}_{k}^{H} \overline{\mathbf{H}}_{2} \mathbf{\Phi} \overline{\mathbf{h}}_{k} \tilde{\mathbf{h}}_{k}^{H} \mathbf{\Phi}^{H} \overline{\mathbf{H}}_{2}^{H} \mathbf{A}_{k}\right\} \mathbf{\Phi} \tilde{\mathbf{h}}_{k}\right\} \\
			=2 c_{k}^{2} \delta \varepsilon_{k} e_{k 2}^{2} \overline{\mathbf{h}}_{k}^{H} \mathbf{\Phi}^{H} \mathbb{E}\left\{\operatorname{Tr}\left\{\overline{\mathbf{H}}_{2} \mathbf{\Phi} \overline{\mathbf{h}}_{k} \tilde{\mathbf{h}}_{k}^{H} \mathbf{\Phi}^{H} \overline{\mathbf{H}}_{2}^{H}\right\} \mathbf{\Phi} \tilde{\mathbf{h}}_{k}\right\} \\
			=2 c_{k}^{2} \delta \varepsilon_{k} e_{k 2}^{2} \overline{\mathbf{h}}_{k}^{H} \mathbf{\Phi}^{H} \mathbb{E}\left\{\mathbf{\Phi} \tilde{\mathbf{h}}_{k}\left(\tilde{\mathbf{h}}_{k}^{H} \mathbf{\Phi}^{H} \overline{\mathbf{H}}_{2}^{H} \overline{\mathbf{H}}_{2} \mathbf{\Phi} \overline{\mathbf{h}}_{k}\right)\right\} \\
			=2 c_{k}^{2} \delta \varepsilon_{k} e_{k 2}^{2} \overline{\mathbf{h}}_{k}^{H} \mathbf{\Phi}^{H} \mathbf{\Phi} \mathbb{E}\left\{\tilde{\mathbf{h}}_{k} \tilde{\mathbf{h}}_{k}^{H}\right\} \mathbf{\Phi}^{H} \overline{\mathbf{H}}_{2}^{H} \overline{\mathbf{H}}_{2} \mathbf{\Phi} \overline{\mathbf{h}}_{k} \\
			=2 c_{k}^{2} \delta \varepsilon_{k} e_{k 2}^{2} \overline{\mathbf{h}}_{k}^{H} \mathbf{\Phi}^{H} \overline{\mathbf{H}}_{2}^{H} \overline{\mathbf{H}}_{2} \mathbf{\Phi} \overline{\mathbf{h}}_{k} \\
			=2 c_{k}^{2} \delta \varepsilon_{k} e_{k 2}^{2} M\left|f_{k}(\mathbf{\Phi})\right|^{2}.
		\end{array}
	\end{align}
	
	The tenth term is
	\begin{align}\label{leakage_10Cross_terms_10}
		\begin{array}{l}
			2 \mathbb{E}\left\{\sqrt{c_{k} \varepsilon_{k}} \sqrt{c_{k} \varepsilon_{k}} \, \overline{\mathbf{h}}_{k}^{H} \mathbf{\Phi}^{H} \tilde{\mathbf{H}}_{2}^{H} \mathbf{A}_{k}^{H} \tilde{\mathbf{H}}_{2} \mathbf{\Phi} \overline{\mathbf{h}}_{k}
			\left(\sqrt{c_{k}} \sqrt{c_{k}} \,\tilde{\mathbf{h}}_{k}^{H} \mathbf{\Phi}^{H} \tilde{\mathbf{H}}_{2}^{H} \mathbf{A}_{k}^{H} \tilde{\mathbf{H}}_{2} \mathbf{\Phi} \tilde{\mathbf{h}}_{k}\right)^{H}\right\} \\
			=2 c_{k}^{2} \varepsilon_{k} \overline{\mathbf{h}}_{k}^{H} \mathbf{\Phi}^{H} \mathbb{E}_{\tilde{\mathbf{h}}_{k}}\left\{\mathbb{E}_{\tilde{\mathbf{H}}_{2}}\left\{\tilde{\mathbf{H}}_{2}^{H} \mathbf{A}_{k}^{H} \tilde{\mathbf{H}}_{2} \mathbf{\Phi} \overline{\mathbf{h}}_{k} \tilde{\mathbf{h}}_{k}^{H} \mathbf{\Phi}^{H} \tilde{\mathbf{H}}_{2}^{H} \mathbf{A}_{k} \tilde{\mathbf{H}}_{2}\right\} \mathbf{\Phi} \tilde{\mathbf{h}}_{k} \mid \tilde{\mathbf{h}}_{k}\right\} \\
			=2 c_{k}^{2} \varepsilon_{k} \overline{\mathbf{h}}_{k}^{H} \mathbf{\Phi}^{H} \mathbb{E}\left\{\left(e_{k 1}^{2} M^{2} \mathbf{\Phi} \overline{\mathbf{h}}_{k} \tilde{\mathbf{h}}_{k}^{H} \mathbf{\Phi}^{H}+e_{k 3} M\operatorname{Tr}\left\{\mathbf{\Phi} \overline{\mathbf{h}}_{k} \tilde{\mathbf{h}}_{k}^{H} \mathbf{\Phi}^{H}\right\} \mathbf{I}_{N}\right)  \mathbf{\Phi} \tilde{\mathbf{h}}_{k}  \right\}  \\
			=2 c_{k}^{2} \varepsilon_{k} \mathbb{E}\left\{e_{k 1}^{2} M^{2} \overline{\mathbf{h}}_{k}^{H} \overline{\mathbf{h}}_{k} \tilde{\mathbf{h}}_{k}^{H} \tilde{\mathbf{h}}_{k}+e_{k 3} M \overline{\mathbf{h}}_{k}^{H} \tilde{\mathbf{h}}_{k}\left(\tilde{\mathbf{h}}_{k}^{H} \overline{\mathbf{h}}_{k}\right)\right\} \\
			=2 c_{k}^{2} \varepsilon_{k}\left(e_{k 1}^{2} M^{2} N^{2}+e_{k 3} M N\right).
		\end{array}
	\end{align}
	
	Thus, we have completed the calculation of $10$ cross-terms. After some direct simplifications, we can obtain $\mathbb{E}\left\{\left|\hat{\mathbf{q}}_{k}^{H} \mathbf{q}_{k}\right|^{2}\right\}$  by combining (\ref{leakage_7terms_2}) - (\ref{leakage_7terms_7}), (\ref{leakage_16terms_1}) - (\ref{leakage_16terms_16}) and (\ref{leakage_10Cross_terms_1}) - (\ref{leakage_10Cross_terms_10}) with (\ref{leakage_expanded-sum}). With the aid of $ \mathbb{E}\left\{\left|   \hat{\underline{\mathbf{q}}}_{k}^{H} \underline{\mathbf{q}}_{k}\right|^{2}\right\} $ and (\ref{derive_noise_and_signal}), we can complete the calculation of the signal leakage $E_k^{\rm leak} \left(\mathbf{\Phi}\right) $ using (\ref{leakage_make_difference}).

	\section{}\label{appendix5}
	Recall the definition of $f_{k}({\bf\Phi})$ in (\ref{f_k_Phi_definition}). If $N=1$, we have $ \zeta_{1}^{k}=0 $. Then, for any design of $\theta_1$, we have $ \left|f_{k}({\bf\Phi})\right|=\left|e^{j \theta_{1}}\right|=1 $.
	
	If $N>1$, we aim to prove that $0\leq \left|f_{k}({\bf\Phi})\right|\leq N$. Firstly, by invoking the triangle inequality, we have
	\begin{align}
		\left|f_{k}({\bf\Phi})\right|=\left|\sum\nolimits_{n=1}^{N} e^{j\left(\zeta_{n}^{k}+\theta_{n}\right)}\right| \leq \sum\nolimits_{n=1}^{N}\left|e^{j\left(\zeta_{n}^{k}+\theta_{n}\right)}\right|=N.
	\end{align}
	The equality holds if the phase shifts of all the RIS elements are aligned as
	\begin{align}\label{phase_N}
		\theta_{n}=-\zeta_{n}^{k}+{C_0}, \forall n,
	\end{align}
	where $C_0$ is an arbitrary constant. 
	
	Next, we aim to prove that the minimum value of  $ \left|f_{k}({\bf\Phi})\right| $ is zero. Firstly, if $N$ is even, the minimum value $0$ is obtained when
	\begin{align}\label{phase_0}
		\theta_{2 i-1}+\zeta_{2 i-1}^{k}=\left(\theta_{2 i}+\zeta_{2 i}^{k}\right)+\pi, 1 \leq i \leq \frac{N}{2}.
	\end{align}
	
	Otherwise, if $N$ is odd, the minimum value $0$ is still achievable for
	\begin{align}\label{phase_0_odd}
		\begin{array}{l}
			\theta_{2 i-1}+\zeta_{2 i-1}^{k}=\left(\theta_{2 i}+\zeta_{2 i}^{k}\right)+\pi, 1 \leq i \leq \frac{N-1}{2}-1, \\
			\theta_{N-2}+\zeta_{N-2}^{k}=\frac{\pi}{3}, \\
			\theta_{N-1}+\zeta_{N-1}^{k}=-\frac{\pi}{3}, \\
			\theta_{N}+\zeta_{N}^{k}=\pi.
		\end{array}
	\end{align}
	
	Next, we aim to prove that when the phase shifts of the RIS are designed to maximize $\left|f_{k}({\bf\Phi})\right|$, the corresponding term $\left|f_{i}({\bf\Phi})\right|$ for the user $i$ is bounded when $N\to\infty$. Note that we can prove this result rigorously under the one-dimensional uniform linear array (ULA) model. 
	Since the USPA model is only a two-dimensional extension of the ULA model, we can deduce that the conclusion still holds.
	
	By ignoring the elevation direction in (\ref{uspa_hk}) and (\ref{uspa_H2}) of the USPA model, we can obtain a one-dimensional ULA model for $\overline{\mathbf{h}}_k$ and $\mathbf{a}_N$ with AoA $\varphi_{k r}^{a}$ and AoD $\varphi_{t}^{a}$, respectively. Then, we can rewrite $ f_{k}({\bf\Phi}) $ as
	\begin{align}
		\begin{array}{l}
			\overline{\mathbf{h}}_{k} \triangleq \mathbf{a}_{N}\left(\varphi_{k r}^{a}\right)=\left[1, e^{j2 \pi \frac{d}{\lambda} \sin \varphi_{k r}^{a}}, \ldots, e^{j2 \pi \frac{d}{\lambda}(N-1) \sin \varphi_{k r}^{a}  }\right]  ^{T}, \\
			\mathbf{a}_{N} \triangleq \mathbf{a}_{N}\left(\varphi_{t}^{a}\right)=\left[1, e^{j2 \pi \frac{d}{\lambda} \sin \varphi_{t}^{a}}, \ldots, e^{j2 \pi \frac{d}{\lambda}(N-1) \sin \varphi_{t}^{a}}\right]^{T}, \\
			f_{k}({\bf\Phi})=\mathbf{a}_{N}^{H} {\bf\Phi} \overline{\mathbf{h}}_{k}=\sum_{n=1}^{N} e^{j2 \pi \frac{d}{\lambda}(n-1)
				\left( \sin \varphi_{k r}^{a} -\sin \varphi_{t}^{a}  \right)+j\theta_{n}}.
		\end{array}
	\end{align}
	
	With $ \theta_{n}=2 \pi \frac{d}{\lambda}(n-1)\left(  \sin \varphi_{t}^{a}-\sin \varphi_{k r}^{a}   \right) $, we have $\left|f_{k}({\bf\Phi})\right|=N$. At the same time, for the user $i$, we have
	\begin{align}
		\begin{aligned}
			f_{i}({\bf\Phi})&=\mathbf{a}_{N}^{H} {\bf\Phi} \overline{\mathbf{h}}_{i}=\sum\nolimits_{n=1}^{N} e^{j2 \pi \frac{d}{\lambda}(n-1)\left(\sin \varphi_{i r}^{a}  - \sin \varphi_{t}^{a}  \right)+j\theta_{n}} \\
			&=\sum\nolimits_{n=1}^{N} e^{j2 \pi \frac{d}{\lambda}(n-1)\left( \sin \varphi_{i r}^{a} - \sin \varphi_{t}^{a}  +\sin \varphi_{t}^{a}-\sin \varphi_{k r}^{a}  \right)} \\
			&=\sum\nolimits_{n=1}^{N} e^{j2 \pi \frac{d}{\lambda}(n-1)\left(   \sin \varphi_{i r}^{a}  -  \sin \varphi_{k r}^{a}     \right)}.
		\end{aligned}
	\end{align}
	
	Then, by using the property of geometric progression, we obtain
	\begin{align}
		\begin{aligned}
			\left|f_{i}({\bf\Phi})\right|&=\left|\sum\nolimits_{n=1}^{N}\left(e^{j2 \pi \frac{d}{\lambda}\left(   \sin \varphi_{i r}^{a}  -  \sin \varphi_{k r}^{a} \right)}\right)^{(n-1)}\right|
			= \left|\frac{1-e^{j2 \pi \frac{d}{\lambda} N\left( \sin \varphi_{i r}^{a}  -  \sin \varphi_{k r}^{a}  \right)}}
			{1-e^{j2 \pi \frac{d}{\lambda}\left(   \sin \varphi_{i r}^{a}  -  \sin \varphi_{k r}^{a}   \right)}}\right| \\
			&=\left|\frac{e^{-j\pi \frac{d}{\lambda} N\left( \sin \varphi_{i r}^{a}  -  \sin \varphi_{k r}^{a} \right)}
				-e^{j\pi \frac{d}{\lambda} N\left( \sin \varphi_{i r}^{a}  -  \sin \varphi_{k r}^{a} \right)}}
			{ e^{-j\pi \frac{d}{\lambda}\left( \sin \varphi_{i r}^{a}  -  \sin \varphi_{k r}^{a} \right)}
				-e^{j\pi \frac{d}{\lambda}\left( \sin \varphi_{i r}^{a}  -  \sin \varphi_{k r}^{a} \right)}} 
			\times \frac{e^{j\pi \frac{d}{\lambda} N\left( \sin \varphi_{i r}^{a}  -  \sin \varphi_{k r}^{a} \right)}}
			{e^{j\pi \frac{d}{\lambda}\left( \sin \varphi_{i r}^{a}  -  \sin \varphi_{k r}^{a} \right)}}\right| \\
			&=\frac{\sin \left(\pi \frac{d}{\lambda} N\left( \sin \varphi_{i r}^{a}  -  \sin \varphi_{k r}^{a} \right)\right)}{\sin \left(\pi \frac{d}{\lambda}\left( \sin \varphi_{i r}^{a}  -  \sin \varphi_{k r}^{a} \right)\right)}.
		\end{aligned}
	\end{align}
	
	Therefore, if the user $i$ does not have the same AoA as user $k$, the term $\left|f_{i}({\bf\Phi})\right|$ is bounded when $N\to\infty$. Then, following a similar process, we can prove that the term $\left|\overline{\mathbf{h}}_{k}^{H} \overline{\mathbf{h}}_{i}\right|^{2}$ is bounded when $N\to\infty$.

	
	{\color{blue}
		\section{}\label{appendix7}
		To begin with, we need to derive the first and second order statistical properties for the aggregated channel and the observation vector.  The expectation is $\mathbb{E}\left\{\mathbf{y}_{c, p}^{k}\right\}=\sqrt{\widehat{c}_{k} \delta} \,\overline{\mathbf{H}}_{2} {\bf\Phi} \overline{\mathbf{h}}_{k}$.
		Aided by Lemma \ref{lemma_HWH}, the covariances between $\mathbf{q}_{c, k}$ and $\mathbf{y}_{c, p}^{k}$ is given by
		\begin{align}\label{cov_q_y}
			\begin{aligned}
				&\operatorname{Cov}\left\{\mathbf{q}_{c, k}, \mathbf{y}_{c, p}^{k}\right\}=\mathbb{E}\left\{\left(\mathbf{q}_{c, k}-\mathbb{E}\left\{\mathbf{q}_{c, k}\right\}\right)\left(\mathbf{y}_{c, p}^{k}-\mathbb{E}\left\{\mathbf{y}_{c, p}^{k}\right\}\right)^{H}\right\}\\
				&=\mathbb{E}\left\{\left(\sqrt{\widehat{c}_{k}} \tilde{\mathbf{H}}_{c, 2} \boldsymbol{\Phi} \overline{\mathbf{h}}_{k}+\sqrt{\gamma_{k}} \tilde{\mathbf{d}}_{k}\right)\left(\sqrt{\widehat{c}_{k}} \tilde{\mathbf{H}}_{c, 2} {\bf\Phi} \overline{\mathbf{h}}_{k}+\sqrt{\gamma_{k}} \tilde{\mathbf{d}}_{k}\right)^{H}\right\}\\
				&=\mathbb{E}\left\{\widehat{c}_{k} \tilde{\mathbf{H}}_{2} \mathbf{R}_{r i s}^{1 / 2} \boldsymbol{\Phi} \overline{\mathbf{h}}_{k} \overline{\mathbf{h}}_{k}^{H} \boldsymbol{\Phi}^{H} \mathbf{R}_{r i s}^{1 / 2} \tilde{\mathbf{H}}_{2}^{H}+\gamma_{k} \tilde{\mathbf{d}}_{k} \tilde{\mathbf{d}}_{k}^{H}\right\}\\
				&=\left(\widehat{c}_{k} \overline{\mathbf{h}}_{k}^{H} \boldsymbol{\Phi}^{H} \mathbf{R}_{ris} {\bf\Phi} \overline{\mathbf{h}}_{k}+\gamma_{k}\right) \mathbf{I}_{M}.
			\end{aligned}
		\end{align}
		
		Using Lemma \ref{lemma_HWH}, the definition of $\mathbf{H}_{c, 2}$ in (\ref{H2_cor}), the fact $\mathbf{V}=\mathbf{R}_{e m i}^{1 / 2} \tilde{\mathbf{V}}$, and the independence between channels, noise, and EMI, the covariance of $\mathbf{y}_{c, p}^{k}$ is calculated as
		\begin{align}\label{cov_y_y}
			\begin{aligned}
				&\operatorname{Cov}\left\{ \mathbf{y}_{c, p}^{k}, \mathbf{y}_{c, p}^{k}\right\}=\mathbb{E}\left\{\left(\mathbf{y}_{c, p}^{k}-\mathbb{E}\left\{\mathbf{y}_{c, p}^{k}\right\}\right)\left(\mathbf{y}_{c, p}^{k}-\mathbb{E}\left\{\mathbf{y}_{c, p}^{k}\right\}\right)^{H}\right\}\\
				&=\mathbb{E}\left\{\widehat{c}_{k} \tilde{\mathbf{H}}_{c, 2} \boldsymbol{\Phi} \overline{\mathbf{h}}_{k} \overline{\mathbf{h}}_{k}^{H} \boldsymbol{\Phi}^{H} \tilde{\mathbf{H}}_{c, 2}^{H}+\gamma_{k} \tilde{\mathbf{d}}_{k} \tilde{\mathbf{d}}_{k}^{H}+\frac{\mathbf{H}_{c, 2} \mathbf{\Phi} \mathbf{V} \mathbf{s}_{k} \mathbf{s}_{k}{ }^{H} \mathbf{V}^{H} \boldsymbol{\Phi}^{H} \mathbf{H}_{c, 2}^{H}}{\tau p}+\frac{\mathbf{N s}_{k} \mathbf{s}_{k}{ }^{H} \mathbf{N}^{H}}{\tau p}\right\}\\
				&=\left(\widehat{c}_{k} \overline{\mathbf{h}}_{k}^{H} \boldsymbol{\Phi}^{H} \mathbf{R}_{ris} \boldsymbol{\Phi} \overline{\mathbf{h}}_{k}+\gamma_{k}+\frac{\sigma^{2}}{\tau p}\right) \mathbf{I}_{M}+\mathbb{E}\left\{\frac{\frac{\beta \delta}{\delta+1} \overline{\mathbf{H}}_{2} \boldsymbol{\Phi} \mathbf{V} \mathbf{s}_{k} \mathbf{s}_{k}{ }^{H} \mathbf{V}^{H} \boldsymbol{\Phi}^{H} \overline{\mathbf{H}}_{2}^{H}}{\tau p}\right\}\\
				&\;\;\;\;+\mathbb{E}\left\{\frac{\frac{\beta}{\delta+1} \tilde{\mathbf{H}}_{c, 2} \boldsymbol{\Phi} \mathbf{V}  \mathbf{s}_{k} \mathbf{s}_{k}{ }^{H} \mathbf{V}^{H} \boldsymbol{\Phi}^{H} \tilde{\mathbf{H}}_{c, 2}^{H}}{\tau p}\right\}\\
				&=\left(\widehat{c}_{k} \overline{\mathbf{h}}_{k}^{H} \boldsymbol{\Phi}^{H} \mathbf{R}_{ris} \boldsymbol{\Phi} \overline{\mathbf{h}}_{k}+\gamma_{k}+\frac{\sigma^{2}}{\tau p}\right) \mathbf{I}_{M}+\frac{\sigma_{e}^{2} \beta \delta \overline{\mathbf{H}}_{2} \boldsymbol{\Phi} \mathbf{R}_{e m i} \boldsymbol{\Phi}^{H} \overline{\mathbf{H}}_{2}^{H}}{\tau p(\delta+1)}+\mathbb{E}\left\{\frac{\sigma_{e}^{2} \beta \tilde{\mathbf{H}}_{c,2}  \mathbf{\Phi} \mathbf{R}_{e m i} \mathbf{\Phi}^{H} \tilde{\mathbf{H}}_{c,2}^{H}}{\tau p(\delta+1)}\right\} \\
				&=\left(\widehat{c}_{k} \overline{\mathbf{h}}_{k}^{H} \boldsymbol{\Phi}^{H} \mathbf{R}_{ris} \boldsymbol{\Phi} \overline{\mathbf{h}}_{k}+\gamma_{k}+\frac{\sigma^{2}}{\tau p}\right) \mathbf{I}_{M}+\frac{\sigma_{e}^{2} \beta \delta \overline{\mathbf{H}}_{2} \boldsymbol{\Phi} \mathbf{R}_{e m i} \mathbf{\Phi}^{H} \overline{\mathbf{H}}_{2}^{H}}{\tau p(\delta+1)}+\frac{\sigma_{e}^{2} \beta \operatorname{Tr}\left\{\mathbf{R}_{e m i} \mathbf{\Phi}^{H} \mathbf{R}_{r i s} \boldsymbol{\Phi}\right\}}{\tau p(\delta+1)} \mathbf{I}_{M}.
			\end{aligned}
		\end{align}
		
		Then, the LMMSE channel estimate for channel $\mathbf{q}_{c, k}$ is given by
		\begin{align}\label{lmmse_cor}
			\hat{\mathbf{q}}_{c, k}=\mathbb{E}\left\{\mathbf{q}_{c, k}\right\}+\operatorname{Cov}\left\{\mathbf{q}_{c, k}, \mathbf{y}_{c, p}^{k}\right\} \operatorname{Cov}^{-1}\left\{\mathbf{y}_{c, p}^{k}, \mathbf{y}_{c, p}^{k}\right\}\left(\mathbf{y}_{c, p}^{k}-\mathbb{E}\left\{\mathbf{y}_{c, p}^{k} \right\}\right).
		\end{align}
		Combining (\ref{lmmse_cor}) with (\ref{cov_q_y}) and (\ref{cov_y_y}) completes the proof.


		\section{}\label{appendix8}
		Apply Lemma \ref{lemma_HWH} and \ref{lemma_HCHWHCH}, the proof can be done following a similar process as in Appendix \ref{appendix4}. Using the orthogonal property, the noise term is given by
		\begin{align}\label{noise_deriv}
			\begin{aligned}
				&E_{c, k}^{\mathrm {noise }} = \mathbb{E}\left\{\left\|\hat{\mathbf{q}}_{c, k}\right\|^{2}\right\} 
				= \mathbb{E}\left\{\hat{\mathbf{q}}_{c, k}^{H} \mathbf{q}_{c, k}\right\}\\
				&=\widehat{c}_{k} \delta \overline{\mathbf{h}}_{k}^{H} \boldsymbol{\Phi}^{H} \overline{\mathbf{H}}_{2}^{H} \overline{\mathbf{H}}_{2} \boldsymbol{\Phi} \overline{\mathbf{h}}_{k}+\mathbb{E}\left\{\widehat{c}_{k} \overline{\mathbf{h}}_{k}^{H} \boldsymbol{\Phi}^{H} \tilde{\mathbf{H}}_{c, 2}^{H} \mathbf{\Upsilon}_{k}^{H} \tilde{\mathbf{H}}_{c, 2} \boldsymbol{\Phi} \overline{\mathbf{h}}_{k}\right\}+\gamma_{k} \mathbb{E}\left\{\tilde{\mathbf{d}}_{k}^{H} \mathbf{\Upsilon}_{k}^{H} \tilde{\mathbf{d}}_{k}\right\}\\
				&=M \widehat{c}_{k} \delta \overline{\mathbf{h}}_{k}^{H} \boldsymbol{\Phi}^{H} \mathbf{a}_{N} \mathbf{a}_{N}^{H} \boldsymbol{\Phi} \overline{\mathbf{h}}_{k}+\widehat{c}_{k} \overline{\mathbf{h}}_{k}^{H} \boldsymbol{\Phi}^{H} \mathbf{R}_{r i s}^{1 / 2} \mathbb{E}\left\{\tilde{\mathbf{H}}_{2}^{H} \mathbf{\Upsilon}_{k}^{H} \tilde{\mathbf{H}}_{2}\right\} \mathbf{R}_{r i s}^{1 / 2} \boldsymbol{\Phi} \overline{\mathbf{h}}_{k}+\gamma_{k} \mathbb{E}\left\{\tilde{\mathbf{d}}_{k}^{H} \mathbf{\Upsilon}_{k}^{H} \tilde{\mathbf{d}}_{k}\right\}\\
				&=M \widehat{c}_{k} \delta\left|f_{k}({\bf\Phi})\right|^{2}+\widehat{c}_{k} \operatorname{Tr}\left\{  { \bf\Upsilon}_{k}^{H}\right\} \overline{\mathbf{h}}_{k}^{H} {\bf\Phi}^{H} \mathbf{R}_{r i s} {\bf\Phi} \overline{\mathbf{h}}_{k}+\gamma_{k} \operatorname{Tr}\left\{{\bf\Upsilon}_{k}^{H}\right\}.
			\end{aligned}
		\end{align}
		
		By substituting $ \mathbf{H}_{c, 2}=\sqrt{\frac{\beta}{\delta+1}}\left(\sqrt{\delta} \overline{\mathbf{H}}_{2}+\tilde{\mathbf{H}}_{c, 2}\right) $, the EMI term is calculated as
		\begin{align}\label{EMI_deriving}
			\begin{aligned}
				&\mathbb{E}\left\{\hat{\mathbf{q}}_{c, k}^{H} \mathbf{H}_{c, 2} \mathbf{\Phi} \mathbf{R}_{e m i} \mathbf{\Phi}^{H} \mathbf{H}_{c, 2}^{H} \hat{\mathbf{q}}_{c, k}\right\} =\frac{\beta}{\delta+1}  \bigg( \delta \mathbb{E}\left\{\hat{\mathbf{q}}_{c, k}^{H} \overline{\mathbf{H}}_{2} \boldsymbol{\Phi} \mathbf{R}_{e m i} \boldsymbol{\Phi}^{H} \overline{\mathbf{H}}_{2}^{H} \hat{\mathbf{q}}_{c, k}\right\} \\
				&+2\sqrt{\delta} \mathbb{E}\left\{\hat{\mathbf{q}}_{c, k}^{H} \overline{\mathbf{H}}_{2} \mathbf{\Phi} \mathbf{R}_{e m i} \boldsymbol{\Phi}^{H} \tilde{\mathbf{H}}_{c, 2}^{H} \hat{\mathbf{q}}_{c, k}\right\} + \mathbb{E}\left\{\hat{\mathbf{q}}_{c, k}^{H} \tilde{\mathbf{H}}_{c, 2} \mathbf{\Phi} \mathbf{R}_{e m i} \boldsymbol{\Phi}^{H} \tilde{\mathbf{H}}_{c, 2}^{H} \hat{\mathbf{q}}_{c, k}\right\} \bigg).
			\end{aligned}
		\end{align}
		
		(\ref{EMI_deriving}) can be derived by inserting the definition of $\hat{\mathbf{q}}_{c, k}$ from (\ref{channel_estimation_cor}), using Lemma \ref{lemma_HWH} and \ref{lemma_HCHWHCH}, and utilizing the independence between $\tilde{\mathbf{H}}_{c, 2}^{H} $, $\mathbf{V}$, and $\mathbf{N}$. Details of the proof are omitted here for brevity.
		
		Next, we discuss the derivation of the interference term. For notational simplicity, define 
		\begin{align}
			\begin{aligned}
				&\hat{\underline{\mathbf{q}}}_{c, k}^{H} = \sqrt{\widehat{c}_{k} \delta} \overline{\mathbf{h}}_{k}^{H} \boldsymbol{\Phi}^{H} \overline{\mathbf{H}}_{2}^{H}+\sqrt{\widehat{c}_{k}} \overline{\mathbf{h}}_{k}^{H} \boldsymbol{\Phi}^{H} \tilde{\mathbf{H}}_{c, 2}^{H} \mathbf{\Upsilon}_{k}^{H}, \\
				&\underline{\mathbf{q}}_{c, i}=\sqrt{\widehat{c}_{i} \delta} \overline{\mathbf{H}}_{2} \boldsymbol{\Phi} \overline{\mathbf{h}}_{i}+\sqrt{\widehat{c}_{i}} \tilde{\mathbf{H}}_{c, 2} \boldsymbol{\Phi} \overline{\mathbf{h}}_i.
			\end{aligned}
		\end{align}
		
		Then, based on the independence, the interference term can be divided by
		\begin{align}
			\mathbb{E}\left\{\left|\hat{\mathbf{q}}_{c, k}^{H} \mathbf{q}_{c, i}\right|^{2}\right\} 
			= \mathbb{E}\left\{\left|   \hat{\mathbf{q}}_{c, k}^{H} \underline{\mathbf{q}}_{c, i}\right|^{2}\right\} 
			+ \mathbb{E}\left\{\left|\sqrt{\gamma_{i}} \hat{\mathbf{q}}_{c, k}^{H} \tilde{\mathbf{d}}_{i}\right|^{2}\right\},
		\end{align}
		where
		\begin{align}
			\mathbb{E}\left\{\left|\sqrt{\gamma_{i}} \hat{\mathbf{q}}_{c, k}^{H} \tilde{\mathbf{d}}_{i}\right|^{2}\right\}
			=\mathbb{E}\left\{\gamma_{i} \hat{\mathbf{q}}_{c, k}^{H} \tilde{\mathbf{d}}_{i} \tilde{\mathbf{d}}_{i}^{H} \hat{\mathbf{q}}_{c, k}\right\}
			=\gamma_{i} \mathbb{E}\left\{\left\|\hat{\mathbf{q}}_{c, k}^{H}\right\|^{2}\right\} 
			= \gamma_{i}  E_{c, k}^{\mathrm {noise }},
		\end{align}
		and
		\begin{align}\label{interference_deriving}
			\begin{aligned}
				&\mathbb{E}\left\{\left|   \hat{\mathbf{q}}_{c, k}^{H} \underline{\mathbf{q}}_{c, i}\right|^{2}\right\} 
				=\mathbb{E}\left\{\left|   \hat{\underline{\mathbf{q}}}_{c, k}^{H} \underline{\mathbf{q}}_{c, i}\right|^{2}\right\} 
				+\mathbb{E}\left\{\left|\left(\sqrt{\gamma_{k}} \tilde{\mathbf{d}}_{k}^{H} \mathbf{\Upsilon}_{k}^{H}+\frac{\mathbf{s}_{k}^{H} \mathbf{V}^{H} \mathbf{\Phi}^{H} \mathbf{H}_{c, 2}^{H} \boldsymbol{\Upsilon}_{k}^{H}}{\sqrt{\tau p}}+\frac{\mathbf{s}_{k}^{H} \mathbf{N}^{H} \mathbf{\Upsilon}_{k}^{H}}{\sqrt{\tau p}}\right) \underline{\mathbf{q}}_{c, i}\right|^2\right\}\\
				&=\mathbb{E}\left\{\left|   \hat{\underline{\mathbf{q}}}_{c, k}^{H} \underline{\mathbf{q}}_{c, i}\right|^{2}\right\} 
				+\mathbb{E}\left\{\left(\gamma_{k}+\frac{{\sigma}^{2}}{\tau p}\right) \underline{\mathbf{q}}_{c, i}^{H} \boldsymbol{\Upsilon}_{k}^{2} \underline{\mathbf{q}}_{c, i}\right\} 
				+\mathbb{E}\left\{\frac{{\sigma}_{e}^{2}}{\tau {p}} \underline{\mathbf{q}}_{c, i}^{H} \mathbf{\Upsilon}_{k} \mathbf{H}_{c, 2} \mathbf{\Phi} \mathbf{R}_{e m i} \mathbf{\Phi}^{H} \mathbf{H}_{c, 2}^{H} \boldsymbol{\Upsilon}_{k}^{H} \underline{\mathbf{q}}_{c, i}\right\}.
			\end{aligned}
		\end{align}
		By utilizing Lemma \ref{lemma_HWH} and Lemma \ref{lemma_HCHWHCH} and following a similar procedure in Appendix \ref{appendix4}, the calculation of interference term can be completed by respectively calculating three expectations in (\ref{interference_deriving}). The detailed process is omitted to save the space.
		
		Finally, we will tackle the signal leakage term. Recall that $ E_{c, k}^{\mathrm {leak }}=\mathbb{E}\left\{\left|\hat{\mathbf{q}}_{c, k}^{H} \mathbf{q}_{c, k}\right|^{2}\right\}-\left|\mathbb{E}\left\{\hat{\mathbf{q}}_{c, k}^{H} \mathbf{q}_{c, k}\right\}\right|^{2}  $  and $ \mathbb{E}\left\{\hat{\mathbf{q}}_{c, k}^{H} \mathbf{q}_{c, k}\right\}   $ has been derived in (\ref{noise_deriv}). Therefore, we only need to derive $\mathbb{E}\left\{\left|\hat{\mathbf{q}}_{c, k}^{H} \mathbf{q}_{c, k}\right|^{2}\right\}$, which can be divided by
		\begin{align}\label{leakge_deriv1}
			\begin{aligned}
				\mathbb{E}\left\{\left|\hat{\mathbf{q}}_{c, k}^{H} \mathbf{q}_{c, k}\right|^{2}\right\}
				&=
				\mathbb{E}\left\{\left|\left(\hat{   \underline{\mathbf{q} }   }_{c, k}^{H}+\sqrt{\gamma_{k}} \tilde{\mathbf{d}}_{k}^{H} \mathbf{\Upsilon}_{k}^{H}\right) \mathbf{q}_{c, k}\right|^{2}\right\}\\
				&+\mathbb{E}\left\{ \left|\frac{\mathbf{s}_{k}^{H} \mathbf{V}^{H} \boldsymbol{\Phi}^{H} \mathbf{H}_{c, 2}^{H} \mathbf{\Upsilon}_{k}^{H}}{\sqrt{\tau p}} \mathbf{q}_{c, k}\right|^{2}\right\}
				+\mathbb{E}\left\{      \left|\frac{\mathbf{s}_{k}^{H} \mathbf{N}^{H} \mathbf{\Upsilon}_{k}^{H}}{\sqrt{\tau p}}   \mathbf{q}_{c, k}\right| ^2    \right\},
			\end{aligned}
		\end{align}
		where
		\begin{align}\label{leakge_deriv2}
			\begin{aligned}
				&\mathbb{E}\left\{\left|\left(\hat{ \underline {\mathbf{q}}}_{c, k}^{H}+\sqrt{\gamma_{k}} \tilde{\mathbf{d}}_{k}^{H} \mathbf{\Upsilon}_{k}^{H}\right) \mathbf{q}_{c, k}\right|^{2}\right\}\\
				& =\mathbb{E}\left\{\left|  \hat{ \underline {\mathbf{q}}}_{c, k}^{H}    \mathbf{q}_{c, k}\right|^{2}\right\}
				+\mathbb{E}\left\{\left|\sqrt{\gamma_{k}} \tilde{\mathbf{d}}_{k}^{H} \mathbf{\Upsilon}_{k}^{H} \mathbf{q}_{c, k}\right|^{2}\right\}
				+2 \operatorname{Re}\left\{\sqrt{\gamma_{k}} \mathbb{E}\left\{   \hat{ \underline {\mathbf{q}}}_{c, k}^{H}  \mathbf{q}_{c, k} \mathbf{q}_{c, k}^{H} \mathbf{\Upsilon}_{k} \tilde{\mathbf{d}}_{k}\right\}\right\}.
			\end{aligned}
		\end{align}
		The calculation of signal leakage can be completed after obtaining expectations in (\ref{leakge_deriv1}) and (\ref{leakge_deriv2}). The details are similar to those in the calculation of interference, and therefore is omitted for brevity.

		\section{}\label{appendix9}
		Recall that $\mathbf{\Phi} = \mathrm{diag}\{\boldsymbol{c}\}$ and $\boldsymbol{c} = e^{j \boldsymbol{\theta}}$. Then, we can re-express $ \operatorname{Tr}\left\{\mathbf{A} \mathbf{\Phi} \mathbf{B} \boldsymbol{\Phi}^{H}\right\} $ as
		\begin{align}
			\operatorname{Tr}\left\{\mathbf{A} \mathbf{\Phi} \mathbf{B} \boldsymbol{\Phi}^{H}\right\} 
			=\sum_{i}\left[\mathbf{A} \boldsymbol{\Phi} \mathbf{B} \boldsymbol{\Phi}^{H}\right]_{i i}
			=\sum_{i} \sum_{a}[\mathbf{A}]_{i a}[\mathbf{\Phi}]_{a a}[\mathbf{B}]_{a i}\left[\boldsymbol{\Phi}^{H}\right]_{i i}
			=\boldsymbol{c}^{H}\left(\mathbf{A} \odot \mathbf{B}^{T}\right) \boldsymbol{c}.
		\end{align}
		
		Applying the chain rule, the gradient of $ \operatorname{Tr}\left\{\mathbf{A} \mathbf{\Phi} \mathbf{B} \boldsymbol{\Phi}^{H}\right\} $ with respect to the $n$-th elements of $\boldsymbol{\theta}$, i.e., $\theta_n$, can be calculated as
		\begin{align}\label{apbp}
			\begin{aligned}
				&\frac{\partial \operatorname{Tr}\left\{\mathbf{A} \boldsymbol{\Phi} \mathbf{B} \boldsymbol{\Phi}^{H}\right\}}{\partial \theta_{n}}=\frac{\partial \boldsymbol{c}^{H}}{\partial \theta_{n}}\left(\mathbf{A} \odot \mathbf{B}^{T}\right) \boldsymbol{c}+\mathbf{c}^{H}\left(\mathbf{A} \odot \mathbf{B}^{T}\right) \frac{\partial \boldsymbol{c}}{\partial \theta_{n}} \\
				&=-j e^{-j \theta_{n}}\left[\left(\mathbf{A} \odot \mathbf{B}^{T}\right) \boldsymbol{c}\right]_{n}+j\left[\mathbf{c}^{H}\left(\mathbf{A} \odot \mathbf{B}^{T}\right)\right]_{n} e^{j \theta_{n}}.
			\end{aligned}
		\end{align}
		
		(\ref{apbp}) is the $n$-th element of $\frac{\partial \operatorname{Tr}\left\{\mathbf{A} \boldsymbol{\Phi} \mathbf{B} \boldsymbol{\Phi}^{H}\right\}}{\partial  \boldsymbol{\theta}}$. Thus, the proof of (\ref{apbp_vec}) is completed by combining (\ref{apbp}) to a vector. The proof of (\ref{apbp_unitary}) can be done by noting that $\left\{\boldsymbol{\Phi}^{T}\left(\mathbf{A}^{T} \odot \mathbf{B}\right) \boldsymbol{c}^{*}\right\}^{*}=\boldsymbol{\Phi}^{H}\left(\mathbf{A}^{H} \odot \mathbf{B}^{*}\right) \boldsymbol{c}=\boldsymbol{\Phi}^{H}\left(\mathbf{A} \odot \mathbf{B}^{T}\right) \boldsymbol{c}$ if $\mathbf{A}$ and $\mathbf{B}$ are unitary.

	}

	\section{}\label{appendix6}
	Instantaneous CSI-based schemes need to estimate the cascaded channel and the direct channel in each coherence interval, and then optimize the phase shifts of the RIS in each coherence interval. 
	In the following, we give a brief introduction of the instantaneous CSI-based scheme in single-user systems, including the system model, channel estimation, problem formulation, and phase shift design.
	
	Assume that only the user $k$ exists in the system. 
	The specific realizations of the channel $\mathbf{H}_2$,  $\mathbf{h}_{k}$, and $\mathbf{d}_k$ in the $i$-th coherence interval are denoted by $\mathbf{H}_2^{(i)}$,  $\mathbf{h}_{k}^{(i)}$, and  $\mathbf{d}_k^{(i)}$, respectively. Besides, the phase shifts matrix $\bf\Phi$  in the $i$-th coherence interval is equal to $\mathbf{\Phi}^{(i)} = \mathrm{diag}\left\{ \mathbf{v}^{(i)}\right\}$, where  $\mathbf{v}^{(i)}=[e^{j\theta_1^{(i)}},\ldots,e^{j\theta_N^{(i)}}]^T$. Then, the equivalent channel in the $i$-th coherence interval can be expressed as 
	\begin{align}
		\mathbf{q}_k^{(i)} = \mathbf{H}_2^{(i)} \mathbf{\Phi}^{(i)} \mathbf{h}_k^{(i)} + \mathbf{d}_k^{(i)}
		=\mathbf{H}_2^{(i)}  \mathrm{diag}\left(\mathbf{h}_k^{(i)}\right) \mathbf{v}^{(i)}+ \mathbf{d}_k^{(i)}.
	\end{align}
	
	Let $\mathbf{G}_k^{(i)} \triangleq \mathbf{H}_2^{(i)}  \mathrm{diag}\left(\mathbf{h}_k^{(i)}\right) $ represent the cascaded channel in the $i$-th coherence interval.
	Next, 
	the instantaneous CSI-based scheme needs to respectively estimate the $M\times N$ cascaded channel matrix $\mathbf{G}_k^{(i)} $ and the $M\times 1$ direct channel vector $\mathbf{d}_k^{(i)}$ in each channel coherence time. 
	
	The estimation of $\mathbf{G}_k^{(i)} $ and $\mathbf{d}_k^{(i)}$ can be performed by using a two-phase pilot-based scheme \cite{9130088}. In the first phase, the direct link $\mathbf{d}_k^{(i)}$ can be estimated by using the MMSE estimator, and the needed pilot length is equal to the number of users, i.e., $1$. In the second phase, using the estimated direct channel, the cascaded channel $\mathbf{G}_k^{(i)} $ can be estimated by using the LMMSE estimator, and the needed pilot length is equal to the number of RIS elements, i.e., $N$. Therefore, the overall pilot needed in the considered instantaneous CSI-based scheme is $N+1$. Then, we denote the estimated cascaded channel and direct channel as $\hat{  \mathbf{G} }_k^{(i)} $ and $\hat{\mathbf{d}}_k^{(i)}$, respectively. The detailed process of the estimation is omitted here, and interested readers can refer to \cite[Section $ \rm\uppercase\expandafter{\romannumeral5} $]{9130088}. 
	
	Based on the estimated channels $\hat{  \mathbf{G} }_k^{(i)} $ and $\hat{\mathbf{d}}_k^{(i)}$, the BS can design the MRC beamforming as $\mathbf{w}^H = \left(  \hat{  \mathbf{G} }_k^{(i)}  \mathbf{v}^{(i)}+ \hat{\mathbf{d}}_k^{(i)}  \right)^H$. Then, the received signal at the BS in the $i$-th coherence interval can be expressed as
	\begin{align}
		\mathbf{y}^{(i)} &= \sqrt{p} \mathbf{w}^H  \left(  {  \mathbf{G} }_k^{(i)}  \mathbf{v}^{(i)}+ {\mathbf{d}}_k^{(i)}  \right) {x}_k^{(i)}+   \mathbf{w}^H \mathbf{n}^{(i)}\nonumber\\
		&= \sqrt{p} \mathbf{w}^H  \left(  \hat{  \mathbf{G} }_k^{(i)}  \mathbf{v}^{(i)}+ \hat{\mathbf{d}}_k^{(i)}  \right) {x}_k^{(i)}
		+\sqrt{p} \mathbf{w}^H  \left(  \tilde{  \mathbf{G} }_k^{(i)}  \mathbf{v}^{(i)}+ \tilde{\mathbf{d}}_k^{(i)}  \right) {x}_k^{(i)}
		+   \mathbf{w}^H \mathbf{n}^{(i)},
	\end{align}
	where $ \tilde{  \mathbf{G} }_k^{(i)} = {  \mathbf{G} }_k^{(i)} - \hat{  \mathbf{G} }_k^{(i)} $ and $\tilde{\mathbf{d}}_k^{(i)} ={\mathbf{d}}_k^{(i)} -\hat{\mathbf{d}}_k^{(i)} $ denote the channel estimation errors.
	
	Then, we can express the effective SNR as
	\begin{align}
		\mathrm{S N R}_k^{(i)} \left(\mathbf{v}^{(i)}\right)= 
		\frac{p \left|   \mathbf{w}^H  \left(  \hat{  \mathbf{G} }_k^{(i)}  \mathbf{v}^{(i)}+ \hat{\mathbf{d}}_k^{(i)}  \right) \right|^2}
		{p \left|  \mathbf{w}^H  \left(  \tilde{  \mathbf{G} }_k^{(i)}  \mathbf{v}^{(i)}+ \tilde{\mathbf{d}}_k^{(i)}  \right)  \right|^2  + \sigma^2 \left\| \mathbf{w}^H\right\|^2},
	\end{align}
	and the effective rate of user $k$ in the $i$-th coherence interval is given by
	\begin{align}
		{R}_k^{(i)}= \left(1- \frac{N+1}{\tau_c}\right) \log_{2}\left(1+\mathrm{S N R}_k^{(i)} \left(\mathbf{v}^{(i)}\right)\right),
	\end{align}
	where the factor $\left(1- \frac{N+1}{\tau_c}\right)$ represents the rate loss due to the pilot estimation overhead.
	
	Next, the instantaneous CSI-based schemes need to optimize the phase shifts $\mathbf{v}^{(i)}$ in the $i$-th coherence interval. We note that the maximization of ${R}_k^{(i)}$ is equivalent to the maximization of $\mathrm{S N R}_k^{(i)} $. However, it is challenging to find an optimal solution for the maximization of the SNR when considering the channel estimation error from imperfect CSI. Therefore, we resort to a low-complexity sub-optimal solution which only uses the RIS to maximize the desired signal power. The optimization problem is formulated as follows
	\begin{subequations}\label{Problem3}
		\begin{align}
			&\max _{\mathbf{v}^{(i)}   }   \;\;\left|   \mathbf{w}^H  \left(  \hat{  \mathbf{G} }_k^{(i)}  \mathbf{v}^{(i)}+ \hat{\mathbf{d}}_k^{(i)}  \right) \right|^2 \\
			&\text { s.t. } \quad  0 \leq \theta_{n}^{(i)} < 2 \pi, \forall n.
		\end{align}
	\end{subequations}
	
	A closed-from solution for the problem in (\ref{Problem3}) can be obtained by using alternating optimization\cite[Section $ \rm\uppercase\expandafter{\romannumeral3} $]{wu2019intelligent}. First, given the phase shifts vector $ \mathbf{v}^{(i)}  $, the MRC beamforming vector is set to $\mathbf{w}^H = \left(  \hat{  \mathbf{G} }_k^{(i)}  \mathbf{v}^{(i)}+ \hat{\mathbf{d}}_k^{(i)}  \right)^H$. Then, given the MRC beamforming vector $\mathbf{w}^H$, the RIS phase shifts are optimized by aligning the phase of the cascaded channel with the phase of the direct channel, i.e.,
	$ \mathrm{arg}\left(   \mathbf{w}^H \hat{  \mathbf{G} }_k^{(i)}  \mathbf{v}^{(i)}  \right)  
	=  \mathrm{arg}\left(   \mathbf{w}^H  \hat{\mathbf{d}}_k^{(i)}   \right) $.
	Then, the solution $ \mathbf{v}^{*(i)} $ is obtained when the alternating optimization algorithm reaches convergence. Based on the optimized solution $ \mathbf{v}^{*(i)} $, the achievable rate in the $i$-th coherence interval is obtained as $ {R}_k^{*(i)}= \left(1- \frac{N+1}{\tau_c}\right) \log_{2}\left(1+\mathrm{S N R}_k^{(i)} \left(\mathbf{v}^{*(i)}\right)\right) $. 
	
	Finally, by repeating the above procedure for $T_{ci}$ coherence intervals, the average achievable rate for the instantaneous CSI-based scheme is given by
	\begin{align}\label{rate_instan1}
		\overline{R}_k^{*}= \left(1- \frac{N+1}{\tau_c}\right)\frac{1}{T_{ci}}\sum_{i=1}^{T_{ci}} \log_{2}\left(1+\mathrm{S N R}_k^{(i)} \left(\mathbf{v}^{*(i)}\right)\right).
	\end{align}
	The rate in (\ref{rate_instan1}), which consists of a rate loss factor equal to $1- \frac{N+1}{\tau_c}$, is plotted in Fig. \ref{figure2.5} in Section \ref{section_6}. It is apparent that the rate in (\ref{rate_instan1}) is negatively affected by the channel estimation overhead. If $N+1>\tau_c$, the rate reduces to zero, since all the symbols in the coherence interval are used for pilot transmission, and no symbol is left for data transmission. To gain more insights, we consider to replace the rate loss factor in (\ref{rate_instan1}) with $1-\frac{1}{\tau_c}$, which is the same as that in the proposed two-timescale scheme. In this case, the rate is given by
	\begin{align}\label{rate_instan2}
		\overline{R}_k^{*}= \left(1- \frac{1}{\tau_c}\right)\frac{1}{T_{ci}}\sum_{i=1}^{T_{ci}} \log_{2}\left(1+\mathrm{S N R}_k^{(i)} \left(\mathbf{v}^{*(i)}\right)\right).
	\end{align}
	The rate in (\ref{rate_instan2}), which is, however,  not achievable, is plotted  in Fig. \ref{figure2.5} in Section \ref{section_6}. Compared with (\ref{rate_instan1}), the only difference in (\ref{rate_instan2}) is that the additional, but necessary, channel estimation overhead is ignored.

	\section{Expressions for Gradient Vectors}\label{appendix_K}
	\begin{thm}
		The gradient of $f(\boldsymbol{\theta})$ with respect to $\boldsymbol{\theta}$ is given by
		\begin{align} 
			&\frac{\partial f(\boldsymbol{\theta})}{\partial \boldsymbol{\theta}}=\frac{\tau^{o}  \sum_{k=1}^{K}\left\{\frac{\exp \left\{-\mu \underline{R}_{ k}(\boldsymbol{\theta})\right\}}{1+\operatorname{SINR}_{ k}(\boldsymbol{\theta})} \frac{\partial \operatorname{SINR}_{ k}(\boldsymbol{\theta})}{\partial \boldsymbol{\theta}}\right\}}{(\ln 2)\left(\sum_{k=1}^{K} \exp \left\{-\mu \underline{R}_{ k}(\boldsymbol{\theta})\right\}\right)},
		\end{align}
		and
		\begin{align} 
			\begin{aligned}
				&\frac{\partial \operatorname{SINR}_{ k}( \boldsymbol{\theta} )}{\partial \boldsymbol{\theta}}=\frac{p \frac{\partial E_{ k}^{\mathrm {signal }}}{\partial \boldsymbol{\theta}}}{p E_{ k}^{\mathrm {leak }}+p \sum_{i=1, i \neq k}^{K} I_{k i}+\sigma^{2} E_{ k}^{\mathrm {noise }}} -p E_{ k}^{\mathrm {signal }} \frac{p \frac{\partial E_{ k}^{\mathrm {leak }}}{\partial \boldsymbol{\theta}}+p \sum_{i=1, i \neq k}^{K} \frac{\partial I_{ k i}}{\partial \boldsymbol{\theta}}+\sigma^{2} \frac{\partial E_{ k}^{\mathrm {noise }}}{\partial \boldsymbol{\theta}}}{\left(p E_{ k}^{\mathrm {leak }}+p \sum_{i=1, i \neq k}^{K} I_{ k i}+\sigma^{2} E_{ k}^{\mathrm {noise }}\right)^{2}}.
			\end{aligned}
		\end{align}
		where  
		\begin{align}
			& 	E_{k}^{(\mathrm {signal })}=2 E_{k}^{(\mathrm {noise })} \frac{\partial E_{k}^{(\mathrm {noise })}}{\partial \boldsymbol{\theta}},\\
			&E_{k}^{(\mathrm {noise })}=(M c_{k} \delta \varepsilon_{k} ) \frac{\partial\left|f_{k}(\boldsymbol{\Phi})\right|^{2}}{\partial \boldsymbol{\theta}},
		\end{align}
		and
		\begin{align}
			\begin{aligned}
				\frac{\partial I_{k i}}{\partial \boldsymbol{\theta}} &=s_{k i 5} \frac{\partial\left|f_{k}(\boldsymbol{\Phi})\right|^{2}}{\partial \boldsymbol{\theta}}\left|f_{i}(\boldsymbol{\Phi})\right|^{2}+s_{k i 5}\left|f_{k}(\boldsymbol{\Phi})\right|^{2} \frac{\partial\left|f_{i}(\boldsymbol{\Phi})\right|^{2}}{\partial \boldsymbol{\theta}} \\
				&+s_{k i 6} \frac{\partial\left|f_{k}(\boldsymbol{\Phi})\right|^{2}}{\partial \boldsymbol{\theta}}+s_{k i 7} \frac{\partial\left|f_{i}(\boldsymbol{\Phi})\right|^{2}}{\partial \boldsymbol{\theta}}+s_{k i 8} \frac{\partial f_{k}^{H}(\boldsymbol{\Phi}) f_{i}(\boldsymbol{\Phi})}{\partial \boldsymbol{\theta}}+s_{k i 9} \frac{\partial f_{i}^{H}(\boldsymbol{\Phi}) f_{k}(\boldsymbol{\Phi})}{\partial \boldsymbol{\theta}},
			\end{aligned}
		\end{align}
		and
		\begin{align}
			\frac{\partial E_{k}^{(\mathrm {leakage })}}{\partial \boldsymbol{\theta}}=s_{k 11} \frac{\partial\left|f_{k}(\boldsymbol{\Phi})\right|^{2}}{\partial \boldsymbol{\theta}},
		\end{align}
		with
		\begin{align}
			\begin{aligned}
				&\frac{\partial\left|f_{k}(\boldsymbol{\Phi})\right|^{2}}{\partial \boldsymbol{\theta}}=2 \operatorname{Im}\left\{\boldsymbol{\Phi}^{H}\left(\mathbf{a}_{N} \mathbf{a}_{N}^{H} \odot\left(\overline{\mathbf{h}}_{k} \overline{\mathbf{h}}_{k}^{H}\right)^{T}\right) \boldsymbol{c}\right\}, \\
				&\frac{\partial\left|f_{i}(\boldsymbol{\Phi})\right|^{2}}{\partial \boldsymbol{\theta}}=2 \operatorname{Im}\left\{\boldsymbol{\Phi}^{H}\left(\mathbf{a}_{N} \mathbf{a}_{N}^{H} \odot\left(\overline{\mathbf{h}}_{i} \overline{\mathbf{h}}_{i}^{H}\right)^{T}\right) \boldsymbol{c}\right\}, \\
				&\frac{\partial f_{k}^{H}(\boldsymbol{\Phi}) f_{i}(\boldsymbol{\Phi})}{\partial \boldsymbol{\theta}}=\boldsymbol{f}_{d}\left(\mathbf{a}_{N} \mathbf{a}_{N}^{H}, \overline{\mathbf{h}}_{i} \overline{\mathbf{h}}_{k}^{H}\right), \\
				&\frac{\partial f_{i}^{H}(\boldsymbol{\Phi}) f_{k}(\boldsymbol{\Phi})}{\partial \boldsymbol{\theta}}=\boldsymbol{f}_{d}\left(\mathbf{a}_{N} \mathbf{a}_{N}^{H}, \overline{\mathbf{h}}_{k} \overline{\mathbf{h}}_{i}^{H}\right),
			\end{aligned}
		\end{align}
		and
		\begin{align}
			\begin{aligned}
				&s_{k i 5}=M^{2} c_{k} c_{i} \delta^{2} \varepsilon_{k} \varepsilon_{i}, \\
				&s_{k i 6}=M c_{k} \delta \varepsilon_{k}\left\{c_{i}\left(M N \delta+N \varepsilon_{i}+N+2 M e_{k 1}\right)+\gamma_{i}\right\}, \\
				&s_{k i 7}=M c_{i} \delta \varepsilon_{i}\left\{c_{k} e_{k 2}\left(M N \delta e_{k 2}+N \varepsilon_{k} e_{k 2}+N e_{k 2}+2 M e_{k 1}\right)+\left(\gamma_{k}+\frac{\sigma^{2}}{\tau p}\right) e_{k 2}^{2}\right\}, \\
				&s_{k i 8}=M^{2} c_{k} c_{i} \varepsilon_{k} \varepsilon_{i} e_{k 1} \delta \overline{\mathbf{h}}_{i}^H \overline{\mathbf{h}}_{k}, \\
				&s_{k i 9}=M^{2} c_{k} c_{i} \varepsilon_{k} \varepsilon_{i} e_{k 1} \delta \overline{\mathbf{h}}_{k}^{H} \overline{\mathbf{h}}_{i},\\
				&s_{k 11}=M c_{k}^{2} \delta \varepsilon_{k}\left\{N\left(M \delta+\varepsilon_{k}+1\right)\left(e_{k 2}{ }^{2}+1\right)+2\left(M e_{k 1}+e_{k 2}\right)\left(e_{k 2}+1\right)\right\}\\
				&+M c_{k} \delta \varepsilon_{k}\left(\gamma_{k}+\left(\gamma_{k}+\frac{\sigma^{2}}{\tau p}\right) e_{k 2}^{2}\right) ,
			\end{aligned}
		\end{align}
	\end{thm}
	
	\begin{thm}
		The gradient of $f_c(\boldsymbol{\theta})$ with respect to $\boldsymbol{\theta}$ is given by
		\begin{align} 
			&\frac{\partial f_{c}(\boldsymbol{\theta})}{\partial \boldsymbol{\theta}}=\frac{\tau^{o}  \sum_{k=1}^{K}\left\{\frac{\exp \left\{-\mu \underline{R}_{c, k}(\boldsymbol{\theta})\right\}}{1+\operatorname{SINR}_{c, k}(\boldsymbol{\theta})} \frac{\partial \operatorname{SINR}_{c, k}(\boldsymbol{\theta})}{\partial \boldsymbol{\theta}}\right\}}{(\ln 2)\left(\sum_{k=1}^{K} \exp \left\{-\mu \underline{R}_{c, k}(\boldsymbol{\theta})\right\}\right)},
		\end{align}
		and
		\begin{align} 
			\begin{aligned}
				&\frac{\partial \operatorname{SINR}_{c, k}( \boldsymbol{\theta} )}{\partial \boldsymbol{\theta}}=\frac{p \frac{\partial E_{c, k}^{\mathrm {signal }}}{\partial \boldsymbol{\theta}}}{p E_{c, k}^{\mathrm {leak }}+p \sum_{i=1, i \neq k}^{K} I_{c, k i}+\sigma_{e}^{2} E_{c, k}^{\mathrm {emi }}+\sigma^{2} E_{c, k}^{\mathrm {noise }}} \\
				&-p E_{c, k}^{\mathrm {signal }} \frac{p \frac{\partial E_{c, k}^{\mathrm {leak }}}{\partial \boldsymbol{\theta}}+p \sum_{i=1, i \neq k}^{K} \frac{\partial I_{c, k i}}{\partial \boldsymbol{\theta}}+\sigma_{e}^{2} \frac{\partial E_{c, k}^{\mathrm {emi }}}{\partial \boldsymbol{\theta}}+\sigma^{2} \frac{\partial E_{c, k}^{\mathrm {noise }}}{\partial \boldsymbol{\theta}}}{\left(p E_{c, k}^{\mathrm {leak }}+p \sum_{i=1, i \neq k}^{K} I_{c, k i}+\sigma_{e}^{2} E_{c, k}^{\mathrm {emi }}+\sigma^{2} E_{c, k}^{\mathrm {noise }}\right)^{2}}.
			\end{aligned}
		\end{align}
		The gradient of signal is
		\begin{align}
			\frac{\partial E_{c, k}^{\mathrm {signal }}}{\partial \boldsymbol{\theta}}=\frac{\partial\left\{\left(E_{c, k}^{\mathrm {noise }}\right)^{2}\right\}}{\partial \boldsymbol{\theta}}=2 E_{c, k}^{\mathrm{noise}} \frac{\partial E_{c, k}^{\mathrm{noise}}}{\partial \boldsymbol{\theta}},
		\end{align}
		and
		\begin{align}
			\frac{\partial E_{c, k}^{\mathrm {noise }}}{\partial \boldsymbol{\theta}}=M \widehat{c}_{k} \delta \boldsymbol{f}_{c, k, 7}^{\prime}(\boldsymbol{c})+\left\{\widehat{c}_{k} f_{c, k, 2}(\boldsymbol{\Phi})+\gamma_{k}\right\} \mathbf{z}_{k}\left(\mathbf{I}_{M}\right)+\widehat{c}_{k} \operatorname{Tr}\left\{\boldsymbol{\Upsilon}_{k}\right\} \boldsymbol{f}_{c, k, 2}^{\prime}(\boldsymbol{c}).
		\end{align}

		The gradient of EMI is $\frac{\partial E_{c, k}^{e m i}}{\partial \boldsymbol{\theta}}=\frac{\beta}{\delta+1} \sum_{\omega=1}^{8} \frac{\partial E_{c, k}^{\omega, e m i}}{\partial \boldsymbol{\theta}}$, where
		\begin{align}
			\begin{aligned}
				\frac{\partial E_{c, k}^{1, e m i}}{\partial \boldsymbol{\theta}}&=M^{2} \widehat{c}_{k} \delta^{2} \mathbf{a}_{N}^{H} \boldsymbol{\Phi} \mathbf{R}_{e m i} \boldsymbol{\Phi}^{H} \mathbf{a}_{N} \boldsymbol{f}_{c, k, 7}^{\prime}(\boldsymbol{\theta}) \\
				&+2 M^{2} \widehat{c}_{k} \delta^{2} f_{c, k, 7}(\boldsymbol{\Phi}) \operatorname{Im}\left\{\boldsymbol{\Phi}^{H}\left(\mathbf{a}_{N} \mathbf{a}_{N}^{H} \odot \mathbf{R}_{e m i}\right) \boldsymbol{c}\right\},
			\end{aligned}
		\end{align}
		and
		\begin{align}
			\begin{aligned}
				\frac{\partial E_{c, k}^{2, e m i}}{\partial \boldsymbol{\theta}} &=\left\{\widehat{c}_{k} \delta \boldsymbol{f}_{c, k, 2}^{\prime}(\boldsymbol{\theta})+\frac{2 \beta \delta \sigma_{e}^{2}}{\tau p(\delta+1)} \boldsymbol{f}_{c, 1}^{\prime}(\boldsymbol{\theta})\right\} f_{c, k, 3}(\boldsymbol{\Phi}) \\
				&+\left\{\widehat{c}_{k} \delta f_{c, k, 2}(\boldsymbol{\Phi})+\frac{2 \beta \delta \sigma_{e}^{2}}{\tau p(\delta+1)} f_{c, 1}(\boldsymbol{\Phi})+\delta\left(\gamma_{k}+\frac{\sigma^{2}}{\tau p}\right)\right\} \boldsymbol{f}_{c, k, 3}^{\prime}(\boldsymbol{\theta}),
			\end{aligned}
		\end{align}
		and
		\begin{align}
			\begin{aligned}
				&\frac{\partial E_{c, k}^{3, e m i}}{\partial \boldsymbol{\theta}} \\
				&=\left[M c_{k} \delta \boldsymbol{f}_{c, k, 7}^{\prime}(\boldsymbol{\theta})+\left\{\widehat{c}_{k} \boldsymbol{f}_{c, k, 2}^{\prime}(\boldsymbol{\theta})+\frac{\beta \sigma_{e}^{2}}{\tau p(\delta+1)} \boldsymbol{f}_{c, 1}^{\prime}(\boldsymbol{\theta})\right\} f_{c, k, 4}(\boldsymbol{\Phi})\right] f_{c, 1}(\boldsymbol{\Phi}) \\
				&+\left\{\frac{\sigma^{2}}{\tau p}+\gamma_{k}+\widehat{c}_{k} f_{c, k, 2}(\boldsymbol{\Phi})+\frac{\beta \sigma_{e}^{2}}{\tau p(\delta+1)} f_{c, 1}(\boldsymbol{\Phi})\right\} f_{c, 1}(\boldsymbol{\Phi}) \boldsymbol{f}_{c, k, 4}^{\prime}(\boldsymbol{\theta}) \\
				&+\left[M \widehat{c}_{k} \delta f_{c, k, 7}(\boldsymbol{\Phi})+\left\{\frac{\sigma^{2}}{\tau p}+\gamma_{k}+\widehat{c}_{k} f_{c, k, 2}(\boldsymbol{\Phi})+\frac{\beta \sigma_{e}^{2}}{\tau p(\delta+1)} f_{c, 1}(\boldsymbol{\Phi})\right\} f_{c, k, 4}(\boldsymbol{\Phi})\right] \boldsymbol{f}_{c, 1}^{\prime}(\boldsymbol{\theta}),
			\end{aligned}
		\end{align}
		and
		\begin{align}
			\frac{\partial E_{c, k}^{4, e m i}}{\partial \boldsymbol{\theta}}=\frac{2 \beta \delta^{2} \sigma_{e}^{2}}{\tau p(\delta+1)}\left\{\begin{array}{l}
				2 \operatorname{Im}\left\{\boldsymbol{\Phi}^{H}\left(\overline{\mathbf{H}}_{2}^{H} \boldsymbol{\Upsilon}_{k} \overline{\mathbf{H}}_{2} \odot\left(\mathbf{R}_{e m i} \boldsymbol{\Phi}^{H} \overline{\mathbf{H}}_{2}^{H} \boldsymbol{\Upsilon}_{k} \overline{\mathbf{H}}_{2} \mathbf{\Phi} \mathbf{R}_{e m i}\right)^{T}\right) \boldsymbol{c}\right\} \\
				+\mathbf{z}_{k}\left(\overline{\mathbf{H}}_{2} \boldsymbol{\Phi} \mathbf{R}_{e m i} \boldsymbol{\Phi}^{H} \overline{\mathbf{H}}_{2}^{H} \boldsymbol{\Upsilon}_{k} \overline{\mathbf{H}}_{2} \boldsymbol{\Phi} \mathbf{R}_{e m i} \boldsymbol{\Phi}^{H} \overline{\mathbf{H}}_{2}^{H}\right),
			\end{array}\right\}
		\end{align}
		and
		\begin{align}
			\begin{aligned}
				&\frac{\partial E_{c, k}^{5, e m i}}{\partial \boldsymbol{\theta}} =2 \widehat{c}_{k} \delta \operatorname{Re}\left\{\overline{\mathbf{h}}_{k}^{H} \boldsymbol{\Phi}^{H} \overline{\mathbf{H}}_{2}^{H} \overline{\mathbf{H}}_{2} \boldsymbol{\Phi} \mathbf{R}_{e m i} \mathbf{\Phi}^{H} \mathbf{R}_{r i s} \boldsymbol{\Phi} \overline{\mathbf{h}}_{k}\right\} \mathbf{z}_{k}\left(\mathbf{I}_{M}\right) \\
				&+\widehat{{c}}_{k} \delta \operatorname{Tr}\left\{\boldsymbol{\Upsilon}_{k}\right\}\left\{\begin{array}{l}
					\boldsymbol{f}_{d}\left(\mathbf{R}_{r i s} \boldsymbol{\Phi} \overline{\mathbf{h}}_{k} \overline{\mathbf{h}}_{k}^{H} \boldsymbol{\Phi}^{H} \overline{\mathbf{H}}_{2}^{H} \overline{\mathbf{H}}_{2}, \mathbf{R}_{e m i}\right)+\boldsymbol{f}_{d}\left(\overline{\mathbf{H}}_{2}^{H} \overline{\mathbf{H}}_{2} \mathbf{\Phi} \mathbf{R}_{e m i} \mathbf{\Phi}^{H} \mathbf{R}_{r i s}, \overline{\mathbf{h}}_{k} \overline{\mathbf{h}}_{k}^{H}\right) \\
					\left.+\boldsymbol{f}_{d}\left(\overline{\mathbf{H}}_{2}^{H} \overline{\mathbf{H}}_{2} \boldsymbol{\Phi} \overline{\mathbf{h}}_{k} \overline{\mathbf{h}}_{k}^{H} \boldsymbol{\Phi}^{H} \mathbf{R}_{r i s}, \mathbf{R}_{e m i}\right)+\boldsymbol{f}_{d}\left(\mathbf{R}_{r i s} \mathbf{\Phi} \mathbf{R}_{e m i} \mathbf{\Phi}^{H} \overline{\mathbf{H}}_{2}^{H} \overline{\mathbf{H}}_{2}, \overline{\mathbf{h}}_{k} \overline{\mathbf{h}}_{k}^{H}\right)\right\}
				\end{array}\right\},
			\end{aligned}
		\end{align}
		and
		\begin{align}
			\begin{aligned}
				&\frac{\partial E_{c, k}^{6, e m i}}{\partial \boldsymbol{\theta}}=\frac{2 \beta \delta \sigma_{e}^{2}}{\tau p(\delta+1)} \operatorname{Tr}\left\{\overline{\mathbf{H}}_{2} \boldsymbol{\Phi} \mathbf{R}_{e m i} \boldsymbol{\Phi}^{H} \mathbf{R}_{r i s} \mathbf{\Phi} \mathbf{R}_{e m i} \boldsymbol{\Phi}^{H} \overline{\mathbf{H}}_{2}^{H} \boldsymbol{\Upsilon}_{k}^{H}\right\} \mathbf{z}_{k}\left(\mathbf{I}_{M}\right)\\
				&+\frac{2 \beta \delta \sigma_{e}^{2}}{\tau p(\delta+1)} \operatorname{Tr}\left\{\boldsymbol{\Upsilon}_{k}\right\}\left\{\begin{array}{l}
					\boldsymbol{f}_{d}\left(\mathbf{R}_{r i s} \mathbf{\Phi} \mathbf{R}_{e m i} \boldsymbol{\Phi}^{H} \overline{\mathbf{H}}_{2}^{H} \boldsymbol{\Upsilon}_{k}^{H} \overline{\mathbf{H}}_{2}, \mathbf{R}_{e m i}\right) \\
					+\boldsymbol{f}_{d}\left(\overline{\mathbf{H}}_{2}^{H} \boldsymbol{\Upsilon}_{k}^{H} \overline{\mathbf{H}}_{2} \mathbf{\Phi} \mathbf{R}_{e m i} \mathbf{\Phi}^{H} \mathbf{R}_{r i s}, \mathbf{R}_{e m i}\right) \\
					+\mathbf{z}_{k}\left(\overline{\mathbf{H}}_{2} \mathbf{\Phi} \mathbf{R}_{e m i} \mathbf{\Phi}^{H} \mathbf{R}_{r i s} \mathbf{\Phi} \mathbf{R}_{e m i} \mathbf{\Phi}^{H} \overline{\mathbf{H}}_{2}^{H}\right)
				\end{array}\right\},
			\end{aligned}
		\end{align}
		and
		\begin{align}
			\begin{aligned}
				&\frac{\partial E_{c, k}^{7, e m i}}{\partial \boldsymbol{\theta}} =\widehat{{c}}_{k} f_{c, k, 6}(\boldsymbol{\Phi}) \boldsymbol{f}_{c, k, 5}^{\prime}(\boldsymbol{\theta})+\widehat{{c}}_{k} f_{c, k, 5}(\boldsymbol{\Phi}) \boldsymbol{f}_{c, k, 6}^{\prime}(\boldsymbol{\theta}),
			\end{aligned}
		\end{align}
		and
		\begin{align}
			\begin{aligned}
				&\frac{\partial E_{c, k}^{8, e m i}}{\partial \boldsymbol{\theta}} \\
				&=\frac{\beta \sigma_{e}^{2}}{\tau p(\delta+1)} \operatorname{Tr}\left\{\left(\mathbf{R}_{r i s} \boldsymbol{\Phi} \mathbf{R}_{e m i} \boldsymbol{\Phi}^{H}\right)^{2}\right\} \boldsymbol{f}_{c, k, 5}^{\prime}(\boldsymbol{\theta})+\frac{2 \beta \sigma_{e}^{2}}{\tau p(\delta+1)} f_{c, k, 5}(\boldsymbol{\Phi}) \boldsymbol{f}_{d}\left(\mathbf{R}_{r i s} \boldsymbol{\Phi} \mathbf{R}_{e m i} \boldsymbol{\Phi}^{H} \mathbf{R}_{r i s}, \mathbf{R}_{e m i}\right).
			\end{aligned}
		\end{align}

		The gradient of interference is $  \frac{\partial  I_{c,ki}}{\partial \boldsymbol{\theta}} =\sum_{\omega=1}^{8} \frac{\partial   I_{c,ki}^{\omega}  }{\partial \boldsymbol{\theta}}$, where
		\begin{align}
			\frac{\partial I_{c, k i}^{1}}{\partial \boldsymbol{\theta}}=\gamma_{i} \frac{\partial E_{c, k}^{\mathrm {noise }}}{\partial \boldsymbol{\theta}}+M^{2} \widehat{c}_{k} \widehat{c}_{i} \delta^{2} f_{c, i, 7}(\boldsymbol{\Phi}) \boldsymbol{f}_{c, k, 7}^{\prime}(\boldsymbol{\theta})+M^{2} \widehat{c}_{k} \widehat{c}_{i} \delta^{2} f_{c, k, 7}(\boldsymbol{\Phi}) \boldsymbol{f}_{c, i, 7}^{\prime}(\boldsymbol{\theta}),
		\end{align}
		and
		\begin{align}
			\begin{aligned}
				&\frac{\partial I_{c, k i}^{2}}{\partial \boldsymbol{\theta}} =\left\{M \widehat{c}_{k} \widehat{c}_{i} \delta \boldsymbol{f}_{c, k, 7}^{\prime}(\boldsymbol{\theta})+\frac{\widehat{c}_{i} \beta \delta \sigma_{e}^{2}}{\tau p(\delta+1)} \boldsymbol{f}_{c, k, 3}^{\prime}(\boldsymbol{\theta})\right\} f_{c, i, 2}(\boldsymbol{\Phi}) \\
				&+\frac{\widehat{c}_{i} \beta \sigma_{e}^{2}}{\tau p(\delta+1)} f_{c, k, 4}(\boldsymbol{\Phi}) f_{c, i, 2}(\boldsymbol{\Phi}) \boldsymbol{f}_{c, 1}^{\prime}(\boldsymbol{\theta}) \\
				&+\left\{\widehat{c}_{i}\left(\gamma_{k}+\frac{\sigma^{2}}{\tau p}\right)+\frac{\widehat{c}_{i} \beta \sigma_{e}^{2}}{\tau p(\delta+1)} f_{c, 1}(\boldsymbol{\Phi})\right\} f_{c, i, 2}(\boldsymbol{\Phi}) \boldsymbol{f}_{c, k, 4}^{\prime}(\boldsymbol{\theta}) \\
				&+\left\{M \widehat{c}_{k} \widehat{c}_{i} \delta\left|f_{k}(\boldsymbol{\Phi})\right|^{2}+\left\{\widehat{c}_{i}\left(\gamma_{k}+\frac{\sigma^{2}}{\tau p}\right)+\frac{\widehat{c}_{i} \beta \sigma_{e}^{2}}{\tau p(\delta+1)} f_{c, 1}(\boldsymbol{\Phi})\right\} f_{c, k, 4}(\boldsymbol{\Phi})+\frac{\widehat{c}_{i} \beta \delta \sigma_{e}^{2}}{\tau p(\delta+1)} f_{c, k, 3}(\boldsymbol{\Phi})\right\} \boldsymbol{f}_{c, i, 2}^{\prime}(\boldsymbol{\theta}),
			\end{aligned}
		\end{align}
		and
		\begin{align}
			\begin{aligned}
				&\frac{\partial I_{c, k i}^{3}}{\partial \boldsymbol{\theta}} \\
				&=\left\{\widehat{c}_{k} \widehat{c}_{i} \delta \boldsymbol{f}_{c, k i, 8}^{\prime}(\boldsymbol{\theta})+\widehat{c}_{k} \widehat{c}_{i} f_{c, i, 2}(\boldsymbol{\Phi}) \boldsymbol{f}_{c, k, 4}^{\prime}(\boldsymbol{\theta})+\widehat{c}_{k} \widehat{c}_{i} f_{c, k, 4}(\boldsymbol{\Phi}) \boldsymbol{f}_{c, i, 2}^{\prime}(\boldsymbol{\theta})\right\} f_{c, k, 2}(\boldsymbol{\Phi}) . \\
				&+\left\{\widehat{c}_{k} \widehat{c}_{i} \delta f_{c, k i, 8}(\boldsymbol{\Phi})+\widehat{c}_{k} \widehat{c}_{i} f_{c, k, 4}(\boldsymbol{\Phi}) f_{c, i, 2}(\boldsymbol{\Phi})\right\} \boldsymbol{f}_{c, k, 2}^{\prime}(\boldsymbol{\theta}),
			\end{aligned}
		\end{align}
		and
		\begin{align}
			\begin{aligned}
				&\frac{\partial I_{c, k i}^{4}}{\partial \boldsymbol{\theta}} \\
				&=\frac{\widehat{c}_{i} \beta \delta \sigma_{e}^{2}}{\tau p(\delta+1)} f_{c, k i, 8}(\boldsymbol{\Phi}) \boldsymbol{f}_{c, 1}^{\prime}(\boldsymbol{\theta})+\left\{\frac{\widehat{c}_{i} \beta \delta \sigma_{e}^{2}}{\tau p(\delta+1)} f_{c, 1}(\boldsymbol{\Phi})+\widehat{c}_{i} \delta\left(\gamma_{k}+\frac{\sigma^{2}}{\tau p}\right)\right\} \boldsymbol{f}_{c, k i, 8}^{\prime}(\boldsymbol{\theta}) ,
			\end{aligned}
		\end{align}
		and
		\begin{align}
			\begin{aligned}
				&\frac{\partial I_{c, k i}^{5}}{\partial \boldsymbol{\theta}} \\
				&=\left\{\begin{array}{l}
					2 \widehat{c}_{k} \widehat{c}_{i} \operatorname{Im}\left\{\boldsymbol{\Phi}^{H}\left(\mathbf{R}_{r i s} \boldsymbol{\Phi} \overline{\mathbf{h}}_{k} \overline{\mathbf{h}}_{k}^{H} \boldsymbol{\Phi}^{H} \mathbf{R}_{r i s} \odot\left(\overline{\mathbf{h}}_{i} \overline{\mathbf{h}}_{i}^{H}\right)^{T}\right) \boldsymbol{c}\right\} \\
					+2 \widehat{c}_{k} \widehat{c}_{i} \operatorname{Im}\left\{\boldsymbol{\Phi}^{H}\left(\mathbf{R}_{r i s} \boldsymbol{\Phi} \overline{\mathbf{h}}_{i} \overline{\mathbf{h}}_{i}^{H} \boldsymbol{\Phi}^{H} \mathbf{R}_{r i s} \odot\left(\overline{\mathbf{h}}_{k} \overline{\mathbf{h}}_{k}^{H}\right)^{T}\right) \boldsymbol{c}\right\}+\frac{\widehat{c}_{i} \beta \sigma_{e}^{2}}{\tau p(\delta+1)} \boldsymbol{f}_{c, i, 6}^{\prime}(\boldsymbol{\theta})
				\end{array}\right\} f_{c, k, 5}(\boldsymbol{\Phi}) . \\
				&+\left\{\widehat{c}_{k} \widehat{c}_{i}\left|\overline{\mathbf{h}}_{k}^{H} \boldsymbol{\Phi}^{H} \mathbf{R}_{r i s} \boldsymbol{\Phi} \overline{\mathbf{h}}_{i}\right|^{2}+\frac{\widehat{c}_{i} \beta \sigma_{e}^{2}}{\tau p(\delta+1)} f_{c, i, 6}(\boldsymbol{\Phi})\right\} \boldsymbol{f}_{c, k, 5}^{\prime}(\boldsymbol{\theta}),
			\end{aligned}
		\end{align}
		and
		\begin{align}
			\begin{aligned}
				&\frac{\partial I_{c, k i}^{6}}{\partial \boldsymbol{\theta}}=2 \widehat{c}_{k} \widehat{c}_{i} \delta \operatorname{Re}\left\{\overline{\mathbf{h}}_{k}^{H} \boldsymbol{\Phi}^{H} \overline{\mathbf{H}}_{2}^{H} \overline{\mathbf{H}}_{2} \boldsymbol{\Phi} \overline{\mathbf{h}}_{i} \overline{\mathbf{h}}_{i}^{H} \boldsymbol{\Phi}^{H} \mathbf{R}_{r i s} \boldsymbol{\Phi} \overline{\mathbf{h}}_{k}\right\} \mathbf{z}_{k}\left(\mathbf{I}_{M}\right)\\
				&+\widehat{c}_{k} \widehat{c}_{i} \delta \operatorname{Tr}\left\{\boldsymbol{\Upsilon}_{k}\right\}\left\{\begin{array}{l}
					\boldsymbol{f}_{d}\left(\mathbf{R}_{r i s} \boldsymbol{\Phi} \overline{\mathbf{h}}_{k} \overline{\mathbf{h}}_{k}^{H} \boldsymbol{\Phi}^{H} \overline{\mathbf{H}}_{2}^{H} \overline{\mathbf{H}}_{2}, \overline{\mathbf{h}}_{i} \overline{\mathbf{h}}_{i}^{H}\right)+\boldsymbol{f}_{d}\left(\overline{\mathbf{H}}_{2}^{H} \overline{\mathbf{H}}_{2} \boldsymbol{\Phi} \overline{\mathbf{h}}_{i} \overline{\mathbf{h}}_{i}^{H} {\bf\Phi}^{H} \mathbf{R}_{r i s}, \overline{\mathbf{h}}_{k} \overline{\mathbf{h}}_{k}^{H}\right) \\
					+\boldsymbol{f}_{d}\left(\overline{\mathbf{H}}_{2}^{H} \overline{\mathbf{H}}_{2} \boldsymbol{\Phi} \overline{\mathbf{h}}_{k} \overline{\mathbf{h}}_{k}^{H} \boldsymbol{\Phi}^{H} \mathbf{R}_{r i s}, \overline{\mathbf{h}}_{i} \overline{\mathbf{h}}_{i}^{H}\right)+\boldsymbol{f}_{d}\left(\mathbf{R}_{r i s} {\bf\Phi} \overline{\mathbf{h}}_{i} \overline{\mathbf{h}}_{i}^{H} \boldsymbol{\Phi}^{H} \overline{\mathbf{H}}_{2}^{H} \overline{\mathbf{H}}_{2}, \overline{\mathbf{h}}_{k} \overline{\mathbf{h}}_{k}^{H}\right)
				\end{array}\right\},
			\end{aligned}
		\end{align}
		and
		\begin{align}
			\frac{\partial I_{c, k i}^{7}}{\partial \boldsymbol{\theta}}=\frac{\widehat{c}_{i} \beta \delta^{2} \sigma_{e}^{2}}{\tau p(\delta+1)} \boldsymbol{f}_{c, k i, 9}^{\prime}(\boldsymbol{\theta}),
		\end{align}
		and
		\begin{align}
			\begin{aligned}
				&\frac{\partial I_{c, k i}^{8}}{\partial \boldsymbol{\theta}}=\frac{2 \widehat{c}_{i} \beta \delta \sigma_{e}^{2}}{\tau p(\delta+1)} \operatorname{Re}\left\{\overline{\mathbf{h}}_{i}^{H} \boldsymbol{\Phi}^{H} \mathbf{R}_{r i s} \boldsymbol{\Phi} \mathbf{R}_{e m i} \boldsymbol{\Phi}^{H} \overline{\mathbf{H}}_{2}^{H} \boldsymbol{\Upsilon}_{k}^{H} \overline{\mathbf{H}}_{2} \boldsymbol{\Phi} \overline{\mathbf{h}}_{i}\right\} \mathbf{z}_{k}\left(\mathbf{I}_{M}\right)\\
				&+\frac{\widehat{c}_{i} \beta \delta \sigma_{e}^{2}}{\tau p(\delta+1)} \operatorname{Tr}\left\{\boldsymbol{\Upsilon}_{k}\right\}\left\{\begin{array}{l}
					\boldsymbol{f}_{d}\left(\mathbf{R}_{r i s} \mathbf{\Phi} \mathbf{R}_{e m i} \mathbf{\Phi}^{H} \overline{\mathbf{H}}_{2}^{H} \boldsymbol{\Upsilon}_{k}^{H} \overline{\mathbf{H}}_{2}, \overline{\mathbf{h}}_{i} \overline{\mathbf{h}}_{i}^{H}\right)\\
					+\boldsymbol{f}_{d}\left(\overline{\mathbf{H}}_{2}^{H} \mathbf{\Upsilon}_{k}^{H} \overline{\mathbf{H}}_{2} \boldsymbol{\Phi} \overline{\mathbf{h}}_{i} \overline{\mathbf{h}}_{i}^{H} \boldsymbol{\Phi}^{H} \mathbf{R}_{r i s}, \mathbf{R}_{e m i}\right) \\
					+\mathbf{z}_{k}\left(\overline{\mathbf{H}}_{2} \boldsymbol{\Phi} \overline{\mathbf{h}}_{i} \overline{\mathbf{h}}_{i}^{H} \boldsymbol{\Phi}^{H} \mathbf{R}_{r i s} \boldsymbol{\Phi} \mathbf{R}_{e m i} \boldsymbol{\Phi}^{H} \overline{\mathbf{H}}_{2}^{H}\right) \\
					+\boldsymbol{f}_{d}\left(\mathbf{R}_{r i s} \boldsymbol{\Phi} \overline{\mathbf{h}}_{i} \overline{\mathbf{h}}_{i}^{H} \boldsymbol{\Phi}^{H} \overline{\mathbf{H}}_{2}^{H} \boldsymbol{\Upsilon}_{k} \overline{\mathbf{H}}_{2}, \mathbf{R}_{e m i}\right)\\
					+\boldsymbol{f}_{d}\left(\overline{\mathbf{H}}_{2}^{H} \boldsymbol{\Upsilon}_{k} \overline{\mathbf{H}}_{2} \boldsymbol{\Phi} \mathbf{R}_{e m i} \boldsymbol{\Phi}^{H} \mathbf{R}_{r i s}, \overline{\mathbf{h}}_{i} \overline{\mathbf{h}}_{i}^{H}\right) \\
					+\mathbf{z}_{k}\left(\overline{\mathbf{H}}_{2} \boldsymbol{\Phi} \mathbf{R}_{e m i} \mathbf{\Phi}^{H} \mathbf{R}_{r i s} \boldsymbol{\Phi} \overline{\mathbf{h}}_{i} \overline{\mathbf{h}}_{i}^{H} \boldsymbol{\Phi}^{H} \overline{\mathbf{H}}_{2}^{H}\right)
				\end{array}\right\}.
			\end{aligned}
		\end{align}

		The gradient of signal leakage is $  \frac{\partial E_{c, k}^{ l e a k}}{\partial \boldsymbol{\theta}} =\sum_{\omega=1}^{8} \frac{\partial E_{c, k}^{\omega, l e a k}}{\partial \boldsymbol{\theta}} $, where
		\begin{align}
			\frac{\partial E_{c, k}^{1, \mathrm { leak }}}{\partial \boldsymbol{\theta}}=M \widehat{c}_{k} \delta \gamma_{k} \boldsymbol{f}_{c, k, 7}^{\prime}(\boldsymbol{\theta}),
		\end{align}
		and
		\begin{align}
			\begin{aligned}
				&\frac{\partial E_{c, k}^{2, l e a k}}{\partial \boldsymbol{\theta}} =\left\{M \widehat{c}_{k}^{2} \delta \boldsymbol{f}_{c, k, 7}^{\prime}(\boldsymbol{\theta})+\widehat{c}_{k}^{2} \delta \boldsymbol{f}_{c, k k, 8}^{\prime}(\boldsymbol{\theta})\right\} f_{c, k, 2}(\boldsymbol{\Phi}) \\
				&+\left\{\widehat{c}_{k}^{2} \boldsymbol{f}_{c, k, 2}^{\prime}(\boldsymbol{\theta}) f_{c, k, 4}(\boldsymbol{\Phi})+\left(\widehat{c}_{k}^{2} f_{c, k, 2}(\boldsymbol{\Phi})+2 \widehat{c}_{k} \gamma_{k}+\frac{\widehat{c}_{k} \sigma^{2}}{\tau p}\right) \boldsymbol{f}_{c, k, 4}^{\prime}(\boldsymbol{\theta})\right\} f_{c, k, 2}(\boldsymbol{\Phi}) \\
				&+\left\{M \widehat{c}_{k}^{2} \delta\left|f_{k}(\boldsymbol{\Phi})\right|^{2}+\widehat{c}_{k}^{2} \delta f_{c, k k, 8}(\boldsymbol{\Phi})+\left(\widehat{c}_{k}^{2} f_{c, k, 2}(\boldsymbol{\Phi})+2 \widehat{c}_{k} \gamma_{k}+\frac{\widehat{c}_{k} \sigma^{2}}{\tau p}\right) f_{c, k, 4}(\boldsymbol{\Phi})\right\} \boldsymbol{f}_{c, k, 2}^{\prime}(\boldsymbol{\theta}),
			\end{aligned}
		\end{align}
		and
		\begin{align}
			\begin{aligned}
				&\frac{\partial E_{c, k}^{3 \mathrm { leak }}}{\partial \boldsymbol{\theta}} =\left\{\frac{\widehat{c}_{k} \beta \delta \sigma_{e}^{2}}{\tau p(\delta+1)} \boldsymbol{f}_{c, 1}^{\prime}(\boldsymbol{\theta})\right\} f_{c, k k, 8}(\boldsymbol{\Phi}) +\left\{\widehat{c}_{k} \delta \gamma_{k}+\frac{\widehat{c}_{k} \beta \delta \sigma_{e}^{2}}{\tau p(\delta+1)} f_{c, 1}(\boldsymbol{\Phi})+\frac{\widehat{c}_{k} \delta \sigma^{2}}{\tau p}\right\} \boldsymbol{f}_{c, k k, 8}^{\prime}(\boldsymbol{\theta}),
			\end{aligned}
		\end{align}
		and
		\begin{align}
			\begin{aligned}
				&\frac{\partial E_{c, k}^{4, \mathrm { leak }}}{\partial \boldsymbol{\theta}} =\frac{\beta \sigma_{e}^{2}}{\tau p(\delta+1)}\left\{\widehat{c}_{k} \boldsymbol{f}_{c, k, 2}^{\prime}(\boldsymbol{\theta}) f_{c, 1}(\boldsymbol{\Phi})+\left(\gamma_{k}+\widehat{c}_{k} f_{c, k, 2}(\boldsymbol{\Phi})\right) \boldsymbol{f}_{c, 1}^{\prime}(\boldsymbol{\theta})\right\} f_{c, k, 4}(\boldsymbol{\Phi}) \\
				&+\left\{\gamma_{k}^{2}+\frac{\gamma_{k} \sigma^{2}}{\tau p}+\frac{\beta \sigma_{e}^{2}}{\tau p(\delta+1)}\left(\gamma_{k}+\widehat{c}_{k} f_{c, k, 2}(\boldsymbol{\Phi})\right) f_{c, 1}(\boldsymbol{\Phi})\right\} \boldsymbol{f}_{c, k, 4}^{\prime}(\boldsymbol{\theta}),
			\end{aligned}
		\end{align}
		and 
		\begin{align}
			\frac{\partial E_{c, k}^{5, \mathrm { leak }}}{\partial \boldsymbol{\theta}}=\frac{\widehat{c}_{k} \beta \delta^{2} \sigma_{e}^{2}}{\tau p(\delta+1)}, \boldsymbol{f}_{c, k k, 9}^{\prime}(\boldsymbol{\theta}),
		\end{align}
		and
		\begin{align}
			\begin{aligned}
				&\frac{\partial E_{c, k}^{6, \mathrm { leak }}}{\partial \boldsymbol{\theta}}=\frac{2 \widehat{c}_{k} \beta \delta \sigma_{e}^{2}}{\tau p(\delta+1)} \operatorname{Re}\left\{\overline{\mathbf{h}}_{k}^{H} \boldsymbol{\Phi}^{H} \overline{\mathbf{H}}_{2}^{H} \boldsymbol{\Upsilon}_{k} \overline{\mathbf{H}}_{2} \boldsymbol{\Phi} \mathbf{R}_{e m i} \mathbf{\Phi}^{H} \mathbf{R}_{r i s} \boldsymbol{\Phi} \overline{\mathbf{h}}_{k}\right\} \mathbf{z}_{k}\left(\mathbf{I}_{M}\right)\\
				&+\frac{c_{k} \beta \delta \sigma_{e}^{2}}{\tau p(\delta+1)} \operatorname{Tr}\left\{  \boldsymbol{\Upsilon}_{k}^{H}\right\}\left\{\begin{array}{c}
					\mathbf{z}_{k}\left(\overline{\mathbf{H}}_{2} \boldsymbol{\Phi} \mathbf{R}_{e m i} \boldsymbol{\Phi}^{H} \mathbf{R}_{r i s} \boldsymbol{\Phi} \overline{\mathbf{h}}_{k} \overline{\mathbf{h}}_{k}^{H} \boldsymbol{\Phi}^{H} \overline{\mathbf{H}}_{2}^{H}\right) \\
					+\boldsymbol{f}_{d}\left(\mathbf{R}_{r i s} \boldsymbol{\Phi} \overline{\mathbf{h}}_{k} \overline{\mathbf{h}}_{k}^{H} \boldsymbol{\Phi}^{H} \overline{\mathbf{H}}_{2}^{H} \boldsymbol{\Upsilon}_{k} \overline{\mathbf{H}}_{2}, \mathbf{R}_{e m i}\right) \\
					+\boldsymbol{f}_{d}\left(\overline{\mathbf{H}}_{2}^{H} \boldsymbol{\Upsilon}_{k} \overline{\mathbf{H}}_{2} \boldsymbol{\Phi} \mathbf{R}_{e m i} \boldsymbol{\Phi}^{H} \mathbf{R}_{r i s}, \overline{\mathbf{h}}_{k} \overline{\mathbf{h}}_{k}^{H}\right) \\
					+\mathbf{z}_{k}\left(\overline{\mathbf{H}}_{2} \boldsymbol{\Phi} \overline{\mathbf{h}}_{k} \overline{\mathbf{h}}_{k}^{H} \boldsymbol{\Phi}^{H} \mathbf{R}_{r i s} \boldsymbol{\Phi} \mathbf{R}_{e m i} \boldsymbol{\Phi}^{H} \overline{\mathbf{H}}_{2}^{H}\right) \\
					+\boldsymbol{f}_{d}\left(\overline{\mathbf{H}}_{2}^{H} \boldsymbol{\Upsilon}_{k}^{H} \overline{\mathbf{H}}_{2} \boldsymbol{\Phi} \overline{\mathbf{h}}_{k} \overline{\mathbf{h}}_{k}^{H} \boldsymbol{\Phi}^{H} \mathbf{R}_{r i s}, \mathbf{R}_{e m i}\right) \\
					+\boldsymbol{f}_{d}\left(\mathbf{R}_{r i s} \boldsymbol{\Phi} \mathbf{R}_{e m i} \boldsymbol{\Phi}^{H} \overline{\mathbf{H}}_{2}^{H} \boldsymbol{\Upsilon}_{k}^{H} \overline{\mathbf{H}}_{2}, \overline{\mathbf{h}}_{k} \overline{\mathbf{h}}_{k}^{H}\right)
				\end{array}\right\},
			\end{aligned}
		\end{align}
		and
		\begin{align}
			\begin{aligned}
				&\frac{\partial E_{c, k}^{7, \mathrm { leak }}}{\partial \boldsymbol{\theta}}=\frac{\beta \delta \sigma_{e}^{2}}{\tau p(\delta+1)} \widehat{c}_{k} \boldsymbol{f}_{c, k, 2}^{\prime}(\boldsymbol{\theta}) f_{c, k, 3}(\boldsymbol{\Phi})+\frac{\beta \delta \sigma_{e}^{2}}{\tau p(\delta+1)}\left\{\gamma_{k}+\widehat{c}_{k} f_{c, k, 2}(\boldsymbol{\Phi})\right\} \boldsymbol{f}_{c, k, 3}^{\prime}(\boldsymbol{\theta}),
			\end{aligned}
		\end{align}
		and
		\begin{align}
			\begin{aligned}
				&\frac{\partial E_{c, k}^{8, l e a k}}{\partial \boldsymbol{\theta}} =\frac{\widehat{c}_{k} \beta \sigma_{e}^{2}}{\tau p(\delta+1)} \boldsymbol{f}_{c, k, 5}^{\prime}(\boldsymbol{\theta}) f_{c, k, 6}(\boldsymbol{\Phi})+\frac{\widehat{c}_{k} \beta \sigma_{e}^{2}}{\tau p(\delta+1)} f_{c, k, 5}(\boldsymbol{\Phi}) \boldsymbol{f}_{c, k, 6}^{\prime}(\boldsymbol{\theta}).
			\end{aligned}
		\end{align}
	\end{thm}
\end{appendices}

\bibliographystyle{IEEEtran}
\vspace{-6pt}
\bibliography{myref.bib}
\end{document}